\documentclass[3p,sort&compress]{elsarticle}
\usepackage{float}

\usepackage{fleqn}
\usepackage{soul}

\usepackage{graphicx}
\usepackage{amssymb,latexsym,amsthm,amsmath,dsfont,mathtools}
\usepackage{color,linegoal}
\usepackage[T1]{fontenc}
\usepackage[polutonikogreek,english]{babel}
\usepackage{mathdots}
\usepackage{multirow}
\usepackage{hyperref}

\usepackage{bm} 
\newcommand{\be}{\begin{equation}}
\newcommand{\ee}{\end{equation}}
\newcommand{\ba}{\begin{eqnarray}}
\newcommand{\ea}{\end{eqnarray}}
\def\bea{\begin{eqnarray}}
\def\eea{\end{eqnarray}}

\newcommand{\df}[1]{{d_f}_{#1}}
\def\degree{\hbox{$^\circ$}}
\newcommand{\el}{El Ni\~{n}o}
\newcommand{\lanina}{La Ni\~{n}a}

\journal{Physics Reports}

\begin{document}

\begin{frontmatter}

\title{Statistical physics approaches to the complex Earth system}

\author[JF,JF2]{Jingfang Fan\corref{corrauthor}}
\ead{jingfang@pik-potsdam.de}
\address[JF]{Potsdam Institute for Climate Impact Research, 14412 Potsdam, Germany}
\address[JF2]{School of Systems Science, Beijing Normal University, 100875 Beijing, China}

\author[JF]{Jun Meng}
\ead{meng@pik-potsdam.de}

\author[JF]{Josef Ludescher}
\ead{joseflu@pik-potsdam.de }

\author[JF2]{Xiaosong Chen}
\ead{chenxs@bnu.edu.cn}

\author[YA]{Yosef Ashkenazy}
\ead{ashkena@bgu.ac.il}
\address[YA]{Department of Solar Energy and Environmental Physics, The Jacob Blaustein Institutes for Desert Research, Ben-Gurion University of the Negev, Midreshet Ben-Gurion 84990, Israel}

\author[JF,JK1,JK2]{J\"urgen Kurths}
\ead{kurths@pik-potsdam.de }
\address[JK1]{Department of Physics, Humboldt University, 10099 Berlin, Germany}
\address[JK2]{Lobachevsky University of Nizhny Novgorod, Nizhnij Novgorod 603950, Russia.}

\author[SH1]{Shlomo Havlin}
\ead{havlin@ophir.ph.biu.ac.il}
\address[SH1]{Department of Physics, Bar Ilan University, Ramat Gan 52900, Israel}
\author[JF]{Hans Joachim Schellnhuber}
\ead{john@pik-potsdam.de}

\begin{abstract}
Global climate change, extreme climate events, earthquakes and their accompanying natural disasters pose significant risks to humanity. Yet due to the nonlinear feedbacks, strategic interactions and complex structure of the Earth system, the understanding and in particular the predicting of such disruptive events represent formidable challenges for both scientific and policy
communities. 
During the past years, the emergence and evolution
of Earth system science has attracted much attention and produced  new concepts and frameworks.  Especially, novel statistical physics and complex networks-based techniques have been developed and implemented to substantially advance our knowledge for a better understanding of the Earth system, including climate extreme events, earthquakes and Earth geometric relief features, leading to substantially improved predictive performances. We present here a comprehensive review on the recent scientific progress in the development and application of how combined statistical physics and complex systems science approaches such as, critical phenomena, network theory, percolation, tipping points analysis, as well as entropy can be applied to complex Earth systems (climate, earthquakes, etc.). Notably, 
these integrating  tools and approaches provide new insights and perspectives for understanding the dynamics of the Earth systems.
The overall aim of this review is to offer readers the knowledge on how statistical physics approaches can be useful in the field of Earth system science.
\end{abstract}

\begin{keyword}
\texttt{Statistical Physics\sep Complex Earth Systems\sep Complex Network \sep Climate Change \sep Earthquake}
\MSC[2010] 00-01\sep  99-00
\end{keyword}

\end{frontmatter}
\tableofcontents


\newpage
\section{Introduction}\label{intro}


\subsection{The Earth as a Complex System}
\label{intro:1}

The Earth behaves as an integrated system comprised of geosphere, atmosphere, hydrosphere, cryosphere as well as biosphere components, with nonlinear interactions and feedback loops between and within them \cite{steffen_global_2005}. 
These components can be also regarded as self-regulating systems in their own right, and further broken down into more specialized subsystems. Nevertheless, the growing understanding of the multi-component
interactions between physical, chemical,  biological and human processes suggests that one should bring different
disciplines together and take into account the Earth
system as a whole. Such studies and results  initiate 
the emergence of a new `science of the Earth'--\textit{Earth System Science} (ESS) \cite{steffen_emergence_2020}. The ESS framework has already demonstrated its potential as a powerful
tool for exploring the dynamical and structural properties of
how the Earth operates as a complex system. 

\begin{figure}[]
\begin{centering}
\includegraphics[width=1.0\linewidth]{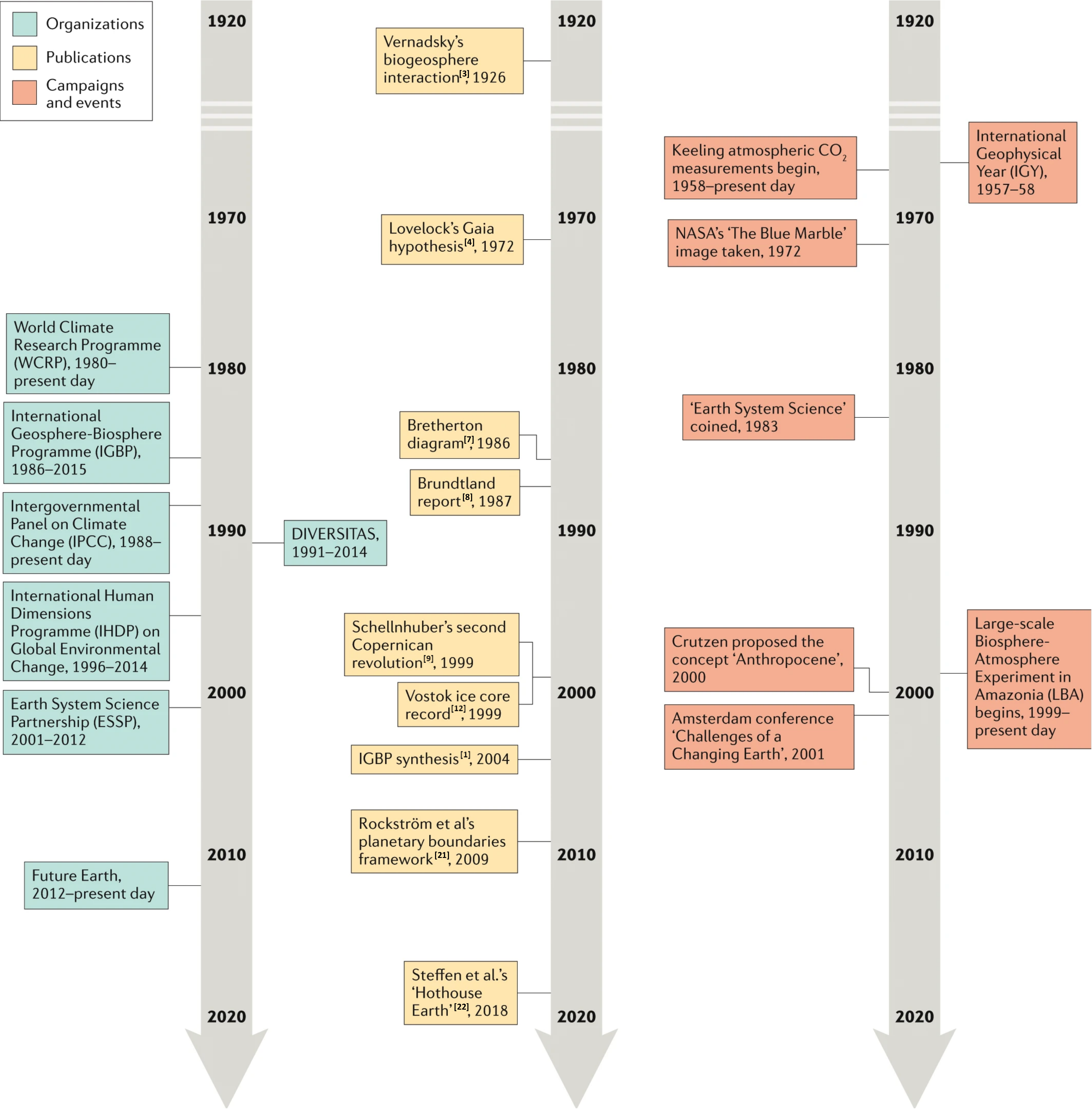}
\caption{\label{Fig1} 
{\bf Timeline illustration of the development of Earth System Science (ESS).}
The figure shows the key organizations, publications, campaigns, events and concepts that have helped to define and develop the ESS. \textit{Source}: Reprinted figure from Ref. \cite{steffen_emergence_2020}. } 
\end{centering}
\end{figure}

The  ESS has  emerged in the early to mid-20th
century,  and has developed rapidly during the last decades. Its historical evolution can be briefly  outlined into four phases \cite{steffen_emergence_2020}: (i) \textit{Precursors and beginnings (pre-1970s)}. The systemic nature of Earth was mainly described and emphasized by some conceptualizations, such as Vernadsky’s biosphere concept that life has a strong influence on the
physical and chemical properties of Earth \cite{vernadsky_biosphere_1998},  and  Lovelock’s Gaia hypothesis that Earth is a synergistic and self-regulating, complex system \cite{lovelock_atmospheric_1974}.
These  conceptualizations play vital roles in the contemporary
understanding of the Earth system.  In particular, the International Geophysical Year (IGY) 1957–1958 promoted the development of  international science and led to the emergence of two contemporary
paradigms, modern climatology and plate tectonics \cite{edwards_vast_2010,oreskes_rejection_1999}.
(ii) \textit{Founding a new science (1980s)}. 
In the 1980s,  the newly emerging recognition of Earth as
an integrated entity: the Earth system,  was called for by a series
of workshops and conference reports. In particular, the Bretherton diagram developed by the National Aeronautics and Space Administration (NASA) \cite{committee_earth_1986} was the first systems–dynamics representation of the
Earth System to couple the physical climate
system and biogeochemical cycle. The Brundtland report published by the World Commission on Environment and Development in 1987 \cite{brundtland1987our}  developed guiding principles for a sustainable development and  recognized the importance of the environmental problems for the Earth system. Some international organizations were also established in this stage, for example, the  World Climate Research Programme (WCRP) aims to determine the predictability of the climate and the effects of human activities on the climate. The International Geosphere-Biosphere Programme (IGBP) was launched in 1987 in order to coordinate international research on global-scale and regional-scale interactions between Earth's biological, chemical and physical processes and their interactions with human systems. The Intergovernmental Panel on Climate Change (IPCC) was created in 1988 to provide policymakers with  scientific assessments on climate change, its implications and potential future risks, as well as to put forward adaptation and mitigation options. (iii) \textit{Going global (1990s–2015)}. Due to great research efforts of
international programmes such as the IGBP, and the
widespread use of the Bretherton diagram,  ESS then developed rapidly, from the `new science of the Earth' movement to a global one.
In particular, the International Human Dimensions Programme (IHDP) on Global Environmental Change was founded with the aim to frame, develop and integrate social science research on global change.
 (iv) \textit{Contemporary ESS (beyond 2015)}. In the 21st century, the ESS framework  was well established, and initiated a new programme, \textit{Future Earth}, which was integrated by several international organizations or programmes,  such as the IGBP and IHDP. Future Earth's mission is to accelerate transformations to global sustainability through research and innovation. Following Ref. \cite{steffen_global_2005}, we highlight the key organizations, publications, campaigns and events
that characterize well  the evolution of ESS in Fig. \ref{Fig1}.

Among the pioneering publications for the evolution of ESS, Hans Joachim Schellnhuber introduced and developed a fundamental  concept for ESS: the dynamic, co-evolutionary relationship between nature
and human factors at the planetary scale \cite{schellnhuber_earth_1999}. He proposed that the Earth system $\boldsymbol{E}$ can be conceptually represented by the following mathematical form:
\begin{equation}
\label{EQ0}
\boldsymbol{E}=(\boldsymbol{N}, \boldsymbol{H}),
\end{equation}
where $\boldsymbol{N}=(\boldsymbol{a}, \boldsymbol{b}, \boldsymbol{c}, \ldots)$; $\boldsymbol{H}=(\boldsymbol{A}, \boldsymbol{S})$.  This model suggests that the overall Earth system contains two main components: $\boldsymbol{N}$ stands the ecosphere and $\boldsymbol{H}$ is interpreted as the human drivers.
$\boldsymbol{N}$ is composed of an alphabet of intricately interacting planetary sub-spheres $\boldsymbol{a}$ (atmosphere), $\boldsymbol{b}$ (biosphere), $\boldsymbol{c}$ (cryosphere), etc. The human factor $\boldsymbol{H}$ is much more subtle: $\boldsymbol{A}$ 
 means the `physical' sub-component and the `metaphysical' sub-component $\boldsymbol{S}$ reflects the emergence of a `global subject'. This work provided the conceptual framework for fully integrating
human dynamics into an Earth system and built a
unified understanding of the Earth.

There exist numerous tools and approaches that support the evolutionary development of the ESS. However, it is worth noting that they can be integrated into three interrelated foci: observations, modeling and computer simulations, assessments and syntheses \cite{steffen_emergence_2020}. The Earth observations are usually referring to the instrumental data, for example, that are collected by the meteorological stations or polar orbiting and geostationary
satellites. However, such data sets only extend, at best,
for about one-to-two centuries into the past. In order to extend our reach beyond thousands or even millions of years ago,  climate proxies fill this gap. There are multiple proxy
records, including coral records \cite{boiseau_climatic_1999}, marine-sediment \cite{taricco_two_2009}, stalagmite time series \cite{cheng_asian_2016}, tree rings \cite{esper_low-frequency_2002} as well as the Vostok \cite{petit_climate_1999} and EPICA \cite{jouzel_orbital_2007} ice core
record  that help us to better understand the past Earth system. 
Mathematical models are currently  key methods for the understanding and projecting of climate and Earth systems. These models include different variants from conceptual climate models--Energy Balance Models (EBMs) \cite{north_energy_1981} to more complex Earth models, the General Circulation Models (GCMs) \cite{stocker2013climate}. In addition, the integrated assessment models (IAMs) were designed to take human dynamics as an integral component into account and aim to understand how human development and societal choices affect each other and the natural world, including climate change \cite{parson_integrated_1997,van_vuuren_how_2011}. Based on the GCMs, the Earth systems Models of Intermediate Complexity (EMICs) were developed  to investigate the Earth's systems on long timescales or at reduced computational cost \cite{flato_earth_2011}. These models provide knowledge-integration to explore the dynamic properties and basic mechanisms  of the Earth system across multiple space and time scales.  The assessments and syntheses also played crucial roles in building new scientific knowledge, linking the scientific and political  communities,  and facilitating new research fields based on the feedback from the political  sector.

Some new concepts and theory have indeed arisen from the evolution of ESS,  including the emerging
concept of sustainability \cite{clark1986sustainable}, Anthropocene \cite{crutzen_geology_2002}, tipping elements \cite{lenton_tipping_2008}, planetary boundaries
framework \cite{rockstrom_safe_2009}, such as planetary
thresholds and state shifts \cite{steffen_trajectories_2018}, etc. Taken together, they create powerful ways to better understand and project the future trajectory of the Earth system.

\subsection{Why Statistical Physics?}

Yet, despite the rapid development of the ESS, the nature and statistical properties of extreme climate events and earthquakes  remain elusive and debated, and furthermore,  the existence of early warning signals of these phenomena is still a major open question. As a consequence, in this review,
we will present several novel approaches  based on or stemmed from statistical physics, that could enhance our understanding of how the Earth system could evolve. In particular, the interdisciplinary perspective on statistical physics and the Earth system yields to an improvement of the prediction skill of high-impact disruptive events within the climate and earthquake systems.

Statistical Physics is a branch of physics that draws heavily on the laws of probability and the statistics of many interacting components. It can describe a wide variety of systems with an inherently stochastic nature, aiming to predict and explain  measurable  properties and behaviours of macroscopic systems. It has been applied to many problems including  fields of physics, biology, chemistry, engineering  and also social sciences. Note that statistical physics does not focus  on the dynamic of every individual particle but on the macroscopic  behaviour of a large number of particles. The basic theory and ideas of statistical physics are depicted in many textbooks, such as \cite{Stanley_1971,chandler1987introduction,huang2009introduction}. 

Sethna motivated  the relationship between statistical physics and  complex systems in his book \cite{sethna2006statistical} as follows: "Many systems in nature are far too complex to analyze directly. Solving for the motion of all the atoms in a block of ice – or the boulders in an earthquake fault, or the nodes on the Internet – is simply infeasible. Despite this, such systems often show simple, striking behavior. We use statistical mechanics to explain the simple behavior of complex systems." A complex system is usually defined as a system composed of many components which  interact with each other. Regarding the large variety of components and interrelations, the Earth system thus can be interpreted as an evolving  complex  system. The concepts and methods of statistical physics can  infiltrate into ESS, in particular, (i)  critical phenomena are analogous to  tipping elements: a system will collapse and follow a breakdown \cite{buldyrev_catastrophic_2010}, if it is close to a phase transition  or tipping point. Critical phenomena and transitions  exist widely in the Earth system \cite{scheffer2009critical}, such as in atmospheric
precipitation \cite{peters_critical_2006} and percolation phase transition in sea ice \cite{golden_percolation_1998}. (ii) The concept of fractal was  used to describe the Earth's relief, shape, coastlines and islands \cite{mandelbrot_stochastic_1975}. (iii) The earthquake process is regarded as a complex spatio-temporal phenomenon, and has been viewed  as a self-organised criticality (SOC) paradigm \cite{sornette_self-organized_1989}. Moreover, the seismic activity  exhibits scaling properties in both temporal and spatial dimensions \cite{de_arcangelis_statistical_2016}.

Complex network theory provides a powerful tool to study the structure, dynamics and function of complex systems \cite{watts_collective_1998,barabasi_emergence_1999,boccaletti_complex_2006,newman2010networks}. Meanwhile, statistical physics is a fundamental framework and has brought theoretical insights for  understanding many properties of complex networks.  From an applied perspective, statistical physics has led to the definition of null models for real-world networks that reproduce global and local features \cite{cimini_statistical_2019}. Complex network theory is an emerging multidisciplinary discipline that has been applied to many fields including mathematics, physics, biology, computer science, sociology, epidemiology and others \cite{barabasi_network_2016}. Ideas from network science have also been successfully applied to the climate system and revealed interesting mechanisms underlying its functions, and leading to the emergence of a new concept, Climate Network (CN) \cite{tsonis_architecture_2004}. The nodes in a CN represent the available, geographically localized, time series, in particular, regular latituted-longitude grids, while the level of similarity and causality between
the nonlinear climate records of different grid points represents the CN's
links \cite{yamasaki_climate_2008,donges_complex_2009}. The development of this data-based approach is providing radical new ways to investigate the patterns and the dynamics of climate variability \cite{dijkstra_2019}.

In this review, we will introduce how to apply statistical physics methods in Earth system science. Particularly, we focus largely on the surface of the Earth system,
including the climate system, Earth's relief as well as the earthquakes system,
but not including  the whole planetary interior.


\subsection{Outline of the Report}

Apart from this introductory preamble, the report is organized along  3 Sections.

In the next Section, we start by offering the overall methodology that will accompany the rest of our discussion. Section (\ref{sec:methodology}) contains several topics that are essential but rather formal, including the CN approach, percolation theory, tipping points analysis, entropy theory and complexity. The reader will find there the attempt to define an overall theoretical framework encompassing the different situations and systems that will be later extensively treated. 

Section (\ref{cap3:App}) gives a comprehensive  review of applications, especially those that were studied in climate systems, the Earth geometric surface relief and earthquake systems. 

Finally, Section (\ref{cap4}) presents our conclusive remarks and perspective ideas.


\section{Methodology}
\label{sec:methodology}

\subsection{Climate Networks}
\label{cap2:CN}
\begin{figure}[H]
\begin{centering}
\includegraphics[width=1.0\linewidth]{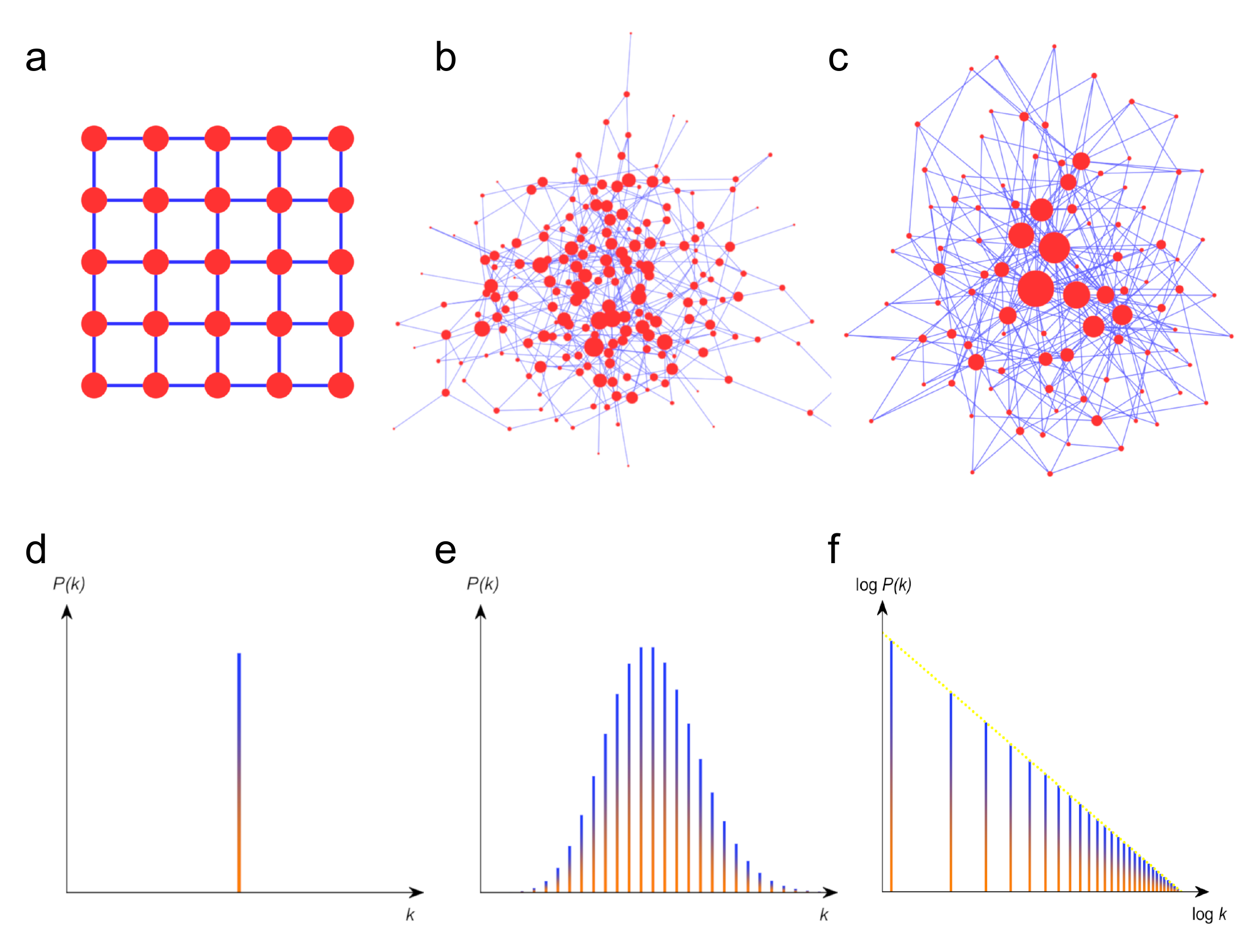}
\caption{\label{Fig2_1} 
{\bf Characteristics
of the basic explored model networks.}
(a) A regular 2D square lattice, is the most
degree-homogeneous network. (b) An Erd\H os-R\'enyi (ER) network, has
a Poisson degree distribution and its degree heterogeneity   is determined by the average degree $\left< k \right>$. (c) A scale--free (SF) network, has a power-law degree distribution, yielding large degree heterogeneity. The degree distributions are shown for (d) the 2D square lattice, (e) the ER network and  (f) the SF network.} 
\end{centering}
\end{figure}

For more than two decades, the complex network paradigm has demonstrated its great potential as a versatile tool for exploring  dynamical and structural
properties of complex systems, from a wide variety of disciplines in physics, biology, social science, economics, and
many other fields \cite{watts_collective_1998,eubank_modelling_2004,caldarelli_scale-free_2007,barrat_dynamical_2008,newman2010networks,kitsak_identification_2010,cohen_complex_2010,barabasi2016network}. 
In the context of network theory, a complex network is a graph with non-trivial topological features which do not occur in simple networks such as regular lattices (Fig. \ref{Fig2_1}a) or random Erd\H os-R\'enyi graphs (Fig. \ref{Fig2_1}b). One of the novelties of complex network theory is that it can relate the topological characteristics to the function and dynamics of the system. Complex network theory  has been successfully applied to many real world systems and revealed fundamental mechanisms underlying their functions. It has been also found that many real world networks  are scale-free networks (Fig. \ref{Fig2_1}c) \cite{albert_statistical_2002,barabasi_emergence_1999}.

\begin{figure}[H]
\begin{centering}
\includegraphics[width=0.5\linewidth]{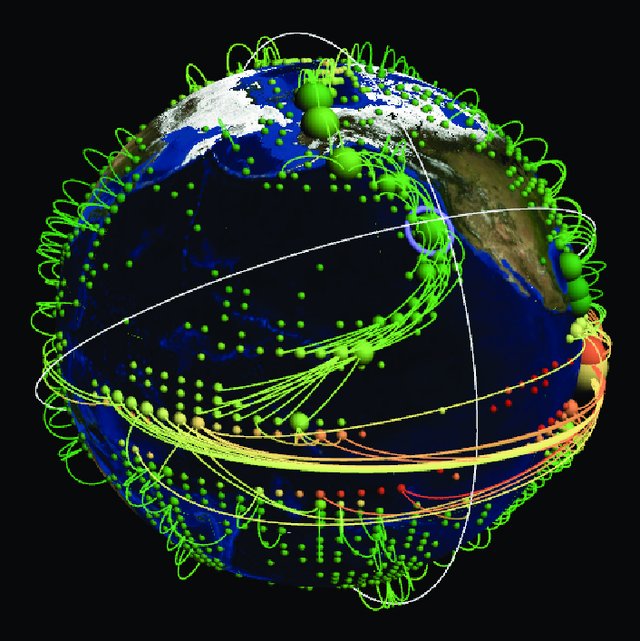}
\caption{\label{Fig2_2} 
{\bf Visualization of a climate network with surface air temperature taking into account the spatial embedding of vertices on the Earth's surface.} \textit{Source}: figure from Ref. \cite{nocke_review_2015}.} 
\end{centering}
\end{figure}

In recent
years, the ideas of network theory have also been implemented in climate sciences
to construct CN \cite{tsonis_architecture_2004,tsonis_what_2006,yamasaki_climate_2008}.
In CNs the geographical locations (or grid
points) are regarded as the nodes of the networks and the
level of similarity (causality) between the records (time series) of two grid points represents the links and their strength.
Various climate data  such as temperature, pressure,
wind and precipitation can be used to construct a network.
The climate networks approach allows to study the interrelationship between the different locations on the globe and
thus represent the global behavior of the climate system, e.g., how energy and matter are transferred from on location to another.
These networks have been used
successfully to analyze, model, understand, and even predict various climate phenomena \cite{tsonis_topology_2008,yamasaki_climate_2008,gozolchiani_pattern_2008,donges_backbone_2009,donges_complex_2009,steinhaeuser_exploration_2010,wang_dominant_2013,guez_global_2013,gozolchiani_emergence_2011,
zhou_teleconnection_2015,fan_network_2017,fan_network_2017_1,meng_percolation_2017,meng_forecasting_2018,fan2018climate,ludescher_improved_2013,mheen_interaction_2013,
feng_deep_2014,dijkstra_2019}. A visualization of a climate network is shown in Fig. \ref{Fig2_2} as an example.

We first provide an introduction of definitions, notations and basic quantities used to describe the topology of a network,  and then present  an overview on how to construct a climate network.

\subsubsection{Network Characteristics}
\label{cap2:NC}

In this Section we review  basic characteristics and metrics used to describe and analyze networks, most of them come from
\textit{graph theory} \cite{west2001introduction}. Graph theory is a large field containing many branches but we present only a small fraction of those results here, focusing on the ones most relevant to the study of the  complex Earth system. A network, also called a graph in the mathematical
literature, is a collection of vertices connected by edges. Vertices
and edges are also called nodes and links in computer science, sites and bonds
in physics, and actors and ties in sociology \cite{newman2010networks}. Throughout this review we will denote the number of vertices by $N$ and the number of edges by $M$, which is also a common notation in the mathematical and physics literature.

\subsubsection*{The Adjacency Matrix}

Most of the networks we will study in this review have at most a single edge
between any pair of vertices, i.e., self-edges or self-loops not allowed.
A simple representation of a network for many purposes is the \textit{adjacency
matrix} \textbf{$A$},  with elements $A_{ij}$
such that
\begin{equation}
A_{i j}=\left\{\begin{array}{l}1 \text { if there is an edge between vertices } i \text { and } j, \\ 0 \text { otherwise. }\end{array}\right.
\label{EQ1}
\end{equation}
We should notice that the diagonal matrix elements are all zero in the adjacency matrix. In some situations it is useful to represent edges as having a strength, weight, or value to them, usually a real number. Such weighted or valued networks can be represented by giving the elements of
the adjacency matrix values equal to the weights of the corresponding connections. A \textit{directed network} or directed graph, also called a \textit{digraph} for short, is a network
in which each edge has a direction, pointing from one vertex to another. Such
edges are themselves called directed edges. For this case, the adjacency matrix is usually not symmetric, and becomes
\begin{equation}
A_{i j}=\left\{\begin{array}{l}1 \text { if there is an edge from  } j \text { to } i, \\ 0 \text { otherwise. }\end{array}\right.
\label{EQ2}
\end{equation}

\subsubsection*{Degree}
The degree of a vertex in a network is the number of edges connected to it. We denote the degree of vertex $i$ by $k_{i}$.
For an undirected network of $N$ vertices the degree can be written in terms of the adjacency matrix (Eq. \ref{EQ2}) as
\begin{equation}
k_{i}=\sum_{j=1}^{N} A_{i j}.
\label{EQ3}
\end{equation}
The number of edges  $M$ is equal to the sum of the degrees of all the vertices divided by 2, so
\begin{equation}
M=\frac{1}{2} \sum_{i=1}^{N} k_{i}=\frac{1}{2} \sum_{i j} A_{i j}.
\label{EQ4}
\end{equation}
The mean degree $\left<k\right>$ of the node in an undirected graph is $\left<k\right>=\frac{1}{N} \sum_{i=1}^{N} k_{i} = 2M/N$.

The concept of degree is more complicated in directed networks. In a directed
network each vertex or node has two degrees. The in-degree $k_{i}^{\mathrm{in}}$ is the number of in-going
edges connected to a node and the out-degree $k_{j}^{\mathrm{out}}$ is the number of out-going edges.
From the  adjacency matrix of a directed network Eq. \ref{EQ2}, they can be written
\begin{equation}
k_{i}^{\mathrm{in}}=\sum_{j=1}^{N} A_{i j}, \quad k_{j}^{\mathrm{out}}=\sum_{i=1}^{N} A_{i j}.
\label{EQ5}
\end{equation}
Bearing in mind that the number of edges in a directed network is equal to the total number of
in-going ends of edges at all vertices, or equivalently to the total number of
out-going ends of edges.

\subsubsection*{Degree Distributions}

One of the most fundamental properties of a network that can be measured directly is the degree distribution, or the fraction $P(k)$ of nodes  having $k$ connections (degree $k$). A well-known result for the Erd\H os-R\'enyi \cite{erdos1960evolution} network is that its degree distribution follows a Poissonian,
\begin{equation}
P(k)=\mathrm{e}^{-z} z^{k} / k !, 
\label{EQ6}
\end{equation}
where $z=\langle k\rangle$ is the average degree. As shown in Fig. \ref{Fig2_1}e, despite
the fact that the position of the edges is random, a typical random graph is rather homogeneous, the maximum number
of the nodes having the same number of edges.

However, direct measurements of the degree distribution for real networks, such as the
Internet \cite{faloutsos_power-law_1999}, WWW \cite{barabasi_emergence_1999}, email network \cite{ebel_scale-free_2002},
metabolic networks \cite{jeong_lethality_2001}, airline networks \cite{barrat_architecture_2004}, neuronal networks \cite{eguiluz_scale-free_2005}, and many more, show that the Poisson law does not apply. But they exhibit an approximate  power-law degree distribution
\begin{equation}
P(k)=C k^{-\lambda},
\label{EQ6_1}
\end{equation}
where $C$ is  a normalization factor. The constant $\lambda$ is known as the exponent of the power law. Values in the range $2 \leq \lambda \leq 3$ are typical, although values slightly outside this range are possible and are observed occasionally \cite{broido_scale-free_2019}.  Networks with power-law degree distributions are  called scale-free networks. The simplest
strategy to determine the scale-free properties is to look at a histogram of the degree distribution on a log-log plot,
as we did in Fig. \ref{Fig2_1}f, to see if we have a straight line. Several models have been proposed for the evolution of scale-free networks, each
of which may lead to a different ensemble. The first proposal was the preferential
attachment model of Barab\'asi and Albert, which is known as the Barab\'asi--Albert
model \cite{barabasi_emergence_1999}. Several variants of this model have been suggested, see, e.g., in  Ref. \cite{krapivsky_connectivity_2000}.

\subsubsection*{Clustering Coefficient}

The extent to which nodes cluster together on very short scales in a network is measured by the clustering coefficient. 
The definition of clustering is related to the number of triangles in the network. The clustering is high
if two nodes sharing a neighbor have a high probability of being connected to each other. The most common way of defining the
clustering coefficient is:
\begin{equation}
C=\frac{(\text { number of triangles }) \times 3}{( \text { number of connected triples) } }.
\label{EQ7}
\end{equation}
Here a ``connected triple''  means three nodes $uvw$ with edges $(u, v)$ and $(v, w)$.
The factor of three in the numerator arises because each triangle is counted three times when the connected triples in the network are counted. 

We can also get a local clustering coefficient $C_i$ for a single vertex by defining 
define
\begin{equation}
C_{i}=\frac{\text { the number of triangles connected to vertex } i}{\text { the number of triples centered on vertex } i}.
\label{EQ8}
\end{equation}
That is, to calculate $C_{i}$, we go through all distinct pairs of vertices that are neighbors of $i$, count the number of such pairs that are connected to
each other, and divide by the total number of pairs. $C_{i}$ 
represents the average probability that a pair of $i$'s friends are friends of one
another.
For vertices with degree $0$ or $1$, for which both numerator and denominator are zero, we assume $C_{i}=0$. Then the clustering coefficient for the whole network \cite{watts_collective_1998} is the average
$C=\frac{1}{N} \sum_{i} C_{i}$.
In both cases, the clustering is in the range $0 \leq C \leq 1$.

\subsubsection*{Subgraphs}
\begin{figure}[H]
\begin{centering}
\includegraphics[width=1.0\linewidth]{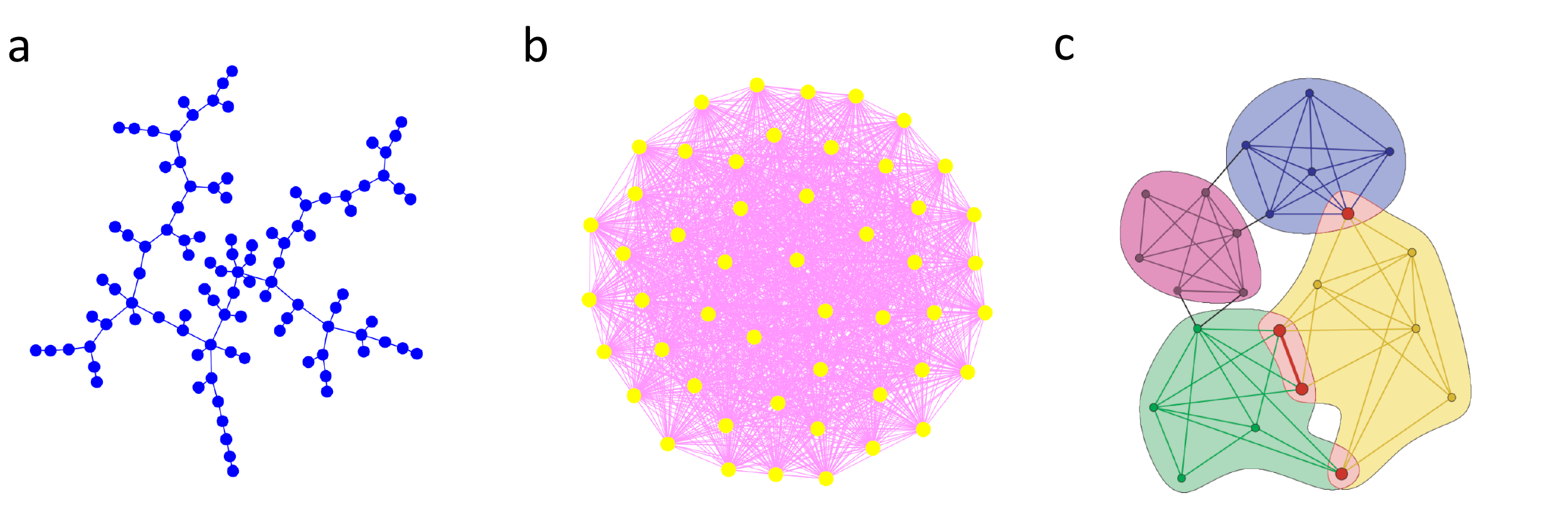}
\caption{\label{Fig2_3_1} 
{\bf Illustration of the concept of subgraphs.} Examples of (a) a tree network,  (b) a fully completed network, (c) overlapping
$k$-clique communities at $k=4$. \textit{Source} (c): Reprinted figure from Ref. \cite{palla_uncovering_2005}.} 
\end{centering}
\end{figure}

A graph $G_{1}$ consisting of a set $P_{1}$ of nodes and a set $E_{1}$ of edges is a \textit{subgraph} of a graph $G=\{P, E\}$ if all nodes in $P_{1}$ are also nodes of $P$ and all edges in $E_{1}$ are also edges of $E$. The simplest examples of subgraphs are cycles, trees, and complete subgraphs \cite{bollobas2001random}. A cycle of order $k$ is a closed loop of $k$ edges such that every two consecutive edges and only those have a common node. That is, graphically a triangle is a cycle of
order 3, while a rectangle is a cycle of order 4. The average degree of a cycle is equal to 2, since every node
has two edges. A tree, as shown in Fig. \ref{Fig2_3_1}a, is a connected, undirected network that contains no closed loops. Here, by ``connected'' we mean that every node in the network is reachable from every other via some path through the network.
A river network is an example of a naturally occurring tree with directed links \cite{fagan_connectivity_2002}. The most important property of trees is that, since
they have no closed loops, there is exactly one path between any pair of nodes. Thus the number of edges in a tree is always exactly $k - 1$ edges.
Complete subgraphs of order $k$, also called $k$-clique, contain $k$
nodes and all the possible $k(k-1) / 2$ edges--in other
words, they are completely connected [see Fig. \ref{Fig2_3_1}b].

Let us consider the ER model, in which we start from $N$ isolated nodes,
then connect every pair of nodes with a probability $p$. Most generally, we can ask
whether there is a critical probability that marks the appearance of arbitrary subgraphs consisting of $k$ nodes
and $l$ edges. A few important special cases that directly give an answer to this question \cite{bollobas2001random} are:
(i) The threshold probability of having a \textit{tree} of order $k$ is $p_{c}(N)=c N^{-k /(k-1)}$;
(ii) The threshold probability of having a \textit{cycle} of order
$k$ is $p_{c}(N)=c N^{-1}$;
(iii) The critical probability of having a \textit{complete subgraph} of order $k$ is $p_{c}(N)=c N^{-2 /(k-1)}$.

Generally,  a network is always composed of many
separate  subgraphs  or \textit{components}, i.e., groups of nodes connected internally, but disconnected from others. In each such component there exists a path between any
two nodes, but there is no path between nodes in different components. A network component with size proportional  to that of the entire
network, $N$, is called a \textit{giant component}.

Similar to component, the concept of \textit{community} is also one of the most fundamental properties in complex networks \cite{girvan_community_2002}. The nodes of the network might be joined together into tightly connected groups, whereas between them  there are still links but less connections.  A single node of the network may belong to more than one community, and most of the actual
networks are made of highly overlapping communities of
nodes \cite{palla_uncovering_2005}. Since community structures are quite common in real networks, community finding and detection are thus  of great importance  for better understanding the function of a network. An example of overlapping $k$-clique communities is shown in Fig. \ref{Fig2_3_1}c.  A good review on the community detection in graphs can be found in Ref. \cite{fortunato_community_2010}.

\subsubsection*{Network Models}

\textit{Erd\H os-R\'enyi model}: It can be interpreted by the following two well-studied graph ensembles,  $G_{N, M}$ the ensemble of all graphs having $N$ vertices and $M$ edges, and $G_{N, p}$ the ensemble consisting of graphs with $N$ vertices, where each possible edge is realized/connected with probability $p$. These two families, initially studied by Erd\H os and R\'enyi \cite{erdos1960evolution}, are known to be similar if $M=\left(\begin{array}{c}N \\ 2\end{array}\right) p$. They are referred to as Erd\H os-R\'enyi (ER) model. These descriptions are quite similar to the microcanonical and canonical ensembles studied in statistical physics \cite{newman_fast_2001}. The ER model has traditionally been the dominant subject of study in the
field of random graphs \cite{bollobas2001random}, with Poissonian degree distributions, see Eq. (\ref{EQ6}).

\textit{Barab\'asi-Albert model}: It is based on two simple assumptions regarding network evolution \cite{barabasi_emergence_1999}.
(i) \textit{Growth}: Starting with a small number $\left(m_{0}\right)$ of nodes, at every time step, a new node with $m\left(\leqslant m_{0}\right)$ edges that link the new node to $m$ different nodes already present in the system is added.
(ii) \textit{Preferential attachment}: This is the heart of the model. When choosing the nodes to which the new node will be connected, the probability $\Pi$ that the new node will be connected to node $i$ depends on the degree $k_{i}$ of node $i$, such that
$\Pi\left(k_{i}\right)=\frac{k_{i}}{\sum_{j} k_{j}}$.
In this process, after $t$ time steps this results in a network with $N=t+m_{0}$ nodes and $m t$ edges. Theoretical and numerical results show that this network model evolves into a power-law degree distribution, see Eq. (\ref{EQ6_1}).

\textit{Watts-Strogatz model}: 

In 1998, Watts and Strogatz \cite{watts_collective_1998} proposed a one-parameter model that interpolates between a regular lattice and a random graph. The details behind the model are the following:
(i) \textit{Start with order}: Start with a ring lattice with $N$ nodes in which every node is connected to its closest $k$ neighbors. In order to have a sparse but connected network at all times, consider $N \gg k$ $ \gg \ln (N) \gg 1$.
(ii) \textit{Randomize}: Randomly rewire each edge of the lattice with probability $p$ such that self-connections and duplicate edges are excluded. This process introduces $pNK/2$ long-range edges. The resulting network properties of this model  are \textit{small-world} and high clustering coefficient.  Specifically, a small-world network  is referring to a network where the characteristic path length $L$ grows proportionally to the logarithm of the number of nodes, $L \propto \log N$ \cite{watts_collective_1998}.

\subsubsection{Pearson Correlation Climate Network}
\label{cap2:PC}
Our climate system is made up of an enormous number of
nonlinear sub-systems having mutual nonlinear interactions and feedback loops active on a wide-range of temporal and spatial scales. Therefore, modeling the Earth climate system from the point of view of complex networks can clearly 
provide critical insights
into the underlying dynamics of the evolving  climate system. As discussed above, in CNs, geographical regions of Earth are regarded as nodes, and the bivariate statistical analysis of similarities between pairs of climatological variables time series represent   the links.

\begin{figure}[H]
\begin{centering}
\includegraphics[width=0.85\linewidth]{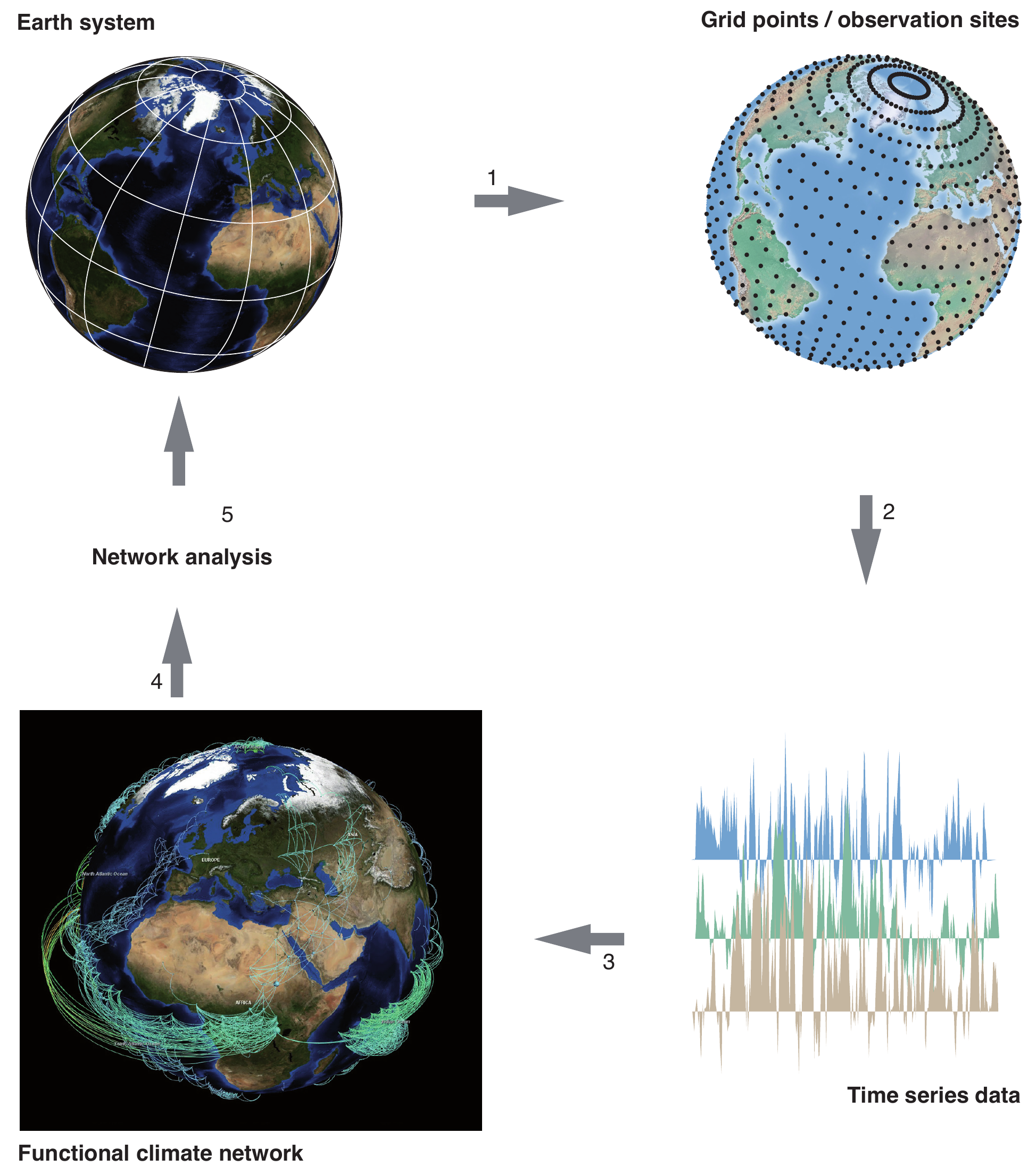}
\caption{\label{Fig2_3} 
{\bf Methodology used to construct a climate network from climatic time series.} \textit{Source}: Reprinted figure from Ref. \cite{donges_unified_2015}.} 
\end{centering}
\end{figure}

In general, there are five steps for the CNs construction and analysis \cite{dijkstra_2019}, the
procedure is displayed in Fig. \ref{Fig2_3}, adapted from \cite{donges_unified_2015}. Step (i): \textit{Nodes}, the nodes in CNs are usually defined as locations in a longitude-latitude spatial grid at various resolutions. (ii): \textit{Climatological variable}, we select the suitable climatological time series to be analyzed,
e.g., surface air temperature, sea surface temperature, precipitation, wind, etc. Some  pre-processing is also often needed. For example, to avoid the strong effect of seasonality, we usually subtract the mean seasonal cycle and divide by the seasonal standard   deviations for each grid point time series.
(iii): \textit{Edges}, in this step we compute the statistical similarity that quantifies  the interdependencies between pairs of time series.
The strength of each edge is correlation based.
There are many measures for quantifying the interdependencies of time series, here we mention the types of CNs by correlation measures. For example, a
Pearson correlation climate network, where we use Pearson correlation to quantify the cross-correlations  between time series; while an event synchronization climate network uses the event synchronization method; and a mutual information climate network is based on mutual information. (iv): \textit{Construction}, in this step we construct the CN which typically  involves some thresholding criterion to select only statistically significant edges. Considering, for example, that the significant links have $\mathcal{W}_{i j}$ values above a given threshold, $\mathcal{W}_{c}$, then the adjacency matrix is determined as
\begin{equation}
\mathcal{A}_{i j}=\mathcal{H}\left(\mathcal{W}_{i j}-\mathcal{W}_{c}\right).
\label{EQ9}
\end{equation}
where $\mathcal{H}$ is the Heaviside function. We then investigate the obtained network by using various network characteristics. Finally, in step (v): \textit{Climatological interpretation}, the results of this analysis are interpreted in terms
of dynamical processes of the climate system (e.g., atmospheric circulation, ocean currents, plenty of waves, etc.).

In the following, we will demonstrate  how to construct a Pearson correlation climate network.  Suppose that the climatic variable is the daily surface air temperature, either from observations, proxy reconstructions, reanalyses, or simulations, gathered at static measurement stations, or provided at grid cells. At each node $i$ of the network, we calculate the daily atmospheric temperature anomalies $T_i(t)$ (actual
temperature value minus the climatological average which  is then divided by the climatological standard
deviation) for each calendar day.

\begin{figure}[H]
\begin{centering}
\includegraphics[width=1.0\linewidth]{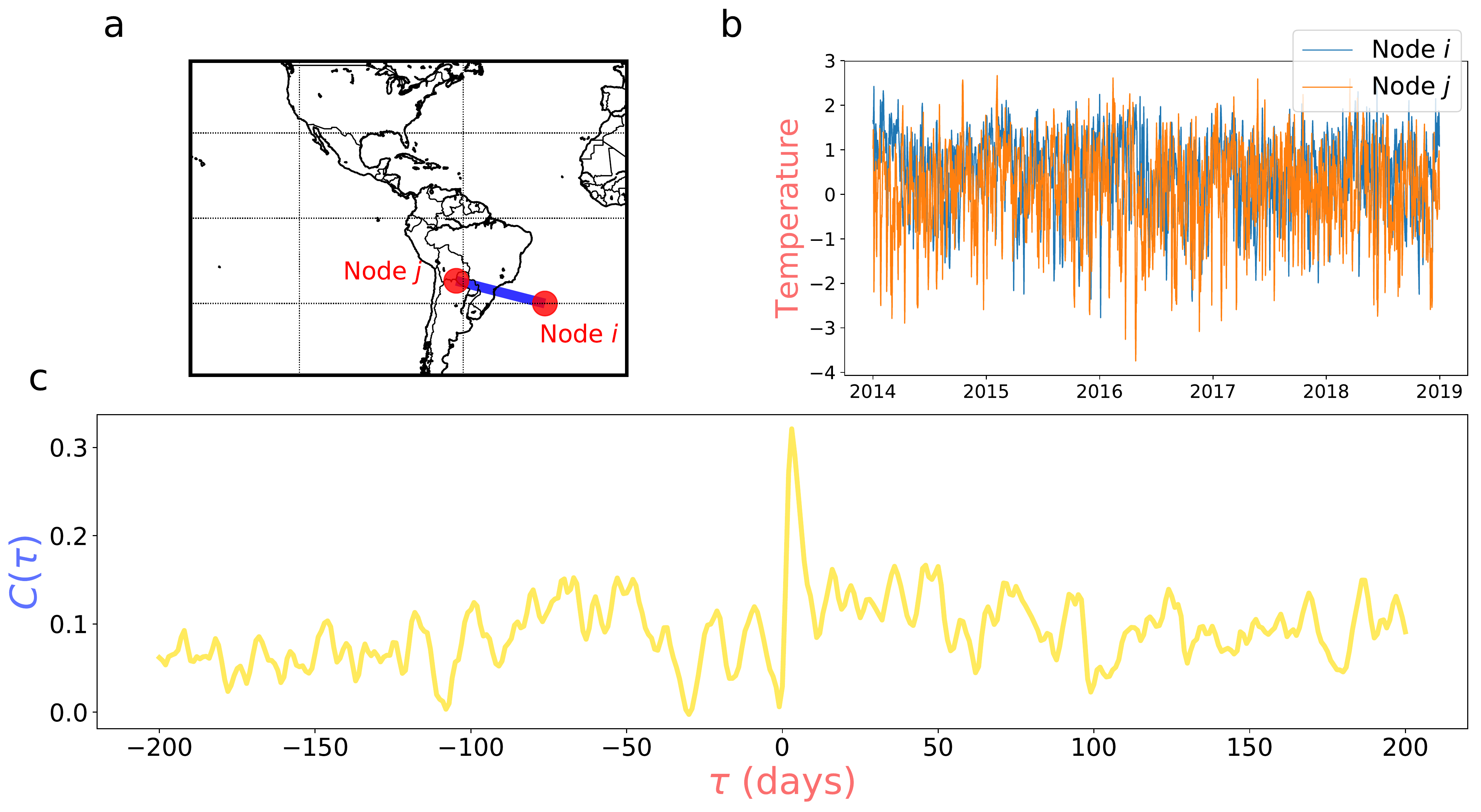}
\caption{\label{Fig2_4} 
{\bf Typical weighted and  directed link in a Pearson Correlation Climate Network.}
(a)  Node $i$ is located on the Southwest Atlantic and node $j$ is in the South American continent. (b) The near surface daily air temperature anomalies for the period [2014,2018]. (c) The cross-correlation function between the time series shown in (b). The direction of this link is from $j$ to $i$ with weight $W_{i,j}^{+} = 5.71$.} 
\end{centering}
\end{figure}

For obtaining the time evolution of the weight of the link between a pair of nodes $i$ and $j$, we follow \cite{meng_forecasting_2018} and compute, for each time windowing (such as month or year) $t$ over the whole time span, the \textit{time-delayed} cross-correlation function defined as
\begin{equation}\label{EQ10}
C^{(t)}_{i,j}(-\tau)=\frac{\langle T_i^{(t)}(t) T_j^{(t)}(t-\tau) \rangle-\langle T_i^{(t)}(t)\rangle \langle T_j^{(t)}(t-\tau) \rangle }{\sqrt{\langle (T_i^{(t)}(t)-\langle T_i^{(t)}(t)\rangle)^2\rangle}\cdot\sqrt{\langle (T_j^{(t)}(t-\tau)-\langle T_j^{(t)}(t-\tau)\rangle)^2\rangle}},
\end{equation}
and
\begin{equation}\label{EQ11}
C^{(t)}_{i,j}(\tau)=\frac{\langle T_i^{(t)}(t-\tau) T_j^{(t)}(t) \rangle-\langle T_i^{(t)}(t-\tau)\rangle \langle T_j^{(t)}(t) \rangle }{\sqrt{\langle (T_i^{(t)}(t-\tau)-\langle T_i^{(t)}(t-\tau)\rangle)^2\rangle}\cdot\sqrt{\langle (T_j^{(t)}(t)-\langle T_j^{(t)}(t)\rangle)^2\rangle}},
\end{equation}
where the brackets denote an average over the past $T$ days, according to
\begin{equation}\label{EQ12}
\langle f(t) \rangle=\frac{1}{T}\sum_{a=1}^{T}f(t-a).
\end{equation}
Here, $\tau$ is the time lag spanning from $[0, \tau_{max}]$ days. The reliable estimate of the background noise level, i.e., the values of the $\tau_{max}$ was discussed  in \cite{guez_influence_2014,martin_interpretation_2013}. Based on the correlation functions, Eqs. (\ref{EQ10})and (\ref{EQ11}), a weighted and directed link between nodes $i$ and $j$ 
was defined in Refs. \cite{gozolchiani_emergence_2011,wang_dominant_2013,fan_network_2017_1}. This is done by identifying  the value
of the highest peak (or lowest valley) of the cross-correlation
function and denote the corresponding time lag of this peak (valley) as $\theta_{i,j}$. The sign of $\theta_{i,j}$ indicates the direction of each link; i.e., when the time lag is positive ($\theta_{i,j} > 0$), the direction of the link is from $j$ to $i$ \cite{fan_network_2017_1}. The positive and negative links and their weights are determined via  $C_{i,j}(\tau)$, and are defined as
\begin{equation}
W_{i,j}^{+} = \frac{\rm max(C_{i,j}(\tau)) - {\rm mean}(C_{i,j}(\tau))}{{\rm std}(C_{i,j}(\tau))},
\label{EQ13}
\end{equation}
and
\begin{equation}
W_{i,j}^{-} = \frac{\rm min(C_{i,j}(\tau)) - {\rm mean}(C_{i,j}(\tau))}{{\rm std}(C_{i,j}(\tau))},
\label{EQ14}
\end{equation}
where ``max'' and ``min'' are the maximum and minimum values of the cross-correlation function,  ``mean'' and ``std'' are the mean and standard deviation.
Typical time series and their cross-correlation functions are shown 
in Fig. \ref{Fig2_4}.  For demonstration, two nodes are selected, a node $i$ is located in the Southwest Atlantic and a node $j$ is in the South American continent (Fig. \ref{Fig2_4}a). The near surface daily air temperature anomalies are shown in Fig. \ref{Fig2_4}b for the time period between 2014 to 2018. The cross-correlation function between the time series is shown in Fig. \ref{Fig2_4}c, where the absolute value of a
maximum cross-correlation function is much larger than the level of noise and the minimum value. Since $\theta_{i,j} =2 > 0$, then the direction of the link is from $j$ to $i$.  According to Eq. (\ref{EQ13}), we calculate its weight to be $W_{i,j}^{+} = 5.71$. This value represents 5.71 standard deviations above the noise level, i.e., highly significant.  

Finally, networks can be constructed by establishing
links between pairs of nodes with weights larger than some significant threshold. The advantage of this method is that it overcomes the problem of strong auto-correlation values in the data \cite{guez_influence_2014}.

\subsubsection{Event Synchronization Climate Network}
\label{cap2:ES}

The event synchronization method provides an alternative way to construct a network from climate observations: event synchronization CN. It was originally proposed by Quian Quiroga \textit{et al.} \cite{quian_quiroga_event_2002}, where they  measured synchronization and inferred time delays
between signals in neuroscience.
Event synchronization is based on the relative timings of events, such as rainfall, in a
time series and is defined, e.g., by the crossing of a threshold or by local maxima or minima, etc. For instance, an extreme rainfall event is defined as the day where its precipitation is above the 99th percentiles of all days.  Event synchronization is especially appropriate for studying extreme events.
The degree of synchronization is obtained from the number of quasi-simultaneous
events and the delay is calculated from the precedence of events in one time series (signal) with respect to the other.

Event synchronization starts by constructing two event series from two discrete univariate time series, $X$ and $Y$. An event $l$ that occurs at $X$ at time $t_{l}^{x}$ is considered to be \emph{synchronized} with an event $m$ that occurs at $Y$ at time $t_{m}^{y}$ within a time $\operatorname{lag} \pm \tau_{l m}^{x y}$, if $0\leq\left|t_{l}^{x}-t_{m}^{y}\right|<\tau_{l m}^{x y}$, where
\begin{equation}
\tau_{l m}^{x y} = \min \left\{t_{l+1}^{x}-t_{l}^{x}, t_{l}^{x}-t_{l-1}^{x}, t_{m+1}^{y}-t_{m}^{y}, t_{m}^{y}-t_{m-1}^{y}\right\} / 2.  
\label{EQ15}
\end{equation}
Here, $l = 1, 2, \ldots, e_{x}$, and
$m = 1, 2, \ldots, e_{y}$. $e_{x}$ and $e_{y}$ are the number of events in the $X$ and $Y$ respectively. The number of times an event appears in $X$ shortly after it
occurs in $Y$ is counted:
\begin{equation}
c(x | y)=\sum_{l=1}^{e_{x}} \sum_{m=1}^{e_{y}} J_{x y}^{l m}
\label{EQ16}
\end{equation}
with 
\begin{equation}
J_{x y}^{l m}=\left\{\begin{array}{ll}
1, & \text { if } 0<t_{l}^{x}-t_{m}^{y} \leq \tau_{l m}^{x y}, \\
1 / 2, & \text { if } t_{l}^{x}=t_{m}^{y}, \\
0, & \text { else, }
\end{array}\right.
\label{EQ17}
\end{equation}
and analogously to Eq. (\ref{EQ16}) we can define $c(x | y)$. Finally, the symmetrical and anti-symmetrical combinations
\begin{equation}
Q_{xy}=\frac{c(y | x)+c(x | y)}{\sqrt{e_{x} e_{y}}},
q_{xy}=\frac{c(y | x)-c(x | y)}{\sqrt{e_{x} e_{y}}},
\label{EQ18}
\end{equation}
are used to measure the synchronization of the events and their delay behavior, respectively. They are normalized to $0 \leqslant Q_{xy}$
$\leqslant 1$ and $-1 \leqslant q_{xy} \leqslant 1$.
It should be noted that if several extreme events are very close in one record, then only the first one is considered. 
Particularly,  $Q_{xy}=1$ if and only if the events of the signals are fully synchronized. In addition, if the events in $X$ always precede those in $Y$, then we get $q_{xy}=1$.

The event synchronization method has been found useful
for the study of electroencephalogram signals \cite{quian_quiroga_event_2002},
neurophysiological signals \cite{pereda_nonlinear_2005}, and the patterns of
extreme rainfall events \cite{malik_analysis_2012,boers_complex_2013,boers_prediction_2014,stolbova_tipping_2016,boers_complex_2019}.

\subsubsection{Mutual Information Climate Network}
\label{cap2:MI}
The mutual information method is also one of the  common tools for quantifying the interdependencies between time series and for constructing climate networks.
The mutual information of two random variables is a measure of the mutual dependence between the two variables. The concept of mutual information is intricately linked to that of entropy of a random variable \cite{cover2012elements}. Let $X$ and $Y$ be a pair of random variables with values over the space $\mathcal{X} \times \mathcal{Y}$. If their joint distribution is $p(x, y)$, and their associated marginal probability distribution functions are $p(x)$ and $p(y)$, then the mutual information is defined as
\begin{equation}
M_{I}=\sum_{x} \sum_{y} p(x, y) \log \left(\frac{p(x, y)}{p(x) p(y)}\right).
\label{EQ19}
\end{equation}
States with zero probability of occurrence are ignored.

Intuitively, $M_{I}$ is a measure of the inherent dependence expressed in the joint distribution of $X$ and $Y$ relative to their corresponding joint distribution under the assumption of independence. Mutual information, therefore, measures dependence and nonlinearity, in fact: $M_I = 0$ if and only if $X$ and $Y$ are independent random variables, i.e., $p(x, y) = p(x) p(y) $. Moreover, mutual information is non-negative, symmetric and can also be
computed with a time lag \cite{deza_assessing_2015}.  After we get the mutual information coefficient, Eq. (\ref{EQ19}), for each pair of nodes, we can construct a mutual information climate network.  It has been shown that the
mutual information CN can well capture the westward propagation of sea surface temperature (SST) anomalies that occur in the Atlantic multidecadal oscillation \cite{feng_are_2014}.

To summarize, in this chapter we have described different network characteristics
and presented various linear and  nonlinear tools of time series analysis, which can
be used to construct,  define and characterize CNs. Linear and nonlinear
methods include Pearson correlation, event synchronization, and
information-theory measures such as entropy and mutual information. There are also some other powerful tools, such as, spectral analysis, empirical orthogonal function analysis and symbolic ordinal analysis that can be used to reconstruct CNs \cite{dijkstra_2019}.

\subsection{Percolation Theory}
\label{cap2:Percolation}

Percolation originally  has been used to describe  the movement and filtering of fluids through porous materials \cite{kirkpatrick_percolation_1973}. The idea of percolation  was also considered by Flory and Stockmayer on polymerization and the sol-gel transition \cite{flory_molecular_1941}.  
In recent years, it has been a cornerstone in the theory of spatial stochastic processes with broad applications to diverse problems in  fields such as statistical physics \cite{fisher_statistical_1961,essam_percolation_1978,isichenko_percolation_1992,sahimi_flow_1993}, phase transitions \cite{aharony2003introduction,bunde_fractals_1996}, materials \cite{vigolo_experimental_2005,grimaldi_tunneling_2006}, epidemiology \cite{newman_random_2001,sander2002percolation,meyers2007contact,romualdo_pastor-satorras_epidemic_2015}, networks \cite{callaway_network_2000,derenyi_clique_2005,achlioptas_explosive_2009,baxter_bootstrap_2010,buldyrev_catastrophic_2010,liu_core_2012,gao_networks_2012,fan_general_2014,morone_influence_2015,newman2010networks,barabasi_emergence_1999}, colloids \cite{safran_percolation_1985,gnan_casimir-like_2014}, semiconductors \cite{shklovskii2013electronic}, traffic \cite{li_percolation_2015}, turbulence \cite{bernard_conformal_2006}, as well as Earth systems \cite{golden_percolation_1998,ali_saberi_percolation_2013,rodriguez-mendez_percolation-based_2016,meng_percolation_2017,lu_percolation_2016,fan2018climate,fan_percolation_2019}. Percolation theory also plays a pivotal role in studying the structure, robustness and functions of complex systems \cite{saberi_recent_2015}.

\begin{figure}[H]
\begin{centering}
\includegraphics[width=1.0\linewidth]{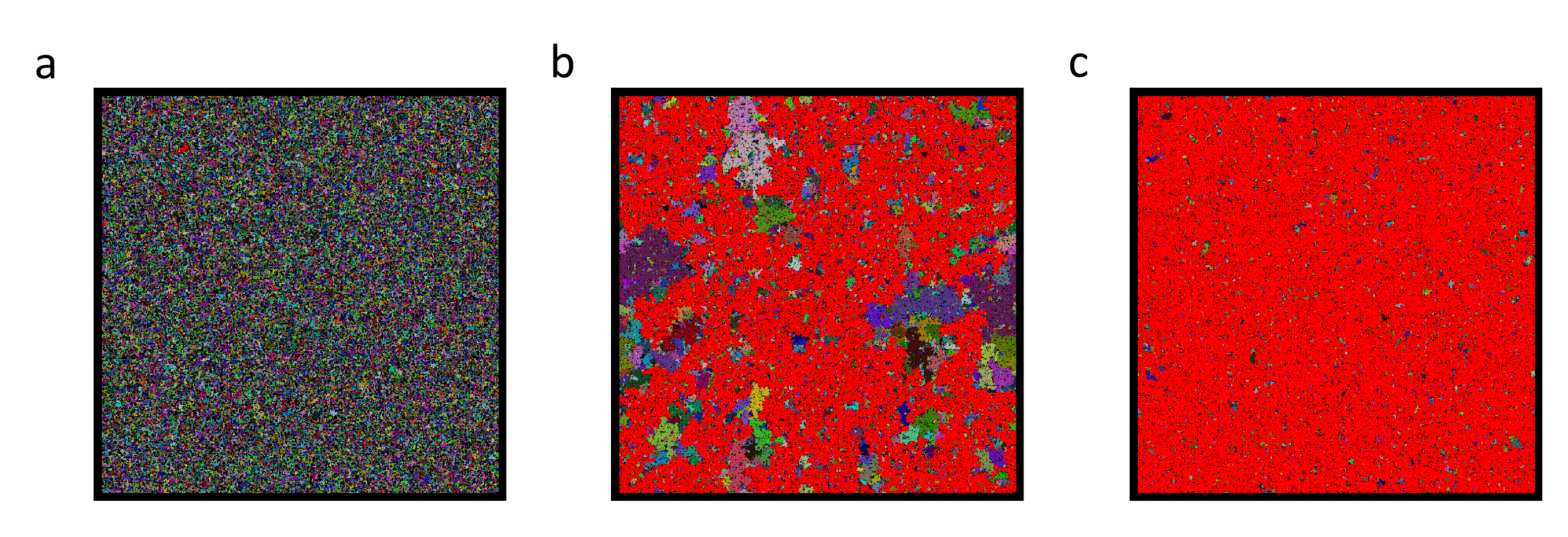}
\caption{\label{Fig2_6} 
{\bf Bond percolation clusters on a 512 $\times$ 512 2D square lattice for $p=$ (a) 0.3, (b) 0.5 and (c) 0.55, respectively.} Different cluster sizes have different colors. The color of the larger clusters depends on their size and varies from blue (small clusters) to red (infinite cluster). The percolation threshold is $p_c = 0.5$.} 
\end{centering}
\end{figure}

It is well known that in lattices and other ordered  networks \cite{essam_percolation_1980,aharony2003introduction,bunde_fractals_1996}, for dimensions greater than one, a percolation phase transition occurs. The percolation process is a simple model in which the nodes (sites) or edges (bonds) are occupied with some probability $p$ and unoccupied with probability $q=1-p$. Take
a regular lattice as an example (see Fig. \ref{Fig2_1}a). A system is regraded as  percolating if there is a path from one side  to the other parallel one, passing only through occupied bonds and sites. When such a path exists, the component or cluster of sites that spans the lattice is called the spanning cluster or the infinite percolation cluster. At low concentration $p$, the sites are either isolated or form small
clusters of nearest-neighbor sites. Two sites belong to the same cluster
if they are connected by a path of nearest-neighbor sites. At large $p$ values, on the other hand, at least one path between opposite sides exist (see Fig. \ref{Fig2_6}).
The percolation phase transition occurs at
some critical density $p_c$ that depends on the type and dimensionality of the lattice.

For many complex networks, the notion of side does not exist. However, the
ideas and tools of percolation theory can still be applied \cite{cohen_complex_2010}. The main
difference is that the condition for percolation is no longer the existence of spanning cluster, but rather having a cluster containing
$\mathcal{O}(N)$ nodes. If such
a component exists, we call it \textit{giant component}, that was discussed in Section \ref{cap2:CN}. Indeed, the condition of the
existence of a giant component applies also to lattice networks, and therefore it can be used as a more general condition  than the spanning property.

There exist two types of percolation, site percolation, where all sites are with probability $p$ occupied and $1-p$ empty. In bond percolation, however, it is the bonds which are occupied with probability $p$  and above $p_c$ they  form a giant component of connected sites. In this review, we will focus on the critical
phenomena of percolation near the percolation threshold $p_c$, where for the first time a giant component is formed or the system collapses \cite{cohen_resilience_2000}. These aspects are called critical phenomena, and the fundamental theory to describe them is the scaling theory from phase transitions.



\subsubsection{Phase Transition}

\begin{figure}[H]
\begin{centering}
\includegraphics[width=1.0\linewidth]{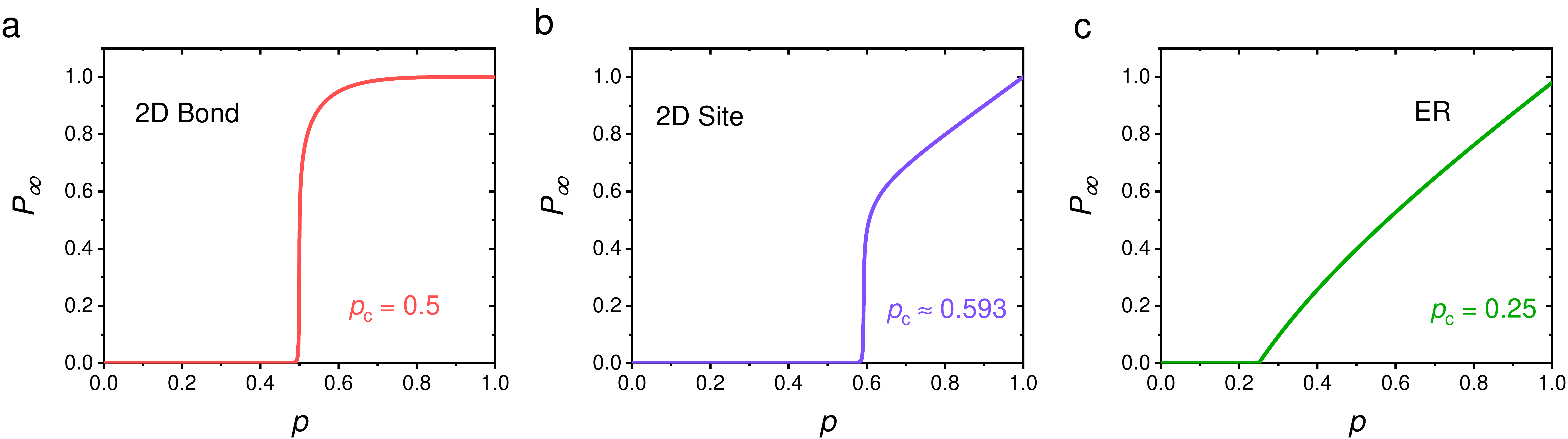}
\caption{\label{Fig2_7} 
{\bf Percolation phase transitions in lattice and network systems.} 
The percolation order parameter $P_{\infty}$ as a function of occupied probability $p$ for (a) 2D bond percolation with $p_c = 0.5$, (b) 2D site percolation with $p_c \approx 0.593$ and (c)  ER network with $p_c = 0.25$. Here the average degree $\left<k\right> = 4$ for ER network.} 
\end{centering}
\end{figure}

The concept of phase transition is usually used to describe transitions between solid, liquid, and gaseous states of matter for thermodynamic physical systems, where an ordered phase (e.g., solid) changes into a disordered phase (e.g., liquid) at some critical temperature $T_c$ \cite{Stanley_1971}. Another classical example of a phase transition is the magnetic phase transition, explained by the Ising model, where a spontaneous magnetization $m >0$ appears, at low temperature without any external field. While  increasing  temperature,  the spontaneous magnetization decreases and vanishes at $T_c$.

Percolation is indeed a simple \textit{geometrical} phase transition, where the percolation threshold $p_c$ distinguishes a phase of finite clusters (disordered phase) from a phase of an infinite cluster (ordered phase). The occupied probability $p$ of sites or bonds plays the same role as the temperature in the thermal phase transition. They are usually called \textit{control parameters} in statistical physics.

\subsubsection*{Percolation Order Parameter}

The percolation order parameter $P_{\infty}$  in the relative size of the infinite cluster which is defined as the fraction of the sites belonging to the infinite cluster. For $p < p_c$, there exist only finite clusters, thus, $P_{\infty} = 0$; For the case $p > p_c$, however, $P_{\infty}$ is analogous to the magnetization below $T_c$ and behaves near $p_c$ as a power law
\begin{equation}
\label{EQ20}
P_{\infty} \sim\left(p-p_{c}\right)^{\beta}.
\end{equation}
It describes the order in the percolation system
and is therefore called the \textit{order parameter}. We show in Fig. \ref{Fig2_7}, how $P_{\infty}$ behaves as a function of $p$, for both bond and site percolation in the 2D square lattice model, as well as, for site percolation in the ER network, respectively. It is worth noting that we adopt $p$ as the control parameter in Fig. \ref{Fig2_7}, 
which results $P_{\infty}$ being a monotonic increasing function. 
When $p$ is gradually  decreased (from 1), it can be regarded as removal (or attack) of nodes or links  and the system collapses after removing a fraction of $1-p_c$. Thus, $1-p_c$ can be a measure of the resilience of the system, i.e., how close it is to collapse.

\subsubsection*{Cluster Size Distribution}

The finite components distribution near criticality follows the scaling form \cite{aharony2003introduction,bunde_fractals_1996}
\begin{equation}
\label{EQ21}
n_{s} \sim s^{-\tau} \mathrm{e}^{-s / s_{\xi}}.
\end{equation}
Where $s$ is the component size, and $n_{s}$ represents the number of components of size $s$. At the percolation threshold $s_{\xi} \sim\left|p-p_{\mathrm{c}}\right|^{-\sigma}$ diverges and the tail of the distribution behaves as a power law with the critical exponent $\sigma$.

\subsubsection*{Average Cluster Size}

For any given site, the probability that it belongs to a cluster of size $s$ is $s n_{s}$. Let us define $\rho(p)$ as the probability that any given site is part of a finite cluster,
\begin{equation}
\label{EQ22}
\rho(p) = \sum_{s=1}^{\infty} s n_{s}(p)=M_{1}(p),
\end{equation}
which is the first moment of the cluster size distribution.
Hence, for any given site of any given finite cluster, the probability $w_{s}(p)$ that the cluster is of size $s$ is
\begin{equation}
\label{EQ23}
w_{s}(p)=\frac{1}{\rho(p)} s n_{s}(p),
\end{equation}
with $\sum_{s=1}^{\infty} w_{s}(p)=1$.
For any given site of any given finite cluster, the average size $\langle s(p)\rangle$ of the cluster is
\begin{equation}
\label{EQ24}
\langle s(p)\rangle=\sum_{s=1}^{\infty} s w_{s}(p)=\frac{1}{\rho(p)}\sum_{s=1}^{\infty} s^{2} n_{s}(p)=\frac{M_{2}(p)}{M_{1}(p)}.
\end{equation}
Note that, the average size $\langle s(p)\rangle$,  excluding the infinite cluster, diverges near the critical point as, 
\begin{equation}
\label{EQ25}
\langle s \rangle \sim\left|p-p_{c}\right|^{-\gamma}, \quad\left(p \rightarrow p_{c}\right).
\end{equation}

\subsubsection*{Correlation Length}

The correlation function $g(\mathbf{r})$ is the probability that a site at position $\mathbf{r}$ from an occupied site in a finite cluster belongs to the same cluster.

Typically, for large $r \equiv|\mathbf{r}|$, there is an exponential cutoff, i.e., $g(\mathbf{r}) \sim e^{-r / \xi}$, at the correlation length $\xi$. The correlation length $\xi$ is defined as
\begin{equation}
\label{EQ26}
\xi^{2}=\frac{\sum_{\mathrm{r}} r^{2} g(\mathbf{r})}{\sum_{\mathbf{r}} g(\mathbf{r})}.
\end{equation}
$\xi$ measures the mean distance
between two sites on the same finite cluster. When $p$ approaches $p_c$,
$\xi$ increases as
\begin{equation}
\label{EQ27}
\xi \sim\left|p-p_{c}\right|^{-\nu}.
\end{equation}

The quantities $P_{\infty}$ and $\langle s \rangle$  are analogous to the magnetization $m$ and the susceptibility $\chi$ in magnetic systems. $\beta$, $\gamma$ and $\nu$ are  called the critical exponents and describe the critical behavior of the percolation phase transition.

\subsubsection{Structural Properties}

Next, we will briefly introduce some fundamental measurements that are used to  characterize the structural properties of a percolation cluster. For more details, the readers can refer to Ref. \cite{aharony2003introduction}.

\subsubsection*{Fractal Dimension}

The fractal concept was  introduced into the physical world by Mandelbrot \cite{mandelbrot_fractal_1982} and applied to percolation  by Stanley \cite{stanley_cluster_1977} to describe the cluster shapes at the percolation threshold $p_c$. The infinite cluster is self-similar
on all length scales,
and can be regarded as a fractal. The fractal dimension $d_f$ is defined as how, on
average, the mass $M$ of the cluster within a sphere of radius $r$ from a site of the cluster changes with $r$,
\begin{equation}
\label{EQ28}
M(r) \sim r^{d_{f}}.
\end{equation}
For length scales  smaller than the correlation length $\xi$,  a fractal structure exists. For length
scales larger than $\xi$, the cluster becomes homogeneous. This can be summarized as
\begin{equation}
\label{EQ29}
M(r) \sim\left\{\begin{array}{ll}
r^{d_{f}}, & r \ll \xi \\
r^{d}, & r \gg \xi.
\end{array}\right.
\end{equation}

The relative size of the giant component, $P_{\infty}$ can also be expressed as
\begin{equation}
\label{EQ30}
P_{\infty} \sim \frac{r^{d_{f}}}{r^{d}} = \frac{\xi^{d_{f}}}{\xi^{d}}, \quad r<\xi.
\end{equation}
The fractal dimension $d_f$ of percolation clusters can be related to the critical exponents $\beta$ and $\nu$. Substituting Eqs. (\ref{EQ20}) and (\ref{EQ27}) into  Eq. (\ref{EQ30}), yields $d_{f}=d-\frac{\beta}{\nu}$.

Fractal analysis was also applied in the study of complex networks \cite{song_self-similarity_2005}. Generally, there are two basic methods to calculate the fractal dimensions of a given system, i.e.,  using either the box counting method \cite{song_origins_2006} or the cluster growing method \cite{daqing_dimension_2011}.
(i) The box counting method: Let $N_{B}$ be the number of boxes of linear size $l_{B}$ that are needed to cover the object, the fractal dimension $d_{f}$ is then given by $N_{B} \sim l_{B}^{-d_{f}}$.
It means that the average number of nodes $\left\langle M_{B}\left(l_{B}\right)\right\rangle$ within a box of size $l_{B}$ is,
\begin{equation}
\label{EQ31}
\left\langle M_{B}\left(l_{B}\right)\right\rangle \sim l_{B}^{d_{f}}. 
\end{equation}
The fractal dimension   $d_{f}$ can be obtained by a power law fit.
(ii) The cluster growing method: Similar to Eq. (\ref{EQ28}), the dimension $d_{f}$ can be calculated by
\begin{equation}
\label{EQ32}
\left\langle M_{C}\right\rangle \sim l^{d_{f}}
\end{equation}
where $\left\langle M_{C}\right\rangle$ is the average mass of the cluster, defined as the average number of nodes within linear size $\ell$ in the cluster. Note that, the above methods are difficult to  apply directly to networks, since networks are generally not embedded in space.  In order to measure the fractal dimension of networks one usually combines the concept of renormalization \cite{song_self-similarity_2005}, i.e.,  for each chemical  size $l_B$, boxes are placed  until the network is covered. Then each box is replaced by a node (renormalization), and the renormalized nodes are connected if there is at least one link between the no-renormalized boxes. This procedure is repeated and the number of boxes $N_B$ scale with $\ell_B$ as $N_B\sim\ell_B^{-d_f}$.

\subsubsection*{Graph Dimensions $d_{min}$ and $d_{\ell}$}

The fractal dimension $d_{min}$ has been used to describe the structural properties of the shortest
path between two arbitrary sites in the same cluster. Let  $\ell$  be the path length, which is often called the "chemical distance" \cite{havlin_chemical_1985}, it scales  with $r$ as
\begin{equation}
\label{EQ33}
\ell \sim r^{d_{\min }},
\end{equation}
and  the number of sites within $\ell$ is
\begin{equation}
\label{EQ34}
M(\ell) \sim \ell^{d_{\ell}},
\end{equation}
where $d_{\ell}$ is often called the ``graph dimension'' or ``chemical'' dimension.
Combining Eqs. (\ref{EQ28}), (\ref{EQ33}) and (\ref{EQ34}), we obtain the relation between $d_{min}$, $d_{\ell}$ and $d_f$, $d_{\ell} = \frac{d_f}{d_{min}}$.

Besides the fractal dimension exponents $d_{min}$, $d_{\ell}$ and $d_f$  discussed above, there are also some other exponents such as the backbone exponent and the red bonds exponents to describe the fractal dimensions of the substructures composing percolation clusters \cite{Stanley_1971,bunde_fractals_1996}. 

\subsubsection{Scaling Theory}

The scaling  hypotheses of phase transitions was developed by Kenneth G. Wilson during the last century and honoured by the 1982 Nobel prize. 
The scaling theory of percolation clusters relates the critical exponents of the percolation transition to the cluster size distribution, $n_s(p)$. According to the scaling hypotheses, it is possible to state the following relation for $n_s(p)$,
\begin{equation}
\label{EQ35}
n_{s}(p) \sim s^{-\tau} f\left(\left|p-p_{c}\right|^{1 / \sigma} s\right),
\end{equation}
where $\tau$ is the Fisher exponent \cite{fisher_statistical_1961}, and $f(x)$ is scaling function following $f(x) = exp (-x)$,  which rapidly decays to zero.

The correlation length  $\xi$ is defined
as the root mean square distance between occupied sites on the same
finite cluster for all clusters, see Eq. (\ref{EQ26}). For clusters with $s$ sites, the root mean square distance between all pairs of
sites on each cluster, is 
\begin{equation}
\label{EQ36}
R_{s}^{2}=\frac{2}{s(s-1)} \sum_{i=1}^{s} \sum_{j=1}^{i} \overline{\left(\mathbf{r}_{\mathbf{i}}-\mathbf{r}_{\mathbf{j}}\right)^{2}},
\end{equation}
and thus,
\begin{equation}
\label{EQ37}
\xi^{2}=\sum_{s=1}^{\infty} R_{s}^{2} s^{2} n_{s} / \sum_{s=1}^{\infty} s^{2} n_{s}.
\end{equation}
Close to the percolation threshold $p_{c}$, $R_{s} \sim s^{1 / d_{f}}$, and we obtain from the above equation
\begin{equation}
\label{EQ38}
\xi^{2} \sim \sum_{s=1}^{\infty} s^{2 / d_{f}+2-\tau} f\left(\left|p-p_{c}\right|^{1 / \sigma} s\right) / \sum_{s=1}^{\infty} s^{2-\tau} f\left(\left|p-p_{c}\right|^{1 / \sigma} s\right).
\end{equation}
To calculate the sums in Eq. (\ref{EQ38}), we transform  them into integrals, and obtain
\begin{equation}
\label{EQ39}
\xi^{2} \sim\left|p-p_{c}\right|^{-2 /\left(d_{f} \sigma\right)},
\end{equation}
which yields the scaling relations between $\nu, \sigma$ and $\tau$
\begin{equation}
\label{EQ40}
\nu=\frac{1}{d_{f} \sigma}=\frac{\tau-1}{d \sigma},
\end{equation}
where $d$ is the system's dimension.

Consider the $k$-th  moment of the cluster size distribution
\begin{equation}
\label{EQ41}
M_{k}=\sum_{s=1}^{\infty} s^{k} n_{s}(p) \sim \sum_{s=1}^{\infty} s^{k-\tau} f\left(s / \xi^{d_{f}}\right),
\end{equation}
which scales in the critical region as,
\begin{equation}
\label{EQ42}
M_{k} \sim \xi^{d_{f}(k-\tau+1)} \sim\left|p-p_{c}\right|^{(\tau-1-k) / \sigma}.
\end{equation}

Next we consider the percolation order parameter  $P_{\infty}$, which behaves as \cite{aharony2003introduction},
\begin{equation}
\label{EQ43}
P_{\infty} = 1 - \frac{1}{p}\sum_{s} s n_s \sim \sum_{s} s (n_s(p_c) - n_s(p))\sim( p- p_c)^{(\tau-2)/\sigma}.
\end{equation}
Combining with Eq. (\ref{EQ20}), we have, 
\begin{equation}
\label{EQ44}
\beta = \frac{\tau-2}{\sigma}.
\end{equation}

Similarly, we
obtain for $k = 2$, the relation $M_{k} \sim \left|p-p_{c}\right|^{(3 -\tau) / \sigma}$, yielding
\begin{equation}
\label{EQ45}
\gamma = \frac{\tau-3}{\sigma}.
\end{equation}

Thus, the critical exponents  are not independent of each other but satisfy two sets of scaling and hyperscaling relations.
The scaling relations can be easily expressed as by the Eqs. (\ref{EQ40}), (\ref{EQ44}) and (\ref{EQ45}).  

\subsubsection{Universal Gap Scaling}

An interesting property of percolation is \textit{universality}, which is a fundamental  principle of behavior at or near a phase transition  critical point. As a result, the critical exponents depend  only on the dimensionality of the system, and are independent
of the microscopic interaction details of the system. The behavior of a system is characterized
by a set of critical exponents, as discussed in the last section. If two systems have the same values of critical exponents, they belong to the same universality class. The universality property is a general feature of phase transitions. A phase transition is also characterized by scaling functions
that govern the finite-size behaviours \cite{privman_universal_1984}. The concept of finite-size scaling provides a versatile tool to study the percolation transition.

\begin{figure}[H]
\begin{centering}
\includegraphics[width=1.0\linewidth]{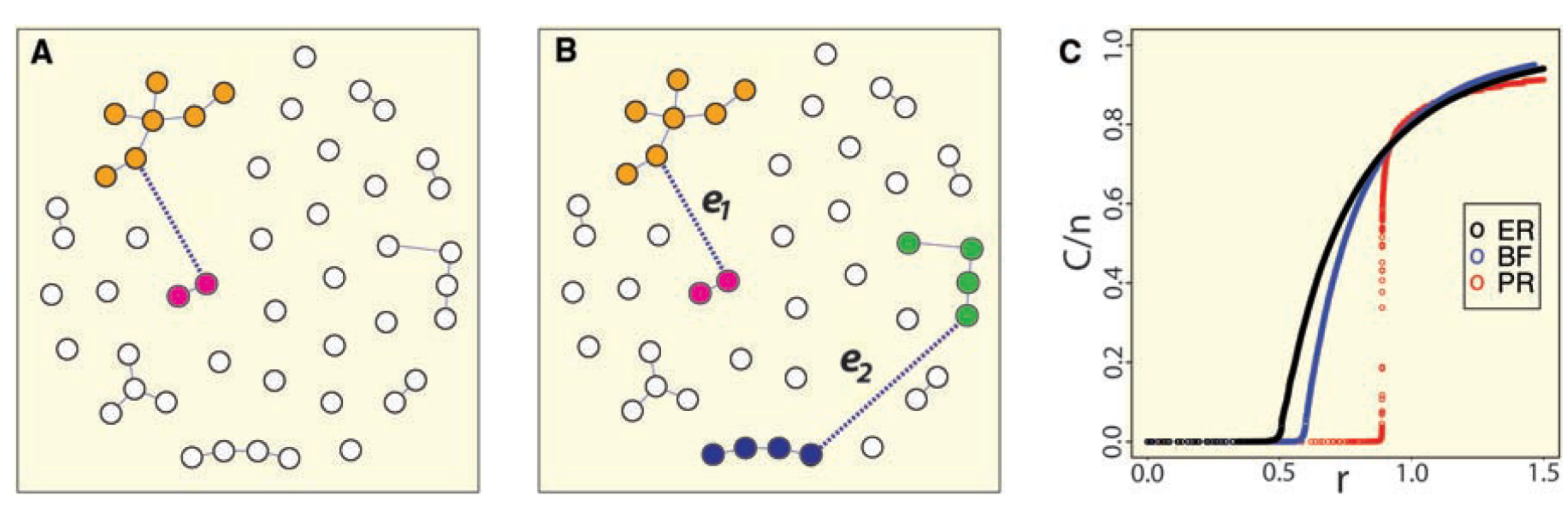}
\caption{\label{Fig2_8} 
{\bf Schematic  of  the  explosive percolation model.} (A) For each
step two vertices are chosen
randomly and connected by an edge. (B) At  each  time  step  of  the  Product  Rule (PR) process,  two edges, $e_1$ and $e_2$, compete for addition. The selected edge is the one minimizing the product of the sizes of the components it merges. Here $e_1$ is accepted and $e_2$ is rejected. (C) Typical evolution of an ER, Bohman Frieze (BF), and PR process on a system of size $N = 512,000$. \textit{Source}: figure from Ref. \cite{achlioptas_explosive_2009}.} 
\end{centering}
\end{figure}

Usually, a lattice or network
is expected to undergo a continuous percolation phase transition
during a random occupation or failure process \cite{bollobas2001random}. Explosive, hybrid and genuinely discontinuous percolation for processes that are not
random but competitive link-addition processes has attracted much
attention in recent years \cite{achlioptas_explosive_2009,friedman_construction_2009,radicchi_explosive_2009,ziff_explosive_2009,cho_cluster_2010, 
da_costa_explosive_2010,riordan2011explosive,nagler_impact_2011,grassberger_explosive_2011,fan_continuous_2012,cho2013avoiding,dsouza_anomalous_2015,dsouza_explosive_2019}. The description of the explosive percolation model is shown in Fig. \ref{Fig2_8}.

Instead of performing a finite-size scaling analysis at the critical
phase transition point,  the finite-size scaling analysis on
the percolation properties  and critical scaling of the size of the largest gap in the
order parameter is considered. For instance, Fan \textit{et al.} \cite{fan_universal_2020} developed a new universal gap scaling theory and propose six gap critical exponents to describe the  universality of percolation phase transitions. This theoretical framework can be applied for both continuous and discontinuous percolation in various lattice and network models by using finite-size scaling functions.

\subsubsection*{Gap Exponents}

Starting with an empty lattice, or network system with $N$ isolated nodes, bonds or links are added randomly or by competitive link-addition processes one by one.
The control parameter is denoted as $r$, which represents the link density $r = T/M$,  with $M$ being the maximal number of edges. The order parameter is the size of the largest cluster, given by
the largest connected component in the entire system. During the evolution of the system, we record the size of the largest cluster $S(T)$  at time step $T$, and calculate its largest one-step gap $\Delta$
\begin{equation}
\Delta \equiv \frac{1}{N}\max_T \left[S(T+1)-S(T) \right]. 
\label{EQ46}
\end{equation}
The step with the largest jump defines $T_c$ and its relative transition point as $r_c$. Moreover, the percolation strength is defined as the size  of the largest connected component at $T_c$, i.e., $S_c = S(T_c)$.

\begin{figure}[H]
\begin{centering}
\includegraphics[width=0.9\linewidth]{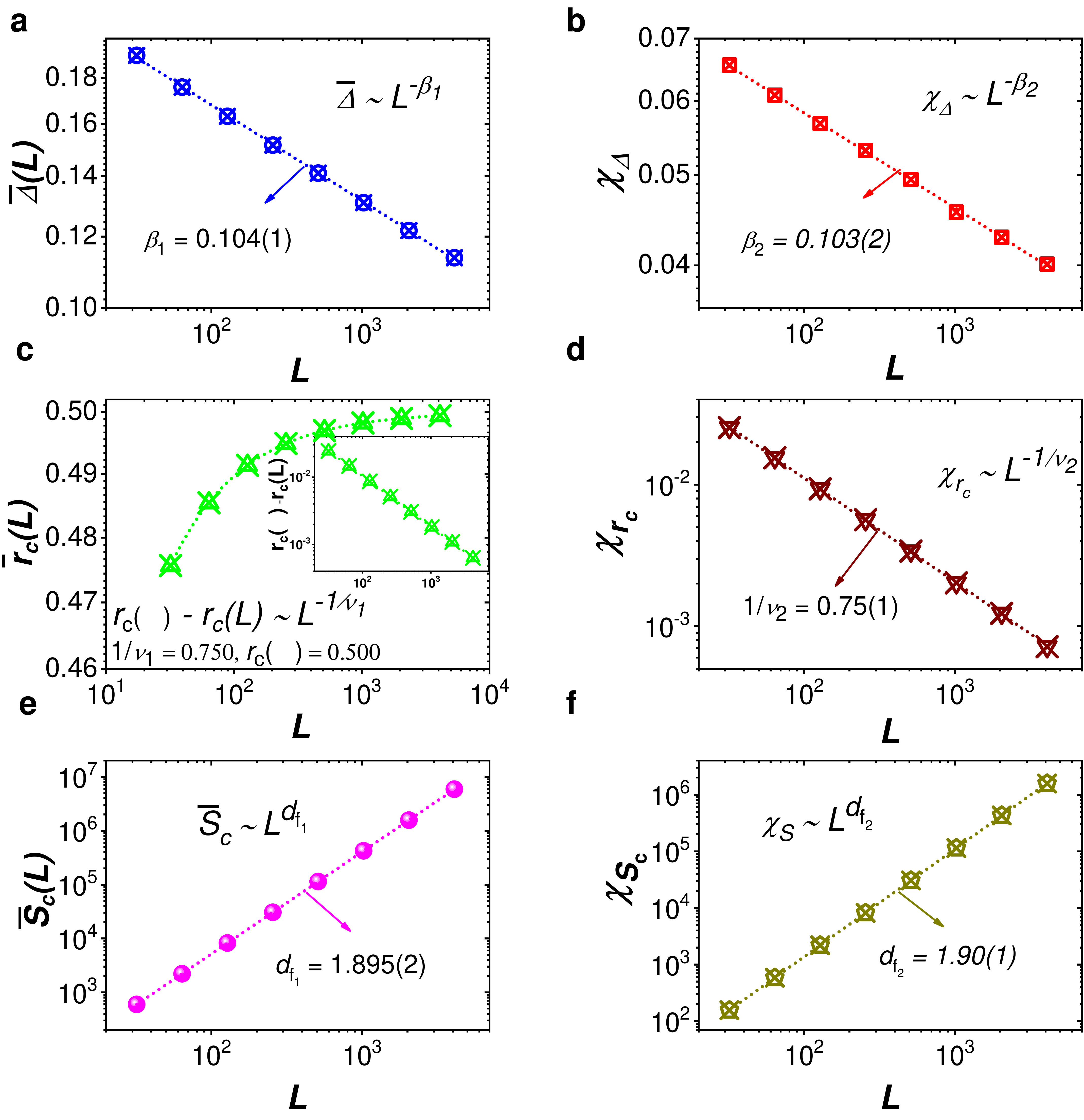}
\caption{\label{Fig2_9} 
{\bf Gap critical exponents for bond percolation on 2D square lattice systems.} (a) and (b), $\bar{\Delta}(L)$ and $\chi_\Delta$ as functions of 
$L$. (c) and (d), log-log plot of the percolation threshold $\bar{r}_c (L)$ and $\chi_\Delta$ versus $L$. 
(e) and (f). The power-law relations of $\bar{S_{c}}(L)$ and $\chi_{S_c}$ with $L$. Inset in (c) shows the log-log plot of $r_c (\infty) - \bar{r}_c (L)$ as a function of $L$. \textit{Source}: Reprinted figure from Ref. \cite{fan_universal_2020}.} 
\end{centering}
\end{figure}

Critical phenomena such as percolation exhibit scale-free behaviors that are quantified by scaling relations, see  discussions in previous sections.
Similarity, the averages 
$\bar{\Delta}$, $\bar{r}_c$ and $\bar{S}_c$ are anticipated to
exhibit the following power-law relations \cite{fan_universal_2020}, as a function of
$L$\footnote{The size of the system is $N = L^d$.},
\begin{equation}
\label{EQ47}
\bar{\Delta}(L) \sim L^{-\beta_{1}},
\end{equation}
\begin{equation}
\label{EQ48}
\bar{r}_c (L)-r_c (\infty) \sim L^{-1/\nu_1},
\end{equation}
\begin{equation}
\label{EQ49}
\bar{S_{c}}(L)  \sim L^{\df{1}},
\end{equation}
where $\beta_{1}$, $\nu_1$ and $\df{1}$ are three critical exponents describing the universal class of the percolation, 
and $r_c(\infty)$ is the percolation threshold in the thermodynamic infinity limit, $L \to \infty$. Their corresponding 
fluctuations $\delta \Delta^{(i)} = \Delta^{(i)}(L) -\bar{\Delta}(L)$, $\delta r_c^{(i)} = r_c^{(i)}(L) - \bar{r}_c (L)$ and $\delta S_c^{(i)} = S_{c}^{(i)}(L) - \bar{S_{c}}(L)$, are also investigated, 
\begin{equation}
\label{EQ50}
\chi_\Delta = \sqrt{\frac{1}{D}\sum_{i=1}^{D}[\delta \Delta^{(i)}]^2},\\
\end{equation}
\begin{equation}
\label{EQ51}
\chi_{r_c} = \sqrt{\frac{1}{D}\sum_{i=1}^{D}[\delta r_c^{(i)}]^2},\\
\end{equation}
\begin{equation}
\label{EQ52}
\chi_{S_c} = \sqrt{\frac{1}{D}\sum_{i=1}^{D}[\delta S_c^{(i)}]^2},
\end{equation}
where  $D$ is the number of independent realizations.
We expect that $\chi_{\Delta}$, $\chi_{r_c}$ and $\chi_{S_c}$ decay algebraically with $L$ with the following  scaling relations  
\begin{equation}
\label{EQ53}
\chi_{\Delta}  \sim   L^{-\beta_2},\\
\end{equation}
\begin{equation}
\label{EQ54}
\chi_{r_c}  \sim   L^{-1/\nu_2},\\
\end{equation}
\begin{equation}
\label{EQ55}
\chi_{S_c}  \sim   L^{\df{2}},
\end{equation}
where $\beta_{2}$, $\nu_2$ and $\df{2}$ constitute another set of 
critical exponents.  The universality class of the percolation
is characterized by the new six gap critical exponents. 
Note that they are highly related to  the standard percolation
exponents, and not fully independent from each other. The relationships between the gap exponents and the standard
percolation critical exponents are derived as \cite{fan_universal_2020}, 
\begin{equation}
\beta_{1} = \beta_{2} = \beta/\nu,\;\;\; \\
\nu_{2} = \nu, \;\;\; \\
\df{1} = \df{2} = d_{f}.
\label{EQ56}
\end{equation}
In particular, the values of $\beta_{1}$ can imply the order of the percolation. That is, $\beta_{1} = 0$ indicates the percolation is a first order transition; whereas models with $0 < \beta_{1} < 1$ are continuous \cite{nagler_impact_2011,fan_general_2014}. We present the numerical results
for bond percolation on a 2D square lattice in Fig. \ref{Fig2_9}, where we obtain $\beta_1 = \beta_2 \approx 0.104$, $1/\nu_1 = 1/\nu_2 \approx 1/\nu$, and $r_c(\infty) \approx 0.5$, and $\df{1} = \df{2} \approx 1.895$. All results are in agreement with the known theoretical  
values \cite{essam_percolation_1978}.

\subsection{Tipping Points Analysis}
\label{subsec:Tipping}

The notion of tipping point: "\textit{little things can make a big difference}'', was first  published by Malcolm Gladwell in his book \cite{gladwell2006tipping}. It means, at a particular moment in time, a small change can result a large and long-term  consequences in a complex system.  Tipping points  are usually associated with bifurcations \cite{strogatz_nonlinear_2015}. 
A tipping point is defined as ``the moment of critical mass, the threshold, the boiling point''.  Many complex systems experience sudden shifts in behavior, often referred to as tipping points or critical transitions, ranging from climate \cite{rahmstorf_ocean_2002,dakos_slowing_2008,lenton_tipping_2008,hirota_global_2011,lenton_arctic_2011,caesar_observed_2018,lenton_climate_2019}, ecosystems \cite{scheffer_catastrophic_2001,kefi_spatial_2007,drake_early_2010,hirota_global_2011,dai_generic_2012,lever_sudden_2014,jiang_predicting_2018}, to social science \cite{centola_experimental_2018,otto_social_2020}, financial markets \cite{haldane_systemic_2011,majdandzic_multiple_2016}, medicine \cite{litt_epileptic_2001,mcsharry_prediction_2003,venegas_self-organized_2005} and event macroeconomic agent-based models \cite{gualdi_tipping_2015}.

Complex systems can shift abruptly from one state to another at tipping points, which may imply  growing a threat and risk of abrupt and irreversible changes \cite{lenton_climate_2019}. It is thus of great practical importance to understand the theoretical mechanisms and predict the tipping phenomena. Although predicting such critical points before they occur 
is notoriously difficult. Theory proposes the existence of generic early-warning signals (EWS) that
may indicate for a wide class of systems if a critical threshold is approaching \cite{scheffer_early-warning_2009,lenton_early_2011,scheffer_anticipating_2012}. EWS is currently  one of
the most powerful tools for predicting critical transitions.
The tipping point analysis technique provides a vital tool to anticipate, detect and predict tipping points in complex dynamical systems. The
methodology usually combines monitoring  memory in time series, includes dynamically derived lag-1 autocorrelation \cite{held_detection_2004}, power-law scaling exponent of detrended fluctuation analysis (DFA) \cite{peng_mosaic_1994,livina_modified_2007}, and power-spectrum-based analysis \cite{prettyman_generalized_2019}.


\subsubsection{Basic Concepts}

\textbf{Defining tipping points}: Following Ref. \cite{lenton_early_2011}, a tipping point is the corresponding critical point at which the future state of the system is qualitatively changed. A single control parameter is identified as $(\rho)$, for which there exists a threshold $\left(\rho_{\text {c}}\right)$, from which a small perturbation $(\delta \rho>0)$ leads to a qualitative change $(\hat{F})$ in a crucial feature of the system $(F)$, after some time $(T>0)$. The actual change $(\Delta F)$ is defined as:
\begin{equation}
\label{EQ63}
|\Delta F|=\left|F\left(\rho \geq \rho_{\text {c}}+\delta \rho | T\right)-F\left(\rho_{\text {c}} | T\right)\right| \geq \hat{F}>0
\end{equation}
In this definition, the critical threshold $\left(\rho_{\text {c}}\right)$ is considered as the tipping point, beyond which a qualitative change occurs. Note that such changes may occur immediately or even much later after the cause.

\begin{figure}[]
\begin{centering}
\includegraphics[width=0.85\linewidth]{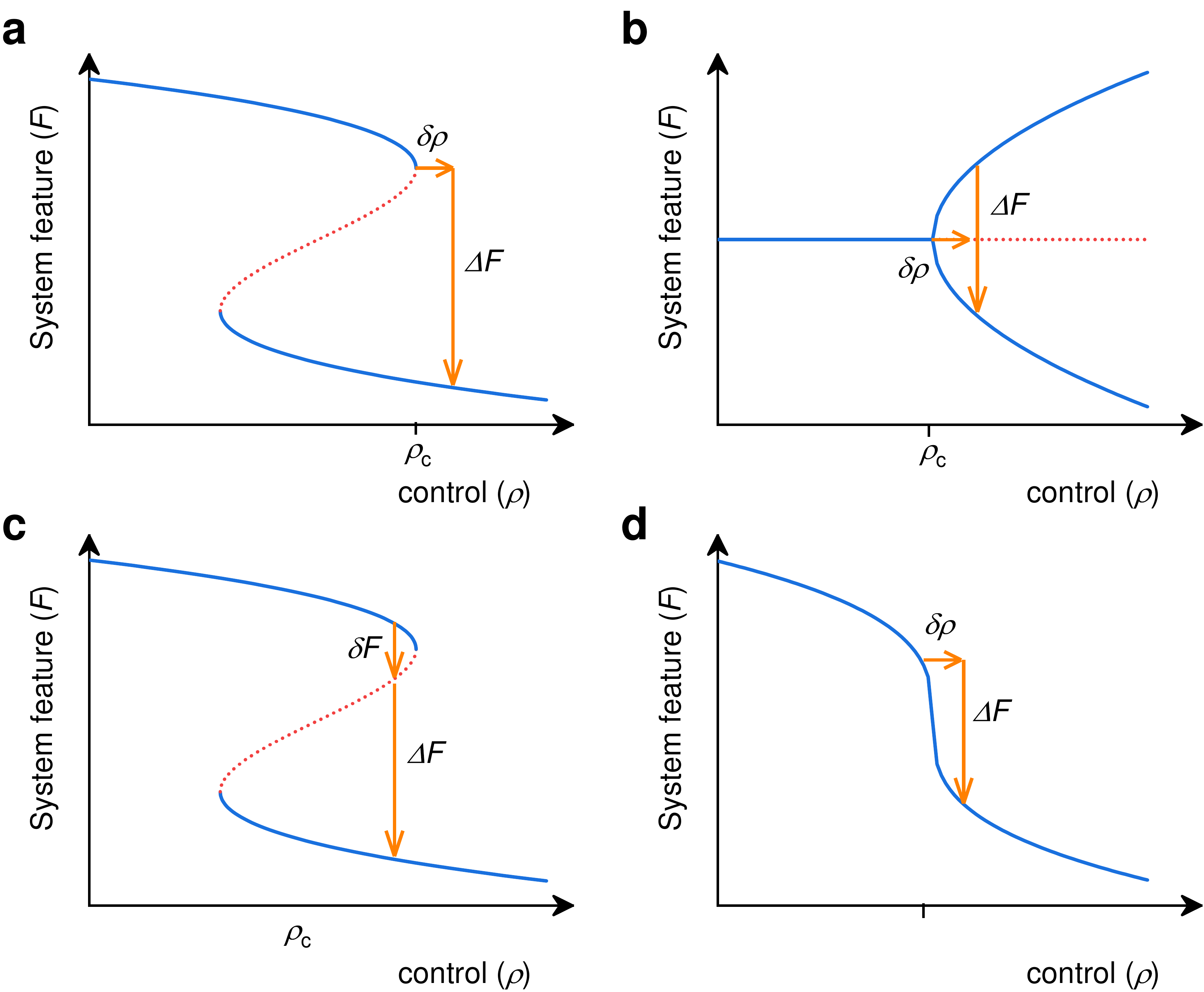}
\caption{\label{Fig2_11} 
{\bf Different sources of cascading tipping point types.} (a) and (b) Bifurcation-induced, (c)  noised-induced transitions, (d) reversible tipping points.  Solid lines are stable steady states, dashed lines are unstable steady states.} 
\end{centering}
\end{figure}

\textbf{Types of tipping points}:
The theoretical mechanisms behind tipping phenomena in a complex system can be effectively divided into
three distinct categories: \textit{bifurcation-induced}, \textit{noise-induced} and \textit{rate-dependent} tipping  \cite{ashwin_tipping_2012}.

\textit{Bifurcation}, means that a small change in forcing $(\rho)$ past a critical threshold causes a large, nonlinear change in the system state, thus meeting the tipping point definition in Eq. (\ref{EQ63}). Given a conceptualized open system \cite{lenton_early_2012},
\begin{equation}
\label{EQ64}
\frac{\mathrm{d} F}{\mathrm{d} t}=f(F, \rho(t)),
\end{equation}
where $\rho(t)$ is in general a time-varying input. In the case that $\rho$ is constant, we refer Eq. (\ref{EQ64})  as the parameterized system with parameter $\lambda$ and to its stable solution as the quasi-static attractor. If $\rho(t)$ passes through a bifurcation point of the system where a quasi-static attractor loses stability, it is intuitively clear that a system may `tip' directly as a result of varying that parameter. We present two bifurcation-induced tipping point examples in Fig. \ref{Fig2_11}. Where the system's states are described by
\begin{equation}
\label{EQ65}
\frac{\mathrm{d} F}{\mathrm{d} t}= \rho + F - F^3,
\end{equation}
and 
\begin{equation}
\label{EQ66}
\frac{\mathrm{d} F}{\mathrm{d} t}= \rho F - F^3,
\end{equation}
for Fig. \ref{Fig2_11} a and b, respectively. In general, as a system approaches a bifurcation tipping point, where its current state becomes unstable, it leads to a shift to an alternative attractor. 

\textit{Noise-induced}, noise-induced transitions between existing stable states  of a  complex system (Fig. \ref{Fig2_11}c), can also be regarded as tipping points \cite{ditlevsen_tipping_2010}, however, they do not meet the definition of forced changes, as in Eq. (\ref{EQ63}). Noise-induced tipping points mean that noisy fluctuations result in the system departing from
a neighbourhood of a quasi-static attractor. For example, Pikovsky and Kurths studied the coherence resonance in a noise-driven  excitable Fitz Hugh-Nagumo system and uncovered that the effect of coherence resonance is explained by different noise dependencies of the activation and the excursion times \cite{pikovsky_coherence_1997}.
The abrupt warming events during the last ice age, known as Dansgaard–Oeschger events, provide a noise-induced tipping real-world example \cite{ditlevsen_tipping_2010}. In addition, Sutera \cite{sutera_stochastic_1981} studied noise-induced tipping points in a simple global zero-dimensional energy balance
model with ice-albedo and greenhouse feedback \cite{fraedrich_catastrophes_1979}. Their results indicate a
characteristic time of $100, 000$ years for  random transitions between  `warm' and  `cold' climate states, which match very well with the observed data.  The noise-induced tipping points approach has also successfully been used for modelling changes in other 
climate models and phenomena \cite{sura_noise-induced_2002}. In contrast to approaching bifurcations, it was found that noise-induced transitions are fundamentally unpredictable and  show none of the EWS \cite{lenton_early_2011}.

\textit{Rate-dependent},
in which the system fails to track a continuously changing
quasi-static attractor. To better understand the phenomenon
of rate-dependent tipping, Ashwin \textit{et al.} introduced a linear model with a tipping radius and discussed three examples where rate-dependent tipping appears \cite{ashwin_tipping_2012}. Given a system for $F \in \mathbb{R}^{n}$ and the parameter $\rho$ has a quasi-static equilibrium $\tilde{F}(\rho)$ with a tipping radius $R>0$, then for some initial $F_{0}$ with $\left|F_{0}-\tilde{F}(\rho)\right|<R$, the evolution of $F$ with time is given by
\begin{equation}
\label{EQ67}
\frac{\mathrm{d} F}{\mathrm{d} t}=M(F-\tilde{F}(\rho)) \quad \text { for } \quad|F-\tilde{F}(\rho)|<R,
\end{equation}
where $M$ is a fixed stable linear operator (i.e. $\left|\mathrm{e}^{M t}\right| \rightarrow 0$ as $t \rightarrow \infty$ ). $\rho(t)$  is a time-varying parameter, that represents the input to the subsystem. The tipping radius may be related to the basin of attraction boundary for the nonlinear problem. This model  shows only rate-dependent tipping- because $M$ is fixed and there is no bifurcation in the system and no noise is present. In particular, the model can be generalized to include $M$ and $R$ that vary with $\rho(t)$.  Eq. (\ref{EQ67}) can be solved with the initial condition $F(0)=F_{0}$ to give
\begin{equation}
\label{EQ68}
F(t)=\mathrm{e}^{M t} F_{0}+\int_{s=0}^{t} \mathrm{e}^{M(t-s)} M \tilde{F}(\rho(s)) \mathrm{d} s.
\end{equation}
Note that the rate-dependent tipping was also observed in the zero-dimensional global energy balance model \cite{fraedrich_catastrophes_1979}.

Besides the above three tipping types, there is potentially another type,  a \textit{reversible} tipping point. In Fig. \ref{Fig2_11}d, we show an example of a reversible tipping point, in which a mono-stable system exhibits  non-linear but reversible change  \cite{lenton_arctic_2011}.


\subsubsection{Tipping Elements in the Earth's Climate System}

\begin{figure}[]
\begin{centering}
\includegraphics[width=1.0\linewidth]{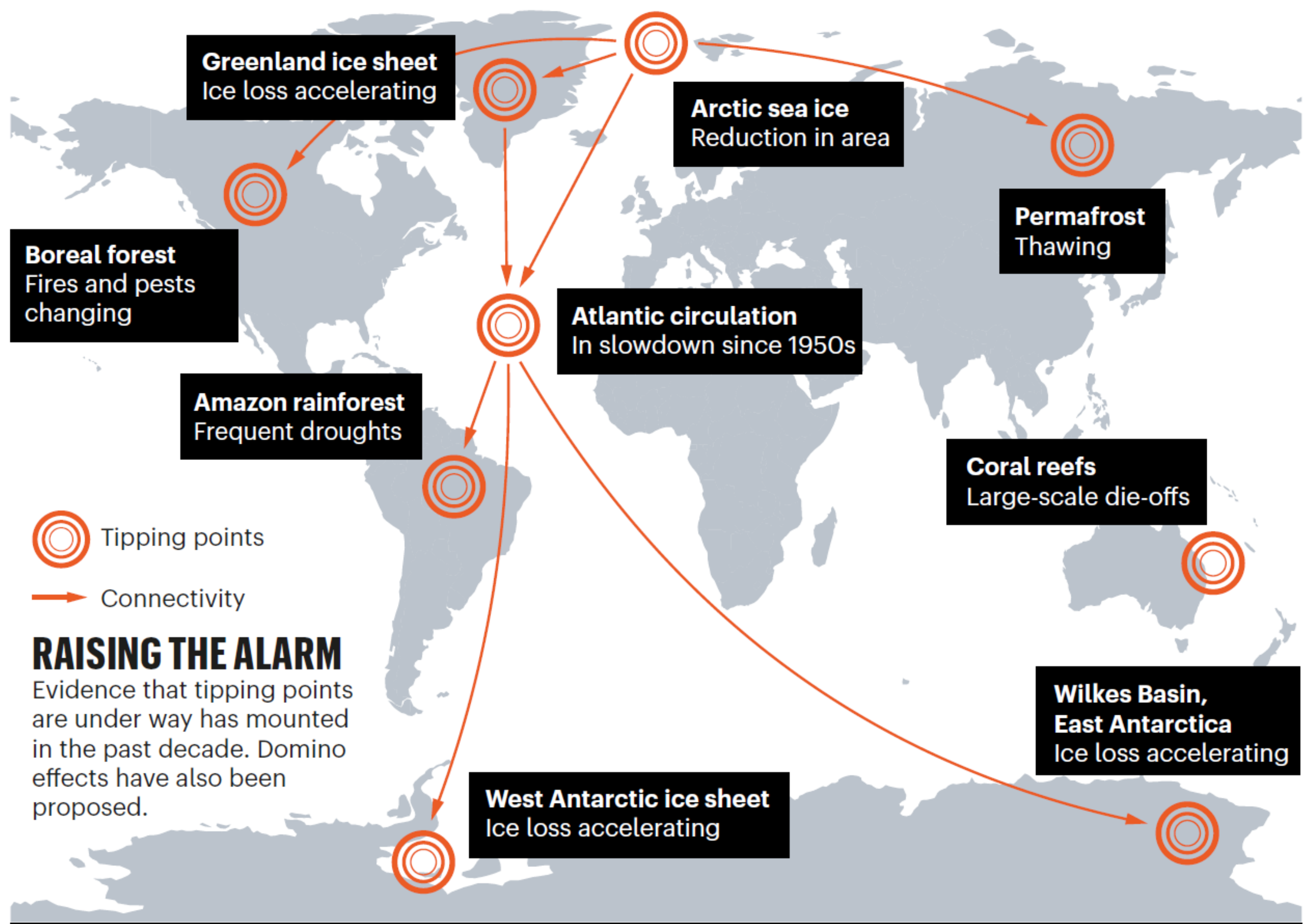}
\caption{\label{Fig2_12} 
{\bf Some basic main tipping elements in the Earth's climate system.} \textit{Source}: figure from Ref. \cite{lenton_climate_2019}. }
\end{centering}
\end{figure}

In Ref. \cite{lenton_tipping_2008} Lenton \textit{et al.} introduced the term tipping elements to describe large-scale components of the Earth system that
may pass a tipping point, and assessed where their tipping
points lie. They emphasized that the human activities may play a vital role to push the components
of the Earth system past their tipping points,  resulting in 
profound impacts on our social and natural systems. They also defined the subset of policy-relevant tipping
elements having the following conditions:
(i) The first condition is described in Eq. (\ref{EQ63}); (ii) the second one is related to the ``political time horizon'' $T_P$; (iii) the third one is called ``ethical time horizon'' $T_E$; (iv) the last one is that a qualitative change should correspondingly be defined in terms of impacts.
According to these four aforementioned conditions, we review here some policy-relevant potential tipping elements in the climate system, as shown in Fig. \ref{Fig2_12}.
For each tipping element, we consider its (1) feature of the system $F$ (see Eq. (\ref{EQ63})) or direction of change, (2) control parameter(s) $\rho$, (3) critical value(s) $\rho_{c}$, (4) transition
timescale $T$ as well as (5) key impacts. 

\textit{Arctic Sea-Ice}. For both summer
and winter Arctic sea-ice, the area coverage is declining and the ice has thinned significantly over a large area \cite{stroeve_arctic_2007}. 
In the IPCC projections with ocean-atmosphere GCMs, half of them will become ice-free in September during this century \cite{holland_future_2006}, at a polar temperature $-9 ^{\circ}$C \cite{winton_does_2006}.
The decline in the areal extent can be regarded as the feature of
the system, denoted by  $F$; the local air temperature $\Delta T_{air}$ and ocean heat
transport can be regarded as control parameters; a numerical value of the critical threshold is still lacking in literature; the transition
timescale is about $T \sim 10$ years; the declining of the Arctic Sea-Ice can amplify warming and cause ecosystem changes.

\textit{Greenland ice sheet}. It has been reported that the Greenland ice sheet is melting at an accelerating rate \cite{portner2019ipcc}, which could add additional  7 meters to sea level over thousands of years if it passes a particular threshold. The decrease of the ice volume can be regarded as $F$, a feature of
system; the local air temperature $\Delta T_{air}$ is a control parameter; the critical value $\rho_c \sim 3$\degree; the transition
timescale $T$ can reach  about 300 years.

\textit{West Antarctic ice sheet}.
According to the IPCC report \cite{portner2019ipcc}, the Amundsen Sea embayment of West Antarctica might have already passed a tipping point, i.e., the `grounding line' (where ice, ocean and bedrock meet) is retreating irreversibly. This could also destabilize the rest of the West Antarctic
ice sheet \cite{feldmann_collapse_2015}.  It has been found, using paleoclimatology data, that such widespread collapse of the West Antarctic ice sheet  occurred repeatedly in the past. Similar to the Greenland ice sheet, the decrease of the Ice volume can be regarded as $F$; the local air temperature $\Delta T_{air}$ is  the control parameter; the critical
value $\rho_c \sim 5-8$\degree; the transition
timescale $T$ can reach  300 years. The collapse of the West Antarctic ice sheet may lead to 
about 5 meters of sea-level rise on a timescale of centuries to millennia \cite{lenton_tipping_2008}.

\textit{Atlantic circulation}. 
The Atlantic meridional overturning circulation (AMOC) is one of Earth’s major ocean circulation systems, redistributing
heat and affecting the climate. Research has  provided evidence for a weakening of the AMOC by about $3  \pm 1$ sverdrups (around 15 per cent)\footnote{In oceanography, a sverdrup (symbol: Sv) is a non-SI metric unit of flow, with 1 Sv equal to 1 million cubic metres per second} since
the mid-twentieth century \cite{caesar_observed_2018}. The AMOC is also considered as one of the main tipping elements of the
Earth system \cite{rahmstorf_ocean_2002,lenton_tipping_2008}. The overturning
of the Atlantic circulation can be regarded as $F$; the additional North Atlantic freshwater input is the control parameter; the critical
value $\rho_c \sim 0.1-0.5$ Sv; the transition
timescale $T$ can reach 100 years. A slowdown of the AMOC is associated with a southward shift of
the tropical rainfall belt by influencing the Intertropical Convergence Zone,  and a warming of the Southern Ocean and
Antarctica \cite{stocker2013climate}.

\textit{\el–Southern Oscillation (ENSO)}. 
ENSO, the interannual
fluctuation between anomalous warm and cold conditions in
the tropical Pacific, is one of the most influential coupled ocean–
atmosphere climate phenomena on Earth \cite{dijkstra_nonlinear_2005,clarke2008introduction,sarachik2010nino}. It has been reported that extreme \el~ events are projected to likely increase in frequency in the 21st century \cite{cai_increasing_2014}. The tipping point behavior of ENSO in a warming world was discussed in Ref. \cite{latif_ninosouthern_2009}.
The amplitude of ENSO  is regarded as $F$; the zonal mean
thermocline depth, thermocline sharpness in the east equatorial Pacific, and the
strength of the annual cycle are the control parameters; the transition
timescale $T$ can  reach  100 years. A stronger \el~
usually causes more extreme events (e.g., floods, droughts, or
severe storms), which have serious consequences for economies,
societies, agriculture  and ecosystems.

\textit{Indian summer monsoon}.
The Indian summer monsoon rainfall (ISMR) has a decisive influence on India's agricultural output
and economy. The monsoon season (from June to September) can bring drought and food shortages or severe flooding, depending on how much rain falls. The land-to-ocean pressure gradient drives the monsoon circulation, related to the moisture-advection feedback \cite{zickfeld_is_2005}.  The ISMR shows a declining trend since 1953 \cite{kumar_recent_2020}. It has been reported that under some
plausible decadal-scale scenarios of land use, greenhouse gas
and aerosol forcing, the Indian summer monsoon switches  between two 
metastable regimes corresponding to
the ``active’’ and ``weak’’ monsoon phases \cite{lenton_tipping_2008}. Thus, the Indian summer monsoon can be also regarded as a tipping element.
In particular, the ISMR is regarded as $F$; the planetary albedo over
India is the control parameter; the critical
value $\rho_c = 0.5$ Sv; and the transition timescale $T$ is 1 years. 

\textit{Amazon rainforest}.
Tropical forests play a vital role in the global carbon cycle \cite{quere_global_2015} and
are the home of more than half of the known species worldwide \cite{wright_tropical_2005}. Deforestation and climate change are
destabilizing the tropical forests with annual deforestation
rates of around 0.5\% since the 1990s, with a strong increase in recent years \cite{achard_determination_2014}.
An empirical finding suggests that the observed tropical forest fragmentation
is near to the critical point in three continents, including the Americas, Africa and Asia–Australia \cite{taubert_global_2018}. In particular, we focus on the Amazon rainforest, since it is the world’s largest
rainforest and is home to one in ten known
species \cite{lenton_climate_2019}. If forests
are close to tipping points, the Amazon dieback
could release 90 gigatonnes Carbon dioxide. Finding
the tipping point of the  Amazon rainforest is thus essential for us to stay within the emissions budget. Here the rainforest in the  Amazon is regarded as $F$; the precipitation and dry season length are the control parameters; the critical
value $\rho_c = 1,100$ mm/year; its transition
timescale $T$ is about 50 years. 

Besides the seven above mentioned tipping elements, there are also some potential policy-relevant tipping elements in the climate system, such as, Sahara/Sahel and West
African monsoon, boreal forest, Antarctic bottom water, tundra, permafrost, marine methane
hydrates, ocean anoxia and Arctic ozone. The readers who are interested can find them in Ref. \cite{lenton_tipping_2008}. Strong evidence indicates that ``tipping points are under way has mounted in the past decade. Domino
effects have also been proposed" \cite{lenton_climate_2019}.


\subsubsection{Early Warning  of Tipping Points}
\label{WATP}
As discussed  above, many complex dynamical systems, in particular  climate systems, can have tipping points and imply risks of unwanted collapse. 
Although predicting such  tipping points before they are reached
is a big challenge, the existence of generic EWS provide useful
indicators for anticipating such critical transitions \cite{scheffer_early-warning_2009,scheffer_anticipating_2012}. Hence, if an early warning of a tipping point can be identified, then it could help broader society, scientists and policymakers to perform early actions to reduce system collapsed related damages. Thus, numerous studies have been dedicated to detecting and predicting these critical transitions, often making use of the EWS.
In this review, we will highlight the tipping point analysis techniques that are used to anticipate, detect and forecast critical transitions in a dynamical system.

\textit{Critical slowing down near tipping points}

\begin{figure}[H]
\begin{centering}
\includegraphics[width=1.0\linewidth]{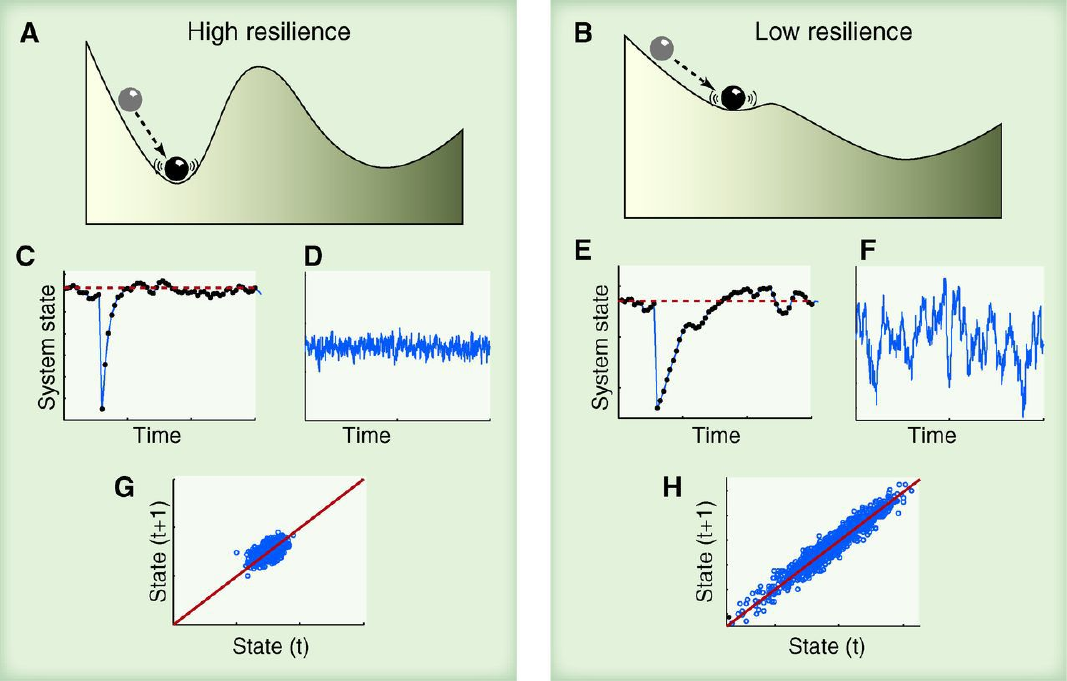}
\caption{\label{Fig2_13} 
{\bf Critical slowing down as an indicator for foreseeing tipping points.} \textit{Source}: figure from Ref. \cite{scheffer_anticipating_2012}. }
\end{centering}
\end{figure}
The ``critical slowing down" phenomenon has been suggested as indicators of whether a dynamical system is getting close to a critical threshold  \cite{wissel_universal_1984}.
This happens, for instance, at the 
fold bifurcation (Fig. \ref{Fig2_11}), often associated with tipping points.
This indicates that the rate at which a system recovers
from small perturbations becomes slow, and the slowness can be inferred from
rising ``memory” in small fluctuations in the state of a system. For example, Fig. \ref{Fig2_13} demonstrates that the critical slowing down has an indicator that the system has lost resilience and may be
tipped more easily into an alternative state (as reflected by lag-1 autocorrelation) \cite{scheffer_anticipating_2012}.

In order to illustrate the relation between the critical slowing down phenomenon, increased
autocorrelation and increased variance, here we show a simple example that reveals the underlying mechanism. We consider a simple autoregressive model of first order,
\begin{equation}
\begin{array}{c}
\label{EQ69}
    x_{n+1}-\bar{x}=e^{\lambda \Delta t}\left(x_{n}-\bar{x}\right)+\sigma \varepsilon_{n} \\
    y_{n+1}=e^{\lambda \Delta t} y_{n}+\sigma \varepsilon_{n},
\end{array}
\end{equation}
where $\bar{x}$ is the stable
equilibrium of the model, $\Delta t$ is the period and $\lambda$ is the recovery speed. Here $y_{n}$ is the deviation of the state variable $x$ from the equilibrium, $\varepsilon_{n}$ is
a random number obtained from a standard normal distribution and $\sigma$ is the standard deviation. If $\lambda$ and $\Delta t$ are independent of $y_{n}$, Eq. (\ref{EQ69}) can  be written as:
\begin{equation}
\label{EQ70}
y_{n+1}=\alpha y_{n}+\sigma \varepsilon_{n}.
\end{equation}
The lag-1 autocorrelation $\alpha \equiv \mathrm{e}^{\lambda \Delta t}$ is zero for white noise and close to one for red noise. The expectation value of a first-order autoregressive process  $y_{n+1}=c+\alpha y_{n}+\sigma \varepsilon_{n}$ is
\begin{equation}
\label{EQ71}
\mathrm{E}\left(y_{n+1}\right)=\mathrm{E}(c)+\alpha \mathrm{E}\left(y_{n}\right)+E\left(\sigma \varepsilon_{n}\right) \Rightarrow \mu=c+\alpha \mu+0 \Rightarrow \mu=\frac{c}{1-\alpha}.
\end{equation}
Let's consider $c=0$, then the expectation value equals zero and the variance to be
\begin{equation}
\label{EQ72}
\operatorname{Var}\left(y_{n+1}\right)=\mathrm{E}\left(y_{n}^{2}\right)-\mu^{2}=\frac{\sigma^{2}}{1-\alpha^{2}}.
\end{equation}
The return speed to equilibrium decreases, when the system is close to the critical point. This implies that $\lambda$ tends to zero and $\alpha$ approaches to one. According to Eq. (\ref{EQ72}), the variance tends to infinity. In summary, the critical slowing
down leads to an increase in lag-1 autocorrelation ($\alpha$) and in the resulting
pattern of fluctuations (variance).

Slowing down causes the intrinsic rates of change
to decrease, and thus the state of the system 
becomes more like its past state, i.e., the autocorrelation increases.
The resulting increase in ``memory'' can be measured in a variety of ways from the frequency
spectrum of the system.

\textit{Autocorrelation function}.
The lag-1 autocorrelation function (ACF(1)) indicator is a simple way to provide an EWS for an impending tipping event. For instance, Held and Kleinen have shown that the autocorrelation increases in the vicinity of a bifurcation in
a model of the thermohaline circulation \cite{held_detection_2004}; Dakos \textit{et al.}  found that the autocorrelation increases before eight well-known
climate transitions in the past  data \cite{dakos_slowing_2008}. 

The coefficient of the correlation between two values in a time series is called the ACF. For example the ACF for a time series $y_{n}$, see Eq. (\ref{EQ70}) in the autoregressive model,  is given by: $Corr(y_{n}, y_{n-s}), s = 1, 2, ...,
$
where $s$ is the time gap and is called the lag. In particular, 
a lag-1 autocorrelation is the correlation between values that are one time step apart. More generally, a lag $s$ autocorrelation is the correlation between values that are $s$ time steps apart.

The ACF scaling exponent, $\gamma$, is the power-law decay of the autocorrelation function with increasing lag $s$ \cite{kantelhardt_detecting_2001}.  Let's denote $C(s)$ as the autocorrelation with lag $s$ of time series, then the scaling is defined as
\begin{equation}
\label{EQ74}
C(s) \sim s^{-\gamma},
\end{equation}
for long-range correlations. Notably,  for short-range correlated records, $C(s)$ decays exponentially and only $\mathrm{ACF}(1)$ is indicative of a data variability close to a tipping point.

\textit{Detrended fluctuation analysis}.
Slowing down causes an increase in memory, which can also be measured using detrended fluctuation analysis (DFA) \cite{peng_mosaic_1994}.  See \cite{peng_mosaic_1994} for the detailed method.
DFA  is often used to detect long-range correlations or the  persistence of diverse time series including DNA sequences \cite{buldyrev_long-range_1995}, heart rate \cite{peng_long-range_1993,ashkenazy_discrimination_1998,bunde_correlated_2000},
earthquakes \cite{fan_possible_2019}, and also climate records \cite{koscielny-bunde_indication_1998}. If the time series is long-range  correlated, the fluctuation function,
$F(n)$, increases according to a power-law relation:
\begin{equation}
\label{EQ75}
F(n) \sim n^\alpha,
\end{equation}
where $n$ is the window size and $\alpha$ the DFA scaling exponent. Here, $F( n ) = \sqrt{\frac{1}{n}\sum_{t = 1}^n \left( X_t - Y_t^z \right)^2}$, where $X_t=\sum_{i=1}^t (x_i-\langle x\rangle)$ is the cumulative sum or the ``profile'' of a time series ${x_i}$, and $Y_t^{z}$ is the fitting polynomial, $z$ stands for the (polynomial) order of the DFA. The DFA exponent $\alpha$ is calculated as the slope of a linear fit to the log-log graph of $F(n)$ vs. $n$. It has been found that the DFA exponent in the temporal range $10 \leq n \leq 100$ is sensitive to changes in a dynamical system, which is  similar to $\mathrm{ACF}(1)$ \cite{livina_modified_2007}.

\textit{Power spectrum}.
Recently,  Prettyman \textit{et al.} introduced a novel scaling indicator based on the decay rate of the power spectrum (PS) \cite{prettyman_novel_2018}.
PS analysis partitions the amount of
variation in a time series into different frequencies. When a system
is close to a critical transition, it tends to show spectral reddening, i.e.,  higher
variation at low frequencies \cite{kleinen_potential_2003}.
The PS scaling exponent $\beta$ is calculated by estimating the slope of the power spectrum $S(f)$ of the data, from the scaling relationship
\begin{equation}
\label{EQ76}
S(f) \sim f^{-\beta}.
\end{equation}
Analytically, the three scaling exponents have the linear relationship:
$\alpha=\frac{1+\beta}{2}=1-\frac{\gamma}{2}$. It has been reported that the PS-indicator is a useful
technique which behaves similarly to the related ACF(1)-
and DFA-indicators. In addition, it also shows
signs of providing  an EWS  for  a real geophysical
system, tropical cyclones, whereas the ACF(1)-
indicator fails \cite{prettyman_novel_2018}.

Besides the slower recovery from perturbations, increased autocorrelation and memory [see Eq. (\ref{EQ71})], increased variance [measured as standard deviation, see Eq. (\ref{EQ72})] is another possible indicator 
of a critical slowing down as a critical transition is
 approached. For example, Carpenter and Brock have shown that the variance increases in the vicinity of a bifurcation in a lake
model \cite{carpenter_rising_2006}.

\textit{Flickering before transitions}.
Another notable EWS is a system's back and forth oscillation between two stable states in the vicinity of a critical transition. This oscillation has been called flickering and was observed  on the model of lake eutrophication \cite{wang_flickering_2012} and trophic cascades \cite{carpenter_leading_2008}.

\textit{Skewness and Kurtosis}.
In addition to the aforementioned EWS before a catastrophic bifurcation, 
two further precursors, observed when approaching a critical transition and thus suggested as EWS, are changes in the skewness and kurtosis of the distribution of states \cite{guttal_changing_2008,biggs_turning_2009,dakos_methods_2012}. While skewness indicates asymmetry in the distribution - with a negative skew indicating a right-sided concentration and a positive skew indicating the opposite. Skewness is the standardized third moment around the mean of a distribution,
\begin{equation}
\label{EQ77}
\gamma=\frac{\frac{1}{n} \sum_{i=1}^{n}\left(y_{i}-\mu\right)^{3}}{[\frac{1}{n} \sum_{i=1}^{n}\left(y_{i}-\mu\right)^{2}]^{3/2}}.  
\end{equation}
 
Similarly,  kurtosis is a measure of the ``peakedness" of the distribution - with positive kurtosis indicating a peak higher than the one of a normal distribution and negative kurtosis indicating a lower peak. 
Kurtosis is the standardized fourth moment around the mean of a distribution estimated as,
\begin{equation}
\label{EQ78}
\kappa=\frac{\frac{1}{n}\sum_{t=1}^{n}\left(z_{t}-\mu\right)^{4}}{(\sqrt{\frac{1}{n} \sum_{t=1}^{n}\left(z_{t}-\mu\right)^{2}})^{2}}.
\end{equation}

\subsubsection{Precursors of Transitions in Real Systems}
In the last section, we first highlighted the theoretical background of tipping points
that may occur in non-equilibrium dynamics before critical transitions.
Nonetheless, it poses much of a challenge  to
detect EWS in real complex  systems such as, climate, social and ecological
systems. Developing reliable predictive systems based
on these generic properties can strengthen our capacity to navigate systemic failure and guide us
for designing more resilient  systems.

We will briefly review emerging precursors of transitions in different real systems, including climate, ecosystems, medicine and finance.
An abrupt climate change occurs when the Earth system is forced to cross a     threshold to a new climate state \cite{alley_abrupt_2003}. Large, abrupt, and widespread climate changes include the Carboniferous Rainforest Collapse \cite{sahney_rainforest_2010}, Younger Dryas \cite{broecker_was_2006}, greenhouse–icehouse transition \cite{liu_global_2009},  Dansgaard-Oeschger events, Heinrich events and Paleocene–Eocene Thermal Maximum \cite{board2002abrupt}. 
All these rapid
climate changes could be explained as critical transitions \cite{scheffer_early-warning_2009}. For instance, a significant increase in autocorrelation is regarded as a precursor  of eight well-known
climate transitions \cite{dakos_slowing_2008}. A flickering phenomenon was found to precede the abrupt end
of the Younger Dryas cold period, which can be also  regarded as a precursor \cite{bakke_rapid_2009}.

In ecology, tipping point analysis has also become a major focus
of research. For example, an EWS was found in the destabilization of
exploited fish stocks, where harvesting leads
to increased fluctuations in fish populations \cite{hsieh_fishing_2006}. The emerging of EWS has been demonstrated in lake and marine systems \cite{scheffer_catastrophic_2003,scheffer_shallow_2007}. In physiology, abrupt transitions were found in epileptic seizures and asthma attacks \cite{venegas_self-organized_2005}. For example, flickering may occur before epileptic
seizures. In finance, market dynamics may
contain information indicating an abrupt change. For example, increased trade volatility may occur  before a main shock \cite{bates_crash_1991}. For other precursors of transitions in real systems, the reader is referred to Ref.  \cite{scheffer_early-warning_2009}.

\subsection{Entropy  and Complexity}

\subsubsection{Introduction}
Entropy is an important concept arising from statistical mechanics.  It is a characteristic that describes the state of a system composed of smaller components, and it has been used to be a general measure of complexity, with
widespread applications. In classical thermodynamics, entropy is related to the loss of energy during an irreversible process,  developed by Rudolf Clausius in the early 1850s. The thermodynamic entropy $S$ is derived from the heat flow $\delta Q$ at a fixed temperature $T$, 
\begin{equation}
\label{EQ79}
S = \int \frac{\delta Q}{T}.
\end{equation}
Note that the entropy of a system is defined only if it is in a thermodynamic equilibrium.
The statistical mechanics definition of entropy was developed by Ludwig Boltzmann in 1870s via analyzing the statistical behavior of the microscopic components of a system as
\begin{equation}\begin{array}{l}
\label{EQ80}
S = k_{B} \ln \Omega.
\end{array}\end{equation}
Here $S$ is the entropy of the macrostate, $k_B$ is the Boltzmann's constant, and $\Omega$ standing the total number of possible microstates that yields the macrostate. When viewed in terms of information theory,  Claude Shannon developed the very general concept of \textit{information entropy}, a fundamental cornerstone of information theory, to describe an analogous loss of information \cite{shannon_mathematical_1948}. It is a measure of the amount of information that is missing before reception.
The definition of the information entropy is,
\begin{equation}
\label{EQ81}
S =-k_{S} \sum_{i} p_{i} \log p_{i},
\end{equation}
where $k_{S}=1 / \log (2)$ (in bits), $p_i$ is the probability of each state.
In quantum statistical mechanics, the concept of entropy has been  developed by John von Neumann and is generally referred to  as ``von Neumann entropy''.
For a quantum-mechanical system described by a density matrix $\rho$, the von Neumann entropy is,
\begin{equation}
\label{EQ82}
S=-k_{B} \operatorname{Tr}(\rho \log \rho),
\end{equation}
where $\operatorname{Tr}$ denotes the trace operator.

There exist also many other types of entropy, such as Gibbs, Residual, Approximate, Sinai-Kolmogorov, Sample, Multiscale. Entropy has been proven useful in many real-world systems, including  analysis of DNA sequences \cite{thanos_entropic_2018}, cosmology and astrophysics \cite{bekenstein_black_1973,hawking_particle_1975,srednicki_entropy_1993}, economics \cite{georgescu1993entropy,ayres_practical_2007}, and climate systems \cite{stephens_entropy_1993,obrien_entropy_1995,meng2019complexity}. Each definition of entropy could give  better results for some systems but fails for others. We will discuss in the following that entropy has three related interpretations \cite{sethna2006statistical} as a measure  (i) irreversible changes, (ii)  disorder and (iii) uncertainty.


\subsubsection{Entropy as Irreversibility}
The idea of \textit{irreversibility} is central to the understanding of entropy. A process that is not reversible is usually called irreversible. This concept arises in thermodynamics. It has been reported that all complex dynamic natural processes are irreversible \cite{lucia_probability_2008}. For an isolated system with an irreversible process, the entropy never decreases. This is known as the second law of thermodynamics. 

\begin{figure}[H]
\begin{centering}
\includegraphics[width=1.0\linewidth]{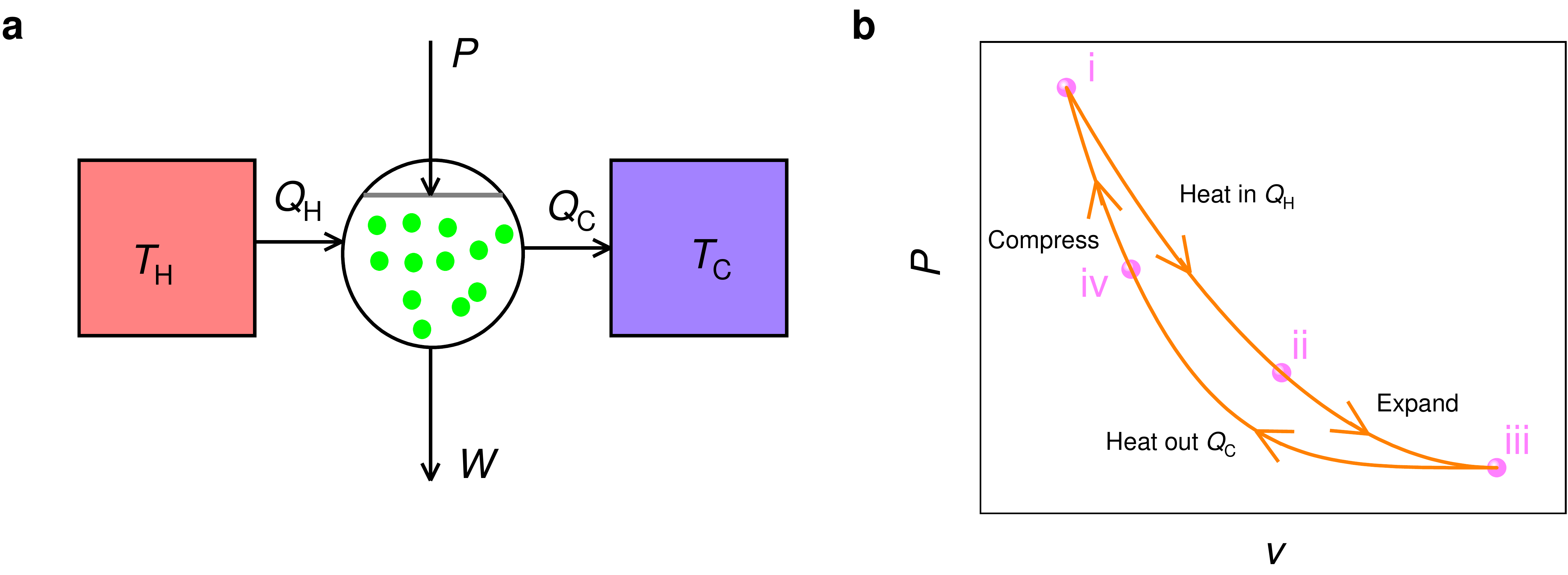}
\caption{\label{Fig2_14} 
{\bf Entropy as Irreversibility.} a. Prototype of Heat Engine. b. Carnot Cycle P-V Diagram.}
\end{centering}
\end{figure}

In order to better understand the behavior of entropy in an irreversible process,  we consider the Carnot heat engine and the corresponding Carnot cycle. As shown in Fig. \ref{Fig2_14}a, given by a piston with pressure $P$ (external), and two heat baths, one at a hot temperature $T_{H}$ and the other at a cold temperature $T_{C}$ with some material (a monatomic ideal gas) inside the piston. During one Carnot cycle, $Q_{H}$ heat flows out of the hot bath, heat
$Q_{C}$ flows into the cold bath, resulting in a net work $W=Q_{H}-Q_{C}$. 
Carnot imagined a fully reversible heat engine, and the Carnot cycle moves the piston in and out with the following four steps (See, Fig. \ref{Fig2_14} b, the Carnot cycle P-V diagram). (1) $(\mathrm{i} \rightarrow \mathrm{ii})$: The gas is connected to the hot bath, and the piston moves outward at a varying pressure and a fixed temperature $T_{H}$. In this step, heat $Q_{H}$ flows  into the piston. (2) $(\mathrm{ii} \rightarrow \mathrm{iii})$: The piston expands at varying pressure $P$. (3) $(\mathrm{iii} \rightarrow \mathrm{iv})$: The expanded gas is then connected to the cold bath
and compressed at a fixed temperature at $T_C$. In this step,  the heat $Q_C$ flows out. (4) $(\mathrm{iv} \rightarrow \mathrm{i})$: The piston is compressed and returns to the original state. In this step, there is no heat
transfer.

The net work $W$ done by the piston is the area inside P-V Loop (integration) in Fig. \ref{Fig2_14}b, 
\begin{equation}
\label{EQ83}
W=\int_{\mathrm{cycle}} P d V=\text { Area inside } \mathrm{PV} \text { Loop. }
\end{equation}
According to the  ideal gas law,
\begin{equation}
\label{EQ84}
P V=N k_{B} T
\end{equation}
where $N$ is the number of particles, the total energy of the ideal gas inside the piston is given by the equipartition theorem
\begin{equation}
\label{EQ85}
E = 3/2 N k_{B} T = 3/2 P V,
\end{equation}
where $E$  stands for the kinetic energy. During the first step, we get
\begin{equation}
\label{EQ86}
  \begin{array}{l}
    Q_{H} = E_{ii} - E_{i} + W_{i,ii} = \int_{i}^{ii} P d V = \int_{i}^{ii} \frac{N k_{B} T_{H}}{V} d V= N k_{B} T_{H} \log \left(V_{ii} / V_{i}\right).\\ 
  \end{array}
\end{equation}
Similarly, for the third step, we have
\begin{equation}
\label{EQ87}
Q_{C}=N k_{B} T_{C} \log \left(V_{iv} / V_{iii}\right).
\end{equation}
For the other two steps, there is no any heat flow, based on 
$d E= -P d V = -\frac{N k_{B} T}{V} d V = 3 / 2 N k_{B} d T$, and we have
\begin{equation}
\label{EQ88}
\int_{ii}^{iii} \frac{d V}{V}=\log \left(V_{iii} / V_{ii}\right)=\int_{ii}^{iii}-3 / 2 \frac{d T}{T}=-3 / 2 \log \left(T_{C} / T_{H}\right).
\end{equation}
The above equation gives $V_{iii} / V_{ii}=\left(T_{H} / T_{C}\right)^{3 / 2}$ and $V_{iv} / V_{i}=\left(T_{H} / T_{C}\right)^{3 / 2}$.  Thus $V_{iii} / V_{ii}=V_{iv} / V_{i}$ and hence
\begin{equation}
\label{EQ89}
\frac{V_{iii}}{V_{iv}}=\frac{V_{ii}}{V_{i}}.
\end{equation}
Substituting Eq. (\ref{EQ89}) into the heat flow, Eqs. (\ref{EQ86}) and (\ref{EQ86}), we yield Carnot's fundamental result,
\begin{equation}
\label{EQ90}
\frac{Q_{H}}{T_{H}}=\frac{Q_{C}}{T_{C}}.
\end{equation}
Let us define the heat flow $\Delta E = Q$ at
a fixed temperature $T$. Then the entropy change $\Delta S$ is
\begin{equation}
\label{EQ91}
\Delta S = \frac{Q}{T}.
\end{equation}
Note that the above is
equivalent to Eq. \ref{EQ79}.  This means that for a reversible Carnot engine, the entropy flows from the hot bath $Q_{H}/T_{H}$ equals the entropy flows from the piston $Q_{C}/T_{C}$, i.e., no entropy is created or destroyed.
However, for any irreversible real engines, entropy will increase.

\subsubsection{Entropy as Disorder}
Traditionally, another interpretation of entropy is described  as a measurement of the disorder or randomness of a system. 
In thermodynamics, the entropy of mixing is the increase of entropy when two or more different types of particles are mixed without chemical reactions.
Consider that $N$ molecules of an ideal gas are separated  in a vessel with two equal volumes
 $V$. The total unmixed entropy $S_{\text {un}}$ is \cite{sethna2006statistical} 
\begin{equation}
\label{EQ92}
S_{\text {un}}=2 k_{B} \log \left[V^{N / 2} /(N / 2)!\right].
\end{equation}
Here we assume that the two separated ideal gases have the same
masses and the same total energy. Let the partition between the gases be removed and they are allowed to mix. Then the mixed  entropy $S_{\text {m}}$ becomes
\begin{equation}
\label{EQ93}
S_{\text {m}}=2 k_{B} \log \left[(2V)^{N / 2} /(N / 2)!\right].
\end{equation}
According to the Eqs. (\ref{EQ92}) and (\ref{EQ93}), we obtain the increased entropy,
\begin{equation}
\label{EQ94}
\Delta S = S_{\text {m}}-S_{\text {un}}= N k_{B} \left[\log (2V) - \log V\right] =N k_{B} \log 2.
\end{equation}
The above equation means that there will be a gain $k_{B} \log 2$ in entropy per molecule  during the mixing process. This is since the temperatures and pressures are equal, and
removing the partition of the vessel does not involve any heat transfer, 
thus mixing of gases (e.g., by diffusion), always results in an increasing  entropy. The mixing is spontaneous!
The diffusion of initially separated gases results in an increase in entropy.  The process has increased the random distribution of molecules.
Therefore, it is appropriate to suggest a relationship between  entropy and  complexity (disorder).

\subsubsection{Entropy as Uncertainty}
The third interpretation of entropy is as a measure of the uncertainty or ignorance of a complex system. This interpretation is more general and strongly related to the information and
memory. In this interpretation, the entropy is not an intrinsic property 
but representing our knowledge about the system.

Next, we will focus on the non-equilibrium entropy and the information entropy to insulate  how entropy can be interpreted as uncertainty. The second law of thermodynamics states that the entropy of an isolated system  never decreases, and is unchanged if and only if all processes are reversible.
What we are interested in is how to define the entropy for non-equilibrium systems. Generally, for both non-equilibrium and equilibrium systems, we use  a probability distribution $\rho$, to define an ensemble
of states. Given a probability distribution of a discrete set of
states, the entropy is 
\begin{equation}
\label{EQ95}
S_{\mathrm{discrete}}=-k_{B}\left\langle\log p_{i}\right\rangle=-k_{B} \sum_{i} p_{i} \log p_{i},
\end{equation}
where $p_{i}$ is the probability
of $i$ state. In the case of continuum distributions, 
the entropy becomes
\begin{equation}
\label{EQ96}
S_{\mathrm{continuum}}=-k_{B}\langle\log \rho\rangle=-k_{B} \int \rho \log \rho.
\end{equation}
We consider a microcanonical ensemble, in which $\rho_{\text {equil }}=1 /\left(\Omega(E) \delta E\right)$, the non-equilibrium of the entropy is shifted from the equilibrium (Eq. (\ref{EQ80})), and the entropy is  
\begin{equation}
\label{EQ97}
\begin{aligned}
S_{\text {micro }} &=-k_{B} \log \rho_{\text {equil }}=k_{B} \log (\Omega(E) \delta E) \\
&=k_{B} \log (\Omega(E))+k_{B} \log (\delta E).
\end{aligned}
\end{equation}
For quantum systems, the non-equilibrium entropy can be described by the density matrix $\rho$, as in Eq. (\ref{EQ82}).

For non-thermodynamic systems, the temperature variable does not exist. Therefore we do
not need to use  Boltzmann’s constant $k_B$, but  the constant $k_{S}=1 / \log (2)$.  Then Eq. (\ref{EQ81}) becomes
\begin{equation}
\label{EQ98}
S_{S}=-\sum_{i} p_{i} \log _{2} p_{i},
\end{equation}
here the entropy is measured in bits. This is called information entropy, which was  introduced by Shannon \cite{shannon_mathematical_1948}.
Information entropy represents an ensemble of possible
messages or images, and  has important implications in communication
technologies (message passing) and computer science (data compression). Note that the non-equilibrium Shannon entropy satisfies
the following properties: (1) it is maximum for equal probabilities; (2) it is unaffected by extra states of zero probability; and (3) it changes for conditional probabilities. For more details, we refer the reader to Ref. \cite{sethna2006statistical}.

\subsubsection{Approximate Entropy, Sample Entropy and System Sample Entropy}

In the following, we will highlight several types of entropy, including 
Approximate Entropy ($ApEn$), Sample entropy ($SampEn$) and System Sample Entropy ($SysSampEn$), which have been developed to quantify the complexity in non-linear time-series. 

\subsubsection*{ApEn}

The $ApEn$ was introduced by Pincus in 1991 \cite{pincus_approximate_1991} motivated by the  exact regularity statistics, Kolmogorov–Sinai (KS) entropy \cite{eckmann_ergodic_1985} and the $K_2$ entropy \cite{grassberger_estimation_1983}. It is used as a quantification of regularity in time-series data, in particular for relatively short and noisy data sets.
$ApEn$ has a wide range of applications in areas from medical data, such as heart rate  \cite{pincus_approximate_1991}, to finance system \cite{pincus_irregularity_2004},  psychology \cite{pincus_physiological_1994} and human factors engineering \cite{mckinley_evaluation_2011}.

Definition of $ApEn$:
Given a time series of data  $u(1), u(2), \ldots, u(N)$. There are $N$ raw data equidistant in time. It forms a sequence of vectors $x(1)$ through $x(N-m+1)$ defined by $x(i)=[u(i), \ldots, u(i+m-1)]$. Here, $m$ is an integer representing the length of the data. Define the embedding  distance $d[x(i), x(j)]$ \cite{takens_detecting_1981} between vectors $x(i)$ and $x(j)$ as the maximum difference in their respective scalar components. Then use $x(1), x(2), \ldots, x(N-m$ +1) to construct, for each $i \leqslant N-m+1$,  $C_{i}^{m}(r)=$ (number of $j \leqslant N-m+1 \text { such that } d[x(i), x(j)] \leqslant r) /(N-m+1)$. Here,
\begin{equation}
\label{EQ99}
d[x(i), x(j)]=\max _{k=1,2, \ldots, m}(|u(i+k-1)-u(j+k-1)|),
\end{equation}
where $r$ is a positive real number which specifies a filtering level. 
Based on  the $C_{i}^{m}(r)$, the $ApEn$ is defined as 
\begin{equation}
\label{EQ100}
\Phi^{m}(r)=(N-m+1)^{-1} \sum_{i=1}^{N-m+1} \log C_{i}^{m}(r).
\end{equation}
For fixed $m$ and $r$,  Eq. (\ref{EQ100}) becomes
\begin{equation}
\label{EQ101}
\operatorname{ApEn}(m, r)=\lim _{N \rightarrow \infty}\left[\Phi^{m}(r)-\Phi^{m+1}(r)\right].
\end{equation}
In particular, 
given $N$ data points, the statistic formula is implemented 
\begin{equation}
\label{EQ102}
\operatorname{ApEn}(m, r, N)=\Phi^{m}(r)-\Phi^{m+1}(r).
\end{equation}
One should note that a similar entropy, called 
Eckmann-Ruelle (E-R)
entropy \cite{eckmann_ergodic_1985}, has been defined as,
\begin{equation}
\label{EQ103}
\text { E-R entropy }=\lim _{r \rightarrow 0} \lim _{m \rightarrow \infty} \lim _{N \rightarrow \infty}\left[\Phi^{m}(r)-\Phi^{m+1}(r)\right].
\end{equation}
Despite their
algorithms being  very similar, $ApEn(m, r)$ is not intended to be an
approximate value of the  $E-R$ entropy.
Compared to the $K-S$,  $E-R$ and $K_2$ entropy, $ApEn(m, r)$ has the following advantages \cite{pincus_approximate_1995}: (i) Less affected by  noise; (ii) It is robust to outliers; (iii) Lower computational demand, with good confidence; (iv) $ApEn(m, r)$ is finite for stochastic, noisy deterministic and composite processes; (v) Increasing $ApEn(m, r)$ corresponds to intuitively increasing process complexity.
$ApEn(m, r)$ has been applied to classify the electroencephalogram (EEG)  in psychiatric diseases, such as schizophrenia \cite{sabeti_entropy_2009}, epilepsy \cite{yuan_epileptic_2011}, and addiction \cite{yun_decreased_2012}.

\subsubsection*{SampEn}
Motivated by the concept of $ApEn$, Richman and Moorman developed the sample entropy ($SampEn$)  to assess the complexity of physiological time-series signals, diagnosing diseased states \cite{richman_physiological_2000}. $SampEn$ is a modification of $ApEn$, but has three advantages: (i) $SampEn$ agrees much
better than $ApEn$ statistics with the theory for random
numbers having known probabilistic character over a
broad range of operating conditions; (ii) maintains relative
consistency and (iii) has a residual bias for very short record lengths.

Definition of $SampEn$:
Similar to  the definition of $ApEn$:
given a time series of $N$ points, $\{u(j): 1 \leq j \leq N\}$ forms the $N-m+1$ vectors $\mathbf{x}_{m}(i)$ for $\{i | 1 \leq i \leq N-m+1\}$,  where $\mathbf{x}_{m}(i)=\{u(i+k): 0 \leq k \leq m-1\}$ is the vector of $m$ data points from $u(i)$ to $u(i+m-1)$. The distance between two vectors is  $d[X(i), X(j)]=\max \{|u(i+k)-u(j+k)|:$
$0 \leq k \leq m-1\},$ the maximum difference of their corresponding scalar components. Let $B_{i}$ be the number of vectors $\mathbf{x}_{m}(j)$ within $r$ of $\mathbf{x}_{m}(i)$ and let $A_{i}$ be the number of vectors $\mathbf{x}_{m+1}(j)$ within $r$ of $\mathbf{x}_{m+1}(i)$. Here $r$ represents the tolerance for
accepting matches. It is convenient to set the tolerance as $r \times SD$, the standard deviation of the data set. Then the sample entropy is defined as:
\begin{equation}
\label{EQ104}
\text {SampEn}=-\log \frac{A}{B}.
\end{equation}
It is clear that $A$ will always be smaller or equal to $B$. Therefore, $SampEn$ will be always either zero or positive. A smaller value of $SampEn$ indicates more self-similarity in the data set or less noise \cite{richman_physiological_2000}.  Note that the parameters $N$, $m$, and $r$ must be fixed for each calculation.

\subsubsection*{SysSampEn}
\label{SysSampEn}

One limitation of the aforementioned entropy is that it can only be applied to univariate or bivariate time series (cross-ApEn) \cite{richman_physiological_2000}. For a  complex system with  multivariate  time series and spatial-temporal structures, Meng \textit{et al.} proposed the so called $SysSampEn$ and applied it to study the climate system. Based on the  $SysSampEn$, they could measure the complexity (disorder) of a system
composed of temperature anomaly time series and forecast
the magnitude of an \el~ event with a prediction horizon of
$1$ year and high accuracy \cite{meng2019complexity}. 

Definition of $SysSampEn$:
Given $N$ interdependent time series $x_\alpha(t)\ (\alpha=1,2,...,N; t = 1,2,...,l)$ of length $l$ composing the system: 
\begin{enumerate}
\item We select sub-records $k$ of length $m < l$, starting at each $q$-th data point, i.e.,
starting at $t= k\times q+1 = 0\times q + 1, 1\times q+1, 2\times q+1,...$, as long as $k\times q+m \leq l$.
Thus a specific sub-sequence is denoted as $X_\alpha^k(m,q)=\{x_\alpha(k\times q+1),x_\alpha(k\times q+2),...,x_\alpha(k\times q+m)\}$. Then we select $n$ sub-sequences from each time series and construct a set of $N\times n$ template vectors from the system, i.e., $\Theta(m,q,n)=\{X_\alpha^k(m,q):0\leq k \leq n-1, 1\leq \alpha \leq N\}$. We assume that two vectors are close (similar) if their Euclidean distance $d(X_\alpha^i(m,q),X_\beta^j(m,q))<\gamma\times max\{\sigma_\alpha,\sigma_\beta\}$ (if $\alpha=\beta$, then $i\neq j$), where $\sigma_\alpha$ and $\sigma_\beta$ are the standard deviations of the time series $x_\alpha(t)$ and $x_\beta(t)$ respectively. $\gamma$ defines the similarity criterion and is a nonzero constant.

\item To examine the probability that two time series, which are close at $m$ data points, still will be close at the next $p$ data points,
we construct analogously another set $\Theta(m+p,q,n)$ by selecting sub-records of length $m+p$. To make the number of template vectors of length $m$ equal to that of length $m+p$, we choose $n\leq \frac{l-m-p}{q}+1$. In order to reduce the parameter degrees of freedom and save calculation time, we take $p=q$, then $n\leq \frac{l-m}{p}$. We assume that two template vectors from the set $\Theta(m+p,q,n)$ are close if $d(X_\alpha^i(m+p,q),X_\beta^j(m+p,q))<\gamma\times max\{\sigma_\alpha,\sigma_\beta\}$ (if $\alpha=\beta$, then $i\neq j$).

\begin{figure}[H]
\begin{centering}
\includegraphics[width=1.0\linewidth]{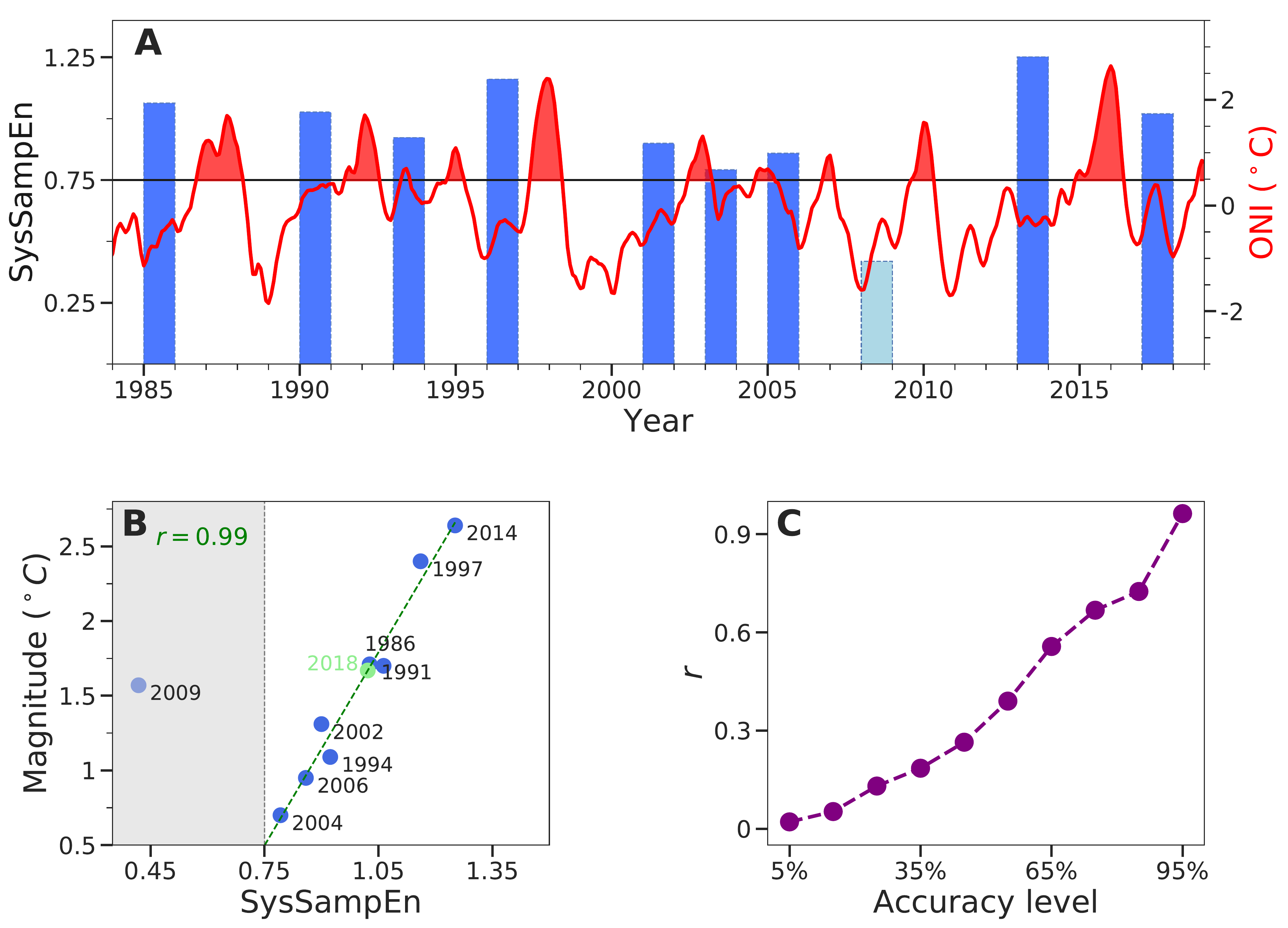}
\caption{\label{Fig2_15} 
{\bf Correlation between SysSampEn and \el~ magnitude.} \textit{Source}: Reprinted figure from Ref. \cite{meng2019complexity}.}
\end{centering}
\end{figure}

\item The $SysSampEn$ of the system is defined as
\begin{equation}
\label{EQ105}
\text {SysSampEn}\left(N, m, p, l_{e f f}, \gamma\right)=-\log \left(\frac{A}{B}\right),  
\end{equation}
where $A$ is the number of close vector pairs from the set $\Theta(m+p,q,n)$, $B$ is the number of close vector pairs from the set $\Theta(m,q,n)$, and $l_{eff}(n)=n*p+m$, is the number of days that is used in the calculation of the $SysSampEn$.
\end{enumerate} 

Note that, when $N=1$, $p=1$, and $l_{eff}=l$, the $SysSampEn$ is equivalent to the classical $SampEn$ \cite{richman_physiological_2000}.   Appropriate parameter values have to
be identified since only certain value combinations can be used
to estimate a system's complexity with considerable accuracy.  In order to  achieve this, 
Meng \textit{et al.}  introduced two  tests \cite{meng2019complexity}: \textit{Spatial asynchrony test} and \textit{temporal disorder test}.
The $SysSampEn$ was calculated  for the climate system composed of the near surface air or sea surface temperature anomaly time series in the Ni\~{n}o 3.4 region. It was found that  a strong positive correlation between the \el ~magnitudes and the values $SysSampEn$ of its previous calendar year exists (Fig. \ref{Fig2_15}A,B). Fig. \ref{Fig2_15}C demonstrates  the calculation of the accuracy level, which allows  to choose effective parameter
combinations.

\begin{figure}[H]
\begin{centering}
\includegraphics[width=0.85\linewidth]{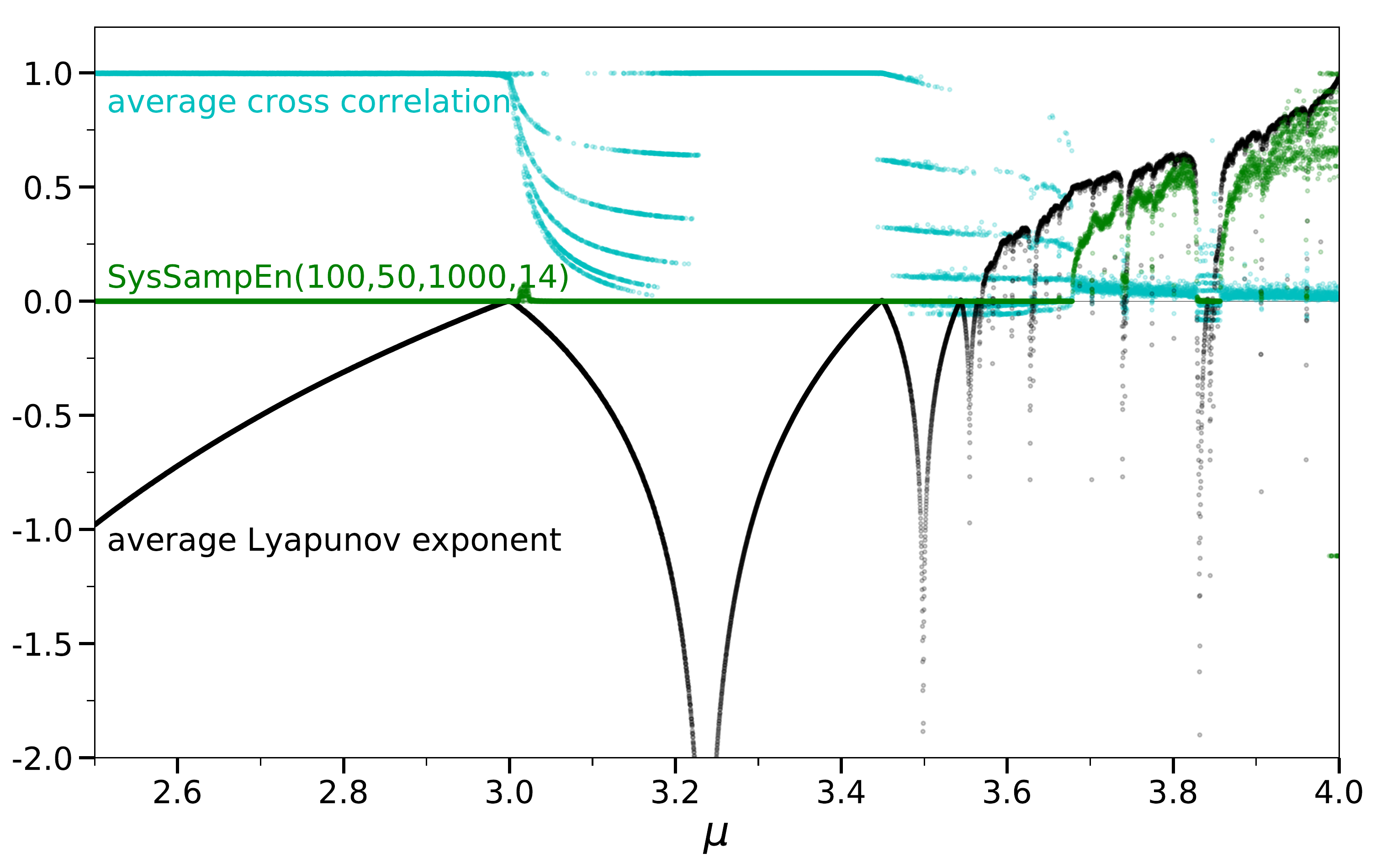}
\caption{\label{Fig2_16} 
{\bf One logistic map example of $SysSampEn$ that gives accurate estimation of the system complexity, i.e., higher values of the $SysSampEn$ indicate  higher complexity of the system.} For each system, $10$ time series are generated from the logistic equation $x_{n+1}=\mu x_n(1-x_n)$ with the same value of $\mu\in[2.5,4]$, but for different initial conditions $x_0 \in [0.0001,0.0002]$. \textit{Source}: Reprinted figure from Ref. \cite{meng2019complexity}.}
\end{centering}
\end{figure}

In addition, to reveal the mathematical meaning of the $SysSampEn$, Fig. \ref{Fig2_16} shows the logistic map as an example of applying the $SysSampEn$ to estimate the system complexity and compare it with the Lyapunov exponents. It is found that higher (lower) values of the $SysSampEn$ are strongly associated with higher (lower) Lyapunov exponents, which indicates that the $SysSampEn$ can well capture the complexity of the system.

\section{Applications}
\label{cap3:App}

\subsection{Climate System}
\label{cap3:CS}

\begin{figure}[]
\begin{centering}
\includegraphics[width=1\linewidth]{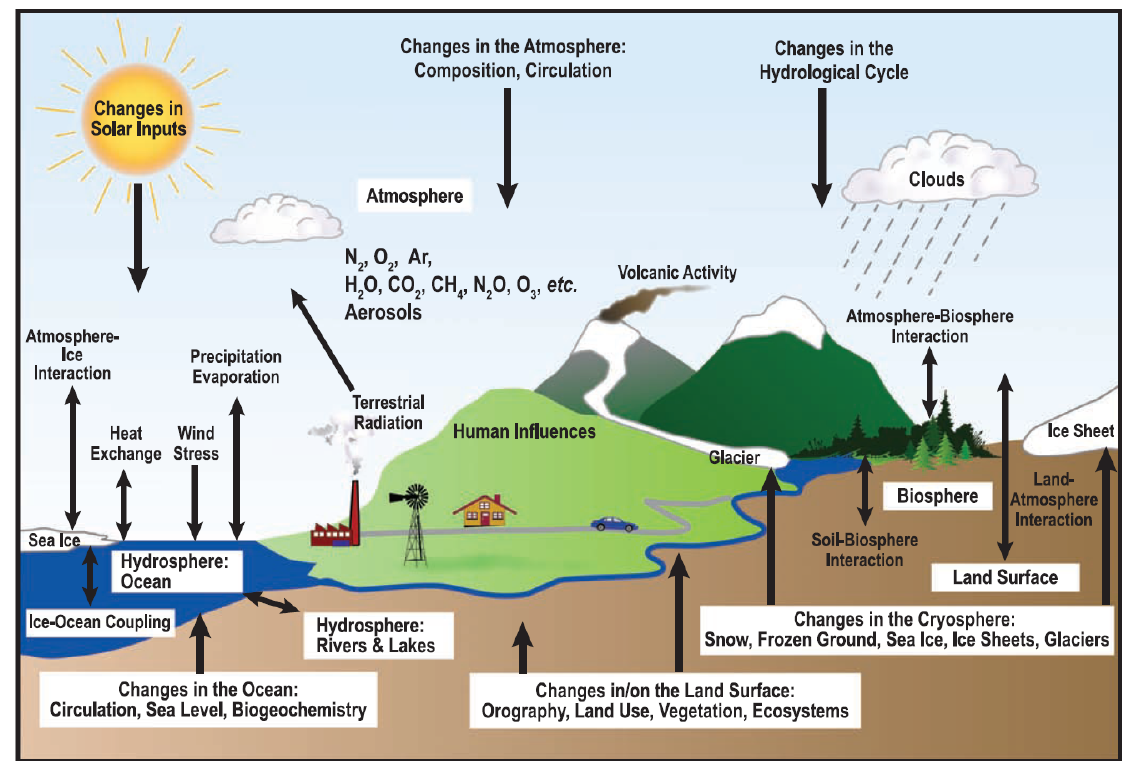}
\caption{\label{Fig3_1} 
{\bf Schematic view of the most important components and associated processes of the climate system on a global scale.} Figure is from the IPCC’s Fourth
Assessment Report (AR4) \cite{solomon_climate_2007}. copyright IPCC 2007. Reproduced with permission.}
\end{centering}
\end{figure}

The Earth's Climate System is a highly complex and interactive system 
as defined by the Global Atmospheric
Research Programme (GARP) of the World Meteorological Organization (WMO) in 1975, as being composed of five major components:  the atmosphere, the hydrosphere, the cryosphere, the land surface and the biosphere. And in 1992, the United Nations’ Framework Convention on Climate Change (FCCC) defined the climate system as `the totality of the atmosphere, hydrosphere, biosphere and geosphere and their interactions’. Fig. \ref{Fig3_1} shows a schematic representation of the most important components of the climate system and  their potential changes.  
The \textit{atmosphere} is the gaseous envelope surrounding the solid
planet and provides oxygen to most animal life at the Earth's surface. It is the most unstable and rapidly changing component. 
Currently, the atmosphere is made up primarily of nitrogen (78.1\%) and oxygen (20.9\%). There are also a number of trace gases that cause of great concern for the future of the planet, including carbon dioxide ($\text{CO}_2$), methane ($\text{CH}_4$) and  ozone ($\text{O}_3$). They are referred to as greenhouse gases (GHGs) and, along with water vapor, provide the Earth
with the greenhouse effect. Their effect is to keep our Earth warm, but not too warm.
The \textit{hydrosphere} is the water on a planet, including oceans, seas, rivers, lakes and underground water. The \textit{cryosphere} is the portion of the Earth's surface where water is in solid form, consisting of land ice  (including
ice shelves and glaciers), snow and sea ice.  It impacts the climate system greatly through its high albedo. Variations in the volume of the ice sheets are regarded as one of the  major factors for sea level rise.
\textit{Land Surface}, i.e., the ``solid'' Earth, creates the distribution of continents, ocean basins, mountain ranges, etc. It influences the transformation
of short-wave to long-wave radiation, reflectivity  of the Earth's surface, reservoir of dust, transfer of momentum and
energy of the land surface. The \textit{biosphere} is the  total sum  of all living things
on the planet, including the organic cover of the land masses (vegetation, soil) and marine organisms. It plays a vital role in exchanging of carbon between the different reservoirs, such as the concentration of $\text{CO}_2$ in the atmosphere, as well as the balances of other gases \cite{solomon_climate_2007}. There are  numerous interactions between the components, by exchanging mass, heat and momentum. For instance, the ocean-atmosphere interaction is a strongly-coupled system exchanging water vapour and heat through evaporation, among others. Each climate system component operates on a rather broad of characteristic temporal
and spatial scales.  Note that the climate system itself is often considered as part of the broader Earth System \cite{steffen_emergence_2020}.

A fundamental factor in  climate science analyses is the climate data, which provides information on changes in the Earth's ecosystem and supplies decision makers with a reliable knowledge base on the climate crisis and its impact. In general, there are two data collection methods to gather quantitative and qualitative climate data (also briefly discussed in Section \ref{intro:1} ):
observations (instrumental, proxies and reanalysis)  and climate models simulations. \textit{Observational} data of the climate system are based on direct measurements and remote sensing from satellites, radar and other platforms.
The first meteorological stations were established in Europe and North America.
Global-scale observations in the instrumental era began in the mid-19th century for temperature and other climatic variables. The increase and quality of the network of observations and the technology supporting the collection and storage of
data have rapidly evolved form 1950 onwards.
Paleoclimate reconstructions extend some \textit{proxy} records back hundreds to millions of years,
including coral  records \cite{boiseau_climatic_1999,karamperidou_response_2015}, tree rings records \cite{esper_low-frequency_2002} for the last few millennia, as well as
stalagmite $\delta^{18} \mathrm{O}$ time series \cite{cheng_asian_2016}, Central-Mediterranean sediment \cite{taricco_two_2009} and ice-core records \cite{noauthor_high-resolution_2004}. 

Climate studies require data of consistent spatial resolution and accuracy 
over long time intervals. To
satisfy this goal, some numerical weather prediction centers have started to produce so-called \textit{reanalyses} that are obtained by subtle data assimilation techniques which efficiently combine observational and numerical data.
For example, the
European Centre for Medium-range Weather Forecasts
(ECMWF) \cite{dee2011era} provides the following reanalysis datasets: ERA5, ERA-Interim, CERA-SAT, CERA-20C, ERA-20CM, ERA-40, ERA-20C and so on\footnote{see  \url{https://www.ecmwf.int/en/forecasts/datasets/browse-reanalysis-datasets}}. The NCEP-NCAR
Reanalysis I/II, 20th Century Reanalysis and North American Regional Reanalysis (NARR)\footnote{see  \url{https://psl.noaa.gov/data/gridded/reanalysis/}} are produced in collaboration by the U.S. National
Centers for Environmental Prediction (NCEP) and
the National Center for Atmospheric Research (NCAR) \cite{kalnay_ncepncar_1996}.
The JRA-25 and JRA-55 reanalysis\footnote{see  \url{https://climatedataguide.ucar.edu/climate-data/jra-55/}} are
produced by the Japan Meteorological Agency (JMA) \cite{onogi_jra-25_2007,kobayashi_jra-55_2015}.
Note that most climatic variables or fields in these reanalysis data 
agree very well with observed data,
such as the geopotential fields over the continents of the Northern Hemisphere. However, 
substantial differences persist in some fields over the
Southern Hemisphere or those that are poorly observed \cite{dellaquila_hayashi_2005,kharin_intercomparison_2005,dellaquila_southern_2007,marques_comparative_2010,kim_examination_2013}. More details are summarized in Ref. \cite{ghil_physics_2020}.

\textit{Climate models} are considered as the primary tools available for investigating the
response of the climate system to various forcings, for making climate
predictions on seasonal to decadal time scales and for making projections of future climate over the coming century and beyond.  Climate models could be defined as a mathematical representation of the climate system based on physical, chemical and biological  principles. 
The models used in climate research range from simple energy balance models to complex Earth System Models (ESMs), which can be slow and costly to use, even on 
high-performance computers, and the results can only be approximations \cite{stocker2013climate}. Here, we provide a brief overview of the basic types of climate
models: (i) \textit{Energy balance models}  propose a highly simplified version of the dynamic of the climate system. They are zero- or one-dimensional models describing 
the surface  temperature as a function of the energy balance
of the Earth, where the number of dimensions, from zero to three,
refers to the number of independent space variables used
in  the model domain. For instance, the global mean surface temperature $\bar{T_s}$ can be expressed as: 
\begin{equation}
\begin{aligned}
c \frac{\mathrm{d} \bar{T_s}}{\mathrm{d} t} &=R_{\mathrm{i}}-R_{\mathrm{o}} \\
R_{\mathrm{i}} &=\mu Q_{0}\{1-\alpha(\bar{T_s})\} \\
R_{\mathrm{o}} &=\sigma m(\bar{T_s})(\bar{T_s})^{4}.\\
\end{aligned}
\end{equation}
Here $R_{\mathrm{i}}$ and $R_{\mathrm{o}}$ are the absorbed shortwave radiation and outgoing longwave radiation. $C$ is the heat capacity of Earth system, in units of $\mathrm{J}\cdot \mathrm{m}^{-2} \cdot \mathrm{K}^{-1}$, $\alpha$ is the planetary albedo, $m$ is the transmissivity of the atmosphere, a number less than 1 that represents the greenhouse effect of the Earth's atmosphere, $\sigma \approx 5.67 \times 10^{-8} ~\mathrm{W}\cdot \mathrm{m}^{-2} \cdot \mathrm{K}^{-4}$ is the Stefan–Boltzmann constant, $\mu$ is an insolation parameter
equal to unity for present-day conditions.
(ii)\textit{ Atmosphere–Ocean General Circulation Models} (AOGCMs) are the most comprehensive
standard climate models that are assessed in the IPCC Fourth Assessment Report (AR). They are used to understand the dynamics of the physical components of the climate system (such as, atmosphere, ocean, land and sea ice), and for making projections based on future GHGs and aerosol forcing.
(iii) \textit{Earth System Models} (ESMs) are the current state-of-the-art models, and they expand on AOGCMs by  including representation of various biogeochemical cycles
such as those involved in the carbon cycle, the sulphur cycle, or ozone. These models provide the most comprehensive tools
available for simulating the past and future response of the climate system
to external forcing, in which biogeochemical feedbacks play an important role.
(iv) \textit{Regional Climate Models}
RCMs are limited-area models with representations of climate processes comparable to those in the atmospheric and land surface components of AOGCMs, though typically run without interactive ocean and
sea ice. RCMs are often used to dynamically `downscale’ global model
simulations for some particular geographical region to provide more
detailed information. In addition, based on the GCMs, the \textit{Intermediate Complexity} models were developed to investigate the Earth systems on long timescales or at a reduced computational cost \cite{flato_earth_2011}. A special application is to achieve reliable predictions of extreme events on a sub-seasonal scale.

Taken together, climate models provide a comprehensive view of the variability and
long-term changes in the climate systems. However, assessing and forecasting  climate extreme events, such as \el~events and extreme rainfall,  still poses challenges. The difficulties are mainly due to the intrinsically rare and  disruptive natural conditions. They are strongly
influenced by weather patterns, modes of variability, thermodynamic processes and various feedbacks. In the following, we will provide an in-depth and detailed description of how the	statistical physics methodologies introduced in Section.~\ref{sec:methodology} can be applied to 
investigate and (or) predict various climate phenomena, such as \el–Southern Oscillation (ENSO), Indian Summer Monsoon (ISM), Extreme Rainfall, Atmospheric Circulation and Atlantic Meridional Overturning Circulation (AMOC).

\subsubsection{\el–Southern Oscillation}
One of the most important climate phenomena on year-to-year time scales is the \el ~Southern Oscillation (\textit{ENSO}). As mentioned in section \ref{subsec:Tipping}, ENSO can be regarded as a tipping element of the Earth system. \el~ is the warm phase of the ENSO and is associated with a band of warm ocean water that develops in the central and east-central equatorial Pacific, sometimes called \el~ basin \cite{yamasaki_climate_2008}. The cool phase of ENSO is called \lanina, with SSTs in the eastern Pacific below average. ENSO significantly impacts Earth's ecosystems and human societies, by influencing  temperatures and precipitation around the globe,  including the Americas, India and surrounding tropical continents \cite{ropelewski_north_1986,power_inter-decadal_1999,kumar_weakening_1999}. These remote effects are known as \textit{teleconnections} \cite{gershunov_interdecadal_1998}.

\begin{figure}[]
\begin{centering}
\includegraphics[width=0.85\linewidth]{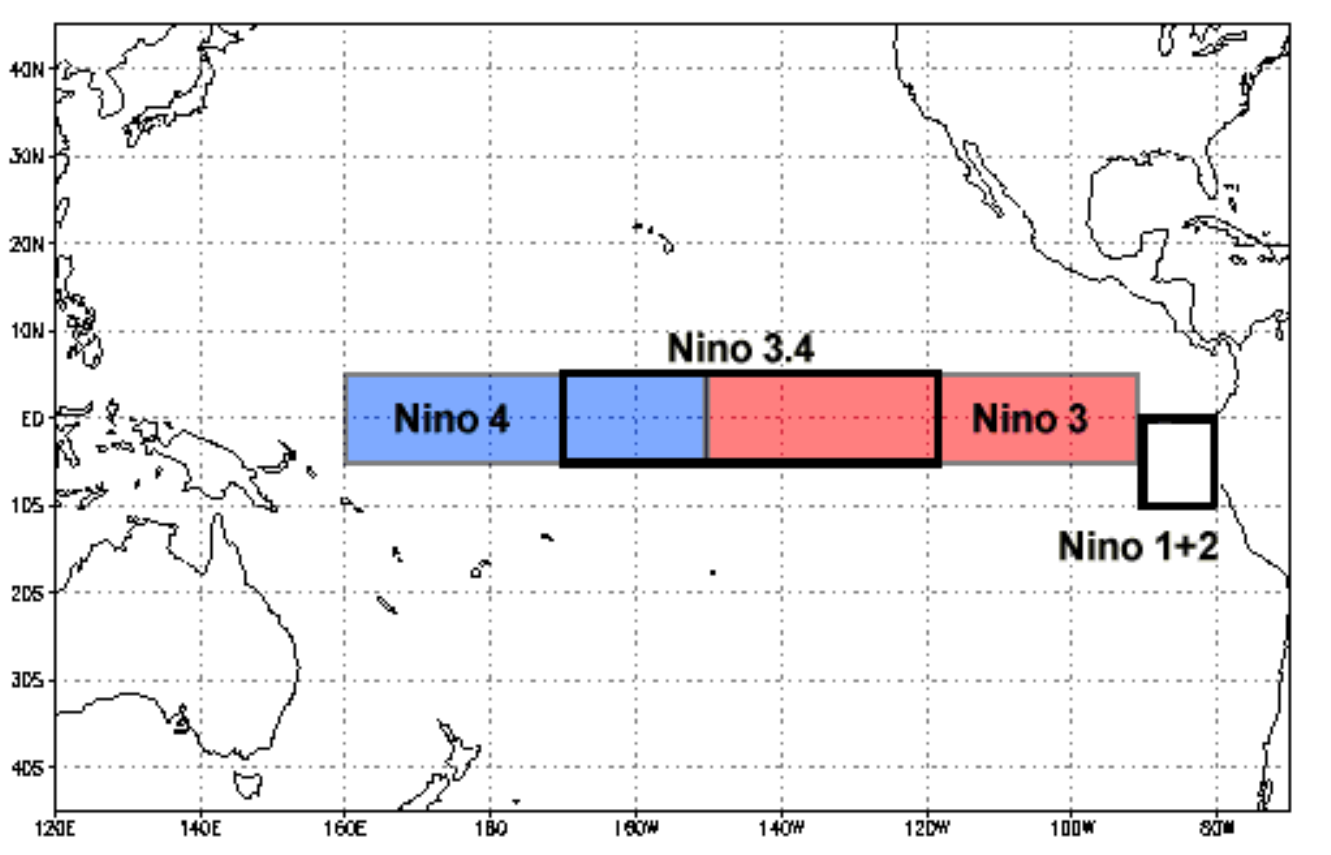}
\caption{\label{Fig3_2} 
{\bf Ni\~no Regions.} The Ni\~no 3.4 region (5 \degree S to 5 \degree N; 170 \degree W to 120 \degree W); The Ni\~no 4 region (5 \degree S to 5 \degree N; 160 \degree E to 150 \degree W); 
The Ni\~no 3 region (5 \degree S to 5 \degree N; 150 \degree W to 90 \degree W); The Ni\~no 1$+$2 region (10 \degree S  to  0 \degree; 90 \degree W to 80 \degree W).
\textit{Source}: figure is from NOAA.}
\end{centering}
\end{figure}

\textit{Definition of \el}: In this review, we use the Oceanic Ni\~no Index (ONI) to define and distinguish the \el ~and \lanina ~events. The ONI is one of the primary indices used to monitor the ONI. It is calculated by  3-month running mean SST anomalies (based on centered 30-year base periods updated every 5 years) in an area of the east-central equatorial Pacific Ocean, which is called the Ni\~no 3.4 region (5 \degree S to 5 \degree N; 170 \degree W to 120 \degree W), see the Ni\~no regions presented in Fig. \ref{Fig3_2}. If the ONI $>$  $+0.5$ \degree C for at least five consecutive months, then the event is defined as an \el; whereas, when ONI is $<$ $-0.5$ \degree C for at least five consecutive months, the corresponding event is regarded as a \lanina.  The ONI is use for  the operational definition of the ENSO phase by NOAA\footnote{The values of ONI are here: \url{https://origin.cpc.ncep.noaa.gov/products/analysis_monitoring/ensostuff/ONI_v5.php}}. There are also some other indices  to measure the \el~ events, such as, Ni\~no 1$+$2, 3, 3.4, 4 and Trans-Ni\~no indexes\footnote{For more details: \url{https://climatedataguide.ucar.edu/climate-data/nino-sst-indices-nino-12-3-34-4-oni-and-tni}}. Besides the SST, ENSO also influences other climate variables like atmosperic pressure gradients, upper ocean currents and the thermocline depth \cite{j_david_neelin_climate_2011}.

\textit{ENSO transition mechanism in brief:}
In the following, we will describe the physical mechanism of ENSO.
Bjerknes first postulated that the ocean–atmosphere interaction is
essential for ENSO and is now referred to as the Bjerknes feedback \cite{bjerknes1969atmospheric}.
This positive feedback mechanism is understood to be a prominent mechanism necessary for the development of ENSO. Consider
an initial  SST anomaly in the central/eastern tropical Pacific. This anomaly reduces the east-west SST gradient and hence weakens the Walker circulation \cite{gill_simple_1980}, resulting in a westerly wind anomaly. The westerly wind anomaly, in turn, drives the ocean circulation change (easterly winds shoal the equatorial
thermocline and induce upwelling in the eastern
Pacific, keeping the east cool) that further reinforces the SST anomaly. As a result of the positive feedback, the tropical Pacific reaches a warm state, i.e., \el.  Tziperman and Yu pointed out that the westerly wind bursts, while partially stochastic, seem an inherent part of the large-scale deterministic ENSO dynamics \cite{tziperman_quantifying_2007}.

Besides the positive
Bjerknes  feedback, the oscillatory nature of ENSO requires a negative feedback to turn the
coupled ocean-atmosphere system from a warm (\el) phase to a cold (\lanina) phase. Different negative feedbacks have been proposed since the
1980s associated with the delayed oscillator, the western Pacific oscillator, the  recharge-discharge
oscillator, and the advective-reflective oscillator. 
The \textit{delayed oscillator} is a mechanism for the oscillatory nature of ENSO, which was proposed by McCreary  based on the Rossby wave reflection at the ocean western boundary \cite{mccreary_model_1983}. Suarez and Schopf introduced the delayed oscillator explaining the oscillatory feature of ENSO to emphasize the delayed effects of oceanic wave reflection at the ocean western boundary \cite{suarez_delayed_1988}. 
Cane and Zebiak developed the  first  coupled ocean–atmosphere model, and used it to predict ENSO \cite{zebiak1987model,cane_experimental_1986}. 
The conceptual delayed oscillator model is represented as follows:
\begin{equation}
\frac{d T}{d t}=A T-B T(t-\eta)-\varepsilon T^{3},
\label{EQ_DOM}
\end{equation}
where $T$ represents the SST anomaly in the equatorial eastern Pacific,  $A, B, \eta$ and $\varepsilon$ are constants. The first term on the right hand side of Eq. (\ref{EQ_DOM}) stands for the Bjerknes positive feedback between the ocean and the atmosphere. The second term represents the delayed negative feedback of wave reflection at the ocean western boundary. The last term is a cubic damping term. It has been found that the delayed oscillator overlooks the role of the ocean–atmosphere interaction in the western Pacific and also assumes that wave reflection at the ocean eastern boundary is not important \cite{wang_review_2018}. 
 
Since the delayed oscillator does not consider the western-Pacific anomaly pattern, Weisberg and Wang \cite{weisberg_western_1997} developed  a \textit{western-Pacific oscillator} model for ENSO. This model stresses that the equatorial wind anomalies in the far western Pacific play an important role in the evolution of ENSO. The western-Pacific oscillator model is represented as:
\begin{equation}
\frac{d T}{d t}=a \tau_{1}+b_{2} \tau_{2}(t-\delta)-\varepsilon T^{3}, 
\end{equation}
\begin{equation}
\frac{d h}{d t}=-c \tau_{1}(t-\lambda)-R_{h} h,
\end{equation}
\begin{equation}
\frac{d \tau_{1}}{d t}=d T-R_{\tau 1} \tau_{1},
\end{equation}
\begin{equation}
\frac{d \tau_{1}}{d t}=d T-R_{\tau 1} \tau_{1},
\end{equation}
\begin{equation}
\frac{d \tau_{2}}{d t}=e h-R_{\tau 2} \tau_{2}
\end{equation}
where $T$ is the SST anomaly in the equatorial eastern Pacific, $h$ is the  thermocline-depth  anomaly in the off-equatorial western Pacific, and $\tau_{1}$ and $\tau_{2}$ are the equatorial zonal wind-stress anomalies in the central Pacific and the western Pacific, respectively. All model parameters are constants.

It has been found that the growth and decrease in sea level over the western Pacific Ocean are related to ENSO \cite{wyrtki_ninodynamic_1975}. Based on these ideas Jin \cite{jin_equatorial_1997,jin_equatorial_1997-1} developed a recharge-discharge oscillator model for ENSO that is represented by:
\begin{equation}
\frac{d T}{d t}=C T+D h-\varepsilon T^{3},
\label{EQ_RDO1}
\end{equation}
\begin{equation}
\frac{d h}{d t}=-E T-R_{h} h.
\label{EQ_RDO2}
\end{equation}
where $T$ is the SST anomaly in the equatorial eastern Pacific and $h$ is the thermocline-depth anomaly in the equatorial western Pacific. The model parameters $C, D, \varepsilon, E$ and $R_{h}$ are constants. In particular, Eqs. (\ref{EQ_RDO1}) and (\ref{EQ_RDO2}) represent the discharge and recharge of tropical Pacific Ocean heat content.

Picaut et al. \cite{picaut_advective-reflective_1997} proposed a
conceptual model of the \textit{advective-reflective oscillator} for ENSO. They found that the positive feedback results from ocean zonal currents that advect the  western-Pacific warm pool (WPWP) toward the east. There are three negative feedbacks that tend to push the WPWP back to the western Pacific, include: (i) anomalous zonal current associated with wave reflection at the ocean western boundary, (ii) anomalous zonal current associated with wave reflection at the ocean eastern boundary and (iii) mean zonal current converging at the WPWP's eastern edge. Unlike the other three oscillator models, the advective-reflective oscillator model  does not have a set of simple and heuristic equations. Instead, they used a linear wind-forced ocean numerical model 
forced by wind anomalies, which were associated
with the zonal current of the first baroclinic Kelvin
and first meridional Rossby waves \cite{dijkstra_nonlinear_2005}.

Moreover, based on the dynamics and thermodynamics of Zebiak and Cane's coupled  model, as well as all previous ENSO oscillator models,  Wang \cite{wang_unified_2001} developed a \textit{unified ENSO oscillator} ocean-atmosphere model with the hypothesis that ENSO may be a multi-mechanism phenomenon and their relative importance may depend on time. The unified ENSO oscillator model is formulated as:
\begin{equation}
\frac{d T}{d t}=a \tau_{1}-b_{1} \tau_{1}(t-\eta)+b_{2} \tau_{2}(t-\delta)
-b_{3} \tau_{1}(t-\mu)-\varepsilon T^{3},
\label{EQ_UNO1}
\end{equation}
\begin{equation}
\frac{d h}{d t}=-c \tau_{1}(t-\lambda)-R_{h} h,
\label{EQ_UNO3}
\end{equation}
\begin{equation}
\frac{d \tau_{1}}{d t}=d T-R_{\tau 1} \tau_{1},
\label{EQ_UNO4}
\end{equation}
\begin{equation}
\frac{d \tau_{2}}{d t}=e h-R_{\tau 2} \tau_{2},
\label{EQ_UNO5}
\end{equation}
where $T, h, \tau_{1}$ and $\tau_{2}$ are variables that stand for the SST anomalies in the equatorial eastern Pacific, the thermocline-depth anomalies in the off-equatorial western Pacific, the zonal wind-stress anomalies in the equatorial central Pacific and the zonal wind-stress anomalies in the equatorial western Pacific, respectively. For a given set of parameters, Eqs. (\ref{EQ_UNO1})--(\ref{EQ_UNO5}) can oscillate on interannual time scales.

\textit{\el~ forecasting:}

\begin{figure}[]
\begin{centering}
\includegraphics[width=0.85\linewidth]{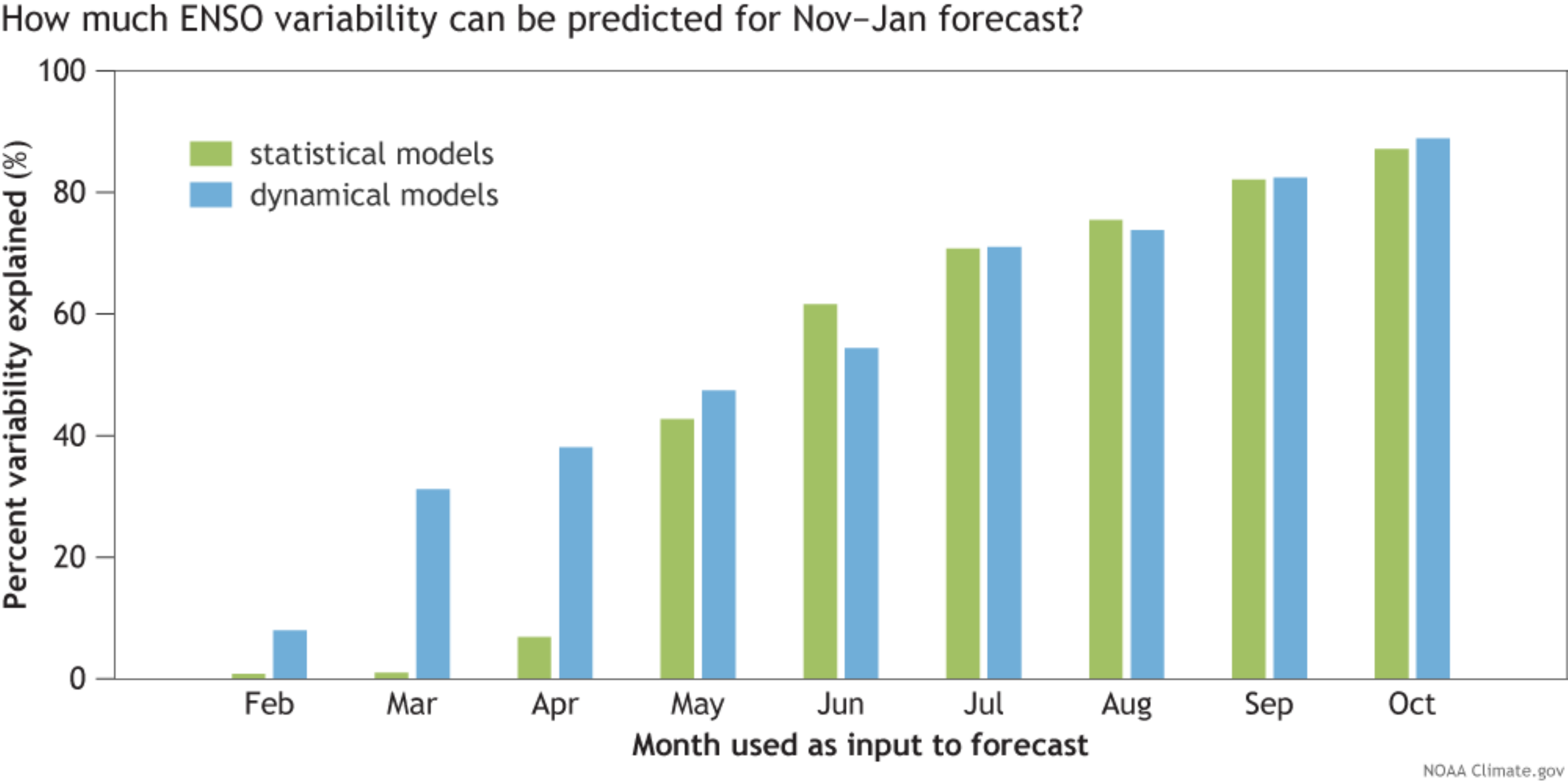}
\caption{\label{Fig3_3} 
{\bf Spring Predictability
Barrier.} The \el~ forecasting skill of models based on February-October observations to predict the November-January average value in the \el-3.4 region (see Fig. \ref{Fig3_2}).  Results are an average correlation coefficient from 20 models between 2002-2011. \textit{Source}: Figure is from NOAA.}
\end{centering}
\end{figure}

Since the 1990s, both dynamical and statistical models have been used to model and  forecast  \el~ events, providing the society the opportunity to prepare for associated hazards such as heavy rains, floods and droughts \cite{chen_predictability_2004,yeh_nino_2009}. Although there are about 17 dynamic and 8 statistical models in CPC/IRI currently providing  ENSO forecasts routinely, high prediction skill is generally limited to about six months ahead, see the description in  Fig. \ref{Fig3_3}. The reason is the presence of the so-called ``spring predictability
barrier'' (SPB),  where errors are greatly amplified due to the coupled feedbacks in the equatorial ocean–atmosphere system \cite{goddard_current_2001,duan_spring_2013}. After the spring, the ability of the models to predict successfully becomes increasingly better.

In general, a poor skill in forecasts can be divided into two categories: imperfections in the forecast system (including, error in model parameterization of sub-grid-scale motions, the relative scarcity of points with data, errors in the measurement
of the data), and fundamental limits of predictability \cite{j_david_neelin_climate_2011}. The fundamental limits are usually  coming from the complex and chaotic nature of a system.

\begin{figure}[]
\begin{centering}
\includegraphics[width=1\linewidth]{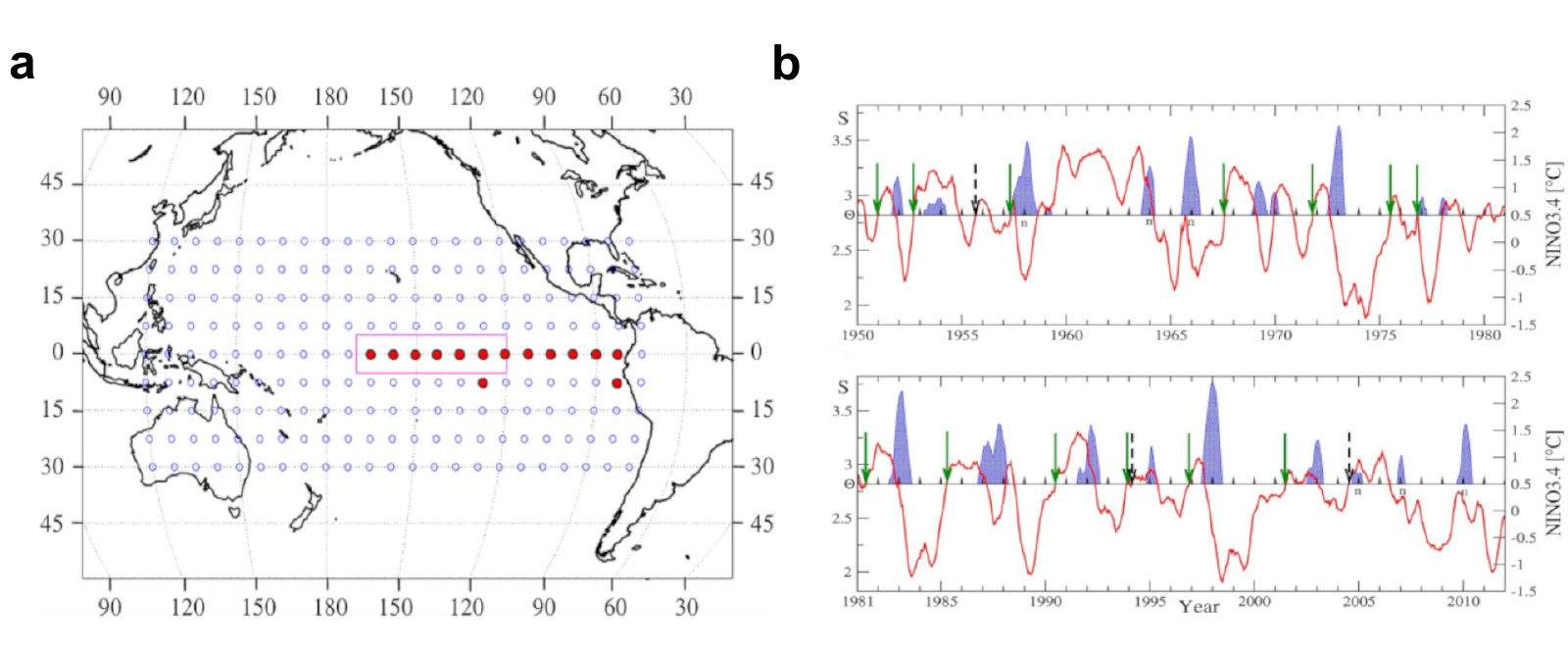}
\caption{\label{Fig3_4} 
{\bf Climate Network and \el~ forecasting algorithm}.  a. The nodes in CN. Each node inside the \el~ basin (solid red symbols) is linked to each node outside the basin (open blue symbols). The red
rectangle denotes the Ni\~no 3.4 region.  b. The red curve is the average link weight $S(t)$ of the CN,
the horizontal line indicates the decision threshold $\Theta = 2.82$, and the blue areas
show the \el~ events. When $S(t)$ crosses the threshold from below, an alarm is given that an \el~
event will start in the following calendar year. \textit{Source}: Reprinted figure from Ref. \cite{ludescher_improved_2013-1}.}
\end{centering}
\end{figure}

To overcome the limits of the SPB in \el~ forecasting, several CN-based approached were developed. For example, by analyzing the CN, Ludescher \textit{et al.} found that a large-scale cooperative mode exists one calendar year before the warming event, linking the \el~ region and the rest of the ocean \cite{ludescher_improved_2013-1}. This mode is 
probably  due to the emergence of \el~ acting
as an autonomous component in the CN \cite{gozolchiani_emergence_2011}. 
All nodes  used in the CN are shown in Fig. \ref{Fig3_4}a, 
where the 14 grid points in the \el~ basin \cite{gozolchiani_emergence_2011} are in red and outside this domain there are 193 grid point in blue. The data is  the daily surface air temperature (SAT) from NCEP/NCAR reanalysis I.
As described  in Section \ref{sec:methodology}, the first step is to remove the seasonal trend  by minus climatological average for each calendar day. And then, the links between the \el~ basin and outside (see Fig. \ref{Fig3_4}a) are considered. The strength $W_{i,j} (t)$ of each link is computed according to Eq. (\ref{EQ13}). The mean strength $S(t)$ of the dynamical
links in the CN is obtained by simply averaging over
all individual link strengths. Fig. \ref{Fig3_4}b, shows the time evolution of $S(t)$ for the learning (top)  and forecasting (bottom) phases. They found that before an \el~ event,  $S(t)$  tends to increase. They marked
an alarm when $S(t)$ crosses a threshold $\Theta = 2.82$ (see Fig. \ref{Fig3_4}b). The prediction accuracy, through Receiver Operating Characteristic (ROC)-type analysis (explained later), is much better than the 12-months forecasts based on climate models.
Based on these findings regarding the temporal evolution of the CN, Ludescher
\textit{et al.} successfully predicted the onset
of the 2014–2016 \el~ event \cite{ludescher_very_2014}.

Motivated by this work, Meng \textit{et al.} developed  a multidisciplinary
approach combining climate, network, and percolation
theory to predict the onset of \el~\cite{meng_percolation_2017}. Their method can forecast \el~ events 1 year-ahead,
with a high prediction accuracy of 70\%, and a low false
alarm of only 4\%. Different from \cite{ludescher_improved_2013-1}, they consider the global CN with initially $726$ isolated nodes.  Links are added one by one
according to the link strength, i.e., first adding the link with
the highest weight, and continue selecting links ordered by decreasing weight. During the evolution of the CN, they found that the structure of the network changes violently approximately one year ahead of \el~events, and thus suggest to use the largest size change of the largest
cluster (percolation cluster) $\Delta$ [see Eq. (\ref{EQ46})] as a percolation-based precursor, to forecast \el~events. In addition, based on finite size
scaling analysis, they found that the percolation process is
discontinuous.

\begin{figure}[]
\begin{centering}
\includegraphics[width=1\linewidth]{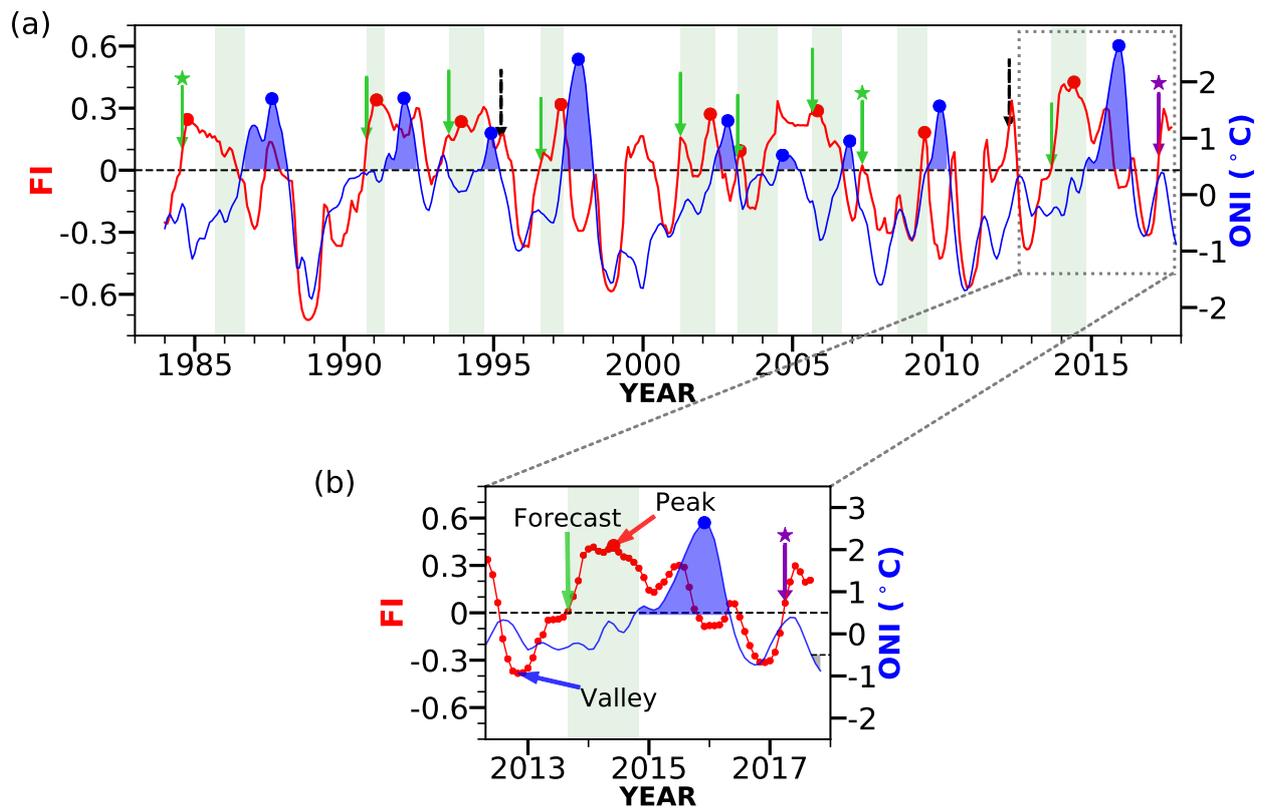}
\caption{\label{Fig3_5} 
{\bf The \el~forecasting algorithm}.  a. The red curve is the  forecasting index defined in Eq. (\ref{EQ108}), the blue line stands for the ONI.
The horizontal black dashed
line indicates $FI =0$, and the blue shades under the ONI curve
indicate the \el~ periods. True positive forecasts are marked by the green arrows, while false alarms are marked in black dashed
arrows. The purple arrow is the alarm for 2018-2019 \el~event. b. A detailed view of $FI(t)$ and the ONI since May 2012. \textit{Source}: Reprinted figure from Ref. \cite{meng_forecasting_2018}.}
\end{centering}
\end{figure}

To forecast both the onset and magnitude  of \el~events, another sophisticated network-based approach was developed by Meng \textit{et al.} \cite{meng_forecasting_2018}. They proposed a new
forecasting index based on CN links representing the similarity of low frequency
temporal temperature anomaly variations between different sites in the Ni\~no 3.4 region. They found that
significant upward trends in the index forecast the onset of \el~approximately 1 year ahead, and
the highest peak since the end of last \el~forecasts the magnitude of the following
event. Their index was also tested on several datasets, including ERA-Interim, 
NCEP Reanalysis I, PCMDI-AMIP 1.1.3 and ERSST.v5. They defined the degree of coherence/disorder of
the Ni\~no 3.4 region as,
\begin{equation}
\label{EQ107}
C(t)=\frac{2}{N(N-1)}\sum_{i=1}^{N-1}\sum_{j=i+1}^{N}{C^{(t)}_{i,j}(\theta)},
\end{equation} 
where $N$ is the number of nodes in the Ni{\~n}o 3.4 region and $C^{(t)}_{i,j}(\theta)$ stands for the max value of the time-delayed cross-correlation function, see Eqs. (\ref{EQ10}) and (\ref{EQ11}). The forecasting index (FI) was proposed as a function of time (months),
\begin{equation}
\label{EQ108}
\rm{FI}(t)=\frac{1}{m+1}\sum_{a=0}^{m}\ln C(t-a)-\ln C(t) 
\end{equation}
where $a=0$ indicates that the average of $ln(C)$ includes the current month, while $m$ is the total number of months preceding $t$ since Jan. 1981. We show the time evolution of the FI and ONI in Fig.~\ref{Fig3_5} by using the ERA-Interim dataset.
Based on the temporal evolution of the FI, they successfully predicted the onset
of the 2018-2019 \el~ event, see the purple star above the purple arrow in Fig.~\ref{Fig3_5}b.

Note that one can use the peak of the FI to predict the magnitude of the following \el~event. This peak value is determined from the end of the last event to the start of a new event, which means that we should wait until the new event begins to occur.
To further improve the forecasting skill, in particular, the magnitude forecasting with a long lead time. Meng \textit{et al.} \cite{meng2019complexity} developed the SysSampEn method, see more details in Sec. (\ref{SysSampEn}). It is defined in Eq. (\ref{EQ105}) and shown in Fig. \ref{Fig2_15}a. The correlation between the 
magnitude of the \el~events and the SysSampEn can  reach to $r = 0.99$ [see Fig. \ref{Fig2_15}b].  They further used this correlation to forecast the magnitude of an \el ~with a prediction horizon of 1 year and high accuracy (i.e., Root Mean Square Error $=0.23^\circ C$ for the average of the individual datasets forecasts).
In particular, for the 2018-2019 \el ~event, they forecasted  a weak \el ~with a magnitude of $1.11\pm 0.23^\circ C$, compared to the observed value $0.9^\circ C$.

Recently, Feng \textit{et al.}  combined  machine learning techniques and climate networks to apply to \el~ predictions, especially for the occurrence of \el~ events and the development of the NINO 3.4 index over time \cite{feng_climatelearn_2016}. In Ref. \cite{petersik_probabilistic_2020}, Petersik and Dijkstra applied  Gaussian density neural network and quantile regression neural network ensembles to predict  ENSO. Classical machine learning models for prediction usually only predict a point value, e.g., the ONI. In contrast, their models forecast directly a probability distribution for the ONI. In particular, they found that their machine learning models have a high-correlation skill ($r = 0.5$) for long lead times (12 months ahead) for an evaluation period between 1963
and 2017. In Ref. \cite{ham_deep_2019}, Ham, Kim and Luo used transfer learning to train a convolutional neural network  on historical simulations and  reanalysis, and developed the heat map analyses. They found that a statistical forecast model employing a deep-learning approach produces skillful ENSO forecasts for lead times of up to one and a half years. A review of ENSO prediction based on  modern machine learning and  artificial neural networks was reported in Ref. \cite{dijkstra_application_2019}.

\textit{ENSO remote impacts: teleconnections}

\begin{figure}[]
\begin{centering}
\includegraphics[width=1\linewidth]{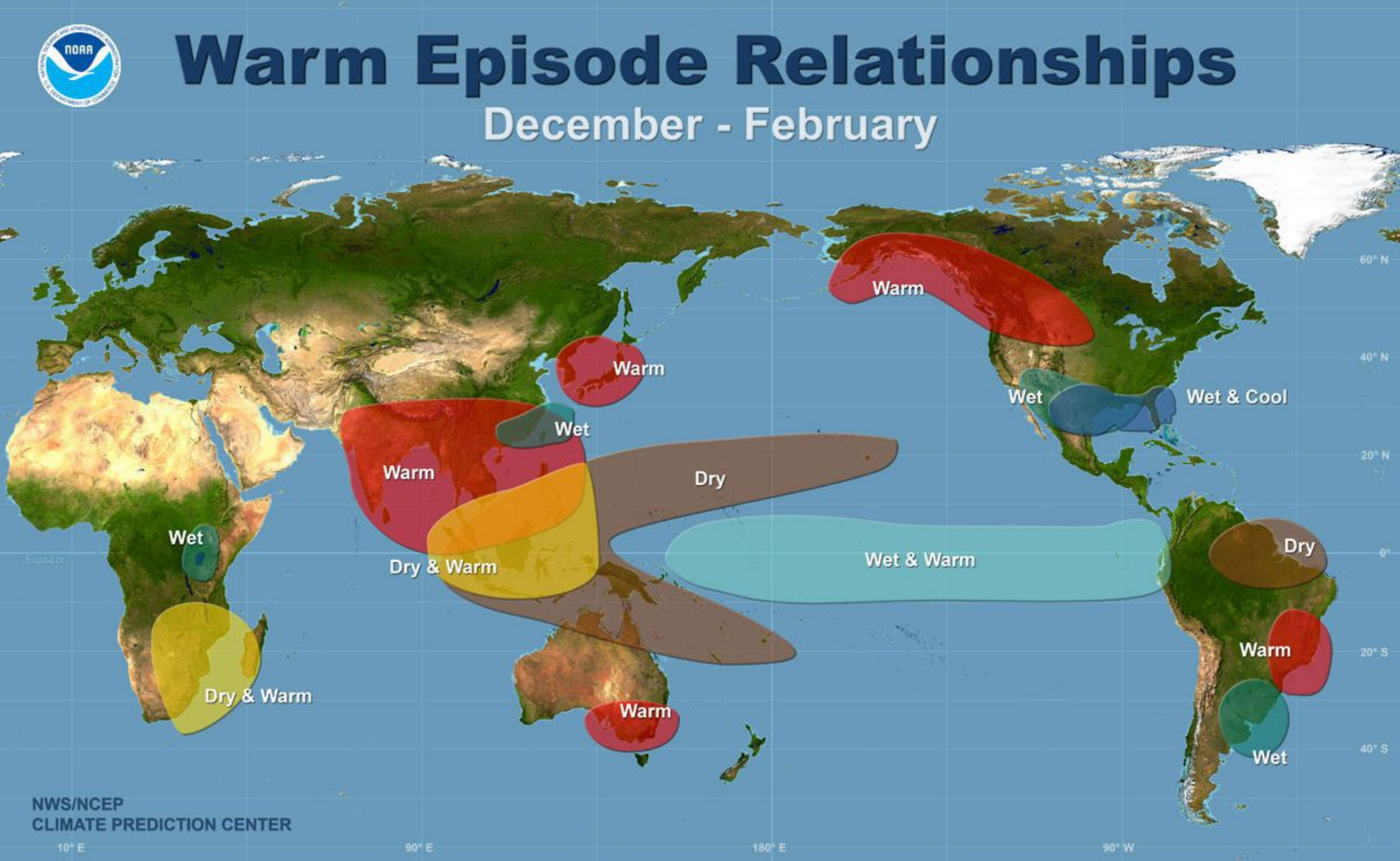}
\caption{\label{Fig3_6} 
{\bf \el~remote impacts: teleconnections, during the boreal winter months
December-January-February}. Regions with statistically reliable relation of precipitation and surface air temperature to a warm ENSO
episode. Figure is reproduced with permission from NOAA.}
\end{centering}
\end{figure}

As  introduced above, the ENSO phenomenon mainly lies in the Pacific ocean. However, its effects are
felt over a large part of the Earth. These remote effects are usually called \textit{teleconnections}, emphasizing that changing climatic conditions in one place can affect areas far from the source \cite{gershunov_interdecadal_1998,alexander_atmospheric_2002}. A teleconnection is usually indicated by the correlation between the values observed at two separate locations or regions. It is related to a pattern of variability, associated with atmospheric wave propagation, the presence of ocean currents, etc.
For example, 
the teleconnections associated with the North Atlantic Oscillation (NAO)  and
ENSO  are main examples of  long-range correlations that can
be found in the atmosphere.
Because of those global teleconnections, \el~ leads to higher precipitation in the central Pacific and dry conditions over Indonesia and Northern Australia, much drier and warmer conditions  in Mozambique, while the western USA tends to be wetter. Fig. \ref{Fig3_6} shows weather patterns statistically associated with \el~conditions during the boreal winter months
December-January-February.  The regions that have
been shown with some degree of reliability to be affected by warm ENSO phases are highlighted, for example, warm-vs.-cold and wet-vs.-dry anomalies. The
effects for cold phases of ENSO, \lanina, are in approximately the same regions, but with the opposite sign.

The basic mechanism of ENSO remote influences are produced in the atmosphere  via large-scale atmospheric waves, such as Rossby waves.
Rossby waves are fundamental for the understanding of weather systems in the atmosphere and the large-scale circulation in the ocean. They depend fundamentally upon the variation of the Coriolis parameter with latitude. In the so called $\beta$-plain approximation, the Coriolis parameter varies linearly with latitude where
\begin{equation}
\label{EQ109}
\beta=\frac{2 \Omega}{R} \cos \phi,
\end{equation}
where $\phi$ is the latitude, $\Omega$ is the angular speed of the Earth's rotation, and $R$ is the mean radius of the Earth. The wavelength of atmospheric Rossby wave is about $6000$ km.
The atmospheric wave dynamics are akin to those observed in the ocean for ENSO in that they involve a transfer of atmospheric mass by anomalous winds, which in turn affect pressure and winds.

In order to assess if a region has an ENSO connection, Fan \textit{et al.} applied the Pearson correlation CNs approach to investigate the global remote impacts of \el ~ and \lanina ~\cite{fan_network_2017_1}. They refer to the Ni\~no 3.4 region as the \el~Basin (ENB). The CNs were constructed by using only directed links from the ENB to regions outside the ENB (``in''-links) based on the global
daily SAT fields of the NCEP/NCAR reanalysis and EAR-Interim datasets.
The links' strengths are computed according to Eqs. (\ref{EQ10}), (\ref{EQ11}) and (\ref{EQ13}). They identified the value of the highest peak of the absolute value of the cross-correlation function $C^{y}_{i,j}(\tau)$ and denote the corresponding time lag as $\theta^{y}_{i,j}$.  The adjacency matrix of a CN became
\begin{equation}
A_{i,j}^{y} = (1 - \delta_{i,j})H(\theta^{y}_{i,j}),
\label{EQ110}
\end{equation}
where $H(x)$ is the Heaviside step function for which $H(x\ge 0)=1$ and $H(x< 0)=0$.  The ``in'' and ``out'' degrees of each node are defined as $I_{i}^{y} = \sum_{j} A_{j,i}^{y}$, $O_{i}^{y} = \sum_{j} A_{i,j}^{y}$ respectively, quantifying the number of links into a node or out from a node. 
The total ``in'' weights for each node outside the ENB using are defined as
\begin{equation}
\label{EQ111}
\begin{array}{l}
\mathrm{IN}\left(C_{i}^{y}\right)=\sum\limits_{j \in \mathrm{ENB}} A_{j, i}^{y} C_{j, i}^{y}(\theta) \\
\mathrm{IN}\left(W_{i}^{y}\right)=\sum\limits_{i \in \mathrm{ENB}} A_{j, i}^{y} W_{j, i}^{y}.
\end{array}
\end{equation}
The values of ${\rm IN}(C_{i}^{y})$ and ${\rm IN}(W_{i}^{y})$ reflect the impacts of the ENB.
%
Specifically, if there are no ``in'' links for a node, both the ``in'' degree and ``in'' weights are zero, indicating no impact of ENB. Based on the ONI, the time is divided into \el~, \lanina~and normal years. The ``in''-weighted degree fields for \el~and \lanina~ are calculated by taking the average of the same type of events 
\begin{equation}
\label{EQ112}
\begin{array}{l}
\mathrm{IN}\left(C_{i}\right)=\sum\limits_{y \in \mathrm{EY} (\mathrm{LY})} \mathrm{IN}\left(C_{i}^{y}\right)/S \\
\mathrm{IN}\left(W_{i}\right)=\sum\limits_{y \in \mathrm{EY} (\mathrm{LY})} \mathrm{IN}\left(W_{i}^{y}\right)/S,
\end{array}
\end{equation}
where $S={\sum\limits_{y\in \mathrm{EY} (\mathrm{LY})}I_{i}^{y}}$, and ``EY'' and ``LY'' refer to the years in which \el ~and \lanina ~start. The number of nodes with zero in-degree is defined as $N^y$ and the average in-weights
per node is 
\begin{equation}
\label{EQ113}
C^{y}=\sum\limits_{i\not\in \mathrm{ENB}}\sum\limits_{j\in \mathrm{ENB}} A_{j,i}^{y}\mid C_{j,i}^{y}(\theta)\mid /N^{y}.
\end{equation}

\begin{figure}[]
\begin{centering}
\includegraphics[width=1\linewidth]{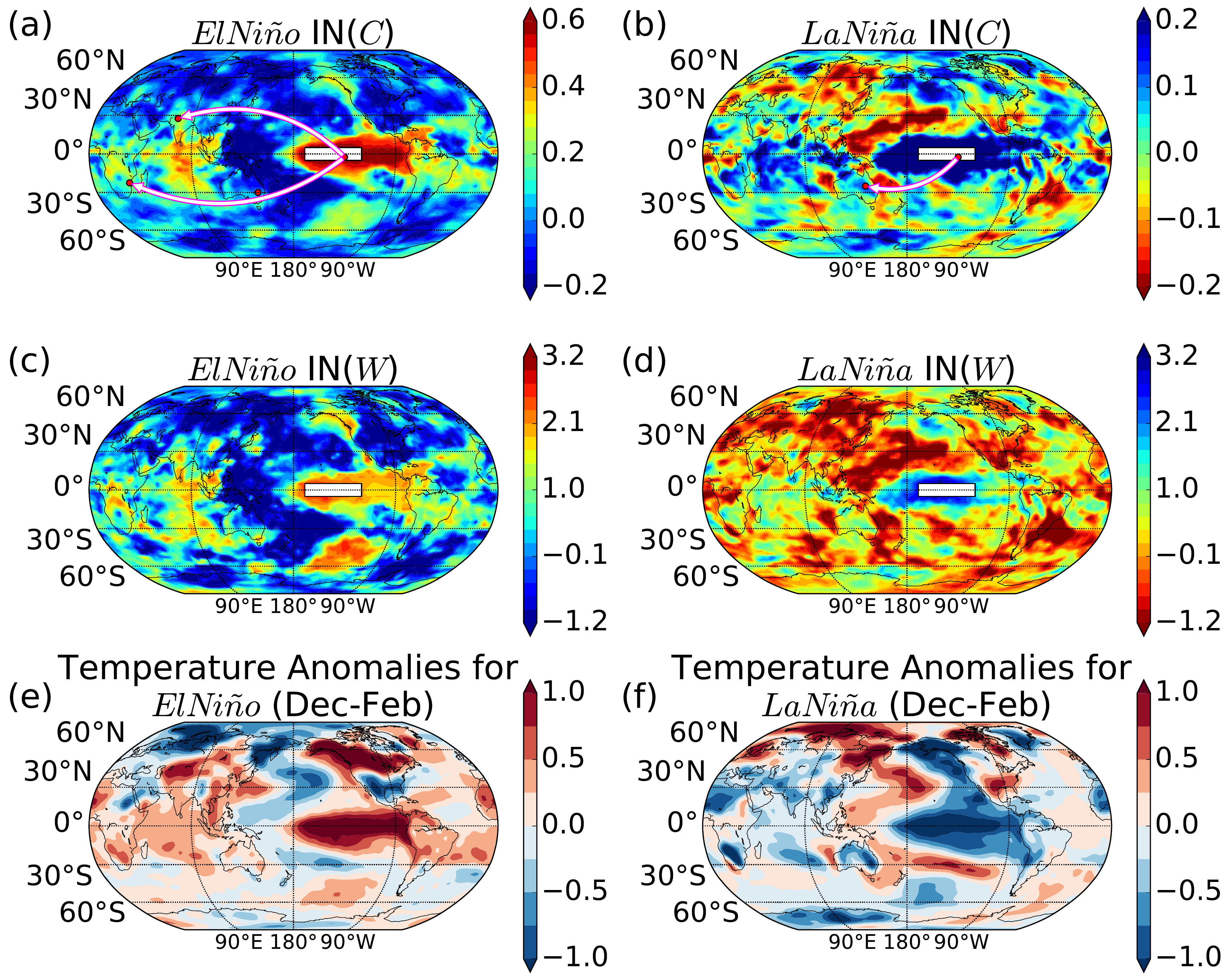}
\caption{ \label{Fig3_7} 
{\bf \el~remote impacts: Climate Networks}. (a), (c) ``In''-weight maps (using $C$ and $W$) for \el ~events. (b),(d), ``In''-weight maps (using $C$ and $W$) for \lanina ~events. (e), (f)  Mean winter (Dec-Feb) temperature anomalies during \el ~and \lanina. \textit{Source}: Reprinted figure from Ref. \cite{fan_network_2017_1}.}
\end{centering}
\end{figure}

\begin{figure}[]
\begin{centering}
\includegraphics[width=1\linewidth]{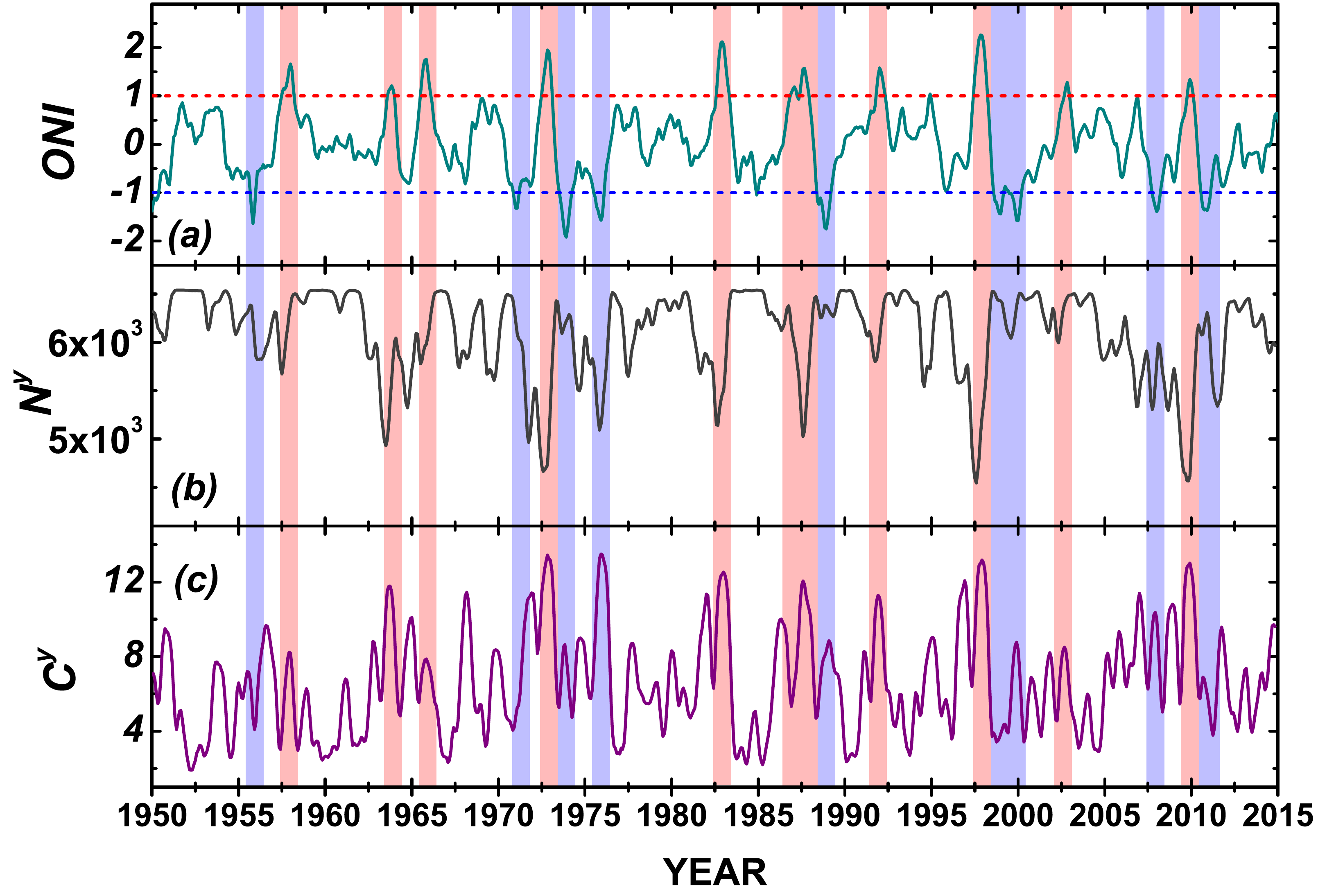}
\caption{\label{Fig3_8} 
{\bf Strongly localized impacts
of ENSO}. (a) The ONI as a function of time. (b) The evolution of the number
of nodes that have in-links with time. (c) The evolution of the average in-weights
per node with time. \el ~and \lanina~periods are respectively emphasized by the red and blue shading. \textit{Source}: Reprinted figure from Ref. \cite{fan_network_2017_1}.}
\end{centering}
\end{figure}

Fig.~\ref{Fig3_7} (a-d) shows that, the affected regions  by \el~ and \lanina,  are characterized by relatively high in-weights (using $C$ and $W$). For comparison, mean winter (Dec-Feb) temperature anomalies during \el ~and \lanina~years are shown in Fig.~\ref{Fig3_7} (e-f).
A quantitative analysis of the area  that is
affected/unaffected during  \el~ and \lanina~years is shown in Fig.~\ref{Fig3_8},   where \el ~and \lanina~years are respectively emphasized by the red and blue shading. The network analysis reveals strongly (high $C^{y}$) localized (low $N^{y}$) impacts of ENSO, i.e., these stronger in-weighted
activities are found to be concentrated in very localized
areas, whereas a large fraction of the globe is not influenced by
the events (low $N^{y}$ means that fewer nodes are influenced).

\textit{Diversity of \el}

It has been  reported that there are  different types of \el~events, 
such as the canonical eastern Pacific (EP) and the Modoki central Pacific (CP) types \cite{ashok_nino_2007,kao_contrasting_2009}.
They are classified based on spatial patterns of the SST anomaly. For instance, the strongest SST anomalies associated with the EP types are located off the coast of South America. However, for CP types, the strongest SST anomalies are located near the International Date Line. In Ref. \cite{wiedermann_climate_2016}, Wiedermann \textit{et al.} proposed an index based on evolving CNs to objectively
discriminate different types of \el~ and \lanina~episodes. In particular,  they find that their network index displays a sharp peak during EP events, whereas during CP events it remains close to
the observed values during normal years. Based on the CNs analysis, Lu \textit{et al.} developed a new approach to estimate the impacts of \el~events in advance, as well as predicted the types of \el~events \cite{lu_impacts_2020}.



\subsubsection{Indian Summer Monsoon}
The Indian Summer Monsoon (ISM) is the most prominent among the  world's monsoon systems and has a decisive influence on India's agricultural output and economy. The ISM delivers more than 70\% of the country's annual rainfall. The prediction of the ISM timing (onset date and withdrawal date) as well as the ISM rainfall is a
vital issue for the Indian subcontinent and strongly affects the gross domestic product of the country, up to 22\% of which is determined by agriculture \cite{subash_statistical_2014}.
A slight deviation of the 
timing manifested as delay (or early arrival) may lead to drastic droughts (floods).
Also, swings in the amount of rainfall, even with variations
of only 10\%, can cause  severe flooding or drought,
causing damages to infrastructure and loss of crops and livelihoods of the population \cite{wang_rethinking_2015}. The onset of the ISM takes place abruptly and the long lead-time operational forecasts for the onset, withdrawal and rainfall amount have shown little skill in the recent decades.

\textit{ISM onset and withdrawal forecasting}

Several statistical climate models have been developed for ISM onset and withdrawal dates
forecasting with different timescales from short-range (up to 4 days), medium-range (4–10 days) \cite{das_skill_2002,durai_prediction_2014} to extended-range (10–30 days) and long-range (more than 30 days) forecasts \cite{alessandri_prediction_2014}. In particular, the Indian Meteorological Department (IMD) provides a forecast of the monsoon
onset 21 days in advance, with an accuracy of $\pm$ 4 days. Note that most of the ISM forecasting models are
based on the input values, including SST, mean sea level pressure, tropospheric moisture, moist static energy, outgoing longwave radiation, wind fields, etc \cite{prasad_onset_2005,wang_objective_2009,taniguchi_comparison_2006,rajagopalan_combining_2014,puranik_index_2013}. However, there are still impending  challenges in the existing forecasting methods:
(i) The false monsoon onsets mostly related to non-monsoonal atmospheric
circulation systems \cite{flatau_dynamics_2001}; (ii)  the limitations in predicting a withdrawal data earlier than 1st September.

To overcome these challenges,  Stolbova \textit{et al.} proposed a novel tipping elements approach to accurately  forecasts the ISM onset and withdrawal dates \cite{stolbova_tipping_2016}.
This approach relies on the combination of two methodologies: the climate network analysis and the nonlinear dynamics. First, the analysis of the network of extreme rainfall events allows to identify and reveal the teleconnection between the most active hubs, the Eastern Ghats (EG) and North Pakistan (NP). Second, the application of the nonlinear dynamics through the analysis of fluctuations allows to detect the critical transitions, which are used to define the tipping elements of the ISM.
Later, the EG and NP were chosen as
optimal observation locations or reference points (RPs) to predict the ISM monsoon onset and withdrawal. It was discovered that there is a tipping point between these RPs for the SAT and relative humidity, which yields a very successful prediction
scheme.

\begin{figure}[]
\begin{centering}
\includegraphics[width=1\linewidth]{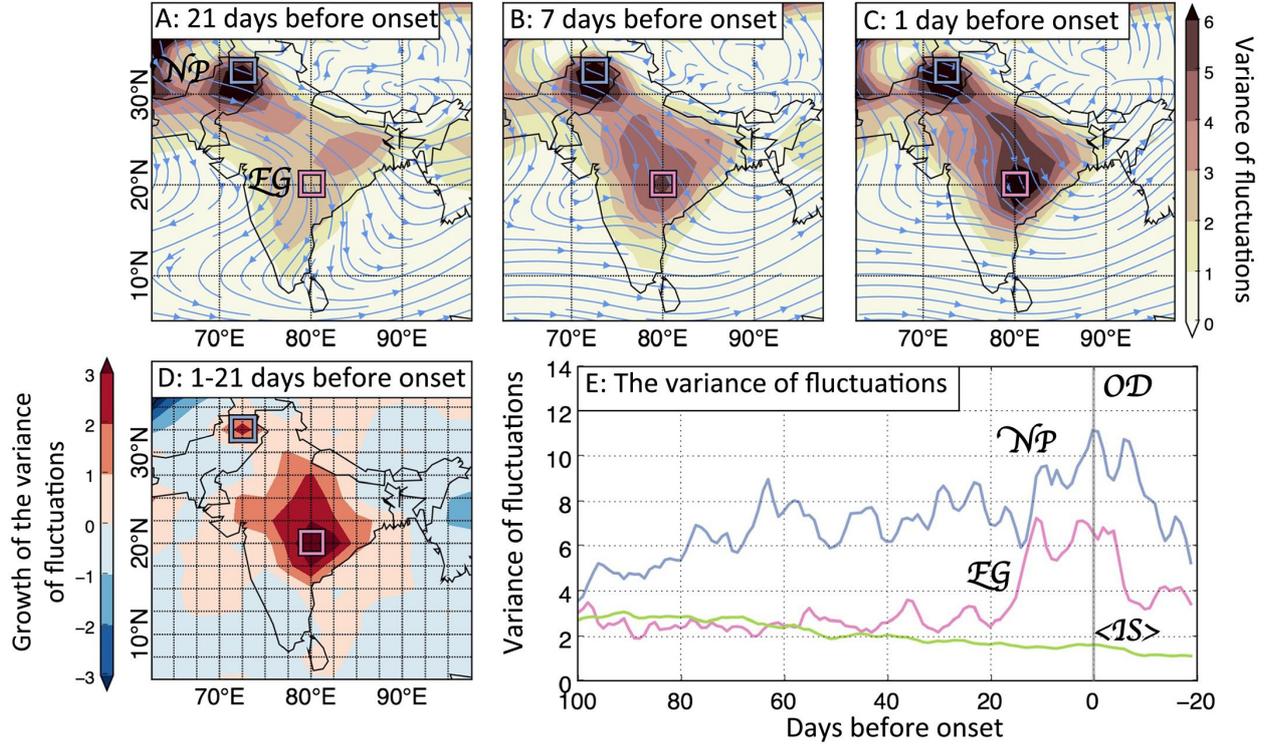}
\caption{\label{Fig3_9}
{\bf Pre-monsoon growth of the variance of fluctuations $\bar{\sigma}^{2}$ defined in Eq. \ref{EQ114} of the weekly mean values of near SAT $T$}. (a) 21 days, (b) 7 days, and (c) 1 day before the ISM onset in the Eastern Ghats (EG), $\mathrm{d} = \mathrm{c} - \mathrm{a}$.
Composites are for the period 1958–2001 and were calculated from the ERA40 reanalysis data set, 700 hPa winds are indicated by the blue lines. Two boxes refer to the RPs: North Pakistan, NP (blue) and EG (pink). (e) Growth of the variance of fluctuations in NP (blue), EG (pink), and averaged over the Indian subcontinent (IS) at approaching the onset date of the monsoon. \textit{Source}: figure from Ref. \cite{stolbova_tipping_2016}.}
\end{centering}
\end{figure}

\begin{figure}[]
\begin{centering}
\includegraphics[width=1\linewidth]{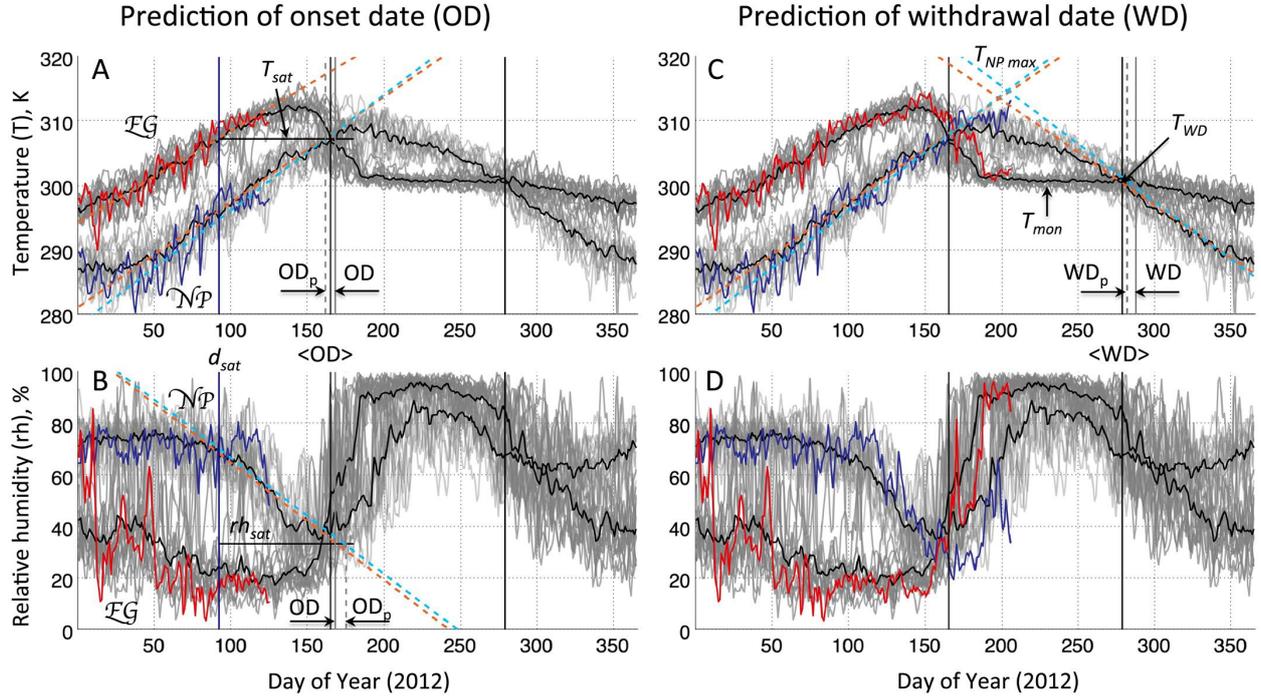}
\caption{\label{Fig3_10} 
{\bf Prediction of onset date ($\mathrm{OD}$) and withdrawal date ($\mathrm{WD}$) of the ISM: case study 2012}. Left column: prediction of the $\mathrm{OD}$; right column: $\mathrm{WD}$ in the Eastern Ghats (EG). (a),
(c): air temperature at 1000 hPa; (b), (d) relative humidity at 1000 hPa. Time series from the reference points: 14-year mean (black) and 2012 values for North Pakistan (NP) (blue) and the EG (red). Gray lines show the time series from  NP and EG for the training period of 14 years. Saturation temperature $T_{sat}(\mathrm{a})$ and saturation humidity $rh_{sat}(\mathrm{c})$ are marked by horizontal black solid lines $(T_{sat}=T_{onset}, T_{onset}$ and $r h_{s a t}$ calculated as the intersection of mean time series for the training period from the EG and $\mathrm{NP}$), and the day of the saturation $\left(d_{sat}\right)$ (when the temperature in the EG in 2012 reaches $T_{sat}$) by vertical dark blue lines. The orange lines indicate trends to the mean time series in  $\mathrm{NP}$ and EG for the training period, light blue are the trends for 2012. Black solid lines indicate mean values of the $\mathrm{OD}~ (<\mathrm{OD}>)$ and $\mathrm{WD}~ (<\mathrm{WD}>)$ for the training period. Vertical dotted  gray lines correspond to the predicted onset $\left(OD_{p}\right)$ and withdrawal dates $\left(WD_{p}\right)$, while  vertical  solid gray lines are the actual onset and withdrawal dates for $2012$. \textit{Source}: figure from Ref. \cite{stolbova_tipping_2016}.}
\end{centering}
\end{figure}

The ISM CN is focused on the monsoon region (62.5–97.5\degree E, 5.0–40.0\degree N) with a spatial resolution of $2.5$ degrees, which results in $15 \times 15 = 225$ grid
points,  see Fig. \ref{Fig3_9}. The near SAT ($T$) at
1000 hPa, relative humidity ($rh$) at 1000 hPa, and wind fields at 700 hPa were used. The variance $\sigma^{2}$ of $(T)$ and $(r h)$ for each grid point are firstly calculated as \cite{stolbova_tipping_2016}
\begin{equation}
\label{EQ114}
\begin{aligned}
\sigma^{2}(x, d, w, y) &=<\left[x\left(t^{*}(y)-d-k\right)-\bar{x}\left(t^{*}(y)-d-k\right)\right]^{2}>_{w}=\\
&=\sum_{k=1}^{w} \frac{1}{w}\left[x\left(t^{*}(y)-d-k\right)-\sum_{i=1}^{w} \frac{x\left(t^{*}(y)-d-i\right)}{w}\right]^{2},
\end{aligned}
\end{equation}
where $x(t)$ stands for a time series, $w$ is the length of the time window, $d$ is the number of days before the onset of the ISM, $y$ is a given year, and $t^{*}(y)$ is the onset date for the given year $y$. The time average value of $\sigma^{2}$ is simply expressed as
\begin{equation}
\label{EQ115}
\bar{\sigma}^{2}(x, d, w)=<\sigma^{2}(x, d, w, y)>_{y}=\sum_{y=1}^{Y} \frac{\sigma^{2}(x, d, w, y)}{Y},
\end{equation}
where $Y$ stands for the total number of years. The results are shown in Fig. \ref{Fig3_9}. It has been confirmed that the EG and NP regions experience the highest growth of $\bar{\sigma}^{2}$ for both $T$ and $rh$,  when approaching the monsoon onset date. 
It has been further discovered that  there exist two intersections in the
RPs for the average time series during the period from 1951 to 2014. At both intersections, it coincides with the mean values of the onset and withdraw as determined by the IMD  with a standard deviation of $\pm 5$ days for  the
EG region, see Fig. \ref{Fig3_10}. Thus, this allows to derive a prediction
scheme for forecasting the onset and withdrawal of the ISM in the EG, based on the trends
of $T$ and $rh$ in the RPs.

Next,  a linear trend estimation
for the two RPs is performed and compared  with the trends of the mean time series of the training period (1951-1965). The slopes of the trends
for the RPs provided an estimation of an early, normal, or late monsoon arrival:
a greater than average slope of $T$ will lead to an earlier than usual onset, and vice
versa. Trends of $rh$ in the RPs in comparison with the average trends for the training
period add up to the predictability of the onset: higher than average values of the relative
humidity lead to a late onset, and vice versa.

It has  also been found that the ISM onset coincides with the date when $T$ in the  EG
and in NP become equal (see Fig. \ref{Fig3_10}a). Therefore, for the forecasting of the onset, one should predict when $T$ for the EG
will abruptly decrease and intersect $T$ for NP.
Finally, the prediction scheme of the withdrawal  is based on the symmetry of $T$ changes
in NP during the year. The withdrawal is estimated as the intersection of
the projected $T$ decrease in NP and the $T$ in the EG during the monsoon season (see Fig. \ref{Fig3_10}c).

The prediction is regarded as successful if the time difference between the predicted onset and the real one is $\leq 7$ days for the onset and $\leq 10$ days for the withdrawal \cite{stolbova_tipping_2016}. The proposed scheme using $T$ results in $74\%$ successful predictions of the onset when made on day 125 of the year (April 10 ). The prediction scheme for the withdrawal succeeds in $84\%$ of the years when made on day 205
(July 25). Based on this prediction scheme, they successfully predicted both the onset and withdrawal since 2016 in the EG region\footnote{For more detailed and updated information have a look at PIK’s information page on the Indian Summer Monsoon Forecast: 
\url{https://www.pik-potsdam.de/services/infodesk/forecasting-indian-monsoon}.}.

\textit{ISM rainfall forecasting}

The ISM is one of the most prominent monsoon systems. It affects the Indian subcontinent, where it is one of the oldest and most anticipated weather phenomena. Several theories have been proposed to explain the underlying mechanism of the monsoon. For example, the \textit{sea breeze} theory: because of the differences in the specific heat capacity of land and water, continents heat up faster than seas. Consequently, the air above coastal lands heats up faster than the air above seas. These create areas of low air pressure above coastal lands compared with pressure over the seas, resulting moisture-laden winds to flow from the seas onto the lands.  On reaching land, these winds rise because of the geographical relief, cooling adiabatically and leading to orographic rains. However,  the understanding and predictability of ISM are still evolving.

The All India Rainfall Index (AIRI) is the total amount of ISM June-to-September rainfall averaged over the entire Indian
subcontinent. It is the IMD's primary indicator for monitoring the ISM rainfall \cite{rajeevan_new_2007}. An accurate long-term  prediction of the ISMR is crucial for taking timely actions  and mitigation activities. 
Over the last few decades, great efforts have been undertaken to
understand the basic physics of the monsoon, and to improve the prediction skill for the AIRI
\cite{shukla_empirical_1987,webster_monsoons:_1998,di_capua_long-lead_2019,gadgil2011seasonal,rajeevan_evaluation_2012,ramu_indian_2016,delsole_climate_2012}.
However, both methods have the challenge of unstable relationships with the
predictors over time, as observed during the recent years due to a  weakened coupling between the boundary forcing and the Indian monsoon \cite{rajeevan2001prediction,gadgil_monsoon_2005}. In particular, the
forecasting skill (cross correlation) of the official operational forecasts made by IMD statistical models is  -0.12 for 1989--2012, for the five
ENSEMBLE models' multi-model ensemble is 0.09 and the four APCC/CliPAS models' MME skill is 0.24 after 1989. 
It was discovered that the recent failure is largely due to the models' inability to capture new predictability sources emerging during the recent global warming \cite{wang_rethinking_2015}.

Traditionally, the April--May SST anomaly in the Ni\~{n}o 3 or Ni\~{n}o 3.4 region was used as a predictor for the ISMR prediction. This is 
since the ISMR variations are primarily driven by the EP-ENSO through tropical teleconnections  \cite{shukla_empirical_1987,webster_monsoons:_1998}.
This inverse relationship between the AIRI and EP--ENSO, however, has weakened in recent decades \cite{kumar_weakening_1999}. It has been found that the CP--ENSO has also distinct teleconnections and affects on the ISMR \cite{kumar_unraveling_2006}. Based on this physical mechanism,  Wang \textit{et al.} \cite{wang_rethinking_2015} proposed  four physical–empirical AIRI predictors, EP–ENSO predictor (EPT), CP–ENSO predictor (CPT), mega-ENSO predictor (PSH) and Anomalous Asian Low (NAT). The EPT is defined as:  May minus March east–west SST dipole
tendency: $\mathrm{DSST}^{*}$ (20 \degree S-5 \degree N, 150 \degree E-170 \degree E)
minus DSST (10 \degree S–10 \degree N, 110 \degree W–80 \degree W); the CPT is defined as:  May minus April SST north–south dipole
tendency: $\mathrm{DSST}^{\dagger}$ (10 \degree–25 \degree S, 170 \degree E–160 \degree W)
minus DSST (5 \degree –20\degree N, 180 \degree–150 \degree W); the PSH  is defined as: April–May mean SLP averaged over (40 \degree
S–10 \degree S, 160 \degree W–90 \degree W) and (10 \degree N–30 \degree N, 180 \degree–130 \degree W); the NAT is defined as: May minus March SLP averaged over
(45 \degree N–60 \degree N, 95 \degree E–125 \degree E). $\mathrm{DSST}^{*}$ means the difference of the SST between May and March (May minus March). $\mathrm{DSST}^{\dagger}$ means the difference of SST between May and April (May minus April). These predictors (EPT)  produce an independent forecast skill of 0.51 for 1989–2012 and a 92-year retrospective forecast skill of 0.64 for 1921–2012. However, the forecasting lead-time for this model is quite short, i.e., starting from May (one month lead-time) or June (IMD's operational model).

In order to achieve the goal of a long-term and reliable prediction of the ISM rainfall,  Fan \textit{et al.}  constructed a series of dynamical and physical CNs based on the global near SAT field \cite{fan_network-based_2020}. It was uncovered that
some characteristics of the directed and weighted CNs can serve
as efficient long-term predictors for ISM rainfall forecasting.
The developed prediction method produces a forecast skill of $0.5$ with a 5-month lead-time in advance by using the previous calendar year's data. The skill of the forecasting, is comparable to the current state-of-the-art models, however, with quite a short (i.e., within one month) lead-time. The new network-based prediction approach is described as follows.

\textit{CN construction}: Based on the NCEP/NCAR reanalysis daily SAT anomalies data,  a CN is constructed for each calendar year from 1948 to the present. In the work, 726 nodes are selected, such that the globe is covered approximately homogeneously \cite{gozolchiani_emergence_2011}. The link strength and direction are determined based on a similarity measure between the temperature anomaly time series of the nodes [See Eqs. (\ref{EQ10})--(\ref{EQ14})]. A pair of nodes is defined to be connected only if their link is within the top $5\%$ positive strength, which is also corresponding to a statistical significance of above the $95\%$ confidence level.

\begin{figure}[]
\begin{centering}
\includegraphics[width=0.85\linewidth]{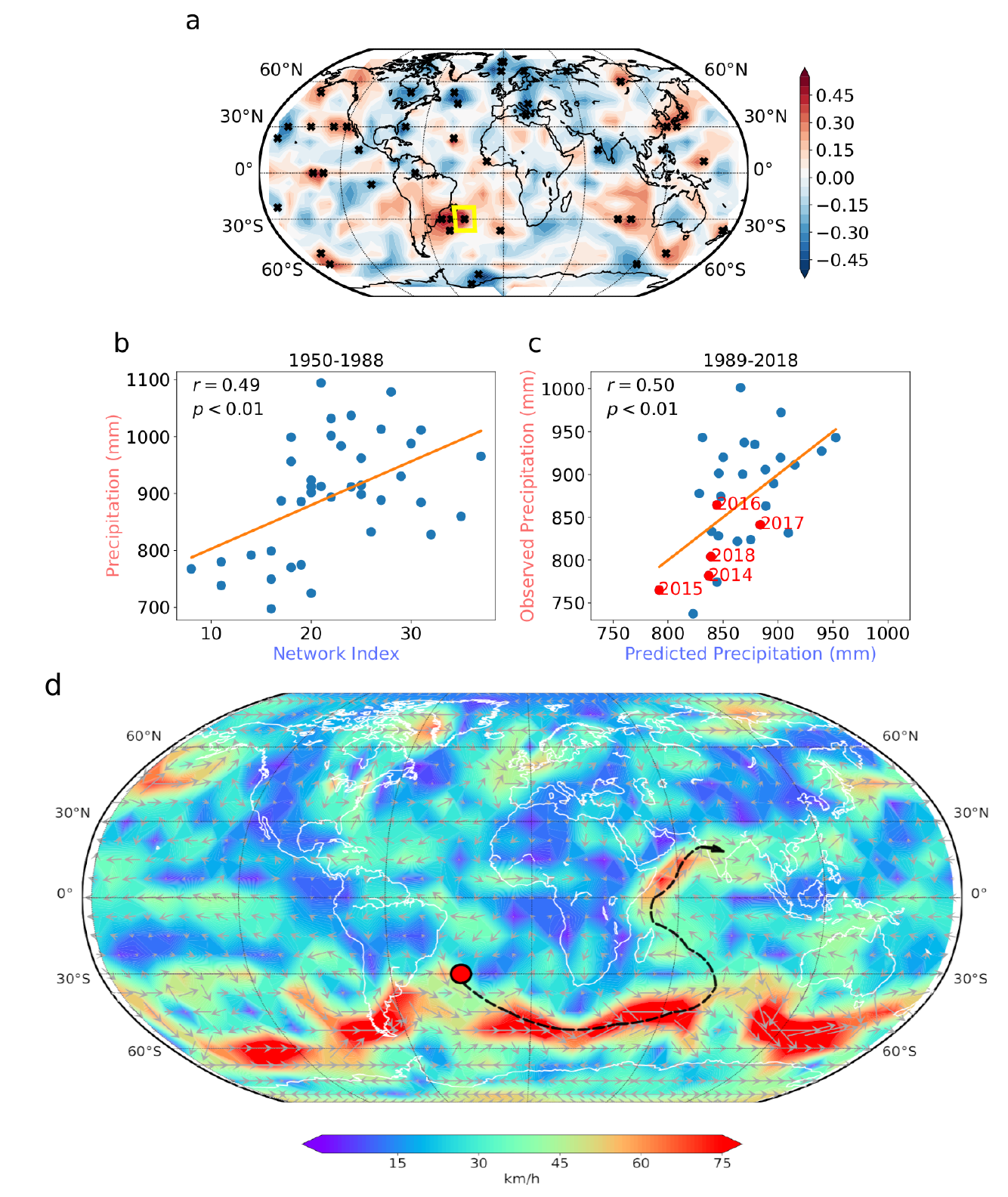}
\caption{\label{Fig3_11} 
{\bf Schematic of ISM rainfall forecasting based on climate networks}. (a)  The correlation coefficients between observed  AIRI and in-degree time series during the training period for all nodes. (b) Scatter plots of the observed  AIRI versus the  optimized network predictor (i.e., the in-degree time series of the key node in the yellow box) in the training period. (c) Scatter plots of the observed AIRI versus the predicted AIRI during the forecast period 1989-2018. (d) Teleconnection path between the key node in South Atlantic and Indian subcontinent.  Black $\times$ in  panel (a)
represent the regions with correlations significant at the 95\% confidence
level (Student's t-test). The recent five years are highlighted
 in red in panel (c). The colors and grey arrows depict the magnitudes and directions of the 850 hPa winds on 12th August 2014 during the Indian summer monsoon season in panel (d). \textit{Source}: Reprinted figure from Ref. \cite{fan_network-based_2020}.}
\end{centering}
\end{figure}

\textit{Network Predictors Mining}: To mine the dynamic network predictors for the AIRI, the cross correlation between the observed AIRI and the time series of network in-degree $K^{y}_{i}$ [see Eq. (\ref{EQ5})] for each node $i$ on the globe is calculated. Because the forecasting skill of the AIRI became very poor in the IMD operational forecasts and other models since the year 1989 \cite{wang_rethinking_2015},  the record is divided into  two consecutive periods, i.e., (i) 1950-1988,  as the training period and (ii) 1989-2018, as the retrospective forecast period.
For the training period, the correlation coefficients are shown in Fig.~\ref{Fig3_11}a. It was found that the in-degree index for some nodes captures well the behavior of the AIRI. Particularly, the node (in the region marked by a yellow box in  Fig.~\ref{Fig3_11}a) having the maximal correlation $r$ is located in the South Atlantic ocean [(30\degree S, 30\degree W)]. Fig.~\ref{Fig3_11}b shows that the value of this maximal correlation is $r=0.49$, with a student t-test significance of  $p<0.01$. Therefore, the in-degree index of this key node is regarded as the optimized network predictor for the AIRI. 

\textit{AIRI Forecasting}: Next, the forecasting capability of the network predictor during the period 1989-2018 is tested.  To obtain the forecasted value of the AIRI for one specific year, the forecasted year's predictor value (i.e., the previous year's network in-degree of the key node) is substituted into the least square linear regression equation. This equation used for prediction is derived using only the past information. The independent forecast skill for
the period 1989-2018 becomes then $\sim0.50$, which is significantly higher than the IMD operational forecast skill [see Fig.~\ref{Fig3_11}c]. The result indicates the good accuracy of the network-based predictor with the long lead-time of 5 months. Finally, based on 2018's data, the AIRI for 2019 is  forecasted as 949.87 mm, a ~7\% departure percentage compared to\footnote{ The observed AIRI data can be found in \url{https://mausam.imd.gov.in/imd_latest/contents/rainfall_time_series.php}.} the long-term  average, which is 887.5 mm. The actual value of the AIRI for 2019 was 968.3 mm.

In the following, a potential physical mechanism is discussed.
From a meteorological perspective, the teleconnection between the key node's location in the mid-latitude of the South West Atlantic Ocean and the Indian subcontinent is no coincidence. The orography of the Eastern coast of the South American continent is prone to the formation of an anticyclone at mid-latitude due to the interaction with mid-latitude Westerlies (winds that blow from the west at the surface level between 30\degree and 60\degree S). Moreover, there is the influence of the South polar jet stream, which appears within the upper air Westerlies. The acceleration/deceleration of the South polar jet stream induces areas of low/high pressure, respectively, which links to the formation of cyclones and anticyclones.  Because the South polar jet stream strengthens and weakens seasonally, and from year to year, it synchronizes with the variability in the strength of the circulation centered around the location of the key node. One of the possible  data-based teleconnection paths is shown in Fig.~\ref{Fig3_11}d. Hence, the key node location is sort of the nexus of the South Atlantic circulation, the south polar jet stream and the main atmospheric circulations patterns over the Indian Ocean. 
It explains why we observe a significant correlation between 
the network characteristics of the key node and the AIRI.

\begin{figure}[]
\begin{centering}
\includegraphics[width=1\linewidth]{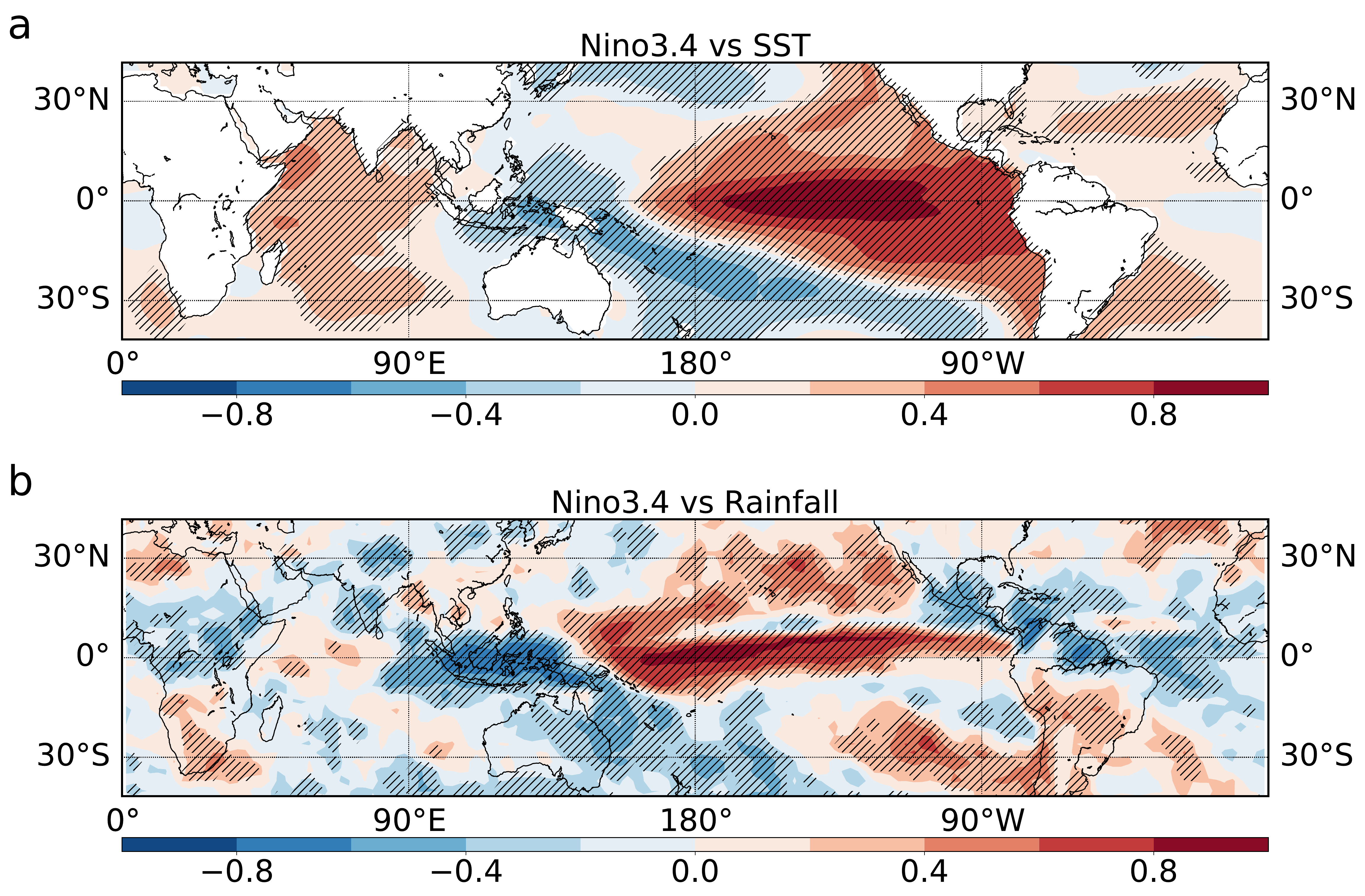}
\caption{\label{Fig3_12} 
{\bf Physical mechanisms of ISM rainfall forecasting}. Correlation between the seasonal mean (Jun-Sep) Ni\~no 3.4 SST and (a) SSTs (b) rainfall at each grid point. Statistically significant (95\% confidence level) correlations are stippled. \textit{Source}: Reprinted figure from Ref. \cite{fan_network-based_2020}.}
\end{centering}
\end{figure}

Further analyses are performed and reveal that the physical mechanisms are related to  (i) the network delayed ENSO teleconnection, i.e., the in-degree index is locally minimized in the previous calendar year of the El Ni\~{n}o onset. (ii) The ENSO-Monsoon teleconnection \cite{webster_monsoon_1992,kumar_weakening_1999,rajeevan_nino-indian_2007}, i.e., the AIRI anomaly and the Ni\~{n}o 3.4 SST anomaly index are negatively correlated.  
The spatial correlation maps between the Ni\~{n}o 3.4 SST anomaly index with both the SST as well as the rainfall index  for all nodes are presented in Fig.~\ref{Fig3_12}. These results indicate that the Ni\~{n}o 3.4 SST anomaly index is significantly anti-correlated
with the SST in the equatorial Indian ocean and also heavily influences the rainfall pattern over the Indian land region.

Moreover, the network-based approach can be also successfully applied to forecast the Indian homogeneous region rainfall as a prediction scheme, including two specific homogeneous regions \cite{fan_network-based_2020}: (1) the Central India Rainfall Index; (2) the East \& Northeast India Rainfall Index.

In addition, there is solid evidence which reveals that climate change affects the forecast skills. The considerable warming trend in the last 70 years - the 850 hPa temperature in the key node area has risen 2.5 \degree C - substantially impacted the key node region.  An increase of the in-degree index and a change in the structure of the network itself are observed. Concurrently with these changes, the prediction skill of this forecasting method for the ISM rainfall amount is improving substantially \cite{fan_network-based_2020}. 

\subsubsection{Extreme Rainfall}
Extreme precipitation events have produced more rain and have become more common since the 1950s in many regions of the world. 
Heavy precipitation may pose threats to our society. The most immediate impact is the threat  of flooding. This risk can be heightened in urban and agricultural areas. In addition to flooding, extreme precipitation also increases the risk of landslides. Excessive precipitation can also degrade water quality, harming human health and ecosystems. Thus is is crucial to predict extreme rainfall events and to establish an early warning system for them. However, the analysis of spatial patterns of co-variability of such events is challenging because traditional techniques based on principal component analysis of the covariance matrix only capture the first two statistical moments of the data distribution and are not suitable to analyze the behavior in the tails of the respective distributions \cite{lakshmanan_machine_2015}.

\textit{Extreme Rainfall over India}

To overcome these challenges, Malik \textit{et al.} applied the CNs approach to analyze the spatial and temporal patterns of extreme rainfall during the ISM \cite{malik_spatial_2010,malik_analysis_2012}.  This approach is based on the event synchronization method (see Section. \ref{cap2:ES}). In Ref. \cite{malik_spatial_2010},
the dataset is a daily precipitation
gridded set for 1961–2004 developed as part of the
project – Asian Precipitation Highly Resolved Observational
Data Integration Towards the Evaluation of Water Resources\footnote{It is freely downloadable from: \url{https://www.chikyu.ac.jp/precip/}.}. Then they extracted the data for the South Asian
region at a 0.5 degree resolution. For calculating the event synchronization,  two different thresholds
of $\alpha = 94$th percentile and $\alpha = 90$th percentile were chosen to obtain the extreme event time series,  referring to very heavy and heavy rainfall events, respectively.
Next, they computed the strength of synchronization $Q_{ij}$ and the delay behaviour $q_{ij}$ between each
pair of the grid sites $i$ and $j$ [see Eq. (\ref{EQ18})].
After normalizing it holds, that $0 \leq Q_{ij} \leq 1$ and $-1 \leq q_{ij} \leq 1$, where $Q_{ij} = 1$ means complete
synchronization, and  $Q_{ij}=0$ stands for the absence of synchronization; and $q_{ij} = 1$
means that events at $i$ precede events in $j$. The CN is then
constructed by selecting only pairs of sites that show strong correlations, and the
time lags between the events are used to define the direction of the links.

\begin{figure}[]
\begin{centering}
\includegraphics[width=1\linewidth]{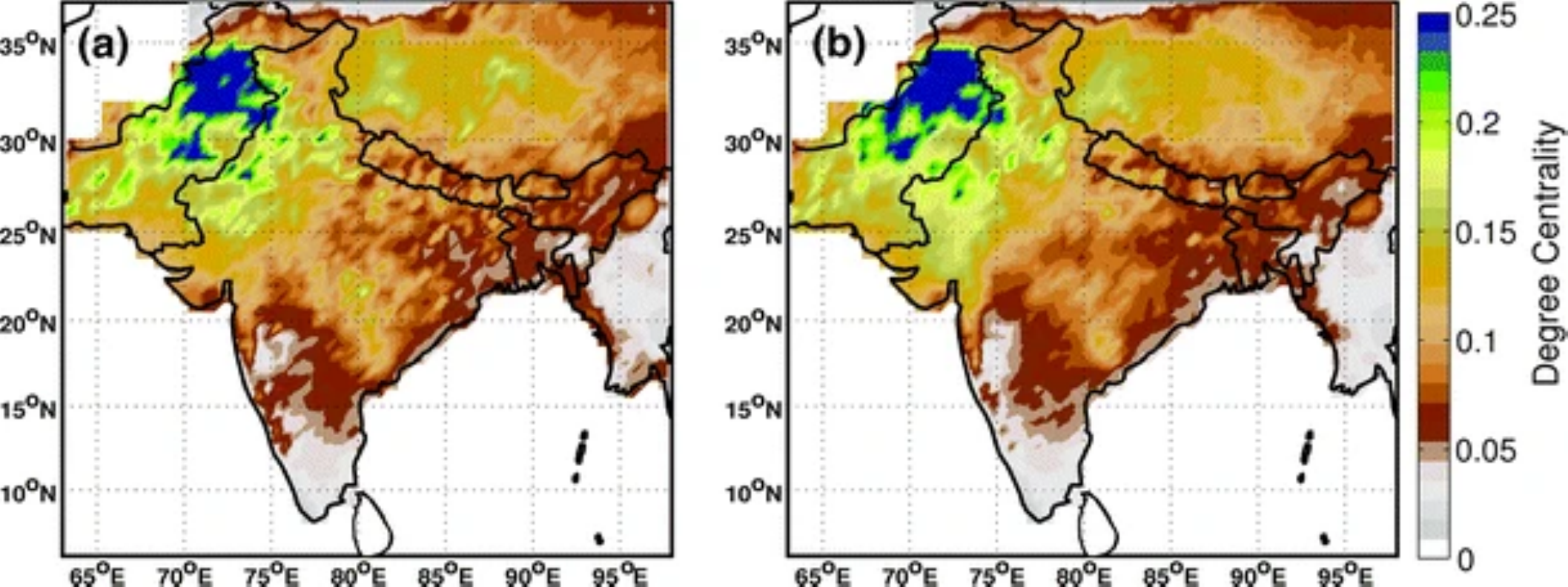}
\caption{\label{Fig3_13} 
{\bf Degree centrality of the event synchronization CN for extreme rainfall over the Indian subcontinent}. The extreme events threshold is (a) $\alpha = 94\%$ and (b) $\alpha = 90\%$.  \textit{Source}: figure from Ref. \cite{malik_analysis_2012}.}
\end{centering}
\end{figure}

The spatial structures, organization, and scales of the extreme
rainfall over India during  the ISM period were investigated. Fig. \ref{Fig3_13} exhibits that the degree centrality  [see Eq. (\ref{EQ3})] of the CNs reveals well-defined spatial structures in the monsoonal extreme precipitation \cite{malik_analysis_2012}. Based on the median of the geographical length  of the links, calculated using the formula for the spherical Earth projected onto a plane, it has been uncovered that extreme rainfall events are synchronized up to 250 km for most of the region.
They further found that such a spatial organization opens the possibility for predictions of a probable spatial extent of monsoonal rainfall activity without delay, e.g., for most of the subcontinent, the accuracy of the prediction $\epsilon$ is above 70\% and even reaches up to 100\% in certain places \cite{malik_analysis_2012}.

\textit{Extreme Rainfall over South America}

This event synchronization CN approach was also  applied to predict extreme rainfall events in the Central Andes of South America \cite{boers_prediction_2014}. Based on the real-time satellite-derived
rainfall data, TRMM 3B42V7 \cite{huffman_trmm_2007}, Boers \textit{et al.} were able to predict more than 60\% of the rainfall events above the 99th percentile in the Central Andes, and even 90\% during \el~ conditions. The
interplay of northward migrating frontal systems and a low-level wind channel from the
western Amazon to the subtropics are considered as the responsible mechanism. 

Extreme rainfall events are defined as times with threshold $\alpha = 99\%$ for all Dec-Jan-Feb seasons in the spatial domain (85 \degree W, 30 \degree W) and (40 \degree S, 15 \degree N),
at a resolution of $0.25 \degree$, and 3-hourly temporal resolution for the
time period from 1998 to 2012. After the construction of the event synchronization CN (see Section \ref{cap2:ES}), the network divergence $\Delta \mathbf{S}$ is defined as 
\begin{equation}
\label{EQ116}
\Delta S_{i}:=S_{i}^{\mathrm{in}}-S_{i}^{\mathrm{out}}:=\sum_{j=1}^{N} A_{i j}-\sum_{j=1}^{N} A_{j i},
\end{equation}
where  $\mathrm{S}^{\text {in }}$  and  $\mathrm{S}^{\text {out }}$ stand the in-strength and out-strength  respectively at each node. Positive values of $\Delta S$ indicate sinks of the network; negative values indicate sources.  The strength out of and into a region $R$ is
\begin{equation}
\label{EQ117}
S_{i}^{\mathrm{in}}(R)=\frac{1}{|R|} \sum_{j \in R} A_{i j},
\end{equation}
\begin{equation}
\label{EQ118}
S_{i}^{\text {out}}(R)=\frac{1}{|R|} \sum_{j \in R} A_{j i},
\end{equation}
where $|R|$ denotes the number of grid  points contained in $R$.

\begin{figure}[]
\begin{centering}
\includegraphics[width=1\linewidth]{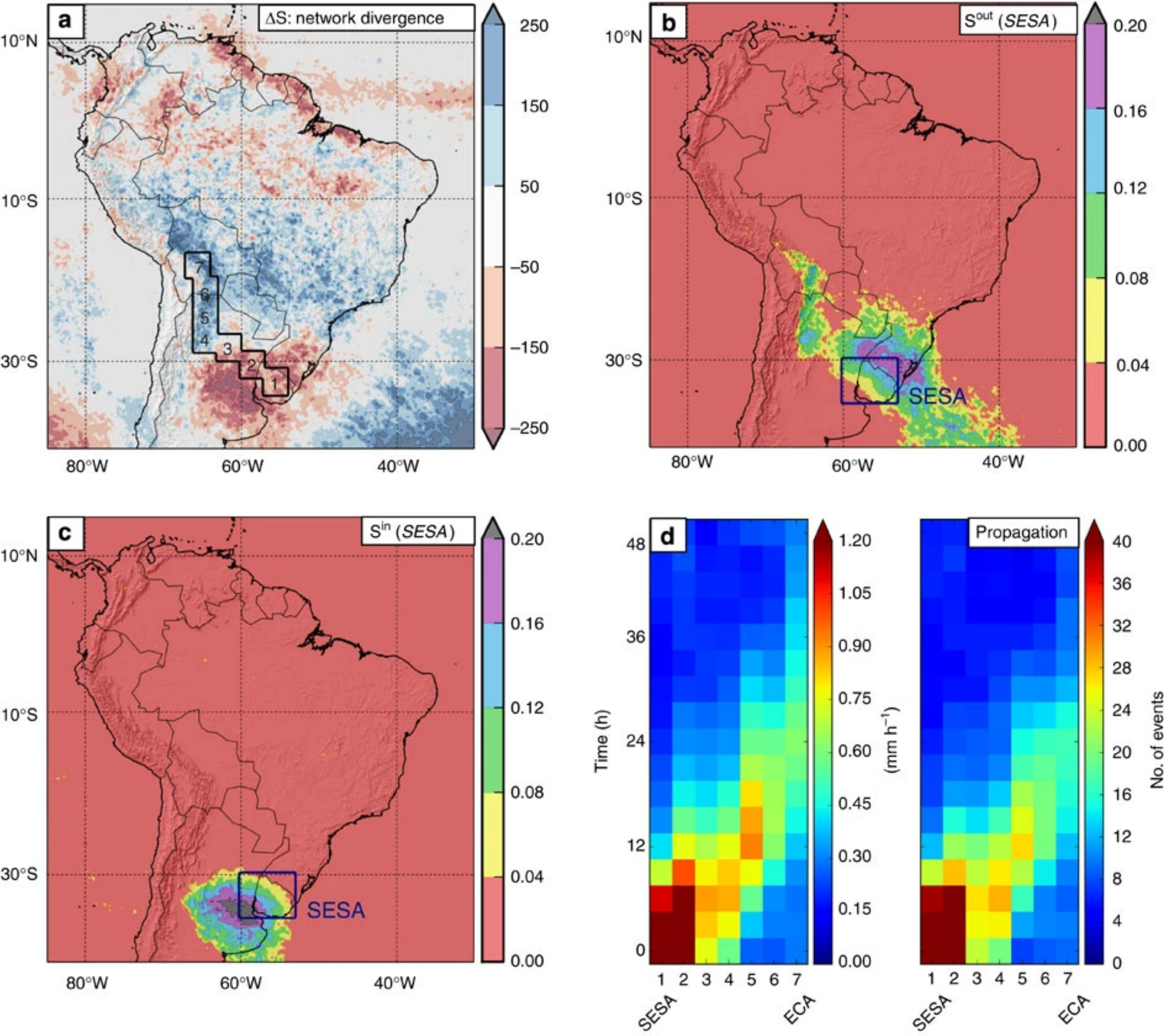}
\caption{\label{Fig3_14} 
{\bf Event synchronization CN for extreme rainfall over South America}. (a) Network divergence, $\Delta S$, is defined as the difference of
in-strength and out-strength at each node (see Eq. \ref{EQ116}). The boxes labelled 1 to 7 are
used for the tracking of extreme events. (b) Strength out of SESA, $S_{i}^{\text {out}}(SESA)$, which is the average out-Strength restricted to SESA ((see Eq. \ref{EQ117})). (c) Strength in of SESA, $S_{i}^{\text {in}}(SESA)$, which is the average in-Strength restricted to SESA ((see Eq. \ref{EQ118})). (d) Temporal evolution of extreme rainfall events from SESA to ECA along the sequence of boxes indicated in (a). \textit{Source}: figure from Ref. \cite{boers_prediction_2014}.}
\end{centering}
\end{figure}

Fig. \ref{Fig3_14}a shows the network divergence $\Delta S$ (defined in Eq. (\ref{EQ116})), which estimates the dynamics and temporal order of extreme
rainfall in South America. The boxes labelled 1 to 7 are
used for the tracking of extreme events. It was found that the
most pronounced source region of the rainfall network is South Eastern South
America (SESA), defined as the box ranging from  (60 \degree W, 53 \degree W) to (35 \degree S, 30 \degree S).  Fig. \ref{Fig3_14}b depicts  where synchronized extreme events occur
within 2 days after extreme events occurred in SESA, measured by $S_{i}^{\text{out}}(SESA)$. For comparison, $S_{i}^{\text{out}}(SESA)$ is shown in Fig. \ref{Fig3_14}c.  This analysis reveals that extreme
events in SESA are followed by extreme events along a narrow
band following the eastern Andean slopes up to western Bolivia, while they are only preceded by extreme events to the
southwest.  Based on the propagation of extreme rainfall from SESA to the eastern Central Andes (ECA), see Fig. \ref{Fig3_14}d,
a prediction for the extreme rainfall  for the ECA  within 2 days is made after they occurred in SESA. It should be noted that these extreme rainfall events could not be forecasted by using other methods.

\textit{Extreme Rainfall over the Globe}

The event synchronization CN method was also applied to investigate the global pattern of
extreme-rainfall by Boers \textit{et al.} \cite{boers_complex_2019}. The gauge-calibrated, satellite-derived rainfall dataset  TRMM
3B42 V731, with daily temporal resolution, from $50$ \degree N to $50$ \degree S for the  period 1998–2016 is used. The extreme rainfall event thresholds are  $80\mathrm{th}, 81\mathrm{th}, \dots, 99\mathrm{th}$ of the wet days (>1 mm) for the Jun-Jul-Aug season.
Note that consecutive days with rainfall above the
threshold are considered as single events and placed on the first day of the occurrence.
To study the robustness of the method,  Global
Precipitation Climatology Project32 and NCEP/NCAR Reanalysis 1 data were also tested.

\begin{figure}[]
\begin{centering}
\includegraphics[width=0.85\linewidth]{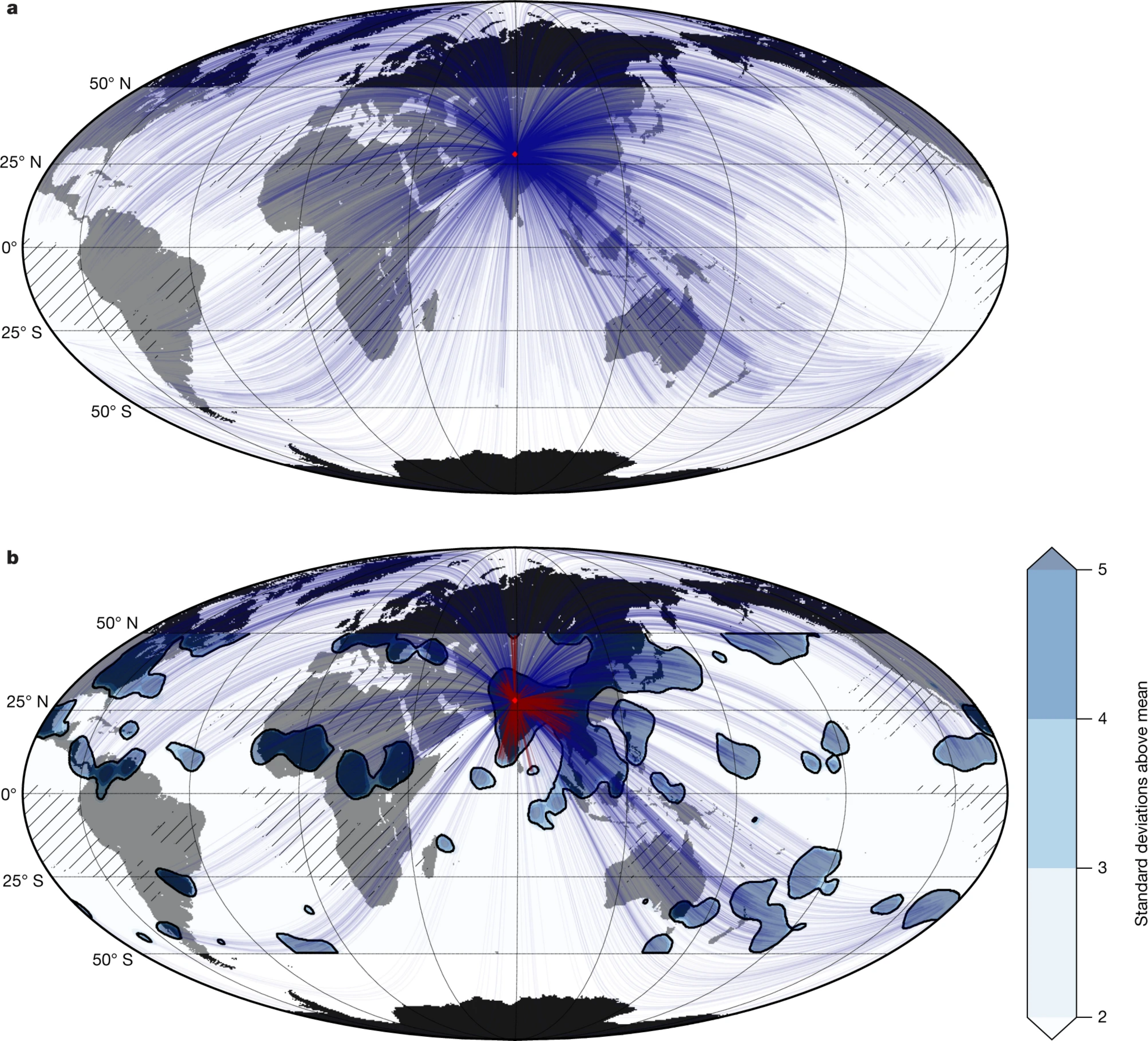}
\caption{\label{Fig3_15} 
{\bf Event synchronization CN for extreme rainfall over globe}. Teleconnection pattern for south-central Asia (SCA) for events above
the 95th percentile. (a) Links connected to SCA, before correcting for the multiple-comparison
bias. (b) Link bundles attached to SCA, after correcting for the multiple-comparison
bias. \textit{Source}: figure from Ref. \cite{boers_complex_2019}.}
\end{centering}
\end{figure}

First, the synchronization of extreme events for each pair of nodes is computed (see Section \ref{cap2:ES}). Network links are added between two nodes if the
corresponding synchronization values are significant,  $p < 0.005$.   The global distribution of the spatial
distances for which significant synchronizations occur was studied. This distribution decays as a power law $p(d) \sim d^{-\alpha}$, with an exponent $\alpha$ very close to 1 for distances $d<$  $2,500~\mathrm{km}$, but exhibits a super-power-law behaviour for longer distances. The scale-break indicates that the significant links can be divided into two distinct classes: (i) links associated with regional weather systems with distances up to $2,500~ \mathrm{km}$, which include mesoscale convective systems and tropical cyclones; and (ii) links associated with global-scale teleconnections. Such teleconnections are generally understood to be caused either by direct signal transport due to large-scale atmospheric circulations or by propagating waves triggered by disturbances of these circulations \cite{boers_complex_2019}. In Fig. \ref{Fig3_15}, the teleconnection pattern for south-central Asia (SCA) for events above the 95th percentile is presented. Pronounced link bundles connecting SCA with eastern Asia, the African tropics, large parts of Europe and the
eastern coast of North America, as well as the Southern Hemisphere extratropics are apparent.
Further analysis reveals that the Rossby waves can be regarded as the physical mechanism underlying these global
teleconnection patterns.

\subsubsection{Atmospheric Circulation and Global Warming}

Earth’s atmosphere, made up essentially of the gases that surround our
planet, consists of circulation patterns that move air from one location to another and
from the surface to top elevations. The atmospheric circulation is controlled by Earth’s rotation, barometric pressure, topography, ocean currents, and differences in temperature, salinity, etc.

\textit{Latitude structure of the circulation}

\begin{figure}[]
\begin{centering}
\includegraphics[width=0.85\linewidth]{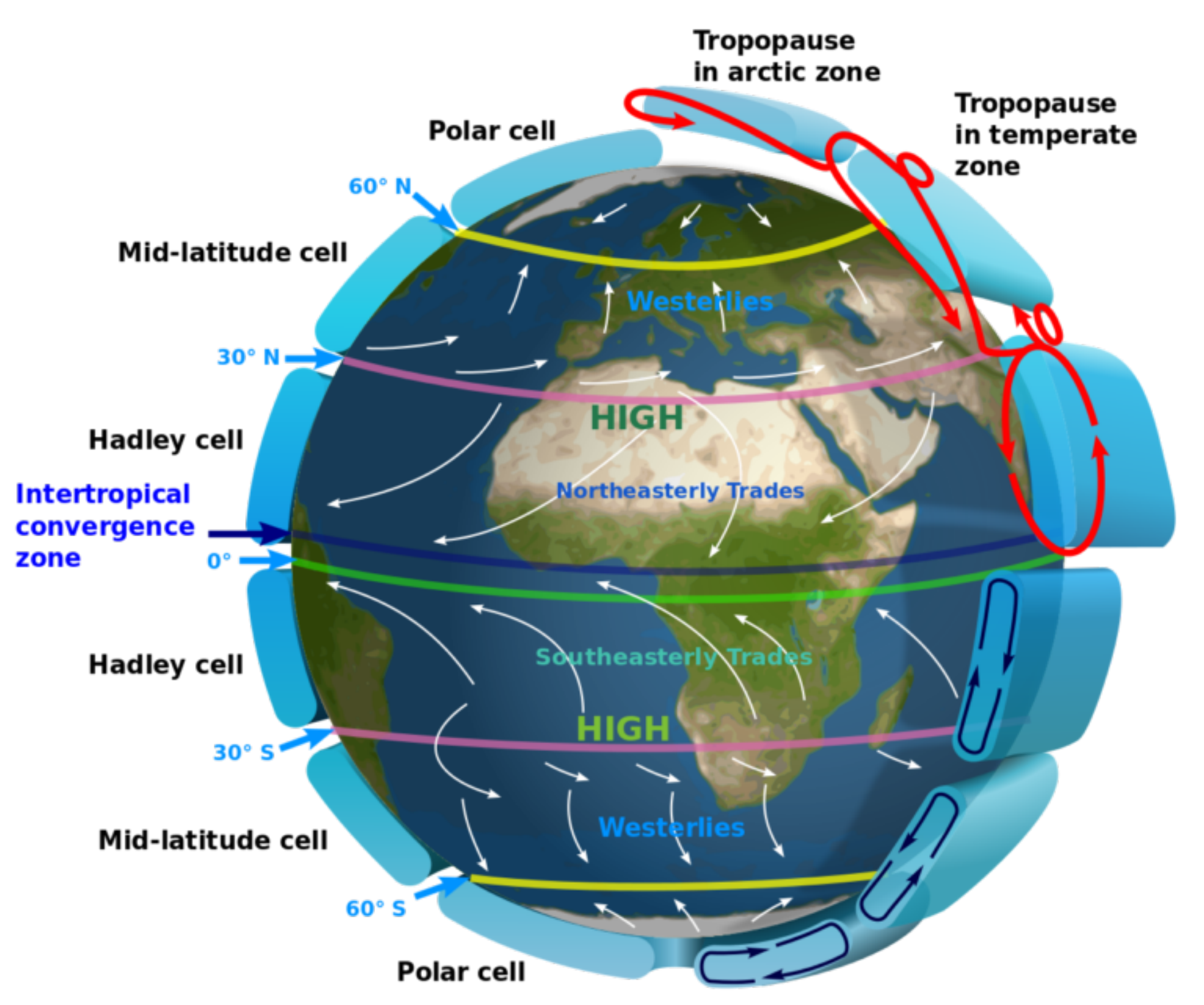}
\caption{\label{Fig3_16} 
{\bf  Schematic of major features of the atmospheric circulation}. The average circulation is idealized as being independent
of longitude and symmetric about the equator. The figure is from \url{https://courses.lumenlearning.com/geophysical/chapter/global-atmospheric-circulations/}.}
\end{centering}
\end{figure}

Since latitudinal gradients in solar energy input are  dominant drivers of the atmospheric circulation,
its primary features depend on the latitude. We show an idealised depiction of the  large-scale atmospheric circulation on Earth in Fig. \ref{Fig3_16}, which schematizes the main features
of the circulation, both in a latitude–height plane and in the horizontal, without yet considering
longitudinal variations due to continents, oceans, etc. This represents the circulation
in an average over all longitudes, known as a zonal average.  As we see,  there are three large-scale convection cells in both hemisphere.

(i) The \textit{Hadley cell} is a thermally driven, overturning circulation that tends to rise in the
tropics and sink at slightly higher latitudes. Warming from the surface near the equator
is transferred upward through a deep layer by convection clouds. The rising air spreads poleward and cools slowly, returning to the surface and then moves toward the equator. Roughly speaking, the Hadley circulation transports heat to about 30\degree ~latitude,
and the average effect of the transient weather disturbances transports the heat further
poleward. 

(ii) The \textit{Ferrel cell} is also called  mid-latitude cell. In the Ferrel cell, air flows poleward and eastward near the surface and equatorward and westward at higher altitudes; this movement is the reverse of the airflow in the Hadley cell. It was the first to account for the westerly winds between latitudes 35\degree ~and 60\degree ~ in both hemispheres. 

(iii) The \textit{Polar cell} is the smallest and weakest cell, which extends from between 60\degree~ and 70\degree ~north and south to the poles. Air in these cells sinks over the highest latitudes and flows out towards the lower latitudes at the surface.
Note that the idealized picture of the atmospheric circulation based  on latitude is a reasonable first approximation, but land–ocean contrasts and other variations in longitude are also important.

\textit{Climate Change Effects on Atmospheric Circulations}

Climate change strongly affects  the mode of atmospheric circulations, especially, the Hadley
cell, which plays a pivotal role in the Earth’s climate by
transporting energy and heat poleward \cite{fu_enhanced_2006,lu_expansion_2007,seo_mechanism_2014}. 
An analysis of satellite observations indicates a poleward
expansion of the Hadley
cell by 2\degree~ latitude from 1979 to 2005 \cite{fu_enhanced_2006}. A possible mechanism
for the changes in the Hadley
cell and its relation to global warming
has been reported \cite{lu_expansion_2007}. Also, a possible mechanism for the
changes in the Hadley
cell’s strength and its relation to global warming
was developed in \cite{seo_mechanism_2014}. Both aforementioned observations and
theories suggest a weakening and poleward expansion of the Hadley
cell under global warming.

As is well known, the locations of the subtropical dry zones and the major tropical/subtropical deserts are associated with the subsiding branches
of the  Hadley
cell \cite{held_robust_2006}. Therefore, the poleward expansion of the Hadley
cell
may result in a drier future in some tropical/subtropical regions \cite{lu_expansion_2007}. In addition, the poleward migration of the location of the tropical cyclone maximum
intensity is also considered as a  consequence of  the  expansion of the Hadley
cell \cite{kossin_poleward_2014}. 

The strength of the Hadley
cell is calculated using the  observed zonal-mean meridional wind in the stream function $\Psi$~ \cite{vallis2017atmospheric},
\begin{equation}
[\overline{V}] =  \frac{g}{2\pi R \cos\phi} \frac{\partial \Psi}{\partial p}, 
\label{EQ119}
\end{equation}
where $V$ is the meridional velocity in pressure coordinates, $R$ is the mean radius of the Earth, and $p$ is the pressure. The operators $\bar{ }$ and $[ ~]$ stand for temporal and zonal averaging, respectively. When computing the $\Psi$ field, we usually assume $\Psi = 0$ at the top of the atmosphere and integrating Eq.~(\ref{EQ119}) downward to the surface. The intensity of the Hadley
cell  is much stronger in winter than in summer.
To quantify the edges of the Hadley
cell, the maximum of the absolute
value of $\Psi$ at 500 hPa is first determined ($\Psi_{500}$), and
then  the edges are identified as the first latitude
poleward of the maximum at which $\Psi_{500}$ becomes zero \cite{lu_expansion_2007}. Theories of the Hadley cell suggest that the
meridional extent should scale with the height of the tropopause $H$ \cite{held_nonlinear_1980},
\begin{equation}
\phi_{H} \sim\left(\frac{g H \Delta_{H}}{\Omega^{2} R^{2}}\right)^{\frac{1}{2}},
\label{EQ120}
\end{equation}
where $H$ was computed from temperature data as the lowest pressure level at
which the lapse rate decreases to 2 \degree C/km \cite{reichler_determining_2003}, $\Delta_{H}$ is the tropospheric mean
meridional potential temperature gradient, and $\Omega$ is the angular velocity of the earth. From the thermodynamic equation and the mass
continuity equation \cite{vallis2017atmospheric}, the meridional overturning stream function $\Psi$ of the circulation scales as,
\begin{equation}
\Psi \sim\left(\frac{g H \Delta_{H}}{\Omega^{2} R^{2}}\right)^{\frac{3}{2}} \frac{R H \Delta_{H}}{\tau \Delta_{V}} \propto \frac{H^{\frac{5}{2}} \Delta_H^{\frac{2}{5}}}{\Delta_{V}},
\label{EQ121}
\end{equation}
where $\Delta_{V}$ is the dry static stability of the tropical troposphere and $\tau$ is the characteristic overturning time scale of the circulation.

Although the conventional analysis of satellite
observations and the climate change simulations  suggests a poleward expansion
of the Hadley cell, there are still two challenges. (i) The latitude–longitude structure of this expansion is not fully
resolved. (ii) In contrast to models and theoretical considerations, which predict a decreasing intensity, an increasing
trend was found in the intensity of the  Hadley cell in reanalysis datasets \cite{mitas_has_2005}. To overcome these unsolved issues, Fan\textit{ et al.} developed an approach based on network
and percolation frameworks to study the impacts of climate
changes on the atmospheric circulations \cite{fan2018climate}. They found an abrupt transition
during the evolution of the climate network, indicating a
consistent poleward expansion of the largest cluster that corresponds
to the tropical area, as well as a weakening of the
strength of link in the tropics. This was found  in both the reanalysis
data and  the CMIP5  simulations. The underlying mechanism for the observed
expansion of the tropical cluster was linked to the weakening and expansion
of the Hadley cell.

\begin{figure}[]
\begin{centering}
\includegraphics[width=1\linewidth]{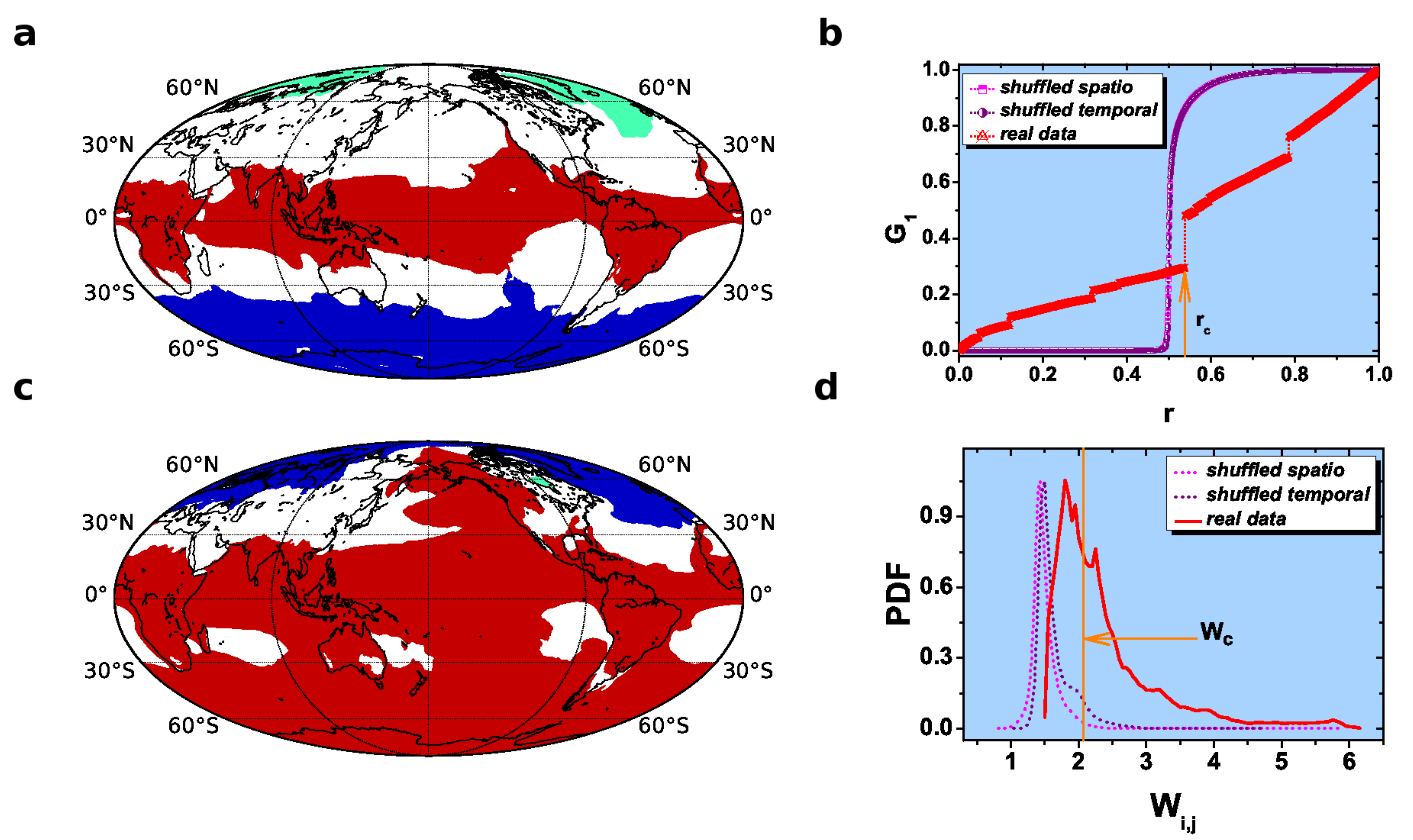}
\caption{\label{Fig3_17} 
{\bf  The network–percolation approach to the atmospheric circulation}. Snapshots of the  component structures of the CN. (a) Just before the percolation threshold. (b) The giant cluster relative size $G_1$ versus the fraction of total number of links $r$, for real (red), spatially shuffled (blue) and temporally shuffled (purple) records. (c) Just before the formation of the spanning cluster (at the second largest jump at $r\approx0.8$).   (d) The PDF of the weight of links $W_{i,j}$ around the globe, in real (solid line) and shuffled (dashed lines) data. The vertical line (orange) indicates the strength of the critical link, $W_c$, at the percolation threshold $r_c\approx0.53$. \textit{Source}: Reprinted figure from Ref. \cite{fan2018climate}.}
\end{centering}
\end{figure}

In Ref. \cite{fan2018climate}, the CNs were embedded into a two-dimensional lattice, where only nearest neighbor links is considered. The strength of each link are calculated based on  Eq. (\ref{EQ13}),
with the month-to-month temperature difference.
The links are sorted in decreasing order of strength and then added one by one according to the decreasing strength. More specifically, the nodes that are more similar are connected first. Existing clusters grow when a new link connects one cluster to another one. Then the CN undergoes an abrupt and statistically significant phase transition, i.e., exhibiting a significant discontinuity in the order parameter $G_1$, the relative size of the largest cluster. Due to the Earth’s spherical shape, the largest component in the climate networks is redefined as
\begin{equation}
G_1(M) = \frac{\max \left[\sum\limits_{i\in S_1 (M)} \cos(\phi_i),\cdots, \sum\limits_{i\in S_m (M)} \cos(\phi_i),\cdots,\right]}{\sum\limits_{i=1}^{N} \cos(\phi_i)}, 
\label{EQ122}
\end{equation}
where $\phi_i$ is the latitude of node $i$.  
The percolation threshold is determined according to Eq. (\ref{EQ46}), i.e., the step with the largest jump is regarded as the phase transition point. Fig.~\ref{Fig3_17}(a) shows the CN component  structure in
the global map at the percolation threshold. Just before this jump, the CN is characterized by three major communities; the largest one is located in the tropical region; the second and third largest are located in the high latitudes of the southern and northern hemispheres. Fig.~\ref{Fig3_17}(b) depicts the relative size of the largest cluster, $G_1$, as a function of the link occupation probability $r$ in the evolution of the CN. It has been found that $G_1$ exhibits an abrupt jump at the percolation threshold $r_c\approx 0.53$. The probability density function (PDF) of the weight $W_{i,j}$ of links is shown in Fig.~\ref{Fig3_17}(d).

To determine the temporal evolution of the largest component $G_c$, and its intensity $W_c$ (the weight of the critical link that leads to the largest transition), a sequence of networks based on successive and non-overlapping temporal windows with lengths of $5$ years each was constructed. The results suggested that the tropical cluster is expanding  poleward, meanwhile, its intensity  decreases significantly with time. The
robust weakening and poleward expansion of the tropical component have been observed both in the reanalysis
data and in 31 CMIP5  models.  The network–percolation approach was also used to identify the climate change response, by comparing the topology of
the tropical component for the first and last twenty years of the
21st century, i.e., 2080–2100 vs. 2006–2026. It was found that some regions, for example,
northern India, southern Africa, and western Australia have
a higher probability to be influenced by the tropical component, whereas the impacts in other regions, e.g., the Northeast
Pacific, will become weaker in the future.

\subsubsection{Atlantic Meridional Overturning Circulation}

\begin{figure}[]
\begin{centering}
\includegraphics[width=1\linewidth]{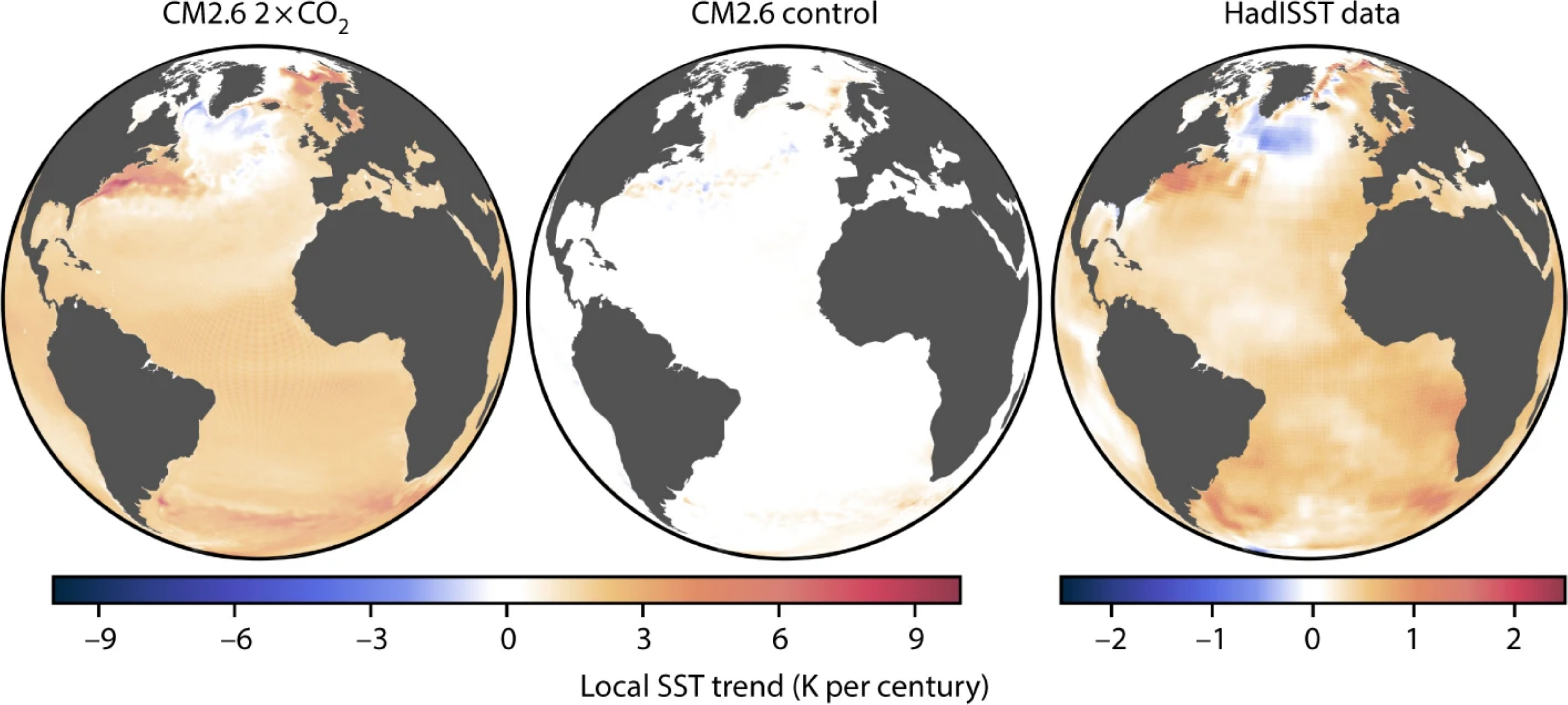}
\caption{\label{Fig3_18} 
{\bf  Slowdown of the Atlantic Meridional Overturning Circulation (AMOC)}. Comparison of SST trends in model and observations. Linear SST trends for the CM2.6 climate model of the
Geophysical Fluid Dynamics Laboratory during a $\mathrm{CO_{2}}$-doubling
experiment (left) and in a control run with fixed $\mathrm{CO_{2}}$ concentrations
(middle). Right, Linear SST trends for the observed HadISST data from 1870 to 2016. The data is shown from the November–May season. \textit{Source}: figure from Ref. \cite{caesar_observed_2018}.}
\end{centering}
\end{figure}

The Atlantic Meridional Overturning Circulation (AMOC) is a crucial part of the climate system in the North Atlantic. It has a major impact on climate  because of its  redistributing meridional heat transport \cite{ganachaud_improved_2000,johns_continuous_2010}. The AMOC has been identified as
one of the important tipping elements  of the
Earth system  \cite{lenton_tipping_2008}. Changes in the AMOC have not only affected the North Atlantic and
surrounding landmasses, but have also had global impacts \cite{rahmstorf_ocean_2002}. For example, negative change leads to increasing storminess
in southern Europe \cite{jackson_global_2015}, connects to above average
sea-level rise at the US east coast \cite{sallenger_hotspot_2012}, and associates with increasing drought
in the Sahel (for both frequency and intensity) \cite{defrance_consequences_2017}. 
Recently, Caesar \textit{et al.} provided  solid evidence for a weakening of the AMOC by about $3 \pm 1$ (around 15\%) sverdrups since the mid-twentieth century \cite{caesar_observed_2018}. This weakening is revealed by  the consisting of a pattern of cooling in the subpolar Atlantic Ocean and warming in the Gulf Stream region (see Fig. \ref{Fig3_18}). The slowdown of the AMOC  was found in both in a high-resolution climate model (CM2.6) in response to increasing atmospheric carbon dioxide concentrations, and in the observed HadISST data since the late 19th century. 
An improved SST-based AMOC index (the subpolar cold patch) was developed, which
is optimized in its regional and seasonal coverage to reconstruct
AMOC changes. In particular, it was found that the AMOC decline since
the 1950s is very likely to be largely anthropogenic. This slowdown of  the AMOC is
mainly caused by the gradually varying freshwater forcing in the
northern North Atlantic.

The AMOC is the zonally averaged volume
transport and its strength at 26\degree~ N in the Atlantic and is now
routinely monitored by the RAPID-MOCHA array \cite{cunningham_temporal_2007}.
Mean patterns of the meridional overturning circulation in the Atlantic 
are determined from the Community Earth System Model (CESM) simulation \cite{hurrell_community_2013,danabasoglu_community_2020}. The AMOC has its maximum at
about 1000 m depth around the separation latitude of the Gulf Stream with
a strength of about 20 Sv. 
Since a future collapse of the AMOC has been
identified as a tipping element, it is
therefore crucial to develop early warning indicators for such a potential collapse. As we discussed in Section \ref{subsec:Tipping}, various techniques have been developed and successfully applied to detect early warning indicators in single time series used based on the concepts of critical slowing down \cite{lenton_early_2011}. In particular, early warning indicators for the tipping
point of the AMOC have been discussed in \cite{held_detection_2004,livina_modified_2007}.

Recently, several CN-based warning-signals have been proposed by analyzing the correlations for the AMOC collapse that efficiently monitor
spatial changes in deep ocean circulation. Specifically, the values
of the mean degree, its kurtosis, assortativity, and clustering increase when approaching a tipping point \cite{mheen_interaction_2013,feng_deep_2014}.
The CN methodology is applied to temperature time
series from an idealized ocean model of the AMOC, as well as to AMOC
strength time series from a coupled atmosphere-ocean GCM.
In addition, the optimal
locations of measurement of the AMOC are formulated to provide early warning
signals of a collapse through the analysis of the performance of this indicator. Next, we will  provide an overview of these CN-based early warning indicators
for the collapse of the  AMOC.

\begin{figure}[]
\begin{centering}
\includegraphics[width=0.85\linewidth]{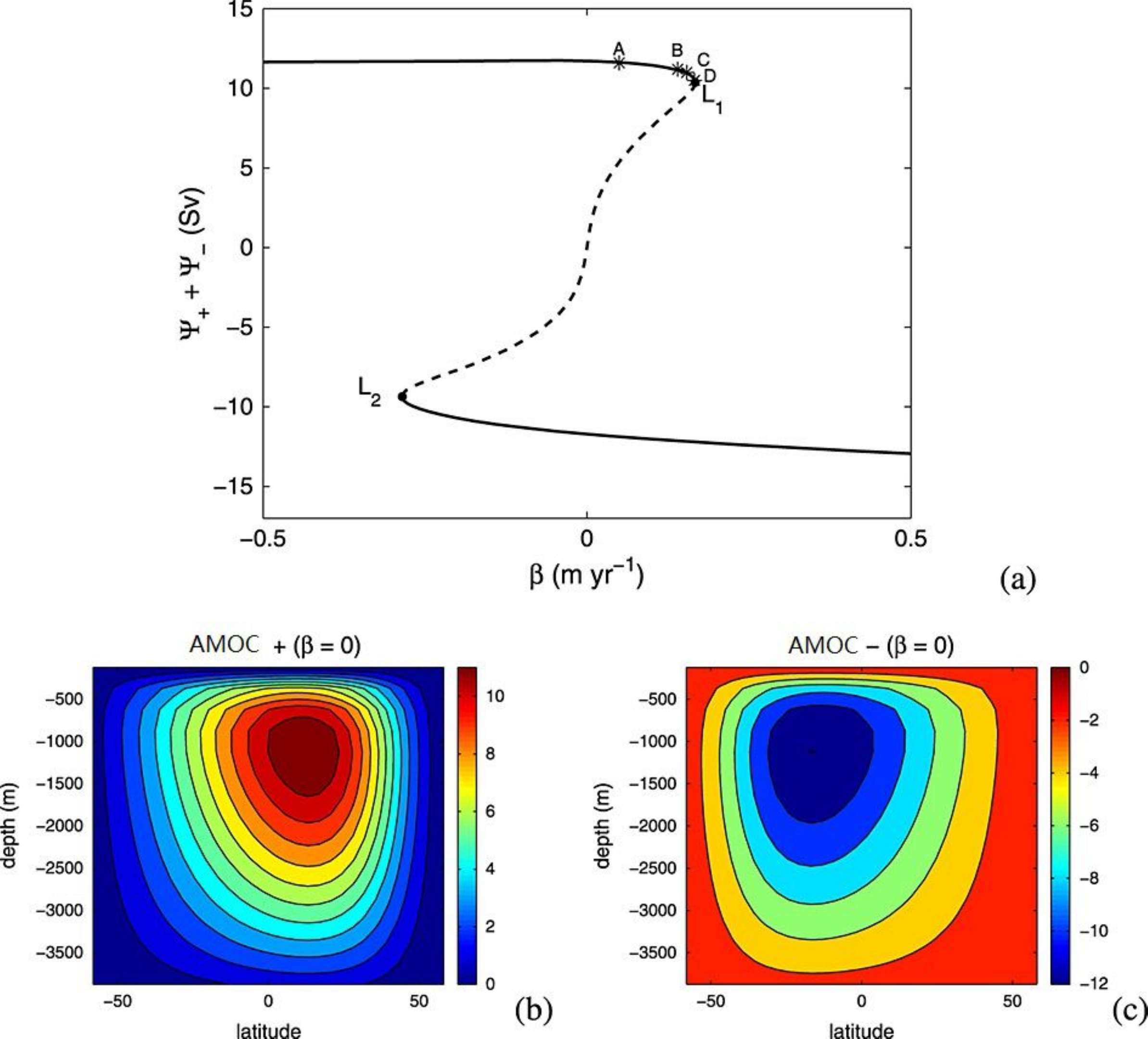}
\caption{\label{Fig3_19} 
{\bf  Tipping points of the AMOC in the idealized ocean model}. (a) Bifurcation diagram with the sum of the maximum ($\Psi_{+}$) meridional overturning stream function and minimum ($\Psi_{-}$) meridional overturning stream function vs. the anomalous freshwater
flux strength $\beta$. (b) Pattern of the meridional overturning
stream function ($\Psi$) at the upper stable branch. (c) Pattern of $\Psi$ at the lower stable branch. \textit{Source}: Reprinted figure from Ref. \cite{mheen_interaction_2013}.}
\end{centering}
\end{figure}

Mheen\textit{ et al.} \cite{mheen_interaction_2013} used a dimensional
(meridional-depth) idealized ocean model of the AMOC, where tipping points can be explicitly
computed. In this model, there are two active tracers: temperature $T$ and salinity $S$, which are related to the density $\rho$ by a linear equation of state \cite{den_toom_spurious_2011}
\begin{equation}
\rho=\rho_{0}\left(1-\alpha_{T}\left(T-T_{0}\right)+\alpha_{S}\left(S-S_{0}\right)\right),
\label{EQ123}
\end{equation}
where $\alpha_{T}$ and $\alpha_{S}$ are the thermal expansion and saline contraction coefficients, respectively, and $\rho_{0}$, $T_{0}$ and $S_{0}$ are reference quantities. The surface freshwater forcing $F_{S}$, which can be applied as a virtual salt flux,  is
\begin{equation}
F_{S}(\phi)=\frac{(\gamma+\eta)}{\cos \phi} \cos \pi \frac{\phi}{\phi_{N}}+\beta F_{p}(\phi)-Q,
\label{EQ124}
\end{equation}
where $\phi$ indicates latitude, $\phi_{N}$ is the northern boundary of the equatorially symmetric domain, $\gamma$ is the strength of the background freshwater forcing, and $\eta$ is a white noise term. $\beta$ is the strength of an anomalous freshwater flux which is only applied over the area $\left[40^{\circ} \mathrm{N}, 60^{\circ} \mathrm{N}\right]$, and $Q$ is a constant used to normalize the surface-integrated salt flux from 0 to 1 for all parameter values. As shown in Fig. \ref{Fig3_19},
the bifurcation diagram of the AMOC in the idealized ocean model indicates the existence of tipping points at $L_1$ and $L_2$. The model was discretized on a 32 $\times$
16 spatial grid, with four values of $\beta$ close to $L_1$, labeled A, B, C and D.
The output of this model is the temperature field time series.

\begin{figure}[]
\begin{centering}
\includegraphics[width=0.85\linewidth]{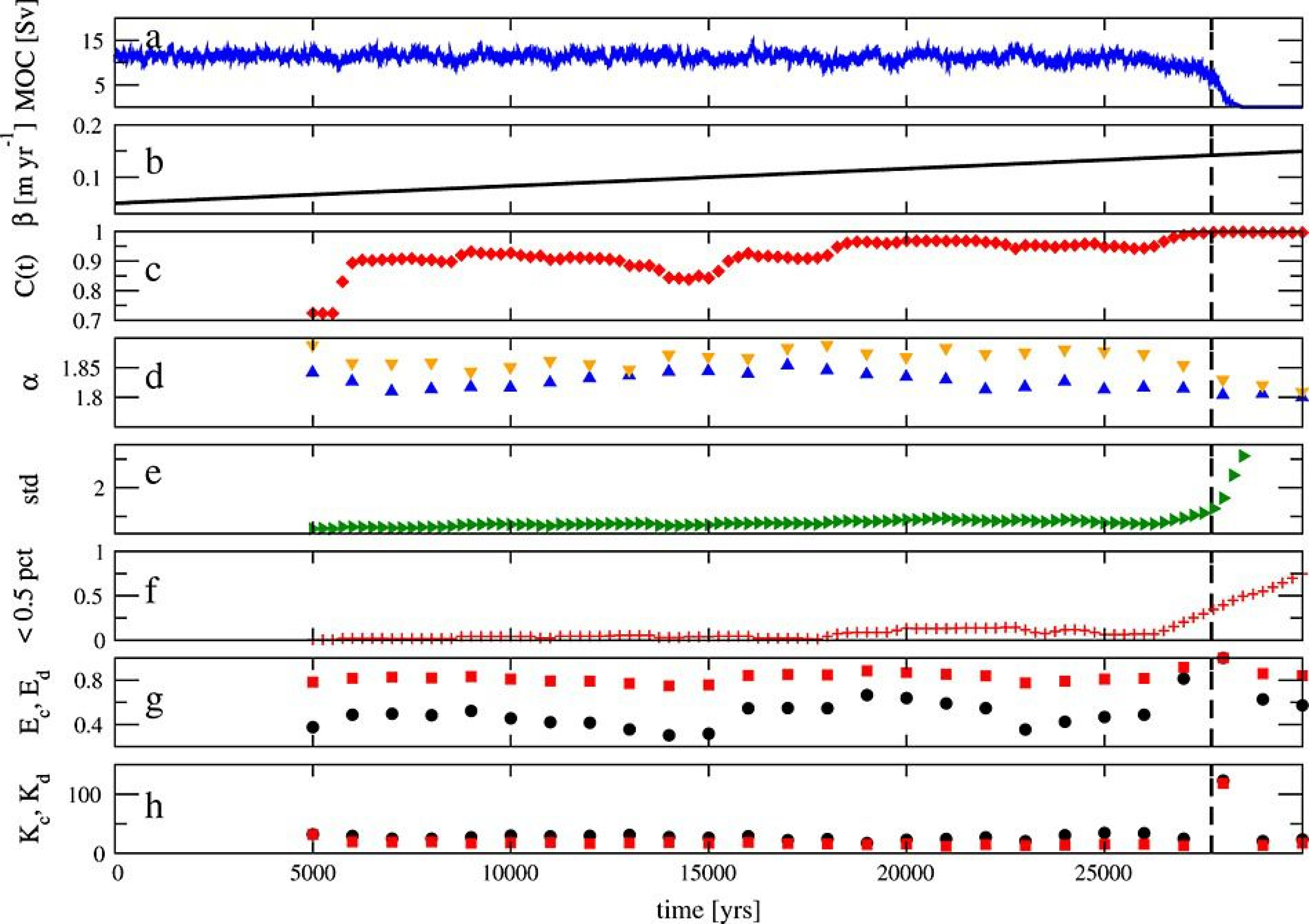}
\caption{\label{Fig3_20} 
{\bf  Early warning signals of the AMOC}. (a) Time series of the maximum value of the AMOC. (b) Transient
change of $\beta$ versus time from 0.05 to 0.17 $myr^{-1}$ during 35,000 years. (c) Th lag–1 autocorrelation of the projection
of the time series onto the first EOF. (d) The DFA exponent, using linear
(up) and quadratic (down) detrending. (e) Standard deviation of the AMOC. (f) Fraction of the time spent below the 0.5 percentile of the AMOC strength. (g) The network-based indicators
$E_d$ (circles) and $E_c$ (squares). (h) Same as (g) but for the kurtosis ($K_d$ (circles) and
$K_c$ (squares). The vertical dashed line indicates the value of $\beta$ at the tipping point. \textit{Source}: figure from Ref. \cite{mheen_interaction_2013}.}
\end{centering}
\end{figure}

Next, an undirected and unweighted CN is constructed based on the Pearson correlation method (see Sec. \ref{cap2:CN}) with no time lags and a threshold $C_c = 0.7$ \cite{mheen_interaction_2013}, i.e, if the
correlation $C_{i,j} >$  $C_c$, the corresponding two nodes $i$ and $j$ are connected. Four network-based indicators
$E_d$, $E_c$, $K_d$ and $K_c$ were developed to study the dynamic evolution of the AMOC. Here
$E_d$ is defined as the
expectation value of the normalized degree distribution $d/d_{max}$; $E_c$ is defined as the
expectation value of the normalized clustering coefficients $c/c_{max}$; $K_d$ is the kurtosis
of the degree distribution; and $K_c$ is the kurtosis  of the clustering coefficients distribution.
These four network indicators are presented in Fig. \ref{Fig3_20}. It was found that $E_d$ increases steeply and smoothly when the tipping point $L_1$ is approached. The
explanation for this increase is based on the increased dominance
of the first EOF in the variance of the solution in a
large part of the domain. The same holds for the indicator $E_c$ (Fig. \ref{Fig3_20}g). 
The kurtosis
distributions $K_d$ and $K_c$ (Fig. \ref{Fig3_20}h) show even more clearly 
a strong increase when the tipping point $L_1$ is approached before the collapse.
Note that the CNs were constructed from the full spatial temperature
field using a sliding window of 5000 years with a shift
of 1000 years.  The results were not sensitive to the sliding window length.
For comparison, previously suggested early warning indicators of
the transition in the same AMOC time series, such as, lag–1 autocorrelation, DFA, standard deviation (see our discussion in Section \ref{subsec:Tipping}) are also plotted in Fig. \ref{Fig3_20} and tend to exhibit a rapid increase/decrease when approaching the tipping point. In particular, the coefficient of the lag-1 autocorrelation (Fig. \ref{Fig3_20}c)
reaches its maximum near the tipping point indicating a critical slowdown of
the AMOC. However, it has been found that the indicator is not monotonic and also increases when the AMOC is still far from the transition. The DFA analysis shown in Fig. \ref{Fig3_20}d does not indicate early warnings of a transition. Fig. \ref{Fig3_20}e depicts the standard deviation of the AMOC record.  It shows a steady increase toward the transition but it is difficult to set a threshold for an alarm. In Fig. \ref{Fig3_20}f, the fraction of the time spent below the 0.5 percentile
(estimated based on the first 5000 years) of the AMOC strength is shown. This indicator displays a much sharper increase, but it is a rather ad-hoc measure which depends on a calibration in the far past \cite{mheen_interaction_2013}.

The same CNs-based analysis was also performed in a more realistic Fast Met Office/UK Universities Simulator (FAMOUS) model by Feng \textit{ et al.} \cite{feng_deep_2014}. The FAMOUS model is a reduction  for Hadley Centre Coupled Model with a lower resolution ocean and atmosphere component. The annual mean of AMOC strength data were obtained from a control simulation and from a freshwater-perturbed (referred to as ``hosing'') simulation of the FAMOUS model. In the hosing simulation, the freshwater flux over the extratropical North Atlantic was increased linearly from zero to 1.0 Sv over 2000 years \cite{feng_deep_2014}. The complete AMOC streamfunction field for each of the six 100-year equilibrium simulations has been considered. It was found that when the freshwater forcing is increased, a high degree in the network – indicating high spatial AMOC correlations – first appears at nodes in the South Atlantic at about 1000 m depth. The kurtosis $K_d$ of the CN was introduced as an effective indicator to capture the changes in the topology of the degree field. For the hosing simulation, there is a strong increase of $K_d$ to values far extending those for the control simulation significantly before the collapse time. However, the critical slowdown indicators variance  and lag-1 autocorrelation do not show any early warning signal. Moreover, the kurtosis $K_d$ can also be used to determine the optimal observation locations of the AMOC, which provides a strong anomalous signal at least 100 years before the transition.

\subsection{Earth Geometric Surface Relief}
\label{cap3:EGSR}

The topography or bathymetry of the Earth shows complex multifractal structures and scaling properties~\cite{mandelbrot_stochastic_1975,gagnon_multifractal_2006,sapoval_self-stabilized_2004,maritan_universality_1996}, which can be regarded as a consequence of plate tectonic processes. Revealing the relations between geometrical features of terrestrial surfaces and their internal geological  processes has long been a fundamental challenge  in
the Earth sciences. The plate tectonic theory is usually used to study most of the major surface topographic features of the Earth \cite{gill2012orogenic}. It has been reported that there exist strong connections between the ocean bottom topography and the Earth's climate \cite{jayne2004connections}. Moreover, the surface topography of the Earth plays a remarkable role in the dynamical evolution of oceans, especially, in regard to  global climate change and sea level rising. Sea level rise is one of the major consequences of global climate warming~\cite{slangen_anthropogenic_2016} and the effects in 
combination with storm surges and other extreme events have been
 observed~\cite{sweet_extreme_nodate}. Decreasing global CO2 emissions is crucial for limiting the risks of sea-level rise. It is estimated that the median sea-level
rise will be between 0.7 and 1.2 m until the year 2300 even within the constraints of the Paris Agreement ~\cite{mengel_committed_2018}. The world’s large ice sheets in Greenland and Antarctica are considered to be the largest possible contributors to sea level rise. If these ice sheets melt completely, the sea level would rise about 7 m, 5 m and 53 m from the Greenland ice sheet, the West Antarctic ice sheet, and the East Antarctic ice sheet, respectively.
There is thus a great interest in clarifying/predicting the influenced areas in response to sea level rise. 

\begin{figure}[]
\begin{centering}
\includegraphics[width=0.85\linewidth]{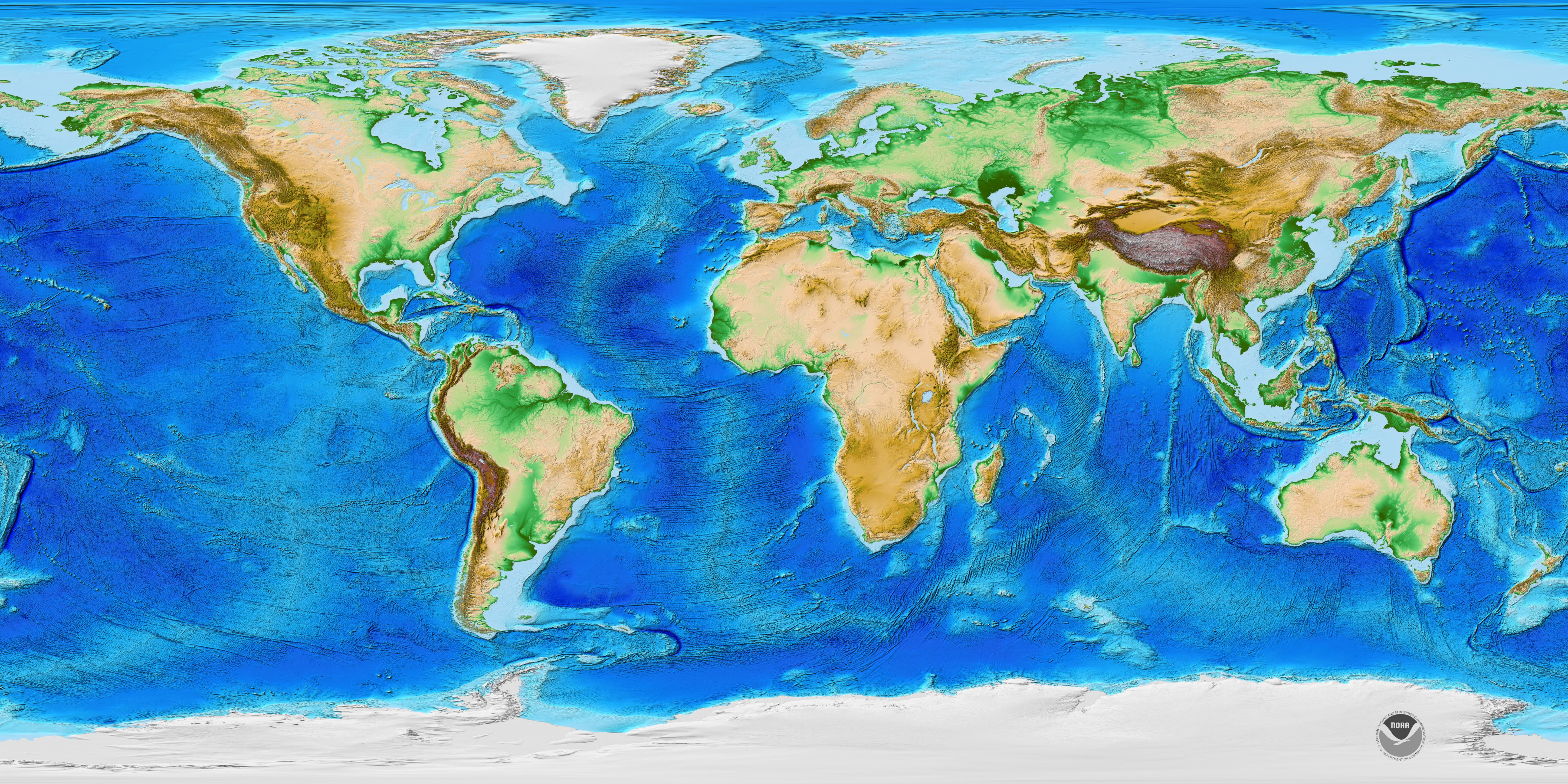}
\caption{\label{Fig4_1} 
{\bf ETOPO1 Global Relief Model}. ETOPO1 is a 1 arc-minute global relief model of the Earth's surface that integrates land topography and ocean bathymetry. The present mean sea level (zero height) is assumed
as a vertical datum of the height relief. Figure is from NOAA \url{https://www.ngdc.noaa.gov/mgg/global/}.}
\end{centering}
\end{figure}

Recently, the ETOPO1 Global Relief Model was released and  provided new opportunities for a better understanding of Earth's surface processes based on geomorphic signatures \cite{amante1noaa}. ETOPO1 is a 1 arc-minute global relief model of the Earth's surface that integrates land topography and ocean bathymetry and was developed by NOAA. It was built from global and regional data sets and used to calculate the volumes of the world's oceans and to derive a hypsographic curve of Earth's surface (see Fig. \ref{Fig4_1}).

In this Section, we will review statistical properties of the
Earth geometric surface relief using percolation theory. Their fractal structure and scaling will also be discussed. In particular, the present mean sea level on Earth will be shown to coincide with the critical threshold in a percolation description of the global topography; strong evidence reveals  abrupt transitions that occurred during the evolution of the Earth’s relief network, indicative of a continental/cluster aggregation. This could help us to identify the critical nodes or locations that will be more exposed to global climate change.

\subsubsection{Self-similarity and Long-range Correlations}

Self-similarity and long-range correlations are  remarkable features of the Earth’s surface topography \cite{mandelbrot_stochastic_1975}. Starting about 30 years ago, new ideas in nonlinear dynamics, particularly fractals and scaling, provoked an explosive growth of research both in modeling and in experimentally characterizing the solid earth geophysics including the topography, for a review see Ref. \cite{lovejoy_scaling_2007}.
The power spectrum $S$ of linear transects of Earth's topography follows the scaling relation $S(k) \sim k^{-\beta_{c}}$, where $k$  stands for  the wave number \cite{mandelbrot_stochastic_1975}. Such a scaling relation was used to measure the self-similarity of the Earth’s surface topography.  Moreover, similar scaling relations have been identified  in Earth's sea-floor topography \cite{bell_statistical_1975},
the topology of river networks \cite{mantilla_testing_2010}, as well as
natural rock surfaces \cite{brown_broad_1985}. 

The long-range correlation features of the Earth’s surface topography can be described in the following way. The exponent $\beta_{c}$ in the power-law spectrum is related to the Hurst exponent $H$ in fractional Brownian motion (fBm) via $\beta_{c}=2 H+1$, which results in $H \simeq 0.5$ for the Earth's topography. Whereas further universal multifractal parameters estimated for specific parts have resulted in a much more complex structure, i.e.,  $H=0.46, 0.66, 0.77$ for bathymetry, continents and continental
margins, respectively \cite{gagnon_multifractal_2006}.

The third characteristic  of the Earth's  surface topography is the well known bimodal distribution, as a consequence of
plate tectonic processes \cite{wegener1966origin}.  It indicates the topographic dichotomy of continents and ocean basins.

\subsubsection{Landmass and Oceanic Clusters}

Although the classical pate tectonic theory provided a framework that might explain most of the major surface topographic features of the Earth,  a  percolation-based  description as discussed in Sec.(\ref{cap2:Percolation})  was developed to  study the geometrical features of the Earth from a statistical physics perspective \cite{ali_saberi_percolation_2013,fan_percolation_2019}. The analysis is
based on high-resolution ETOPO1 global relief records. The resolution is 1 arc-minute, i.e., $N = 10800 \times 21600$ grid points. The present  mean sea level (zero height) is assumed as a vertical datum of the height relief $h(\phi_i,\theta_i)$, where $\phi_i$, $\theta_i$  are the corresponding latitude and longitude of grid point $i$.

\begin{figure}[]
\begin{centering}
\includegraphics[width=0.75\linewidth]{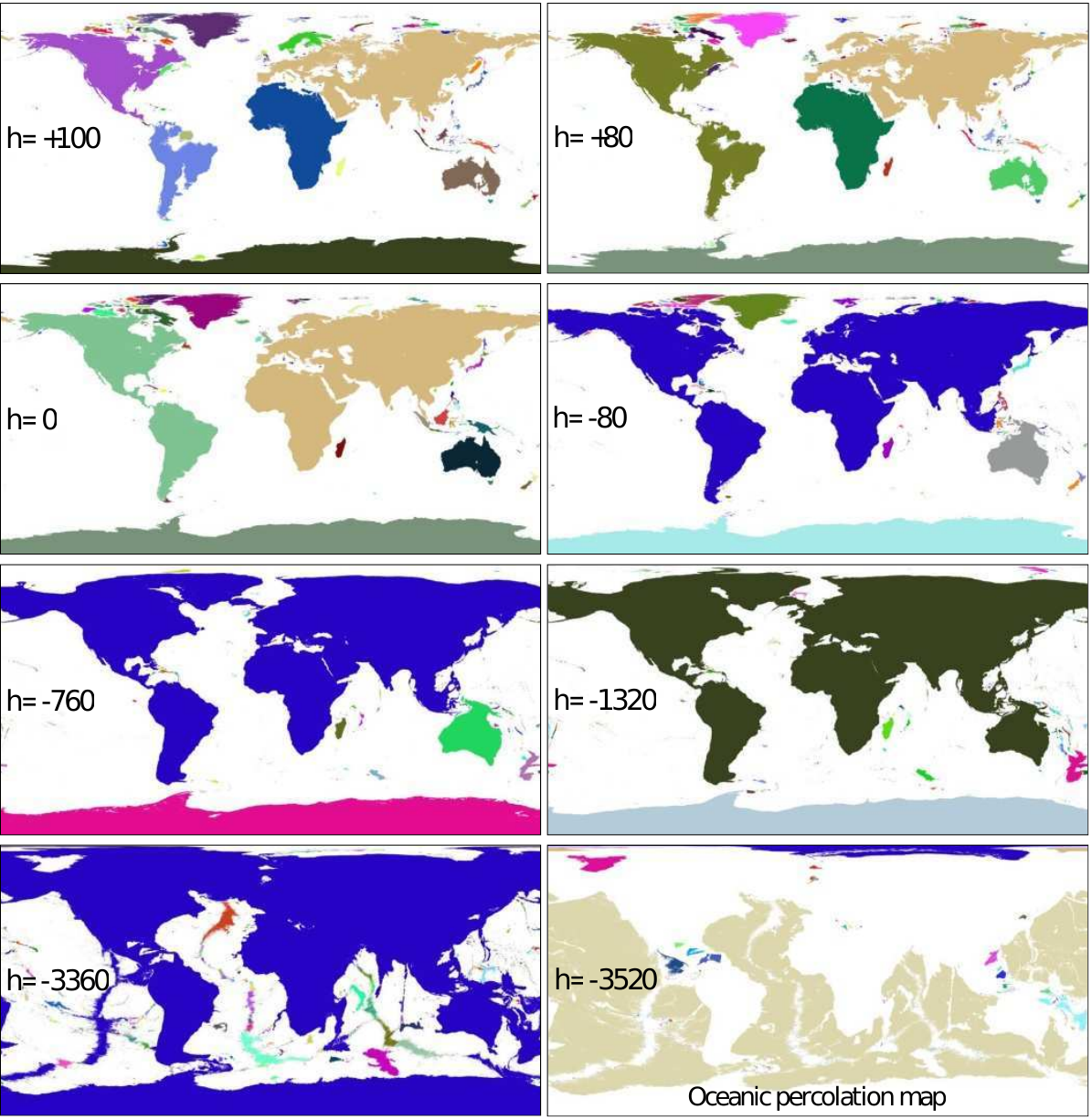}
\caption{\label{Fig4_2} 
{\bf Schematic illustration of the landmass cluster aggregation by decreasing the sea level}. The nodes are occupied if its corresponding height is above the given threshold $h$. Different colours stand for different clusters.   \textit{Source}: figure from Ref.  \cite{ali_saberi_percolation_2013}.}
\end{centering}
\end{figure}
\begin{figure}[]
\begin{centering}
\includegraphics[width=0.75\linewidth]{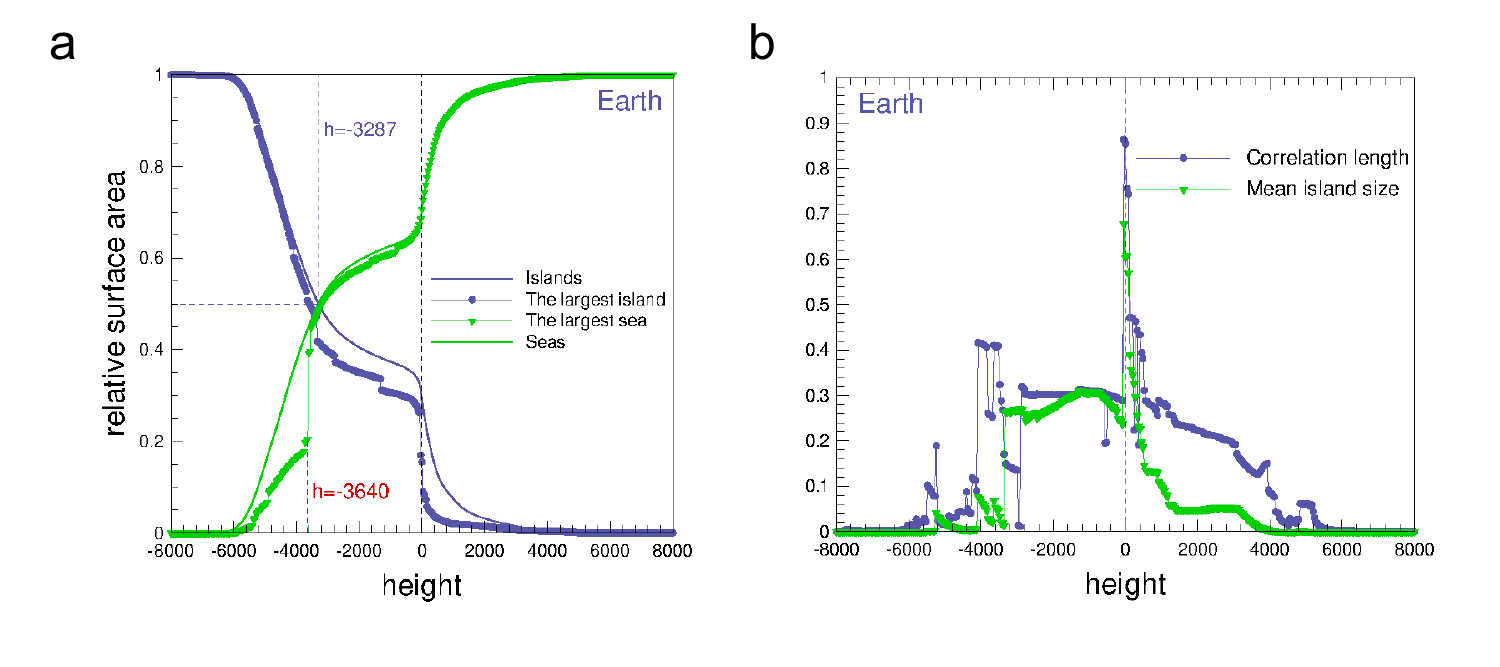}
\caption{\label{Fig4_3} 
{\bf Percolation analysis of the Earth's topography}. (a) Relative surface area of the largest
island (circles) and the largest sea (triangles) as a function of the sea level. The landmass critical level is close to the current sea level $h = 0$, whereas,  the oceanic
critical level is close to the level $h=-3287$ m at which the
total island and oceanic surface areas are equal.  (b) The  correlation length $\xi$ and mean island size $\chi$ as a function of the sea level. The dashed line marks a geometrical percolation phase transition at the present mean sea level. \textit{Source}: Reprinted figure from Ref. \cite{ali_saberi_percolation_2013}.}
\end{centering}
\end{figure}

In Ref. \cite{ali_saberi_percolation_2013}, the hypothetical water level
was considered as the percolation control parameter. When it is decreased from the highest to lowest available heights on Earth, there occurs a geometrical percolation phase transition at a critical level $h_c$ around which  most parts of landmass join together. Fig. \ref{Fig4_2} illustrates the landmass cluster aggregation by decreasing the sea level from $h=100~\mathrm{m}$ (top) to $h=-3520~\mathrm{m}$ (bottom). To determine the percolation threshold, the probability
of any nodes to be part of the largest island was defined as the order parameter; the mean island size  $\chi$ (analogous to the susceptibility of the system) was defined as,
$\chi=\sum_{s}^{\prime} s^{2} n_{s}(h) / \sum_{s}^{\prime} s n_{s}(h),$ where $n_{s}(h)$ denotes the average number of islands of size $s$ at level $h$ and the prime on the sums stands for the exclusion of the largest island; the correlation length $\xi$ is defined as the average distance of nodes belonging to the same island cluster, $\xi^{2}=\sum_{s}^{\prime} 2 R_{s}^{2} s^{2} n_{s}(h) / \sum_{s}^{\prime} s^{2} n_{s}(h),$ where $R_{s}$ is the radius of gyration of a given $s$ cluster. It has been found that the order parameter for islands has an abrupt jump around the zero height level $h=0$, i.e., right at the present mean sea level (Fig. \ref{Fig4_3}a). It was furthermore found, that both quantities $\chi$ and $\xi$ become divergent at the present mean sea level, see Fig. \ref{Fig4_3}b. From these results, the most remarkable observation
is that the critical level $h_c$ coincides with the present mean sea level $h = 0$ on Earth.  This may suggest the important role of water on Earth and shed new light on the tectonic plate motion \cite{ali_saberi_percolation_2013}.

The same percolation analysis was also performed for the oceanic
clusters, i.e., the hypothetical water level was increased from the lowest to highest. Fig. \ref{Fig4_3}b gives rise to a discontinuous
jump in the oceanic order parameter at around $-3640$ m. 
Further investigation in the lunar topography reveals various characteristic features of the Moon. It was found that the critical level for the Moon has
the same amount of land and oceans at the threshold,
indicating a purely geometrical phase transition \cite{ali_saberi_percolation_2013}. 

\subsubsection{Origin of the Discontinuity}

It has been reported that a random network or lattice
system always undergoes a continuous percolation phase transition
and shows standard scaling features \cite{bollobas2001random}. Yet, the order of the percolation transition for the Earth's topography is still an open question. To answer this, Fan \textit{et al.} developed a more sophisticated percolation-based method \cite{fan_percolation_2019}.
The percolation model was defined as follows: starting from an unoccupied lattice, the sites are occupied one by one according to their ranking, i.e., first choosing  the site with the largest height, then the second, etc. At each step, the fraction of occupied sites $p$ increases by the inverse of the total number of sites $N$ in the Earth's relief landscape. By this procedure, a configuration of occupied sites is continually  obtained at every $p$.
Here  all the grid points in the ETOPO1 Global Relief Model were ranked according to their height $h(\phi_i,\theta_i)$, from the largest to the smallest value.

\begin{figure}[]
\begin{centering}
\includegraphics[width=0.75\linewidth]{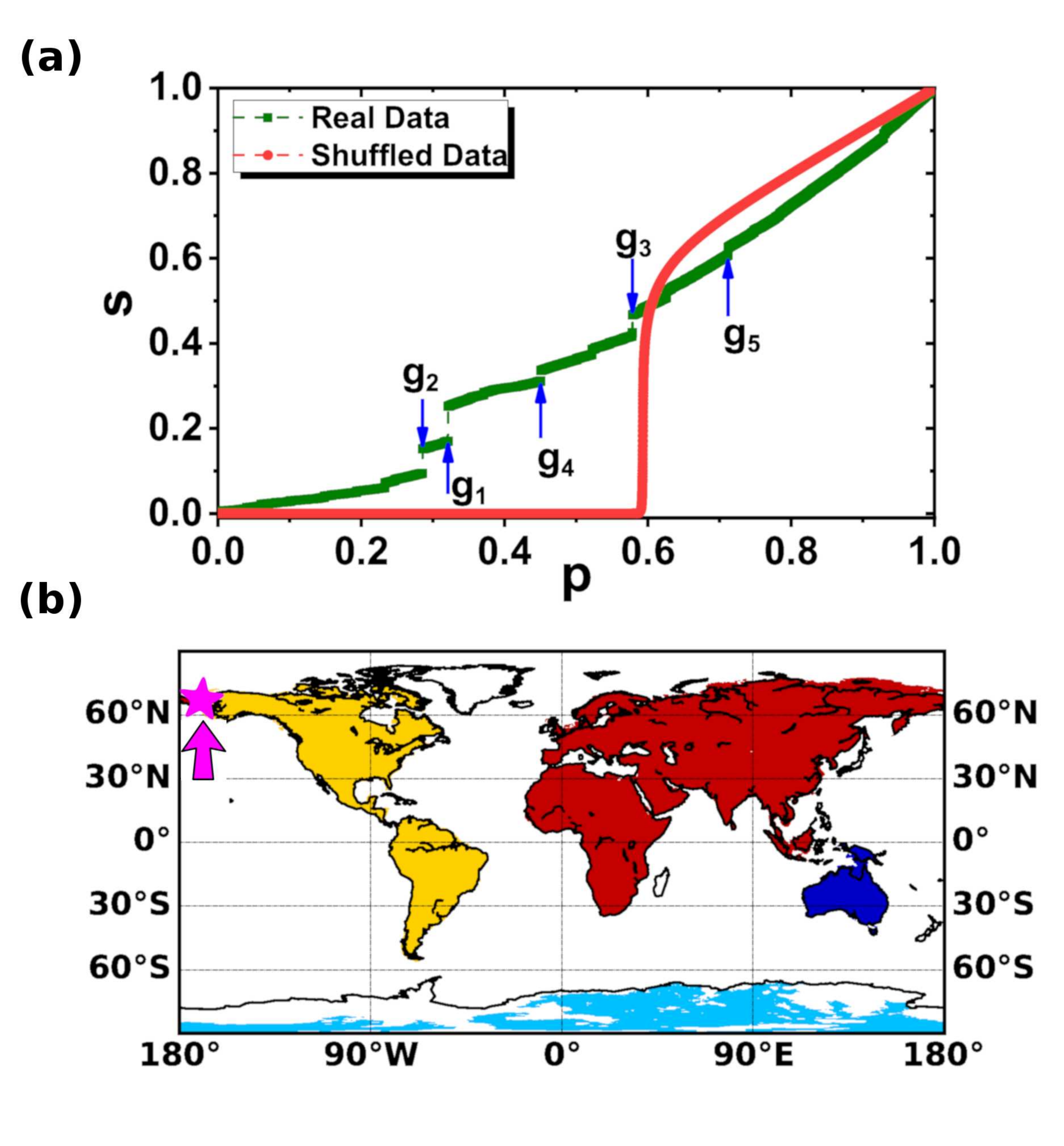}
\caption{\label{Fig4_4} 
{\bf Percolation analysis of the Earth's topography}. (a) The largest landmass cluster relative size $s$ as a function of the fraction  $p$ of the total landmass for real (green) and shuffled (red) Earth relief records.  $g_1$--$g_5$ indicate the five largest gaps, defined in Eq. (\ref{EQ133_1}).  (b) Snapshots of the landmass clusters of the Earth’s surface
topography network at the percolation threshold, $p=0.321$. The star indicates the critical node. \textit{Source}: Reprinted figure from Ref. \cite{fan_percolation_2019}.}
\end{centering}
\end{figure}

In the following, the size of the percolation landmass (oceanic) clusters $s$ were calculated
based on  Eq. (\ref{EQ122}) due to the Earth's spherical nature. The size of
the $i$-th gap $g_i$ at each fraction $p$ is defined as follows:
\begin{equation}
g_{i}(p) \equiv s(p)-s(p-1/N).
\label{EQ133_1}
\end{equation}
Specifically, here $g_1$ denotes the largest gap, $g_2$ indicates the second largest gap, etc. The larger the gap $g_i$ is, the larger are the
two clusters before merging.
The dynamical evolution of the largest landmass cluster $s$ as a function of
the fraction of occupied nodes $p$ is shown in Fig. \ref{Fig4_4}a. It was found that Earth’s relief network undergoes several abrupt
and statistically significant phase transitions, i.e., exhibiting a
 discontinuity in the order parameter $s$. Fig. \ref{Fig4_4}b shows the network  clusters landmass structure at the percolation threshold (just
before the largest gap $g_1$). The results reveal that the network, just before this jump,
is characterized by four major communities: the largest one being
the Afro-Eurasia continental landmass, the
second largest cluster is the Americas, the
third is located in Antarctica, and the fourth is Oceania. Notably, there exists a 
critical node $(64.458333~\hbox{$^\circ$} \mathrm{N},171.141667~\hbox{$^\circ$} \mathrm{W})$ that connects the
largest and second largest cluster at the percolation threshold
$p_c \approx 0.321$, with altitude level $h = - 43$ m, under the current
sea level.

The same analysis has also been  applied for the oceanic clusters, i.e., the nodes are occupied  one by one according to their hight level in increasing order.
It also has given rise to a discontinuous jump in the oceanic order parameter at the percolation threshold $p_c \approx 0.379$ with an altitude level $h= -3621$ m. This is very close to the result $h=-3640$ m reported in Ref. \cite{ali_saberi_percolation_2013}.  Note that the critical node, $(59.908333~\hbox{$^\circ$} \mathrm{S}, 161.308333~\hbox{$^\circ$} \mathrm{E})$, connects the Atlantic+Indian Ocean Plate to the Pacific Plate. It is worth noting that the role of the largest
cluster on the landmass structures is different from that on
the oceanic one. These differences
reveal the complex and different structure of the Earth’s
topography (continents) and bathymetry (oceans). This dichotomy
is manifested in the well-known bimodal distribution
of the Earth’s topography \cite{wegener1966origin}.

\begin{figure}[]
\begin{centering}
\includegraphics[width=0.85\linewidth]{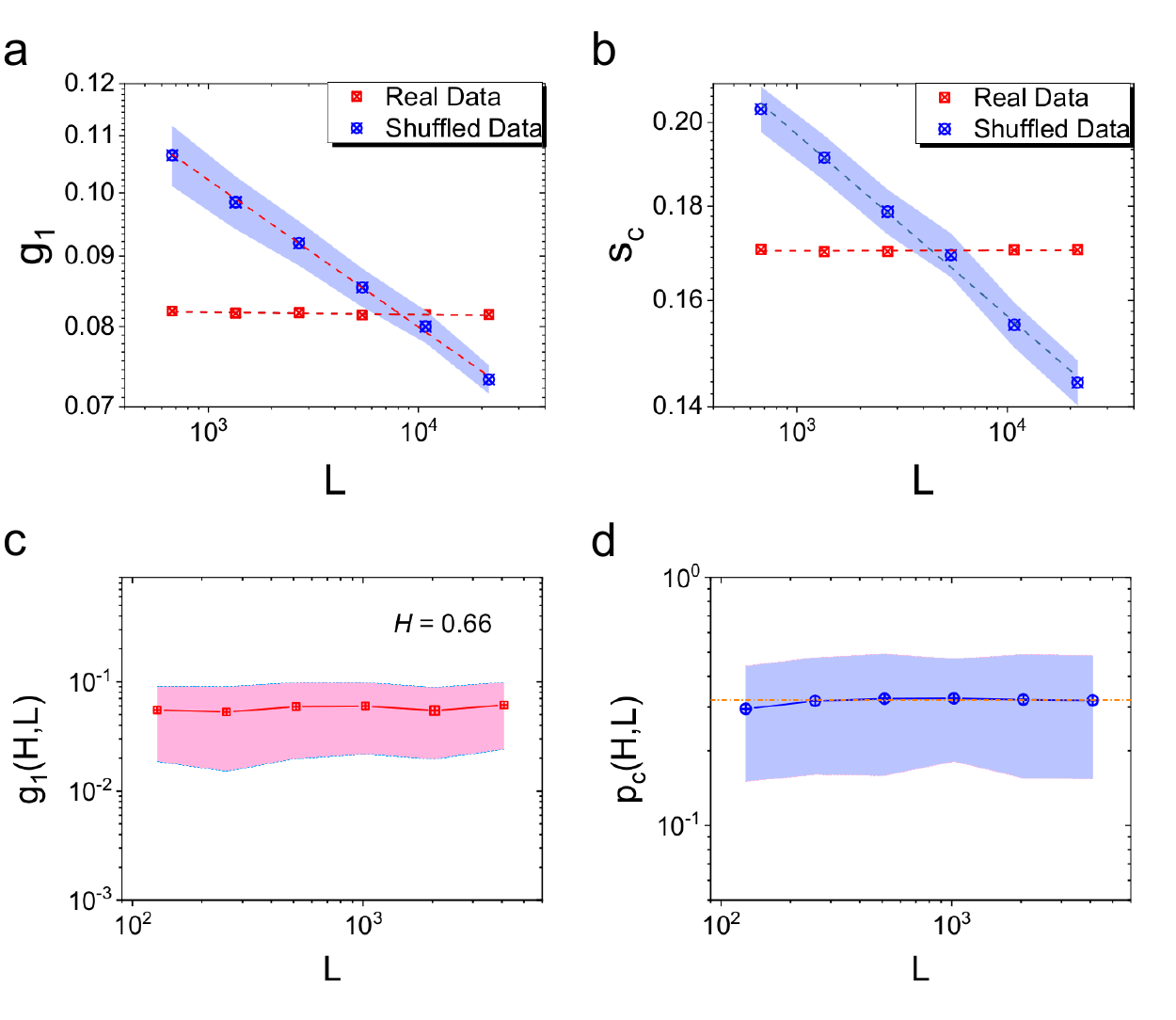}
\caption{\label{Fig4_5} 
{\bf Finite size analysis of the percolation in Earth's relief network}. (a) Log-log plot of the largest gap $g_1$ vs. the network system size $L$ for
original real data (red) and shuffled data (blue).  (b) Log-log plot of the largest landmass cluster relative size $s_c$ at the percolation
threshold vs. $L$ for real data (red) and shuffled data (blue). (c) The average of the largest jump $g_1(H, L)$ as a function of system
size $L$; (d) the corresponding percolation threshold $p_c(H, L)$ as a function of $L$ for percolation on 2D fractional Brownian motion surfaces with the Hurst exponent $H=0.66$. \textit{Source}: Reprinted figure from Ref.  \cite{fan_percolation_2019}.}
\end{centering}
\end{figure}

To demonstrate the order of the percolation phase
transition in Earth's relief network, a finite-size effects analysis was performed by altering the resolution of the nodes. The largest gap $g_1 (L)$ was  calculated for a given system size $L$.  Here, $L$ is defined as the number of nodes in the zonal direction.
According to the Eq. (\ref{EQ47}), if $g_1(L)  \to 0$ as $L \to \infty$, the corresponding giant cluster is assumed to undergo a continuous percolation; otherwise, the corresponding percolation is assumed to be discontinuous \cite{nagler_impact_2011,fan_universal_2020}.
In addition, the size of the order parameter at the percolation threshold [just before the largest jump $g_1$], $s_c (L)$, was also studied [see Eq. (\ref{EQ49})]. 
The results are shown in Fig.~\ref{Fig4_5}a and b, which indicate  a discontinuous percolation since (1) $g_1 (L)$ tends to be a non-zero constant; (2) $d_{f} -d = 0$, also indicates a discontinuous percolation.
For comparison, a randomized
topography obtained from the shuffling (spatial randomizing the height profile) of the original data was also considered (see Fig.~\ref{Fig4_5}a and b). The numerical results for the shuffled case, suggest a continuous percolation transition
with known critical exponents $\beta/\nu = 5/48 \approx 0.104$ and $d_f = 91/48$ \cite{aharony2003introduction,bunde_fractals_1996}, as expected when the  long-range correlations were destroyed.

In order to reveal the origin
of the discontinuity, Fan \textit{et al.} \cite{fan_percolation_2019} further  studied the percolation on 2D fBm surfaces with Hurst exponent $H$ \cite{du_percolation_1996,s1992fractal}. The parameter $H$ is usually between 0 and
1, where 0 is very noisy, and 1 is smoother. The Earth's rough surfaces can be modeled via a fBm \cite{family1991dynamics}, and the estimation for the Earth continents topography is $H = 0.66$ \cite{gagnon_multifractal_2006}. 
Next, the percolation analysis on fBm surfaces with $H = 0.66$ was performed. Similar to the real network, the largest cluster $s$ also exhibited abrupt transitions around $p\approx0.3$. The largest gap $g_1(H,L)$ (average) as a function of system size $L$ is shown in Fig.~\ref{Fig4_5}c. It was found that the percolation on fBm surfaces with $H=0.66$ is discontinuous, since $g_1(H,L)$ tends to be a non-zero constant when $L \to \infty$.  The percolation threshold $p_c (H,L)$ corresponding to the largest gap during the evolution of site percolation is shown in Fig.~\ref{Fig4_5}d. The values are robust and agree well with the real data where $p_c \approx 0.321$ [see Fig.~\ref{Fig4_2}]. This indicates that the combined percolation and fBm methods can be used as an efficient tool to investigate the Earth's surface topography.

\subsection{Earthquake Systems}
\label{cap3:ES}

Earthquakes are one of the most devastating natural disasters that plague society. Thus, reliable and skillful earthquake forecasts over different time scales (from days to decades) are essential for establishing rational seismic risk reduction strategies and to enhance preparedness and resilience.
The earthquake process is a complex spatio-temporal phenomenon,
and has been thought to be an example of  self-organised criticality  systems, in which events occur as cascades on a wide range of sizes, each determined by fine details of the rupture process.
Despite rapid progress in the latter part of the 20th century, the study of earthquakes, like the science of many other complex natural systems, is still in its juvenile stages of exploration and discovery.  The scientific challenge is how to understand the earthquake phenomena that are both profound and practical, in particular, the deterministic prediction of specific event sizes, their locations, and times.
Recently, probabilistic forecasting, based on
statistical patterns of earthquake occurrence, became a much more realistic goal, and has been  actively explored and tested in global initiatives \cite{de_arcangelis_statistical_2016}.

In this section, we will first present a description of the phenomenological
laws of earthquake occurrence, and scale invariance and the intertime distribution is also briefly discussed. Then we will review 
how the statistical mechanics approach (memory analysis) can be applied to improve  the forecasting skill for real earthquake catalogs.

\subsubsection{Scale Invariance and the inter-event Distribution}

First, we introduce some fundamental quantities that characterize earthquake occurrence.  This information, such as occurrence time, hypocentral
location, earthquake magnitude and depth, is usually reported in seismic catalogs  available online for different geographic areas, for instance, the United States Geological Survey (USGS) earthquake catalog\footnote{\url{https://earthquake.usgs.gov/data/}.}.  We  present  here the main phenomenological laws for the statistical distribution of these quantities.

The most common method of describing the size of an earthquake is the Richter magnitude scale $m$, which  takes the logarithm of the ground displacement as measured by a seismograph, and applies a correction which varies with the distance from the earthquake to the seismograph \cite{wells_new_1994}. The characteristic size of the fractured area $L$, as well as the
faulting duration $\tau$, can be both expressed in terms of the magnitude as $L \propto \tau \propto 10^{0.5 \mathrm{m}}$.

\begin{figure}[]
\begin{centering}
\includegraphics[width=1\linewidth]{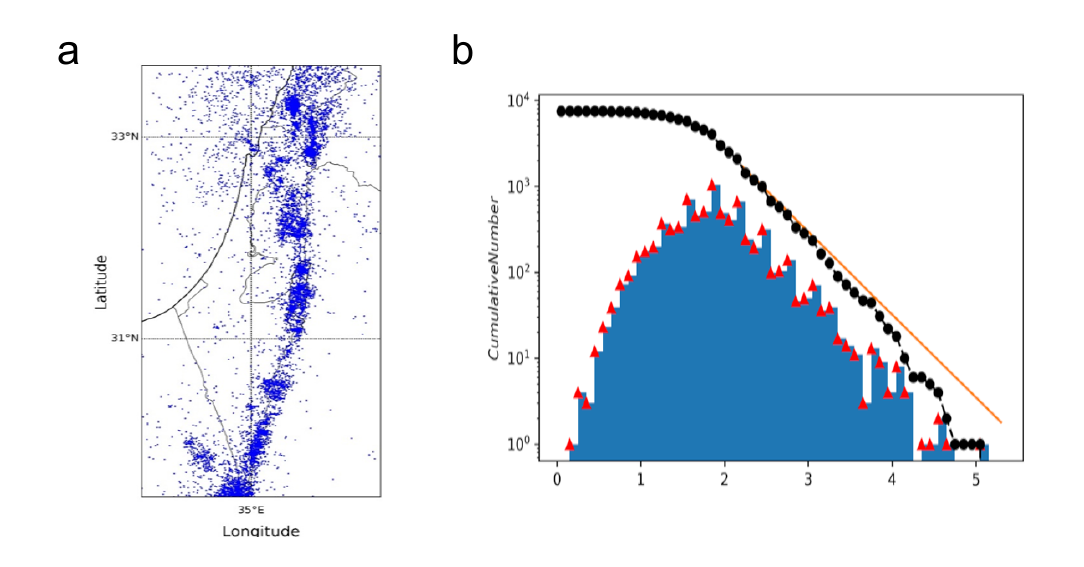}
\caption{\label{Fig5_1} 
{\bf Gutenberg–Richter law for the Israeli earthquake catalog}. (a) Map of the events present in the selected area.  (b) Cumulative number as a function of magnitude. The orange line is the fit of the distribution based on the Eq. (\ref{EQ126}), with $a = 5.36$ and $b = 0.97$.}
\end{centering}
\end{figure}

Next, we will give some well established empirical basic laws of earthquakes.
In seismology, the \textit{Gutenberg–Richter} (GR) law expresses the relationship between the magnitude $m$ and cumulative number of earthquakes $N$ in any given region and time period with magnitude larger or equal to $m$, which exhibits an exponential decay \cite{gutenberg_frequency_1944} 
\begin{equation}
\label{EQ126}
\log N = a - bm,
\end{equation}
where $a$ and $b$ are fitting parameters. We show the Gutenberg–Richter law for the Israeli earthquake catalog as an example, in Fig. \ref{Fig5_1}. 
The $b$-value of the Gutenberg and Richter law has been calculated using the approach described by Marzocchi and Sandri \cite{marzocchi_review_2009}. The value was found to be equal to $0.97 \pm 0.02$. The intercept of the fit line defines the $a$-value that explains the earthquake rate for the region selected. In our case $a = 5.36$.
As expected, different catalogs exhibit different seismicity levels quantified by the
magnitude range and by the total number of events, corresponding to different values of the constant $a$. Conversely,
the parameter $b$ appears to be quite independent of the geographical region and the temporal interval. Typical experimental results provide $b \simeq 1$ for tectonic earthquakes.

The second empirical law, the \textit{Omori law}, was first described  
by the seismologist Omori in 1894 \cite{omori1894after}, 
\begin{equation}
\label{EQ127}
n(t) = K/(c+t)^p,   
\end{equation}
where $n(t)$ is the aftershock rate, $t$ is the time elapsed from the mainshock occurrence; $K$ and $c$ are empirical constants
controlling, respectively, the total number of aftershocks $n(t)$ and the onset of the power law decay. $p$ is a third constant, which modifies the decay rate and typically falls in the range 0.7–1.5.

The magnitude of an event understandably influences the number of aftershocks triggered by it, i.e., the larger the mainshock magnitude, the larger is the total number of triggered
aftershocks. This property is known as the \textit{productivity law}, stating that the number of aftershocks $n_{A}$ [integral of Eq. (\ref{EQ127})] belonging to a
sequence increases exponentially with the mainshock magnitude $m_{M}$, 
\begin{equation}
\label{EQ128}
n_{A} \propto 10^{\alpha_{M} m_{M}},
\end{equation}
The parameter $\alpha_{M}$ is typically small for swarm-type activity,
and large for clear primary aftershock sequences \cite{ogata_detection_1992}.

\begin{figure}[]
\begin{centering}
\includegraphics[width=1\linewidth]{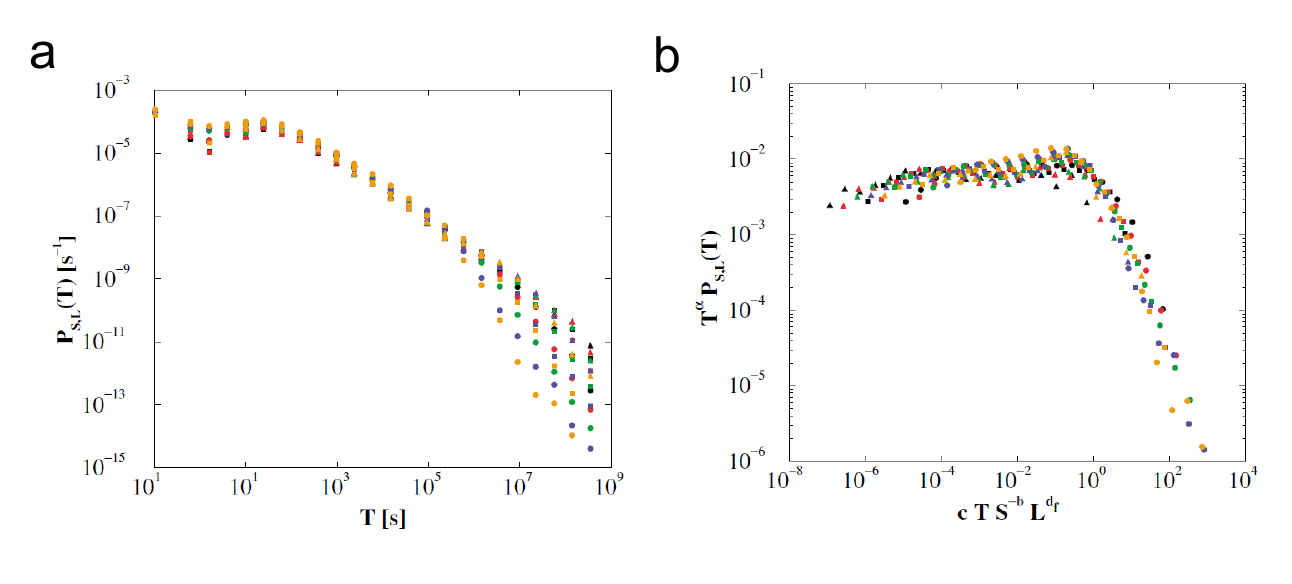}
\caption{\label{Fig5_2} 
{\bf Unified Scaling Law for Earthquakes}. (a) The distribution $P_{S, L}(T)$ of interoccurrence
time $T$ between earthquakes with magnitude $m$ greater than
$\log_{10} S$ within an area of linear size $L$ for the California catalog.  (b) The scaling function, see Eq. (\ref{EQ129}), for the data in (a) with $T>38$ s rescaled with $T^{\alpha} P_{S, L}(T)$ as a function of the variable $x=c T S^{-b} L^{d_{f}}$, with $c=10^{-4}$.  The data collapse implies a unified law for earthquakes. The solid circles, squares, and triangles correspond to magnitude cutoffs $M_c=2,$ $3,$ and $4$, respectively. The color coding represents the linear size $L=0.25^{\circ}$ (black), $0.5^{\circ}$ (red), $1^{\circ}$ (green), $2^{\circ}$ (blue), and $4^{\circ}$ (orange) of the cells. \textit{Source}: Reprinted figure from Ref. \cite{bak_unified_2002}.
}
\end{centering}
\end{figure}

The distribution of waiting times between seismic events has generated much attention and discussion over the last decades, since it was considered to have a universal scaling form \cite{bak_unified_2002}. Bak \textit{et al.}  proposed, for the first time, a unified scaling law for the waiting times between earthquakes, expressing a hierarchical organization
in time, space, and magnitude. They considered the California earthquake catalog and covered the region with a grid of cells of size $L \times L$.  Then they analyzed the inter-event times for each cell and calculated  a histogram of the inter-event times $P_{S, L}$ whose magnitudes are greater than $m = \log (S)$; this was repeated for different values of $L$ and minimum magnitudes (see Fig. \ref{Fig5_2}a). It was found that if they rescaled the inter-event times $T$ by $S^{-b} L^{d_{f}}$ (where $b$ is the Gutenberg-Richter $b$ -value and $d_{f}$ is the spatial fractal dimension), and rescaled $P_{S, L}$ by $T^{\alpha}$, where $\alpha$ is some exponent, the distribution of  waiting times all appeared to collapse onto a single curve (Fig. \ref{Fig5_2}b). The scaling hypothesis was thus expressed as 
\begin{equation}
\label{EQ129}
T^{\alpha} P_{S, L}(T)=f\left(T S^{-b} L^{d_{f}}\right),
\end{equation}
where $\alpha \approx 1$ can be identified as the Omori-law exponent for aftershocks, $b \approx 1$ is the $b$ value in the Gutenberg-Richter law, and $d_{f} \approx 1.2$ describes the $2 d$ fractal dimension of the location of epicenters projected onto the surface of the Earth \cite{bak_unified_2002}.

\begin{figure}[]
\begin{centering}
\includegraphics[width=1\linewidth]{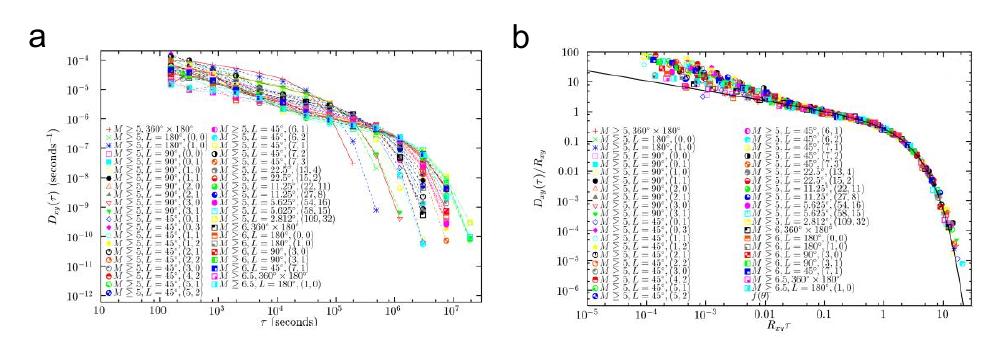}
\caption{\label{Fig5_3} 
{\bf Recurrence-time distributions without
and with re-scaling}. (a) Probability densities from the NEIC
worldwide catalog.  (b) Previous data, after re-scaling, with a fit of
the scaling function $f$, see Eq. (\ref{EQ131}). \textit{Source}: Reprinted figure from Ref. \cite{corral_long-term_2004-1}.}
\end{centering}
\end{figure}

Subsequently, Corral \cite{corral_long-term_2004-1} has discovered that the probability density function $D_{xy}(\tau)$ of the inter-occurrence time $\tau$  can be simplified to the scaling form
\begin{equation}
\label{EQ130}
D_{xy}(\tau) = R_{xy}f(R_{xy}\tau),
\end{equation}
where $R_{xy}$ stands  for the mean seismic rate that refers to the $(x,y)$ region. The scaling function $f$ can be  expressed by a generalized Gamma distribution,
\begin{equation}
\label{EQ131}
f(\theta)=C \frac{1}{\theta^{1-\gamma}} \exp \left(-\theta^{\delta} / B\right),
\end{equation}
with the parameters $\gamma=0.67 \pm 0.05, \delta=0.98 \pm 0.05, B=$
$1.58 \pm 0.15,$ and $C=0.50 \pm 0.10$. Corral verified the above scaling hypothesis by considering data from the global worldwide catalog from the National Earthquake Information Center (NEIC)\footnote{\url{http://earthquake.usgs.gov/contactus/golden/neic.php}.}. and several local catalogs: The Southern California Earthquake Center
(SCEC)\footnote{\url{http://service.scedc.caltech.edu/eq-catalogs/}.},  the Japan University Network Earthquake Catalog (JUNEC)\footnote{\url{http://wwweic.eri.u-tokyo.ac.jp/CATALOG/junec/}.}, the
Bulletins of the IGN (the Iberian Peninsula and the North
of Africa)\footnote{\url{https://www.ign.es/web/ign/portal}.}, and the BGS catalog (the British Islands and the North Sea)\footnote{\url{ http://www.earthquakes.bgs.ac.uk/}.}. The inter-event time  distributions for the different geographic areas  and for different values of the lower magnitude can be collapsed into a universal scaling function, as shown in Fig. \ref{Fig5_3}.

Bak \textit{et al.} and Corral are the main two proponents of the idea of a universal scaling law. The validity of the scaling relation Eq. (\ref{EQ131})  has attracted much interest.
For example, Saichev and Sornette \cite{saichev_universal_2006} carried out an extensive analysis to determine
the distribution of the inter-event time  based on the Epidemic-Type Aftershock Sequences
(ETAS) model (see more details in the following section), and they found that it is only approximately universal and of a gamma form,
assuming that $p$ in Eq. (\ref{EQ127}) has a value close to 1, and
the branching ratio of seismicity, the average number of events each event triggers, is around 0.7–1. Then Molchan \cite{molchan_interevent_2005}  published a rigorous mathematical study on
the earthquake inter-event time, and proved that  if the scaling distribution would be universal, the functional form
must follow an exponential function. Touati \textit{et al.} analyzed  
the variation of the distribution for both real data (SCEC catalog) and ETAS simulation data, and suggested that it is not universal \cite{touati_origin_2009}. They have shown that the distribution is a bimodal mixture
distribution dominated by correlated aftershocks at short waiting times and independent events at longer
times.

\subsubsection{Modeling Seismic Time Series by the Point Process Approach}

There are various  statistical earthquake models for describing the occurrences process
of earthquakes that can be used for real-time
earthquake forecasts \cite{verejones_stochastic_1970}. The principle of these models is to 
evaluate the probabilities of earthquake occurrence by using 
a point process approach. Among the different models, the ETAS model, which describes the features of earthquake clustering of mainshocks,  foreshocks and aftershocks, has become a standard model for testing hypotheses and a starting point for short-term earthquake forecasts \cite{helmstetter_foreshocks_2003,zhuang_analyzing_2004,hainzl_detecting_2005,lombardi_increase_2010}.

Here we will focus on the ETAS model, which is used to generate synthetic earthquake catalogs.  The ETAS model was developed by Ogata \cite{ogata_statistical_1988,ogata_statistical_1989,ogata_detection_1992}  who
observed that seismic activity can be well described by the Gutenberg-Richter law Eq. (\ref{EQ126}), and the Omori law Eq. (\ref{EQ127}). The conditional rate in the ETAS model is given by,
\begin{equation}
\label{EQ132}
\lambda (t|\mathcal{H}_{t}) = \mu + A\sum\limits_{i:t_{i}<t} \exp[\alpha_{M}(m_{i} - M_{z})]\left(1 + \frac{t - t_{i}}{c} \right)^{-p},
\end{equation}
where $t_{i}$ are the times of the past events and $m_{i}$ are their
magnitudes; $\mathcal{H}_{t} = \{(t_{i}, m_{i}); t_{i} <t\}$ is the history of occurrence. $\lambda (t|\mathcal{H}_{t})$ provides the probability to have an earthquake above a threshold magnitude $M_z$ at time $t$, given the earthquake history. Here, $A = K/c^p$ is the occurrence rate of earthquakes in the Omori law at zero lag \cite{touati_origin_2009}, and $\alpha_{M}$ is called the productivity parameter defined in Eq. (\ref{EQ128}). 
The ETAS model can also be written in terms of a stochastic integral \cite{daley_introduction_2003}, 
\begin{equation}
\label{EQ133}
\lambda\left(t | \mathcal{H}_{t} \right)=\mu+\int_{m_{1}}^{\infty} \int_{0}^{t} \frac{K}{(t-s+c)^{p}} \cdot e^{\alpha_M\left(m-M_{z}\right)} N(d s, d m),
\end{equation}
where $N(d s, d m)=1$ if an infinitesimal element $(d s, d m)$ would include an event $\left(t_{i}, m_{i}\right)$ for some $i$, otherwise $N(d s, d m)=0$. The five parameters $\left(\mu, A, c, \alpha_M, p\right)$ represent some characteristics of seismic activity of the region. Therefore, they vary spatially, and also temporally in some cases.

Then the ETAS model was extended to a space-time phase with the conditional intensity function \cite{ogata_space-time_1998},
\begin{equation}
\label{EQ134}
\lambda\left(t, x, y | \mathcal{H}_{t}\right)=\mu(x, y)+\sum_{i: t_{i}<t} \kappa\left(m_{i}\right) g\left(t-t_{i}\right) f\left(x-x_{i}, y-y_{i} | m_{i}\right),
\end{equation}
where $\mu(x, y)$ still is the background intensity, which is a function of space, but not of time; $\kappa(m)$ is the expected number of events triggered from an event of a magnitude $m$ in the form
\begin{equation}
\label{EQ135}
\kappa(m)=A \exp \left[\alpha\left(m-M_{z}\right)\right],
\end{equation}
$g(t)$ is the probability density function of the occurrence times of the triggered events, 
\begin{equation}
\label{EQ136}
g(t)=\frac{p-1}{c}\left(1+\frac{t}{c}\right)^{-p}.
\end{equation}
$f(x, y | m)$ is the spatial distribution of the triggered events, which can be expressed in the following two ways: a short-range Gaussian decay,
\begin{equation}
\label{EQ137}
f(x, y | m)=\frac{1}{2 \pi D^{2} e^{\alpha_M\left(m-M_{z}\right)}} \exp \left[-\frac{x^{2}+y^{2}}{2 D^{2} e^{\alpha_M\left(m-M_{z}\right)}}\right],
\end{equation}
or a long-range inverse power decay,
\begin{equation}
\label{EQ138}
f(x, y | m)=\frac{(q-1) D^{2(q-1)} e^{\alpha_M(q-1)\left(m-M_{z}\right)}}{\pi\left[x^{2}+y^{2}+D^{2} e^{\alpha_M\left(m-M_{z}\right)}\right]^{q}}, \quad q>1.
\end{equation}
The ETAS model has been successfully used for operational
earthquake forecasting, e.g., the complex Amatrice-Norcia
seismic sequence \cite{marzocchi_earthquake_2017}, next-day earthquake forecasts for the Japan region \cite{zhuang_next-day_2011} and California region \cite{field_spatiotemporal_2017}.
More general discussions on modeling seismic time series by the point process approach are given  in Ref. \cite{ogata_seismicity_1999}.

\subsubsection{Memory Analysis}
\begin{figure}[]
\begin{centering}
\includegraphics[width=0.8\linewidth]{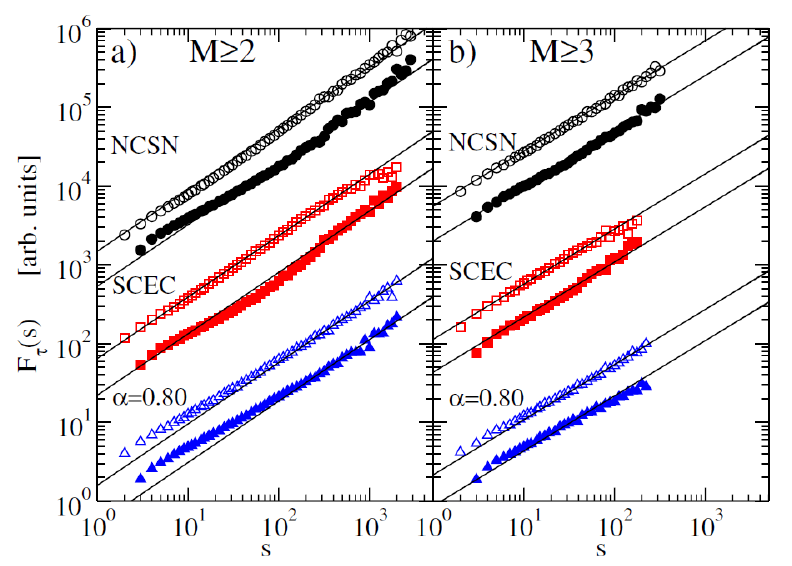}
\caption{\label{Fig5_4} 
{\bf Detrended fluctuation analysis of the
interoccurrence times from the  Californian catalogs}. (a) For all earthquake with magnitude $M \geq 2$, and   (b) $M \geq 3$. Black circles indicate the SCEC catalog (1995–1998) and red squares stand for the the NCSN catalog (1995–1998), blue triangles represent an artificial Gutenberg-Richter distributed correlated data set with $\alpha = 0.8$. The open symbols show the DFA0 results and the filled symbols the DFA1 results. The corresponding slopes are represented by straight lines with $\alpha \approx 0.75$. \textit{Source}: figure from Ref. \cite{lennartz_long-term_2008}.}
\end{centering}
\end{figure}

In this subsection, we will review the memory analysis that has been applied to seismic records.   The first method is the so-called detrended fluctuation analysis (DFA), which is used to detect long-range correlations or long-term memory of diverse time series, such as DNA sequences and climate records \cite{peng_mosaic_1994}. Details of the DFA are presented in  Sec. \ref{WATP}. According to Eq. (\ref{EQ75}), if the time series has a long-term memory, the fluctuation function $F(n)$ increases according to a power-law relation with a  exponent $\alpha$. The DFA has been applied to  earthquake interoccurrence times for the first time by Lennartz \textit{et al.} \cite{lennartz_long-term_2008}. They tested the real and the synthetic
records for long-term correlations by employing the
first two orders of the detrended fluctuation analysis DFA0
and DFA1 on the SCEC catalog (1995–1998) and the NCSN catalog (1995–1998).
They found that there exists a long-term memory between seismic
events in the Californian catalogs, which show up
in characteristic fluctuations in both magnitudes and interocurrence times.
Both DFA0 and DFA1 have resulted in a DFA 
exponent $\alpha \approx 0.75$, independently of the threshold $M_z$ and the catalog, see Fig. \ref{Fig5_4}. 

\begin{figure}[]
\begin{centering}
\includegraphics[width=0.85\linewidth]{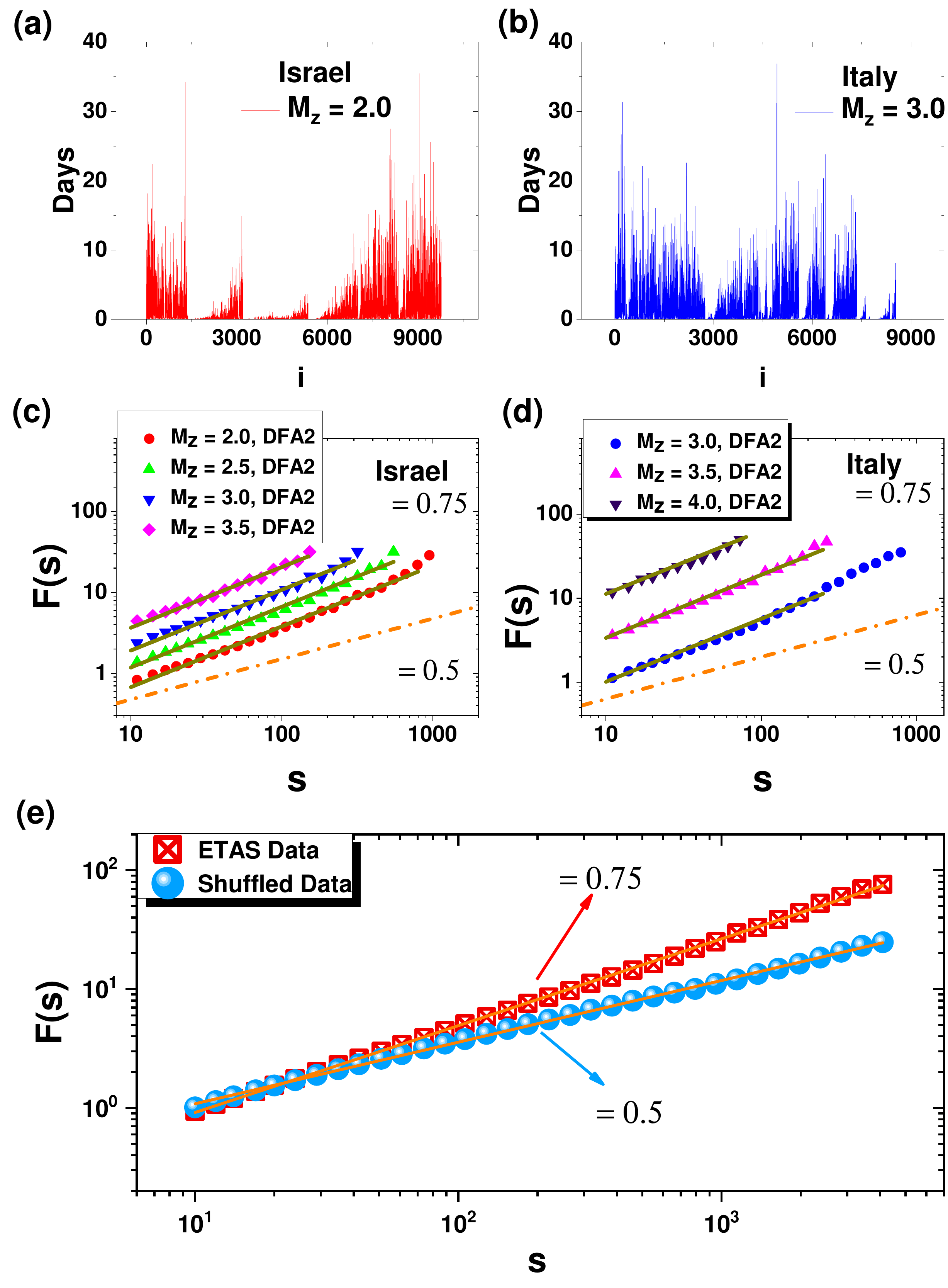}
\caption{\label{Fig5_5} 
{\bf Interoccurrence time series and DFA from the  the Israeli, Italian and ETAS catalogs}. (a) Interoccurrence time series between earthquake events
from the Israeli (a) and Italian (b) catalogs using magnitude threshold $M_z$. DFA of the interoccurrence times from the (c) Israeli and
(d) Italian catalogs with different $M_z$ values. (d) DFA of the interoccurrence times from the ETAS model, with parameters $A = 6.26$, $\mu = 0.2$, $p=1.1$, $\alpha_M = 1.5$ and $c = 0.007$, see Eq. (\ref{EQ132}). The solid line is the best fitting line with slope $\alpha = 0.75$, R-square $>0.99$. For comparison, the shuffled data with with slope $\alpha=0.5$, indicating no memory, are presented in (d). \textit{Source}: Reprinted figure from Ref. \cite{fan_possible_2019}.}
\end{centering}
\end{figure}

A similar analysis, DFA2,  has been  applied to the real Israeli and Italian catalogs by Fan \textit{et al.}  \cite{fan_possible_2019}. They found that the exponent $\alpha$ was also very close to 0.75 for the inter-occurrence time series between earthquake events with different magnitude threshold $M_z$. Figs. \ref{Fig5_5}a-d show the DFA2 results.  They also performed DFA  on other seismic variables, such
as the number of earthquake events and the released energy within a coarse time window $dt$. The energy of earthquakes was defined as $S(t) = \sum_{l=1}^{E(t)} 10^{\frac{3}{2}M_{l}(t)}$,
where $E(t)$ denotes the number of events that occurred between $t$ and $t+dt$, and $M_l(t)$ denotes the magnitude of the event. To deal with more homogeneous time series, Fan \textit{et al.} switched  to $s(t)=\log(S(t))$ for $S(t) > 1$ and zero for $S(t) \leq 1$ and $e(t)=\log(E(t))$ for $E(t) > 1$ and zero for $E(t) \leq 1$ \cite{fan_possible_2019}. Next, applying the DFA analysis for the time series $s(t)$ and $e(t)$ in the Israeli and Italian catalogs, with different time windows $dt$, were tested. It was found that for both regions and for all studied magnitudes, the value of the scaling exponent  $\alpha$ is quite robust $\sim 0.75$, i.e., the size of $dt$ does not affect the memory exponent $\alpha$. Thus, the return intervals, the number of events and the released energy are significantly long-term  correlated and have a very similar scaling exponent. The memory analysis has also been  applied on the ETAS model. The DFA2 of the inter-occurrence times, with the parameters $A = 6.26$, $\mu = 0.2$, $p=1.1$, $\alpha_M = 1.5$ and $c = 0.007$, see Eq. (\ref{EQ132}), yields $\alpha \approx 0.75$ and is shown in Fig. \ref{Fig5_5}e. 

\begin{figure}[]
\begin{centering}
\includegraphics[width=0.85\linewidth]{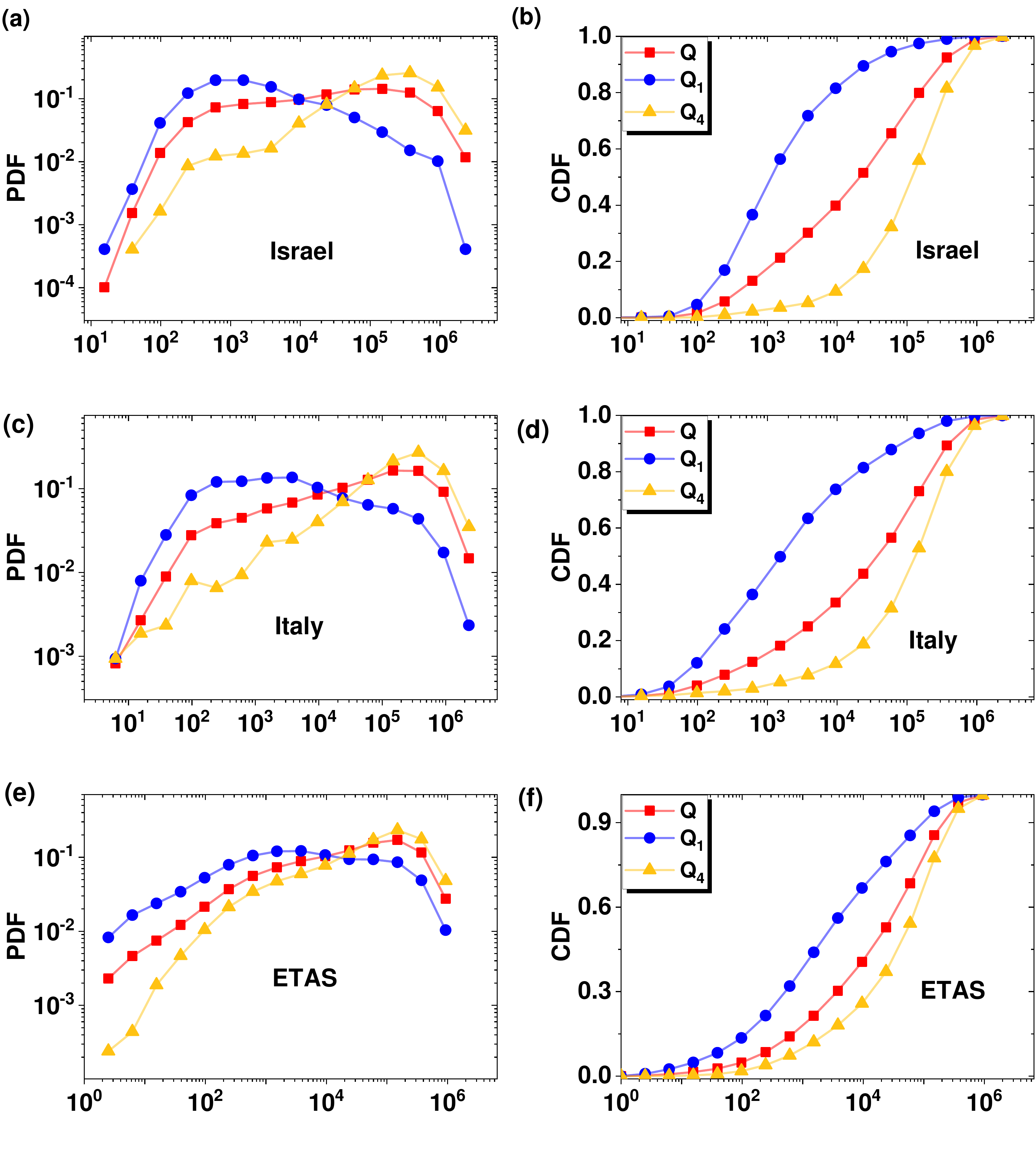}
\caption{\label{Fig5_6} 
{\bf Memory analyses for the Israeli, Italian and ETAS catalogs}. (a, c, e) Conditional PDF and (b, d, f) CDF of the recurrence times $\tau$ for the (a, b) Israeli catalog above the threshold $M_c$ = 2.0, for the (c, d) Italian catalog above the threshold $M_c$ = 3.0 and for the (e, f)  ETAS catalog  above the threshold $M_c$ = 2.0. \textit{Source}: Reprinted figure from Ref. \cite{fan_possible_2019}.}
\end{centering}
\end{figure}

Another common method for studying  memory in the occurrence of earthquakes is called conditional probability (CP), which was  introduced by Livina \textit{et al.} \cite{livina_memory_2005}.  Considering a time series of recurrence intervals, they sorted it in ascending order and divided it into four 25\% quantiles; i.e., the first quantile, $Q_1$, represents the shortest $25\%$ of waiting times, etc. The distribution of recurrence times $\tau$, that follow a prior recurrence time $\tau_0$, $P(\tau|\tau_0)$, was studied, where $\tau_0$ belongs to either one of the quantiles at the extremes, $Q_1$ or $Q_4$. Note that for records without memory, $P(\tau|\tau_0)$ should be independent of $\tau_0$ and should be identical to $P(\tau)$. For the real data, Livina \textit{et al.} found that $P(\tau|\tau_0)$ depends strongly on the previous recurrence time $\tau_0$, such that short recurrence times lead to short ones, and long recurrence times follow long ones (see Figs. \ref{Fig5_6}a, c for the Israeli and Italian catalogs). To quantify and measure the level of memory, Fan \textit{et al.} analyzed the cumulative distribution function (CDF) of the recurrence times and denoted the CDF for $Q$, $Q_1$ and $Q_4$ as $CQ(\tau)$, $CQ_1(\tau)$ and $CQ_4(\tau)$, respectively. The strength of the memory for $Q_1$ is defined as
\begin{equation}
\label{EQ139}
\rho_1 = \int(CQ_1(\tau) - CQ(\tau)) d\tau/{\int d\tau},  
\end{equation}
and similarly, the level of memory for $Q_4$ is
\begin{equation}
\label{EQ140}
\rho_4 = \int(CQ_4(\tau) - CQ(\tau)) d\tau/{\int d\tau}.
\end{equation}
Thus, $0\leq\rho_1\leq 1$ and $-1\leq\rho_4\leq 0$,
and higher $|\rho_1|$ (or $|\rho_4|$), imply stronger memory and $\rho_1 = 0$ (or $\rho_4 = 0$), implies no memory \cite{fan_possible_2019}. The results are shown in Fig. \ref{Fig5_6}b and d  for the Israeli and Italian earthquake catalogs. They found $\rho_1=0.248, \rho_4=-0.193$ for $Q_1$ and $Q_4$ of the Israeli catalog, whereas $\rho_1=0.260, \rho_4=-0.153$ for $Q_1$ and $Q_4$ of the Italian catalog \cite{fan_possible_2019}. Further study suggests that the values of $\rho$ are robust and do not depend on $M_z$. To study the dependence of the correlations ($\alpha$, $\rho_1$, and $\rho_4$) on the geographical location (tectonic setting), they performed the DFA and CP analysis for other earthquake catalogs, including New Zealand, Southern California Earthquake Center, Japan unified hIgh-resolution relocated catalog for earthquakes and Preliminary Determination of Epicenters global earthquake catalogs. It has been found that for all catalogs and for all studied magnitudes the values of $\alpha$, $\rho_1$, and $\rho_4$ are quite robust.

Finally, the memory analyses DFA2 and CP, were performed in ETAS models, by measuring the coefficients $\alpha$, $\rho_1$ and $\rho_4$, as shown in Figs. \ref{Fig5_5} and  \ref{Fig5_6}.
The results indicate that the empirically observed
earthquake memory (for Israeli catalog) can be reproduced only for a narrow range of the model's parameters.  Moreover, the origin of memory in the ETAS model is influenced by: (i) the  model's background (noise) rate parameter $\mu$ which, affects the memory through the interference of temporally overlapping aftershock subsequences, i.e., smaller $\mu$ leads to stronger memory, and (ii) the branching ratio $n' = \frac{Ac}{p - 1} \frac{\beta}{\beta - \alpha_M}$: the exponent relating the production of aftershocks as a function of magnitude, $\alpha_M$, and the power $p$ of Omori's law can also affect the memory through the branching ratio of the ETAS model, i.e., smaller $p$ and larger $\alpha_M$ result in a stronger memory (see Figures. 11 and 12 in Ref. \cite{fan_possible_2019}).

\begin{figure}[]
\begin{centering}
\includegraphics[width=0.75\linewidth]{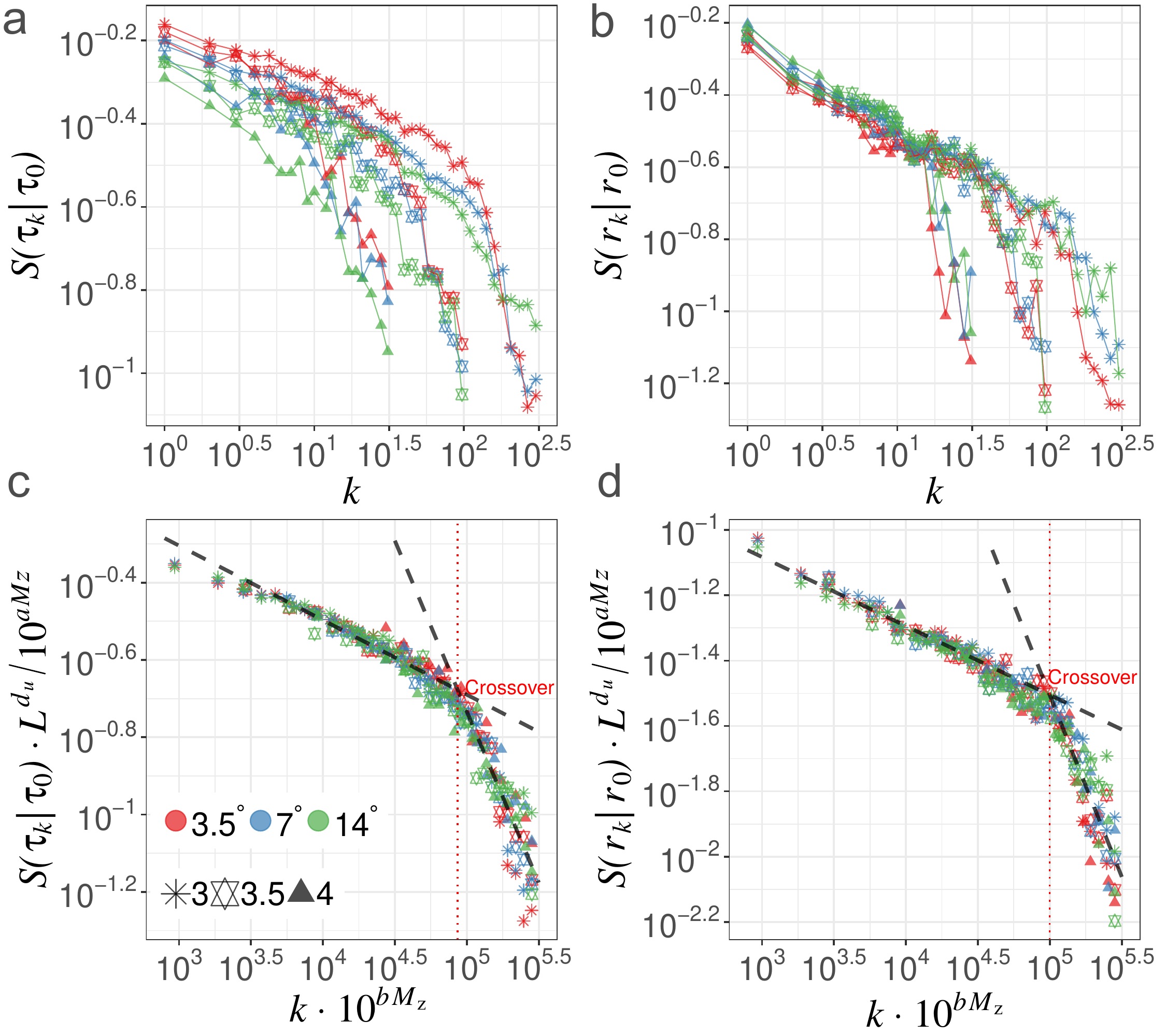}
\caption{\label{Fig5_7} 
{\bf The lagged CP memory analysis for the inter-event times and distances from the Italian catalog}.  The memory measures (a) $S(\tau_{k}|\tau_{0})$ and (b) $S(r_{k}|r_{0})$ as a function of the lag index $k$. Rescaled memory measure (c) for inter-event times and (d) distances. Colors represent different grid sizes $L$. The shapes of the symbols represent different magnitude thresholds $M_z$. \textit{Source}: Reprinted figure from Ref. \cite{zhang_scaling_2020}.}
\end{centering}
\end{figure}

More recently, a lagged CP method was  introduced by Zhang \textit{et al.} to study a possible long-term memory in both the interevent
times and distances of earthquakes \cite{zhang_scaling_2020}. The CP of times and distances were defined as $\rho\left(\tau_{k} | \tau_{0}\right)$ and  $\rho\left(r_{k} | r_{0}\right)$, respectively, where $\tau_{0}\left(r_{0}\right)$ belongs to the $Q_1$ or $Q_3$ group. $\tau_{k}\left(r_{k}\right)$ is the lagged $k$-th inter-event time that follows $\tau_{0}\left(r_{0}\right)$. In order to  quantify the level of memory,  $S\left(\tau_{k} | \tau_{0}\right) \equiv 1-s_{13}$ was introduced, where $s_{13}$ is the common area of the PDF between $Q_{1}$ and  $Q_{3}$. Here $S\left(\tau_{k} | \tau_{0}\right) \in [0,1]$. Similarly, $S\left(r_{k} | r_{0}\right)$ represents the level of memory for the inter-event distances.
Interestingly,  it has been shown that $S(k)$ can be expressed by the following scaling form,
\begin{equation}
\label{EQ141}
S(k) = L^{-d_{u}} F(k\cdot10^{b M_{z}}) 10^{a M_{z}},
\end{equation}
where $d_u$ and $a$ are constants, $b=1$ as  for the
Gutenberg-Richter law. The distribution of $S\left(\tau_{k} | \tau_{0}\right)$, $S\left(r_{k} | r_{0}\right)$ and their corresponding scaling functions for the Italian catalog are presented in Fig. \ref{Fig5_7},  where $d_u=0.14$ and $a=0.09$ for the inter-event times $\tau$; $d_u=-0.08$ and $a=0.24$ for the inter-event distances $r$. The scaling functions shown in Fig. \ref{Fig5_7}c and d suggest 
a crossover between two distinct power-law relations with $F(x) \sim x^{-\gamma}$. Both
scaling functions exhibit a significant crossover at $k\cdot10^{b M_{z}} \simeq 10^{5}$. These results indicate that the memory measure of different grid
sizes and different magnitude thresholds can be rescaled into a single function. However, it was found that the memory function in the ETAS model is very 
different from that of real earthquake catalogs, i.e.,  the model’s memory
is weaker (stronger) on short (long) timescale compared
to the real catalogs. Moreover, the model does not exhibit a
clear crossover observed in the real catalogs.

\subsubsection{Earthquake Forecasting}
Earthquakes are one of the most destructive natural disasters in the  world. 
Skilled and reliable earthquake forecasting remains an ultimate goal.  The term \textit{earthquake prediction} is usually referring to the specification of the
occurrence time, location or magnitude of a future earthquake. However, ``the term \textit{earthquake forecasting} usually refers to the evaluation of the occurrence probability $P_{\Sigma}$ of an earthquake inside a hypercell of volume $\Sigma$ centered in the point $\vec{\omega}=(t, x, y, z, m)$. Therefore, the fundamental quantity is the occurrence probability $P_{\Sigma}(\vec{\omega} | \ell)$ conditioned on the whole set of information $\ell$ available at time
$t$'' \cite{de_arcangelis_statistical_2016}. The probability of having an earthquake at time $t$, in the $k$-th interval conditioned to the previous history is
\begin{equation}
\label{EQ142}
P\left(\zeta_{k}=1 | \mathcal{H}_{t_{k}}\right)=\lambda\left(\vec{\omega}_{k} | \mathcal{H}_{t_{k}}\right) d \Sigma+\mathcal{O}\left(d \Sigma^{2}\right),
\end{equation}
where $\lambda$ is the local conditional rate as defined in Eqs. (\ref{EQ132}) or  (\ref{EQ134}), $\vec{\omega}_{k}$ denotes the center of the $k$-th cell. The forecasting probability, $P_{\Sigma}$ the integral over $\Sigma$ of the occurrence rate $\lambda\left(\vec{\omega}_{k} | \mathcal{H}_{t_{k}}\right)$, depends on the volume of the hypercell $\Sigma$, i.e., the smaller $\Sigma$, the more accurate is the evaluation of $\lambda$.
Secondly, the temporal
extension $T$ of $\Sigma$ also influences the forecasting accuracy. According to the time scale of $T$, one can separate arbitrarily Long-Term from Short-Term forecasting \cite{kagan_statistical_1987}: earthquake Long-Term forecasting, for $T$ in the interval of decades to centuries; Intermediate-Term forecasting, for $T$ of the order of months and Short-Term forecasting, for $T$ in the interval from few seconds up to several weeks. Short-Term forecasting is usually subdivided into post-seismic and pre-seismic forecasting, referring to the aftershock and foreshock of a mainshock. It is notable that the post seismic forecasting is still very important from the point of view of risk management, since aftershocks can have sizes comparable or even larger than their triggering mainshock. Long-Term forecasting, however, is probably the most
relevant from the engineering point of view, such as urban planning risk assessment. Based on the early warning foreshock observations, one could successfully predict some big earthquakes, such as the 1975 Haicheng, China earthquake \cite{wang_predicting_2006} and the 1995 Kozani–Grevena, Greece earthquake \cite{bernard_precursors_1997}. Nevertheless, most earthquakes
do not present significant precursory patterns, e.g.,  the 2004 Parkfield (California) earthquake \cite{bakun_implications_2005}. Thus the evaluation of foreshocks or precursory patterns is still an open question. 
Since most earthquake models are based on the well-established empirical laws, aftershock spatio-temporal occurrence clustering, as the ETAS model, they are able, in some degree, to provide reliable risk assessment only for the Long-Term and the post-seismic Short-Term forecasting but not for the foreshocks. Nevertheless, how to validate a forecasting model is still of great importance. In the following, we will review various  methods  for  evaluating  earthquake  predictions and earthquake forecasts.

\textit{Receiver operating characteristic diagram}

\begin{figure}[]
\begin{centering}
\includegraphics[width=0.4\linewidth]{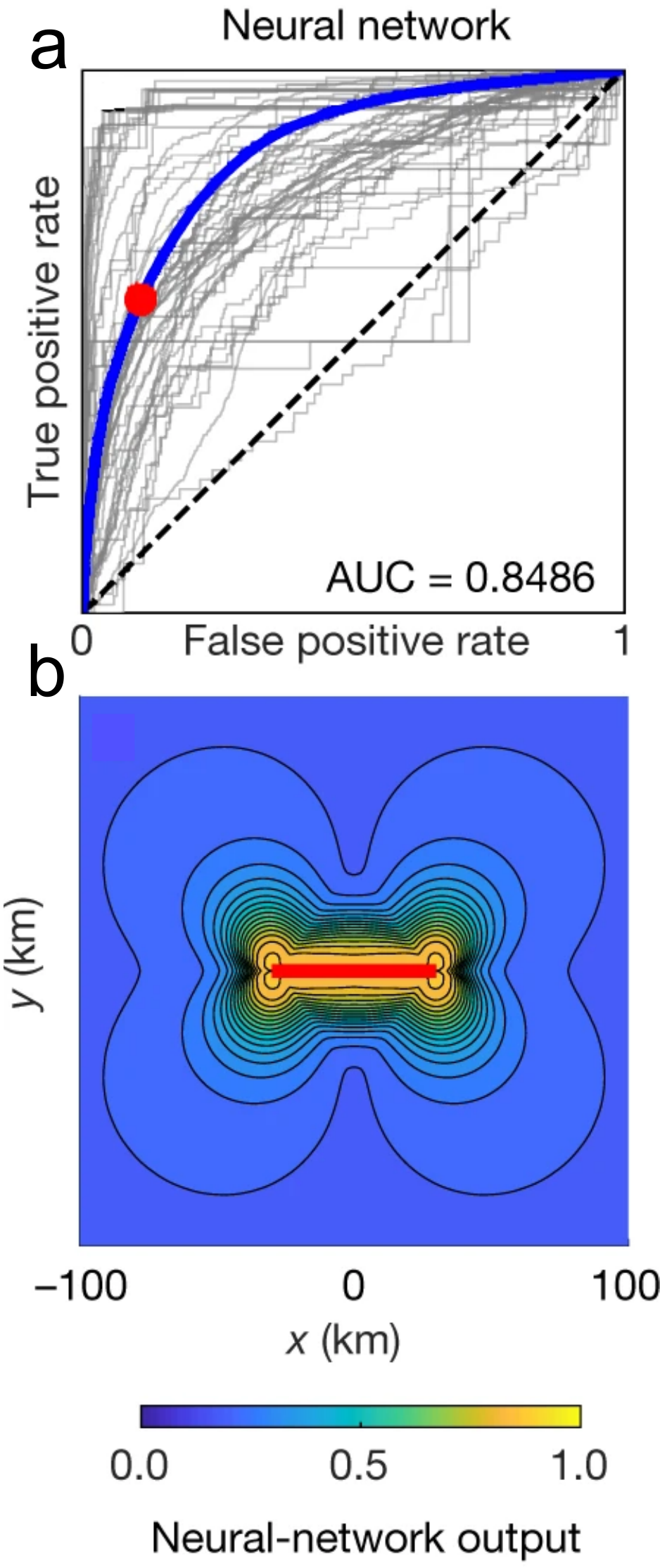}
\caption{\label{Fig5_8} 
{\bf An example of ROC curves based on the neural network}.  (a) The true positive rate is plotted versus the false alarm rate. (b) For a
synthetic case of a 60-km-long, right-lateral strike-slip fault (red lines) at a depth of 10 km. The dashed diagonal represents random
prediction (the null-hypothesis).  \textit{Source}: Reprinted figure from Ref. \cite{devries_deep_2018}.}
\end{centering}
\end{figure}

The receiver operating characteristic (ROC) diagram 
is a graphical plot that illustrates the prediction quality of a binary classifier system as its discrimination threshold is varied. For evaluating earthquake predictions, if an earthquake is expected (``YES'') with a probability $P_{\Sigma}>P_{t}$, where $P_{t}$ is a threshold, we call it a \textit{hit}, if an event has occurred;  if earthquakes are not expected (``NO'') in the cell, we call it a $miss$, if an event has occurred. Each ``YES'' prediction in which no corresponding earthquake is observed is a \textit{false alarm}, and each ``NO'' prediction in which no earthquake occurred is a\textit{ correct rejections}. In this representation, there are four possible combinations of alarm declaration and event observation, see Table. \ref{tab:caption} for the number of each of these contingencies. Here we define $a$ as the number of hits, $b$ as the number of false alarms, $c$ as the number of misses and $d$ as the number of correct rejections, respectively. We define a true positive rate ($TPR$) as
\begin{equation}
\label{EQ143}
T P R=\frac{a}{a+c},
\end{equation}
and false positive rate ($FPR$) as
\begin{equation}
\label{EQ144}
F P R=\frac{b}{b+d}.
\end{equation}
When $TPR$ and  $FPR$ are plotted together on the square [0,1] $\times$ [0,1], the resulting metric is called the ROC, whereas the perfect prediction
corresponds to the point with coordinates $TPR=1$ and  $FPR=0$, the diagonal line represents the long-term behavior of random guessing. One can obtain the full ROC curve by changing  $P_t$ continuously. Here we show an aftershock forecasting ROC curve by using a deep learning method in Fig. \ref{Fig5_8} as an example. 

\textit{The Molchan diagram}

Closely related to the ROC is the Molchan diagram, which is an error diagram that allows to compare the prediction of the test model and another generic reference model \cite{molchan_strategies_1990}. It is a plot of the miss rate, $M = 1 - TPR$, and the fraction of experiment space-time volume, $F$, occupied by alarms or YES predictions. The best prediction in the Molchan diagram is the point ($F = 0,M = 0$). In particular, the probability of obtaining $h$ or more hits by chance, given that there have been $n$ observed target earthquakes, is described by the binomial distribution
\begin{equation}
\label{EQ145}
\Gamma_{M}=\sum_{i=n}^{n}\left[\left(\begin{array}{c}
N \\i
\end{array}\right) F^{i}(1-F)^{n-i}\right]    
\end{equation}
where $\left(\begin{array}{l}n \\ i\end{array}\right)$ stands the binomial coefficient. One can obtain a confidence level curve in the Molchan diagram by  fixing $\Gamma_{M}$ and changing $F$ continuously. Therefore, a very small $\Gamma_{M}$ value suggests that the alarm set has high skill.

\textit{The Number test (N-test)}

\begin{figure}[]
\begin{centering}
\includegraphics[width=0.85\linewidth]{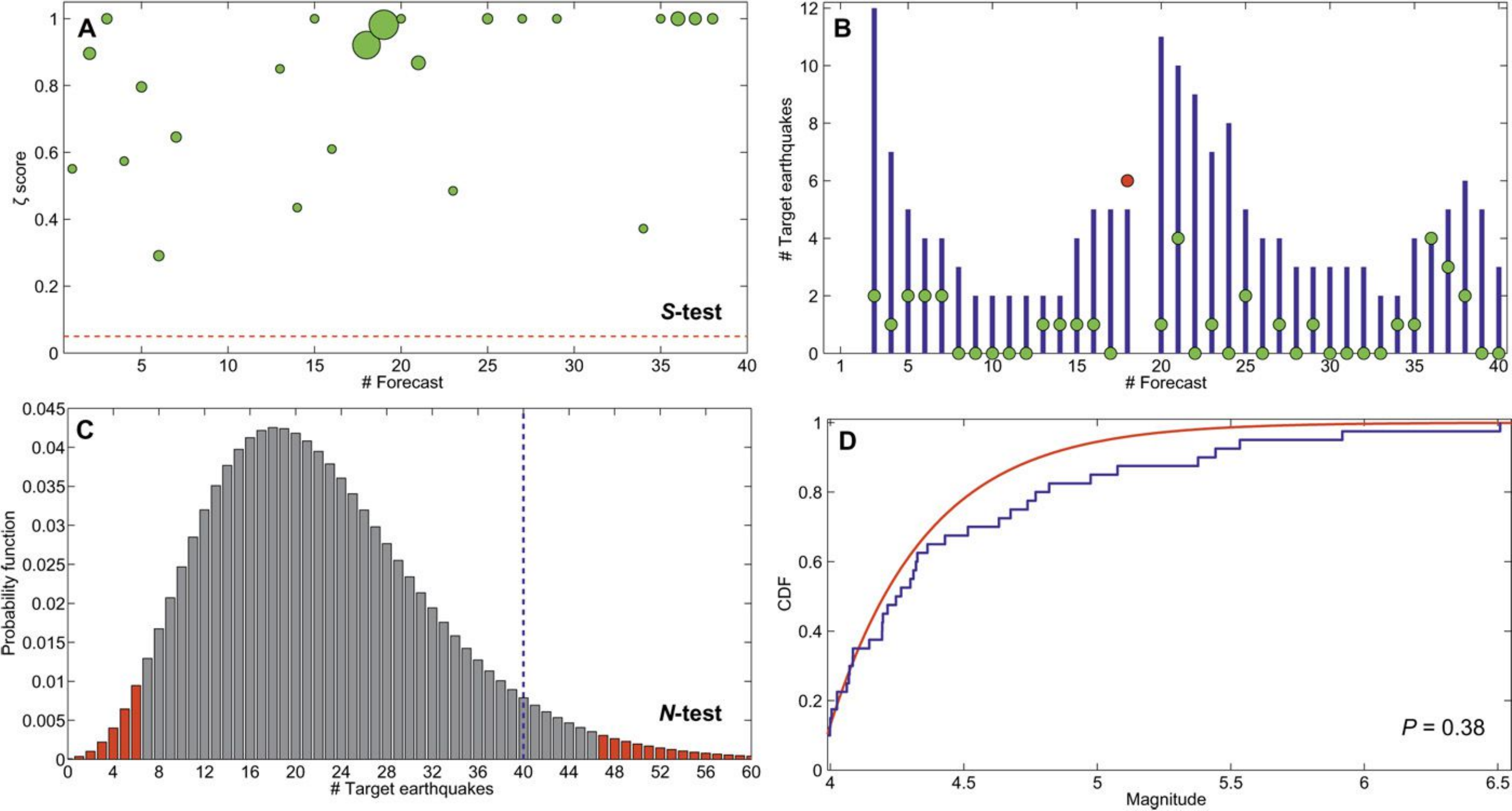}
\caption{\label{Fig5_9} 
{\bf The statistical tests for earthquake forecasting for the complex
Amatrice-Norcia seismic sequence}.  (a) The Space test (S-test). The red dashed horizontal line is the critical value $\zeta = 0.05$ for rejection. (b) Expected number of target earthquakes for each forecast. (c) The Space test (S-test): probability distribution of the expected number of target earthquakes with a 95\% confidence interval (marked by red) and the observed number of target earthquakes
(vertical dashed blue line). (d) The CDF of the forecast and observed frequency-magnitude distribution.  \textit{Source}: figure from Ref. \cite{marzocchi_earthquake_2017}.}
\end{centering}
\end{figure}

The N-test is used to access how well the total number of forecasted earthquakes $N_{forecast}$ matches the number of events observed $N_{obs}$ \cite{schorlemmer_earthquake_2007}. The target earthquakes are assumed to follow a generalized Poisson
distribution, and the mass function of the negative binomial distribution (NBD) can be expressed as \cite{marzocchi_earthquake_2017}
\begin{equation}
\label{EQ146}
f(n ; r, p)=\frac{\Gamma(r+n)}{n ! \Gamma(r)} p^{n}(1-p)^{r},
\end{equation}
where $\Gamma(r)$ is the gamma function, $n$ is the number of target earthquakes, and $r, p$ are the two parameters of the distribution. Then NBD can be derived as 
\begin{equation}
\label{EQ147}
f(n ; r, p)=\int_{0}^{\infty} f_{1}(n ; \lambda) f_{2}\left(\lambda ; r, \frac{1-p}{p}\right) d \lambda,
\end{equation}
where $f_{1}(n ; \lambda)$ is the Poisson distribution and $\lambda$ is assumed to follow a gamma distribution
\begin{equation}
\label{EQ148}
f_{2}(\lambda ; k, \theta)=\frac{\lambda^{k-1} e^{-\frac{\lambda}{\theta}}}{\theta^{k} \Gamma(k)},
\end{equation}
where $k, \theta$ are the parameters of the gamma distribution with $r = k$ and $p = \frac{\theta}{1+\theta}$. Next, we calculate the parameters $k$ and $\theta$ of the gamma distribution. To interpret the N-test results, one can use a one-sided test with an effective significance value $\alpha_{eff}$, which is half of the intended significance value $\alpha=0.05$. Fig. \ref{Fig5_9}c shows an example of the N-test.

\textit{The Space test (S-test)}

The S-test is a likelihood test where only the spatial distributions of the forecast and the observation are considered \cite{zechar_likelihood-based_2010}. The $S$-test can be summarized by a \begin{equation}
\label{EQ149}
\zeta=\frac{n\left\{S_{i} | S_{i} \leq S, S_{i} \in S_{\mathrm{s}}\right\}}{n\left\{S_{\mathrm{s}}\right\}},
\end{equation}
where $S_{i}$ is the $i$-th simulated spatial likelihood, $S_{\mathrm{s}}$ is the set of simulated spatial likelihoods, $S$ is the likelihood of the spatial forecast relative to the observed, $n\{A\}$ is the number of elements in a set $\{A\}$.  If $\zeta$ is less than critical significance value $\alpha$, the observed spatial distribution is inconsistent with the forecast. However, values close to 1 indicate optimal spatial forecasts.  We show an example of the S-test in Fig. \ref{Fig5_9}a.

\begin{table}[]
\centering
\caption{Contingency table for a binary event} 
\begin{tabular}{|l|c|c|c|}
\hline
\multicolumn{2}{|l|}{\multirow{2}{*}{}}               & \multicolumn{2}{c|}{Observation}                        \\ \cline{3-4} 
\multicolumn{2}{|l|}{}                                & YES                  & NO                               \\ \hline
\multicolumn{1}{|c|}{\multirow{2}{*}{Forecast}} & YES & a = number of hits   & b = number of false alarms       \\ \cline{2-4} 
\multicolumn{1}{|c|}{}                          & NO  & c = number of misses & d = number of correct rejections \\ \hline
\end{tabular}
\label{tab:caption}
\end{table}

Besides the aforementioned methods, there are still various other methods that have been developed to evaluate the accuracy of a forecasting model, such as, the Likelihood test (L-test), the Likelihood Ratio test (R-test) and the Magnitude test (M-test). An instructive review on the the evaluation of earthquake predictions and earthquake forecasts can be found in Ref. \cite{zechar2010evaluating}.

As discussed in the last section, we studied the possible origin of memory in earthquakes for both real catalogs and ETAS models. Here, we will show that the memory analysis can significantly
improve the short-term forecasting rate for the real earthquake catalog. In estimating the ETAS model parameters for a given earthquake catalog, the ETAS parameters are commonly inverted from the data based on the point-process maximum likelihood (ML) method, by the Davidon-Fletcher-Powell algorithm \cite{ogata_statistical_1988} or by simulated annealing \cite{lombardi_estimation_2015}. For example, by using  the ML method, one can obtain $\mu=0.2$, $A=6.26$, $\alpha_M = 1.4$, $p=1.13$ and $c=0.007$, see Eq. (\ref{EQ132}),   for the Italian catalog \cite{lombardi_estimation_2015}. When considering the memory, however, the ETAS model
can reproduce the same memory as in real catalogs, only
for a small range of parameter values, i.e., $\mu=0.2$, $A=6.26$, $\alpha_M = 1.5$, $p=1.1$ and $c=0.007$ \cite{fan_possible_2019}. 
As a test case the cumulative number of events, $\mathrm{NC}$, after the Capitignano 5.7 main shock that happened on 18 January 2017 \cite{falcucci_campotosto_2018}, was forecasted by using the ETAS model. The cumulative number at time $t$ is calculated by integrating Eq.~(\ref{EQ132}) from 0 to $t$. This way the forecasting accuracy within 14 days after the main shock can be improved by more than 20\% by considering the memory.  Moreover, a revised and generalized ETAS model was proposed recently by Zhang \textit{et al.}, in which the short- and long-term/distance memory was reproduced accurately \cite{zhang_improved_2020}.  The new ETAS model is also found to significantly improve earthquake forecasting for the Italian and South California earthquake catalogs.

\section{Conclusions and Perspectives}
\label{cap4}

In this article, we have reviewed  statistical physics and complex networks-based techniques that advance our knowledge on the complex Earth
system, a relatively novel branch of geophysics.  These techniques can help to
address the understanding as well as the prediction of
climate variability, Earth geometric relief features and earthquakes. In this section we try to draw a few concluding
remarks and address the  perspectives that remain still open for future research.  We believe that  these novel approaches will attract the attention of scientists in the related fields.

In Section \ref{sec:methodology} we started by describing  the overall methodology based on statistical physics for analyzing  important aspects of complex Earth subsystems. It includes complex networks, percolation theory, tipping points analysis, entropy theory and complexity. Although it was shown that the aforementioned approaches can be successfully applied to Earth systems individually or in a combination, there is still a lot of work ahead in order to develop a proper and comprehensive framework. Among the different open problems to be solved, we highlight the following 
four: {\it i)}  The need of gathering a better knowledge and understanding on the underlying physical mechanisms; {\it ii)} The need of extending  concepts of CNs to a more comprehensive  description, such as multilayer climate networks \cite{boccaletti_structure_2014}. This is since the climate is a coupled and multilayered system; {\it iii)} The need of developing a universal theory of critical phenomena (percolation) in climate systems, in particular for out-of-equilibrium  systems; {\it iv)} The need of proposing a principle  on how to anticipate the tipping points and entropy of the Earth system by means of time series of multiple spatio-temporal variables.

In Section \ref{cap3:App}, we have continued  by reviewing the applications of statistical physics approaches to different Earth subsystems. Specifically, five weather and climate dynamical scenarios were highlighted in Section \ref{cap3:CS}, including,  \el-Southern Oscillation, Indian summer monsoon, extreme rainfall, atmospheric circulation and Atlantic meridional  overturning circulation.  We suggested that the evolving CNs (and other combined approaches) interactions and patterns can serve as novel predictors which significantly improve the predictions. Beside the above mentioned scenarios, the CNs method was also widely applied to other high-impact phenomena, such as the westward propagation of the Atlantic multidecadal  oscillation \cite{feng_are_2014}, the detection of teleconnection paths \cite{zhou_teleconnection_2015}, the identifying of causal gateways and mediators \cite{runge_identifying_2015}, the early prediction of extreme stratospheric polar vortex states \cite{kretschmer_early_2017},  the impacts of carbon dioxide (CO2)
on global SAT \cite{ying_rossby_2020}, etc. In spite of the  great achievements, there are still many unexplored problems related to climate systems that merit further attention. Therefore, further topics are needed  to be investigated, for example,  {\it i)} how human activities have changed and continue to change the Earth’s oceanic and atmospheric composition, some of these changes have a direct or indirect impact on the energy balance of the Earth and are thus drivers of climate change; {\it ii)} The statistical mechanics and CNs behaviours in Earth System
Models, ranging from simple energy balance models to general circulation models; {\it iii)} The projected short-term and long-term abrupt phase transitions in paleoclimate, in particular, the role of sea ice and Snowball-Earth initiations \cite{ashkenazy_dynamics_2013,hoffman_snowball_2017,lucarini_transitions_2019};  {\it iv)} Improving the predictability of severe convective weather processes including thunderstorms, hail and tornadoes by using the CNs approach.

In section \ref{cap3:EGSR}, we have reviewed the current literature on the statistical features and the percolation framework of the Earth’s surface topography. We have pointed out the evidence for abrupt transitions
that occurred during the evolution of the Earth’s relief network, indicative of a continental/cluster aggregation. A further analysis via  fractional Brownian motion models suggests that  long-range correlations may play a key role in the observed discontinuity on Earth.  This research may facilitate the understanding of the geometrical phase transition on Earth, but also can be used to identify the critical nodes for future global  change in Earth’s relief network. 
In particular, {\it i)} the future movement of tectonic plates, as well as {\it ii)}  the dynamic evolution of these critical nodes are of key importance and thus need to be further addressed. 

In section \ref{cap3:ES}, we first presented a description of the well established empirical basic laws of earthquakes, including the Gutenberg-Richter, Omori and productivity laws. We then discussed the scaling
hypothesis of the inter-event distributions. An epidemic-type
aftershock sequence model (ETAS) and various methods for evaluating earthquake predictions and earthquake forecasts have been also reviewed. At last, we have shown that the two proposed memory analysis approaches, DFA and CP, can significantly improve the short-term forecasting rate for  real earthquake catalogs.   Nevertheless,  earthquake  prediction  research  has  been  plagued  by  controversy,  and  it  remains  an  outstanding  challenge.
Besides the improvement of short-term aftershock forecasting, we also emphasize the  potential of new directions. {\it i)} The wide applications of tipping points and memory analysis in  theoretical earthquake models which account for accelerated moment release and foreshock occurrence together with other precursory patterns, where mainshocks were treated  as characteristic earthquakes within a seismic cycle \cite{mignan_functional_2012}. {\it ii)}  The development of an 
earthquake forecasting model which is implemented for describing foreshock organization and memory behaviors. {\it iii)} Motivated by the successful applications of network theory to the prediction of climate phenomena, the  proposing of an \textit{earthquake network} framework  based on the occurrence time series \cite{chorozoglou_earthquake_2019}, or in particular the seismic waves (elastic body and shear waves) and elastogravity signals preceding direct seismic waves \cite{vallee_observations_2017}.

One of the key features of statistical physics-based approaches reviewed here is the ability to better and reliably forecast the complex Earth phenomena, such as  \el~events, extreme rainfall in the Eastern Central Andes, Indian summer monsoon, the collapse of the Atlantic multidecadal oscillation and the occurrence of earthquakes.  Moreover, we observed that  artificial intelligence and deep learning techniques \cite{lecun_deep_2015}  have achieved great success during  recent years in many fields, such as  phase transitions in statistical physics \cite{carrasquilla_machine_2017,nieuwenburg_learning_2017,bapst_unveiling_2020}, data-driven Earth system science \cite{reichstein_deep_2019},  ENSO forecasts \cite{ham_deep_2019,petersik_probabilistic_2020}, Indian monsoon rainfall \cite{saha_deep_2017}, as well as the forecasting aftershock patterns \cite{devries_deep_2018} and detecting earthquakes \cite{perol_convolutional_2018} in seismic systems.  In fact, we are confident that the combination and complement of  network-based and artificial intelligence-based skills will boost each other. For instance,  an application of   machine learning to network attribute vectors (products similarity)  can  predict successful and failing firms with much better accuracy than current state of the art techniques for market forecasting \cite{fan_topology_2019}. In addition,  deep learning systems can be regarded  as complex networks, which thus gain some insights into the structural and functional properties of the computational graph resulting from the learning process \cite{testolin_deep_2020}.

As a final remark, there is still room for many approaches considering the analysis of climate resilience of societies, ecosystems and economies by using the complex network theory. We currently lack appropriate models to better understand or predict the effects of cascading failures \cite{buldyrev_catastrophic_2010} triggered by the increasing adverse effect of extreme climate/weather events on interdependent critical infrastructures. Closing this knowledge gap is crucial for developing means to achieve  both climate and infrastructure resilience. Another very relevant subject of investigation that will certainly attract huge attention is the assessing and quantifying of the complex climate–health relationships. There is an overwhelming consensus that climate conditions, including  temperatures and spatial-temporal distribution of precipitation, has key implications for human health \cite{carleton_social_2016}. Moreover, the morbidity and outbreak of some disease, such as the vector-borne (malaria and dengue fever) disease, the ongoing COVID-19 pandemic \cite{mehta_covid-19_2020}, and influenza are strongly affected by
the climate change or the environment \cite{li_climate-driven_2019,towers_climate_2013,deyle_global_2016,pei_forecasting_2018,wu_exposure_2020}. However, the climatic influences are often excluded from consideration in the traditional epidemiologic models \cite{romualdo_pastor-satorras_epidemic_2015}, e.g., the SIR, SIRS and SEIR  models. What we need to emphasize is that  traditional epidemiologic models with taking into account the climate factor could be a very promising road towards a deeper understanding of epidemic processes and  assess the health consequences of climate change both regionally and globally.

\section*{Acknowledgements}
\addcontentsline{toc}{section}{Acknowledgements}

We would like to acknowledge gratefully all
colleagues with whom we maintained interactions and discussions on the topic in our report.

In particular, we would like to thank A. Agarwala, N. Boers, A. Bunde, D. Chen, R. Cohen, H. A. Dijkstra1, G. Falcone, L. Da, N. B. George, G. Dong, J. F. Donges, Z. Fu,  J. Gao, A. Gozolchiani,  H. Hof,  R. Hofstetter, H. Inoue, V. Livina, T. M. Lenton, Y. Liu, W. Lucht,  N. Marwan, W. Marzocchi, M. McPhaden, J. Nagler, P. Nooteboom, A. A. Saberi, L. M. Shekhtman, H. E. Stanley, V. Stolbova, Y. Sun, E. Surovyatkina, K. Yamasaki, N. Yuan, Y. Zhang, D. Zhou.

This work was partly supported by the  `East Africa Peru India Climate Capacities — EPICC' project,
which is part of the International Climate Initiative (IKI). The Federal Ministry for
the Environment, Nature Conservation and Nuclear Safety (BMU) supports this
initiative on the basis of a decision adopted by the German Bundestag. The Potsdam
Institute for Climate Impact Research (PIK) is leading the execution of the project
together with its project partners The Energy and Resources Institute (TERI) and the
Deutscher Wetterdienst (DWD). S.H. thanks the Italian Ministry of Foreign Affairs and International Cooperation jointly with the Israel Ministry of Science, Technology, and Space (MOST); the Israel Science Foundation, and the EU H2020 project RISE for financial support.

\appendix

\section{Acronyms}{\label{app:acronyms}

This Appendix contains two tables of acronyms used throughout the paper: Table~\ref{tab:glossary1} contains the scientific acronyms and Table~\ref{tab:glossary2} the institutional ones.

\begin{table}[!hpb]

\caption{Scientific acronyms} 

\centering

\begin{tabular}{l c}

\hline

Acronym & Meaning \\

\hline

$ACF$ & Autocorrelation function\\

$AIRI$ & All India Rainfall Index \\

$AMOC$ & Atlantic Meridional Overturning Circulation \\

$ApEn$ & Approximate Entropy\\

$CN$s & Climate networks \\

$DFA$ & Detrended fluctuation analysis\\

$EBM$ & Energy Balance Model \\

$EEG$ & Electroencephalogram \\

$ENB$ & \el~Basin (ENB) \\

$ENSO$ & El Ni\~no--Southern Oscillation \\

$ESMs$ & Earth System Models \\

$ESS$ & Earth System Science \\

$ETAS$ & Epidemic Type Aftershock Sequence \\

$EWS$ & Early warning signals\\

$FAMOUS$ & Fast Met Office/UK Universities Simulator (FAMOUS)  \\

$fBm$ & fractional Brownian motion \\

$GCM$ & General Circulation Model \\

$GHGs$ & Greenhouse Gases \\

$ISM$ & Indian Summer Monsoon\\

$NAO$ &  North Atlantic Oscillation \\

$OD$ & Onset date \\

$ONI$ & Oceanic Ni\~no Index (ONI) \\

$ROC$ & Receiver Operating Characteristic  \\

$SampEn$ & Sample entropy \\

$SLP$  & Sea Level Pressure \\

$SST$ & Sea Surface Temperature\\

$SPB$ & Spring Predictability Barrier\\

$SysSampEn$ & System Sample Entropy \\

$WD$ & Withdrawal date \\

\hline

\multicolumn{2}{l} 

\end{tabular}\label{tab:glossary1}

\end{table}

\begin{table}[!hpb]

\caption{Institutional acronyms}  

\centering

\begin{tabular}{l c}

\hline

Acronym & Meaning \\

\hline

$AR$ & Assessment Report \\

$CMIP$ & Climate Model Intercomparison Project\\

$ECMWF$ & European Centre for \\
 & Mid-range Weather Forecast\\

$JUNEC$ & Japan University Network Earthquake Catalog\\

$IMD$ & Indian Meteorological Department \\

$IPCC $ & Intergovernmental Panel on Climate Change \\

$NCAR$ & National Center for Atmospheric Research\\

$NCEP$ & National Center for Environmental Prediction \\

$NEIC$ & National Earthquake Information Center \\

$PCMDI$ & Program for Climate Model \\

$SCEC$ & Southern California Earthquake Center \\

\hline

\multicolumn{2}{l} 

\end{tabular}\label{tab:glossary2}

\end{table}
}

\newpage
\clearpage

\phantomsection


\bibliography{MyLibrary}

\end{document}